\documentclass[british,american,english]{article}
\pdfoutput=1 
\usepackage[T1]{fontenc}
\usepackage[latin9]{inputenc}
\usepackage{array}
\usepackage{verbatim}
\usepackage{units}
\usepackage{url}
\usepackage{multirow}
\usepackage{amsmath}
\usepackage{amssymb}
\usepackage{graphicx}
\usepackage[authoryear]{natbib}

\makeatletter

\providecommand{\tabularnewline}{\\}

\@ifundefined{showcaptionsetup}{}{%
 \PassOptionsToPackage{caption=false}{subfig}}
\usepackage{subfig}
\makeatother

\usepackage{babel}
\begin{document}
\title{Importance Sampling Methods for Bayesian Inference with Partitioned
Data}
\author{Marc Box\thanks{Email: marc.box@protonmail.com}}
\maketitle
\begin{abstract}
This article presents new methodology for sample-based Bayesian inference
when data are partitioned and communication between the parts is expensive,
as arises by necessity in the context of ``big data'' or by choice
in order to take advantage of computational parallelism. The method,
which we call the Laplace enriched multiple importance estimator,
uses new multiple importance sampling techniques to approximate posterior
expectations using samples drawn independently from the local posterior
distributions (those conditioned on isolated parts of the data). We
construct Laplace approximations from which additional samples can
be drawn relatively quickly and improve the methods in high-dimensional
estimation. The methods are ``\foreignlanguage{british}{embarrassingly}
parallel'', make no restriction on the sampling algorithm (including
MCMC) to use or choice of prior distribution, and do not rely on any
assumptions about the posterior such as normality. The performance
of the methods is demonstrated and compared against some alternatives
in experiments with simulated data.
\end{abstract}

\section{Introduction}

Bayesian sample-based computation is a common approach to Bayesian
inference in non-trivial models where it is infeasible to compute
the \foreignlanguage{british}{normalising} constant of the posterior
distribution. The focus of this article is on performing Bayesian
computation when data are partitioned and communication between the
parts is expensive or impossible. We present new methodology for Bayesian
inference in this context without having to combine the parts and
communicating between the parts only at the end of sampling. Our methods
achieve this with competitive performance and with fewer constraints
or assumptions than some other methods.

Bayesian computation with partitioned data is challenging because
the algorithms for generating samples from the posterior typically
require many calculations involving the whole data set; specifically,
likelihood evaluations. For example, in Markov chain Monte Carlo (\foreignlanguage{british}{MCMC})
algorithms such as the Metropolis-Hastings algorithm, a (typically
large) number of dependent samples are generated from a Markov chain,
each constrained by the previous sample and the data (see e.g. \citet{robert2004monte}).
When used to target a posterior distribution $\pi\left(\theta\mid x\right)$
given data $x$, the probability of accepting proposal $\theta^{\prime}$
given $x$ and previous sample $\theta$ is

\begin{equation}
\min\left(\frac{\pi\left(\theta^{\prime}\mid x\right)}{\pi\left(\theta\mid x\right)}\frac{q\left(\theta\mid\theta^{\prime}\right)}{q\left(\theta^{\prime}\mid\theta\right)},1\right),
\end{equation}

where $q$ is the density of the proposal distribution. The ratio
$\frac{\pi\left(\theta^{\prime}\mid x\right)}{\pi\left(\theta\mid x\right)}$
(which may be of unnormalised p.d.f.s because the normalising constant
cancels) must be evaluated for every proposal $\theta^{\prime}$,
i.e. in every iteration. Even if observations are conditionally independent
given $\theta$, so that likelihoods evaluated using parts of the
data can be multiplied to give the full data likelihood, this can
still be a problem if the necessary data transfer is expensive or
impossible.

There are three important data analysis situations where data are
partitioned:
\begin{enumerate}
\item When a data set is too large to work with in the memory of one computer.
\item When there are several sources or owners of data which are unable
or unwilling to share \foreignlanguage{british}{their} data.
\item When there is the possibility of speeding up sampling by running multiple
instances of the sampling algorithm in parallel on separate parts
of the data. 
\end{enumerate}
The first situation arises in the context of ``big data'': data
that must be stored in a distributed manner with no shared memory
for computing. Big data has become very important in modern science
and business because of the possibility of finding patterns not observable
on a smaller scale and which may lead to deeper understanding or provide
a competitive edge (\citet{bryant2008big,sagiroglu2013big}).  The
open source Apache Hadoop framework is widely used for storing and
computing with big \foreignlanguage{american}{data} on a cluster (\citet{hadoop2018apache,borthakur2007hadoop}).
In the \foreignlanguage{british}{Hadoop} file system (HDFS), data
are partitioned into ``blocks'' and stored across the cluster, in
duplicate for resilience to errors, then processed using \foreignlanguage{british}{parallel}
computation models such as MapReduce (\citet{dean2008mapreduce})
and Spark (\citet{zaharia2010spark}) which operate on data using
memory local to each block. A major source of inefficiency in these
computations arises when data must be communicated between cluster
nodes (\citet{kalavri2013mapreduce,sarkar2015mapreduce}).

One possible approach to Bayesian computation in this context is to
down-sample the data to a size that will fit in the local memory of
one computer, but down-sampling a large data set seems to defeat the
purpose of collecting it in the first place. As pointed out in \citet{scott2016bayes},
some large, complex models genuinely require a large amount of data
for robust estimation.

The second situation arises due to data privacy concerns or in meta-analyses.
Inference in this situation is sometimes known as ``federated inference''
(and the distributed data, ``federated data'') (\citet{xiong2021federated,ma2021federated}).
Current approaches to preserving privacy often rely on the masking
of data or the addition of noise (\citet{dwork2008differential,torra2016big}),
both which imply the loss of information. Meta-analyses use statistical
procedures such as mixed effects models to pool the results of primary
studies using aggregated data when there is no access to raw observational
data (\citet{dersimonian1986meta}). Performing inference in global
models for pooled data without any participants having to share their
data may open up new possibilities in these situations.

In the third situation, we may assume there is ample memory for the
entire data set, but computational parallelism is available such as
through multiple CPU cores, with a GPU or array of GPUs (\citet{lee2010utility}),
or on a cluster, and the time complexity of the sampling algorithm
depends on the number of data points. In this situation there is an
opportunity to generate more samples in a given time by running the
sampler in parallel on subsets of the data. This may result in estimators
with lower bias and variance than would otherwise be possible.

If the data can be contained in the memory of a single node, another
way of taking advantage of computational parallelism is to run multiple
MCMC chains in parallel. Besides the large number of samples that
can be generated (e.g. \citet{lao2020tfp}), there is potential for
improved convergence and new adaptive algorithms (\citet{green2015bayesian}).
 This is a different mode of parallelism and not the concern of this
paper.

Our approach is for each worker node (the cluster node or agent managing
each data part) to run the same sampling algorithm independently on
their local data, resulting in sets of samples from posterior distributions
different from the full data posterior distribution.  We regard
these local posteriors as importance proposal distributions or components
of a mixture proposal distribution targeting the posterior (\citet{robert2004monte,owen2013monte}).
By correctly weighting the samples we can construct Monte Carlo estimators
of a posterior expectation that are asymptotically unbiased in the
sense of approaching zero bias in the limit of infinite samples. There
are two observations that suggest importance sampling-based estimation
in this context may be fruitful. Firstly, the local posteriors should
be similar to the posterior (so long as the parts of data are similar
in distribution); secondly, the tails of the local posteriors should
be fatter than those of the posterior because they are conditioned
on less data (see e.g. \citet{mackay2003information}). We make use
of three importance weighting strategies which fit into the class
of multiple importance sampling (\citet{veach1995optimally,hesterberg1995weighted,owen2013monte,elvira2021advances}).
Two of these strategies we devised by extending the methods of \citet{veach1995optimally}
to the case where both the target density and the proposal densities
are only known up to a constant of proportionality. The third strategy
we believe is novel.

Importance sampling is known to suffer the curse of dimensionality,
leading to poor performance in high-dimensional models (\citet{mackay2003information}).
To address this we include samples from Laplace approximations to
the posterior to complement the samples received from the workers.
These additional samples can be generated easily without additional
iterations of the posterior sampling algorithm, which are often relatively
costly. The approximations are simple constructions from the pooled
samples that provide additional importance proposals which, it is
hoped, cover regions of parameter space not covered by the local posteriors.
We consider three ways of doing this, but find that only one of them
is particularly useful in the majority of examples.

The advantages of our methods can be \foreignlanguage{british}{summarised}
as follows. They appear to perform relatively well (comparing with
some alternative approaches) in terms of approximating posterior expectations
across a range of models, and in particular for non-normal posteriors.
We will provide evidence of this from experiments with synthetic data.
Our methods have no preference of algorithm used for sampling by the
workers using local data, so long as it is approximately unbiased,
and no communication between workers is required until the very end
of the sampling. This means we can perform sampling in an ``embarrassingly
parallel'' fashion (\citet{herlihy2020art}). In fact, no data (i.e.
observations) need be transmitted between nodes at all (after any
initial partitioning of data). This is an essential requirement in
the use case of collaboration between parties who are unable to share
data, and is an advantage to parallel computation on a cluster where
data transfer between nodes is a performance bottleneck. Our methods
can be used without any constraint on the choice of prior distribution.
This is in contrast to some other methods, in which it is necessary
that the prior be amenable to a certain transformation for the methods
to be unbiased. This constrains the choices available for the prior
in those methods, which can have unwanted implications for the analysis,
particularly in analyses of small data sets. There are also no \foreignlanguage{british}{hyperparameters}
that need to be set or tuned in our methods.

We make no distributional assumptions about the posterior. The only
assumptions we need beyond those implied by the model or the sampling
algorithm are that observations which are held in separate parts are
conditionally independent of each other given model parameters, that
the likelihood function is computable and the mild assumptions required
by importance sampling. Conditional independence of all observations
is sufficient but stronger than necessary. However, whilst we make
no explicit assumptions against non-random partitioning of the data,
random partitioning would likely be beneficial for methods based on
importance sampling because it makes the local posteriors more likely
to resemble the posterior. We do not require the size of the data
parts to be equal or the number of samples drawn by each worker to
be equal.

There are two useful performance diagnostics we propose to use with
our methods. First, we derive an effective sample size in the manner
of \citet{kong1992note}. This is a measure of the sampling efficiency
lost due to the use of an approximation. Second, we look at the $\hat{{k}}$
diagnostic arising in the Pareto smoothed importance estimator of
\citet{vehtari2015pareto}. This is an estimate of the shape parameter
in a generalised Pareto model for the tail of the importance weight
distribution, for which \citet{vehtari2015pareto} identified a threshold
which seems to be a valuable indicator of poor performance. We find
that, together, these indicate situations where our estimators perform
poorly.

As of writing, the problem of Bayesian inference with partitioned
data remains an open challenge (\citet{green2015bayesian,bardenet2017markov}),
although there are some notable contributions. Some hierarchical models
have a structure that is particularly amenable to distributed processing,
such as the hierarchical Dirichlet process topic model, for which
\citet{newman2009distributed} devise a distributed Gibbs sampling
algorithm. This approach does not generalise to other models, however.

A number of methods start from the observation that the posterior
density is proportional to the product of local posterior densities,
the \emph{product distribution}, under a conditional independence
assumption for the data. \citet{scott2016bayes} make normal distribution
assumptions for the local posteriors, justified by the Bernstein-von
Mises theorem (\citet{van1998asymptotic}), and pool samples using
a weighted linear combination. \citet{neiswanger2013asymptotically}
propose three methods, one which is similar to \citet{scott2016bayes}
and two using kernel density estimators constructed from the product
distribution and sampled from using an independent Metropolis-within-Gibbs
algorithm.  We will describe these methods in more detail in Section
\ref{subsec:Other-approaches} because we use them for performance
comparisons with our methods. \citet{huang2005sampling} use a similar
approach to \citet{scott2016bayes} but provide specific approaches
to normal models, linear models and hierarchical models. \citet{luengo2015bias}
consider a similar estimator to \citet{scott2016bayes} but pool the
local posterior estimators rather than individual samples. This also
relies on the Bernstein-von Mises theorem, but they propose a bias
correction which helps when the posterior does not follow a normal
distribution or with small data sets. \citet{luengo2018efficient}
propose more refined estimators along these lines. An approach related
to \citet{neiswanger2013asymptotically} is \citet{wang2013parallelizing},
who use the Weierstrass transform of the local posterior densities
and sample from the product of these using a Gibbs sampling algorithm.
This method may perform better when the posterior deviates from normality,
but requires some communication between workers during sampling and
involves some hyperparameters. \citet{nemeth2018merging} use a Gaussian
process prior on the log of each of the local posterior densities,
the sum of which is a Gaussian process approximation to the log posterior
density, from which they sample using Hamiltonian Monte Carlo. Importance
weighting is then used to improve the approximation represented by
these samples (this is a different use of importance sampling from
our methods, although the computation of the unnormalised posterior
density is the same).

There are also approaches that do not start from the product distribution.
\citet{xu2014distributed} use expectation propagation message passing
to enforce agreement to the target among the local samplers. This
involves communication between nodes during sampling, and there are
some hyperparameters. \citet{park2020variational} are concerned with
Bayesian inference in situations with a data privacy concern and use
a variational Bayes approach. Their methods involve the injection
of noise and involve some restriction on the models that can be studied.
\citet{neiswanger2015embarrassingly} use nonparametric variational
inference to widen the scope of models to which these methods can
be applied. \citet{jordan2018communication} construct a pseudo-posterior
from Taylor series approximations of the local log likelihoods and
sample from this using MCMC on a single node. In the approach of \citet{rendell2020global},
auxiliary variables are used as local proxies for the global model
parameters. The hierarchical model relating the auxiliaries to the
parameters involves a set of kernel functions which act to smooth
the local likelihood functions. A Metropolis-within-Gibbs sampler
is used to sample the local proxies and global parameters; the latter
samples can be used in estimators of posterior expectations. \citet{vono2018sparse,vono2019efficient}
take a similar approach but are interested in particular in high-dimensional
models rather than partitioned data; their auxiliary variables constitute
a projection of parameters onto a space of much lower dimension than
in the target posterior. These methods using auxiliary variables require
communication between nodes during sampling (although not every iteration).

Multiple importance sampling was introduced in \citet{veach1995optimally}
for Monte Carlo integration in the form of two importance weighting
schemes they call the ``combined estimator'' and the ``balance
heuristic''. The methods have been studied further e.g. by \citet{medina2019revisiting,elvira2021advances},
but the proposal distributions in these works are assumed to be normalised,
a limitation we needed to address for our methods.

The rest of this article is structured as follows:
\begin{itemize}
\item Section \ref{sec:methods} expands on the mathematical details of
the problem at hand and explains our proposed solution. It also explains
two other approaches, the consensus Monte Carlo algorithm and the
density product estimator, which are used in performance comparisons
in Section \ref{sec:experiments}.
\item Section \ref{sec:experiments} demonstrates the methods on some synthetic
data sets, exhibiting their behaviour under different conditions and
comparing performance.
\item Section \ref{sec:discussion} concludes with some further discussion
of the methods and ideas for further investigation.
\end{itemize}
All our analyses were run in R (\citet{R}). We make available R code
to implement our methods, as well as the methods we compare against
in Section \ref{sec:experiments}, at \url{https://github.com/mabox-source/parallelbayes}.
This repository can be compiled into an R package named \emph{parallelbayes}
which we hope will become available on the CRAN repository network
(\url{https://cran.r-project.org/}).

\section{Bayesian computation with distributed data\label{sec:methods}}

 Suppose we have data $x_{1},\ldots,x_{n}$ with dimension $d$ partitioned
into sets of observations $\mathbf{x}_{j},j=1,2,\ldots,M$, $M<n$.
That is, $\left\{ x_{1},\ldots,x_{n}\right\} =\cup_{j=1}^{M}\mathbf{x}_{j}$
and for all $j,k\in1,2,\ldots,M$ such that $j\ne k$, $\mathbf{x}_{j}\cap\mathbf{x}_{k}=\emptyset$.
We will abbreviate the set of observations with indices $1,2,\ldots,n$
as $x_{1:n}$. Suppose also we have a parametric model for $x_{1:n}$
with a parameter vector $\theta$ of dimension $p$. Under this model,
$\mathbf{x}_{j}$ is conditionally independent of $\mathbf{x}_{k}$,
for all $j\ne k$, given $\theta$ ($x_{1:n}$ being conditionally
iid given $\theta$ is sufficient but not necessary). We posit a prior
distribution with p.d.f. $\pi$ for $\theta$, and the model implies
a form for the likelihood function $p\left(x_{1:n}\mid\theta\right)$,
which we assume we are able to compute.

\subsection{Sample-based Bayesian computation\label{subsec:Sample-based-Bayesian-computatio}}

We are interested in estimating posterior expectations of functions
of $\theta$ given $x_{1:n}$. That is, expectations with respect
to the density

\begin{equation}
\pi\left(\theta\mid x_{1:n}\right)\propto p\left(x_{1:n}\mid\theta\right)\pi\left(\theta\right).\label{eq:full-posterior}
\end{equation}
This is the central task of sample-based Bayesian computation. Given
realisations $\theta_{1},\theta_{2},\ldots,\theta_{N}\sim\pi\left(\theta\mid x_{1:n}\right)$,
the posterior expectation of a real-valued function $f$ of $\theta$
can be approximated using

\begin{equation}
\mathbb{{E}}_{\pi}\left[f\left(\theta\right)\right]\approx\frac{1}{N}\sum_{h=1}^{N}f\left(\theta_{h}\right),\label{eq:posterior-expectation}
\end{equation}
which converges almost surely as $N\to\infty$ by the strong law of
large numbers (\citet{robert2004monte}). We use the $\mathbb{{E}}_{\pi}\left[\cdot\right]$
notation for expectations with respect to $\pi\left(\theta\mid x_{1:n}\right)$
specifically and $\mathbb{{E}}\left[\cdot\right]$ more generally
when the p.d.f. is to be inferred from the argument. Most Bayesian
inference tasks can be performed to an arbitrary degree of precision
with this estimator, such as estimation of posterior quantiles and
sampling from the posterior predictive distribution (\citet{gelman2004bayesian}).
It becomes particularly useful when we do not have an analytic expression
for the posterior distribution because, for instance, computation
of the normalising constant in Equation \ref{eq:full-posterior} is
infeasible. In such a situation, algorithms in the family of Markov
chain Monte Carlo (MCMC), if well designed, can efficiently generate
the required samples even in complicated or high-dimensional models.

This works when the realisations $\theta_{h}$ in Equation \ref{eq:posterior-expectation}
were sampled from the posterior distribution. If instead we had realisations

\begin{equation}
\theta_{j,h}\sim\pi_{j}\left(\theta\mid\mathbf{x}_{j}\right),h=1,2,\ldots,N_{j},
\end{equation}
one set for each $j=1,2,\ldots,M$, the sample mean of the $f\left(\theta_{j,h}\right)$
would be a biased estimator of Equation \ref{eq:posterior-expectation}.
We will refer to this approach as the \emph{naive pooling estimator};
this is the simplest and fastest, yet least accurate approximation
of the posterior, as we will demonstrate in Section \ref{sec:experiments}.
We will call the distribution with density

\begin{equation}
\pi_{j}\left(\theta\mid\mathbf{x}_{j}\right)\propto p\left(\mathbf{x}_{j}\mid\theta\right)\pi\left(\theta\right)\label{eq:local-posterior}
\end{equation}
the $j^{\textrm{{th}}}$ \emph{local posterior distribution}.

\subsection{Multiple importance estimators\label{subsec:Multiple-importance-estimation}}

Our solution to the problem of posterior inference using samples from
the local posteriors is to employ multiple importance sampling. One
of our estimators is based on weighting samples as if the local posteriors
$\pi_{j}\left(\theta\mid\mathbf{x}_{j}\right)$ were individual proposals
in a multiple importance sampling scheme; the other two as if a mixture
distribution consisting of components $\pi_{j}\left(\theta\mid\mathbf{x}_{j}\right)$
was a proposal distribution (mixture importance sampling). In aid
of explanation, consider first the use of $\pi_{j}\left(\theta\mid\mathbf{x}_{j}\right)$
alone as an importance proposal distribution. Define the importance
weighting function as

\begin{equation}
w_{j}\left(\theta\right):=\frac{\pi\left(\theta\mid x_{1:n}\right)}{\pi_{j}\left(\theta\mid\mathbf{x}_{j}\right)}\label{eq:importance-weights}
\end{equation}
and assume that $\pi_{j}\left(\theta\mid\mathbf{x}_{j}\right)>0$
for all $\theta$ such that $\pi\left(\theta\mid x_{1:n}\right)>0$.
Then for $\theta\sim\pi_{j}$ and for any function $f\left(\theta\right)$,

\begin{eqnarray}
\mathbb{{E}}\left[w_{j}\left(\theta\right)f\left(\theta\right)\right] & = & \int w_{j}\left(\theta\right)f\left(\theta\right)\pi_{j}\left(\theta\mid\mathbf{x}_{j}\right)\mathrm{{d}}\theta\nonumber \\
 & = & \int\frac{\pi\left(\theta\mid x_{1:n}\right)}{\pi_{j}\left(\theta\mid\mathbf{x}_{j}\right)}f\left(\theta\right)\pi_{j}\left(\theta\mid\mathbf{x}_{j}\right)\mathrm{{d}}\theta\nonumber \\
 & = & \int\pi\left(\theta\mid x_{1:n}\right)f\left(\theta\right)\mathrm{{d}}\theta\nonumber \\
 & = & \mathbb{{E}}_{\pi}\left[f\left(\theta\right)\right].\label{eq:normalised-importance-sampling}
\end{eqnarray}
This motivates the use of weighted samples of $\theta$ from the $j^{\textrm{{th}}}$
local posterior in a Monte Carlo estimate of $\mathbb{{E}}_{\pi}\left[f\left(\theta\right)\right]$
similar to Equation \ref{eq:posterior-expectation}. However, the
normalising constants of the densities in Equation \ref{eq:importance-weights}
are assumed to be unavailable (hence the need for sample-based estimation).
Define the unnormalised importance weights

\begin{equation}
\tilde{w}_{j}\left(\theta\right):=\frac{\tilde{\pi}\left(\theta\mid x_{1:n}\right)}{\tilde{\pi}_{j}\left(\theta\mid\mathbf{x}_{j}\right)},\label{eq:unnormalised-importance-weights}
\end{equation}
where $\tilde{\pi}\left(\theta\mid x_{1:n}\right)$ is the right hand
side of Equation \ref{eq:full-posterior} and $\tilde{\pi}_{j}\left(\theta\mid\mathbf{x}_{j}\right)$
is the right hand side of Equation \ref{eq:local-posterior}. Define
also the \emph{self-normalised} importance weights

\begin{equation}
\tilde{\mathbf{w}}_{j}\left(\theta\right):=\frac{\tilde{w}_{j}\left(\theta\right)}{\sum_{h=1}^{N}\tilde{w}_{j}\left(\theta_{j,h}\right)}.\label{eq:selfnormalised-importance-weights}
\end{equation}
Then, for samples $\theta_{j,h}\sim\pi_{j},h=1,2,\ldots,N_{j}$,

\begin{equation}
\tilde{\mu}_{j}:=\sum_{h=1}^{N_{j}}\tilde{\mathbf{w}}_{j}\left(\theta_{j,h}\right)f\left(\theta_{j,h}\right)\label{eq:single-importance-estimator}
\end{equation}
is an asymptotically unbiased estimator of $\mathbb{{E}}_{\pi}\left[f\left(\theta\right)\right]$
with the bias going to zero as $N_{j}\to\infty$. This is because
the denominator of Equation \ref{eq:selfnormalised-importance-weights}
in the limit supplies the normalising constants for the densities
in Equation \ref{eq:unnormalised-importance-weights}. This is derived
in more detail in Appendix \ref{subsec:Asymptotic-unbiasedness-snis}.
A useful result for importance sampling theory is

\begin{equation}
\lim_{N_{j}\to\infty}\frac{1}{N_{j}}\sum_{h=1}^{N_{j}}\tilde{w}_{j}\left(\theta_{j,h}\right)=\frac{Z_{\pi}}{Z_{j}},\label{eq:estimator-of-normalising-constants}
\end{equation}
also derived in Appendix \ref{subsec:Asymptotic-unbiasedness-snis}.

Whilst Equation \ref{eq:single-importance-estimator} could be used
as an estimator of $\mathbb{{E}}_{\pi}\left[f\left(\theta\right)\right]$,
we have $M$ sets of samples and thus $M$ estimators of this form.
Our idea is to combine all $M$ sets of samples in one estimator,
using \emph{multiple importance sampling} (\citet{veach1995optimally,hesterberg1995weighted,owen2013monte,elvira2021advances}),
of which we present 3 weighting schemes as well as ways to further
improve the estimators by incorporating samples from Laplace approximations.
We refer to the resulting estimators as the \emph{multiple importance
estimators (MIE)}, and when Laplace approximation samples are used,
as the \emph{Laplace enriched multiple importance estimators (LEMIE)}.

We need to compute $\tilde{\pi}\left(\theta\mid x_{1:n}\right)$ in
each of our weighting schemes. We propose an algorithm for this that
does not involve any transfer of data between nodes. This algorithm
is explained in Section \ref{subsec:In-out-in-algorithm-to}.

\subsubsection{In-out-in algorithm to calculate $\tilde{\pi}\left(\theta\mid x_{1:n}\right)$\label{subsec:In-out-in-algorithm-to}}

Each of our multiple importance estimators uses weight functions that
involve the unnormalised posterior density, $\tilde{\pi}\left(\theta\mid x_{1:n}\right)$.
The prior density cancels in the weights, e.g. in Equation \ref{eq:unnormalised-importance-weights},
but still the likelihoods $p\left(x_{1:n}\mid\theta\right)$ must
be computed for every $\theta$ sampled from each local posterior.
Our approach to this involves a total of three data transfers between
each worker node and a designated master node: pooling the samples
at the master, broadcasting the pooled samples to each of the workers,
which then compute likelihoods for each sample using local data, and
finally collecting the likelihoods. We call this the ``in-out-in''
algorithm:
\begin{enumerate}
\item (``In'') After sampling from the local posteriors has completed,
each worker sends their samples to the master node.
\item (``Out'') Samples are pooled by the master, with the indices denoting
their local posterior of origin retained. These pooled samples are
sent to every worker.
\item (``In'') Worker $j$ computes the likelihoods $L_{j}\left(\theta_{k,h}\right)=p\left(\mathbf{x}_{j}\mid\theta_{k,h}\right)$
for each sample $\theta_{k,h},k=1,2,\ldots,M,h=1,2,\ldots,N_{k}$
and sends these back to the master.
\item At the master, the likelihoods can be combined as
\end{enumerate}
\begin{equation}
p\left(x_{1:n}\mid\theta_{k,h}\right)=\prod_{j=1}^{M}L_{j}\left(\theta_{k,h}\right)
\end{equation}
for each sample $\theta_{k,h}$ due to our assumption of conditional
independence of the data parts given $\theta$. This is essentially
how \citet{nemeth2018merging} compute the unnormalised posterior
density of samples in their method.

\begin{figure}
\begin{centering}
\subfloat[``In'': workers send the samples drawn from the local posteriors
to the master node.]{\includegraphics[scale=0.2]{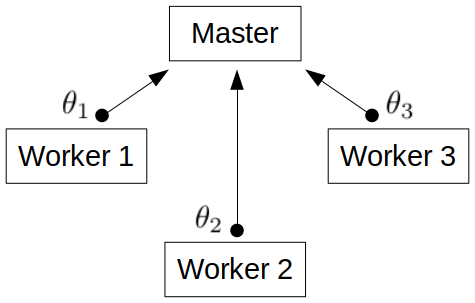}}\hfill{}\subfloat[``Out'': samples are pooled in the master node and broadcast to
each of the workers.]{\includegraphics[scale=0.2]{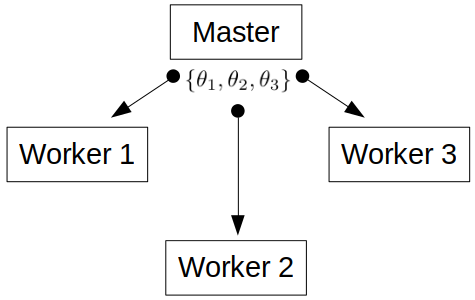}}\hfill{}\subfloat[``In'': workers compute likelihoods for the pooled samples using
local data and send these to the master node.]{\includegraphics[scale=0.2]{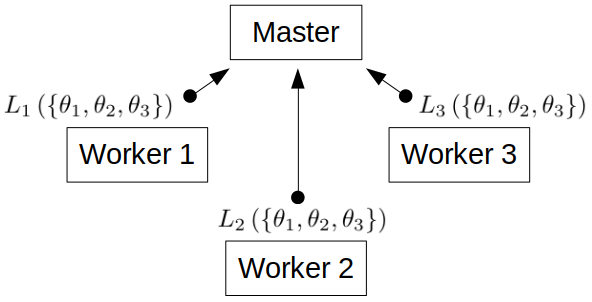}}
\par\end{centering}
\caption{The in-out-in algorithm of Section \ref{subsec:In-out-in-algorithm-to}.
This depicts an example where $M=3$ with data distributed across
3 workers and one master, which could be an edge node on a compute
cluster. $\theta_{j}$ represents the set of samples drawn from the
$j^{\textrm{{th}}}$ local posterior for $j=1,2,3$.\label{fig:in-out-in-algorithm}}
\end{figure}

Figure \ref{fig:in-out-in-algorithm} depicts the first 3 steps of
this algorithm with a simple example in which $M=3$. Samples $\theta_{1},\theta_{2}$
and $\theta_{3}$ are drawn from each of three local posteriors. These
are pooled in a master node and the pooled samples broadcast back
to the workers. Then the workers compute the likelihoods $L_{1}\left(\left\{ \theta_{1},\theta_{2},\theta_{3}\right\} \right),L_{2}\left(\left\{ \theta_{1},\theta_{2},\theta_{3}\right\} \right)$
and $L_{3}\left(\left\{ \theta_{1},\theta_{2},\theta_{3}\right\} \right)$
and send these back to the master. In this example, the master node
does not hold any data or draw any samples. This could be an example
of an edge node on a cluster consisting of itself and 3 compute nodes.
Alternatively, the master node which pools samples and collects likelihoods
could also play the role of a worker node, running its own MCMC sampler
targeting the local posterior using local data.

\subsubsection{Multiple importance estimator 1}

Our first estimator takes a weighted average of local importance estimators.
The weight to use for the $j^{\textrm{{th}}}$ local posterior is

\begin{equation}
q_{j}:=\frac{N_{j}}{N},\label{eq:mie-1-component-weights}
\end{equation}
for $j=1,2,\ldots,M$, where $N=\sum_{j=1}^{M}N_{j}$. This is similar
to the approach \citet{veach1995optimally} called the ``combined
estimator'', but we use self-normalised importance weights Equation
\ref{eq:selfnormalised-importance-weights} where those authors assumed
the densities involved were normalised. The estimator is

\begin{equation}
\tilde{\mu}^{\textrm{{MIE1}}}:=\sum_{j=1}^{M}\frac{N_{j}}{N}\sum_{h=1}^{N_{j}}\tilde{\mathbf{w}}_{j}\left(\theta_{j,h}\right)f\left(\theta_{j,h}\right).\label{eq:mie-1}
\end{equation}
That this is asymptotically unbiased follows from the single self-normalised
importance estimator, Equation \ref{eq:single-importance-estimator},
being unbiased and

\begin{equation}
\mathbb{{E}}\left[\sum_{j=1}^{M}\frac{N_{j}}{N}\tilde{\mu}_{j}\right]=\mathbb{{E}}\left[\tilde{\mu}_{j}\right].
\end{equation}
The variance of this estimator can be approximated as

\begin{equation}
\mathrm{{Var}}\left(\tilde{\mu}^{\textrm{{MIE1}}}\right)\approx\sum_{j=1}^{M}\frac{N_{j}}{N^{2}}\mathbb{{E}}_{\pi}\left[w_{j}\left(\theta\right)\left(f\left(\theta\right)-\mathbb{{E}}_{\pi}\left[f\left(\theta\right)\right]\right)^{2}\right],\label{eq:mie-1-variance}
\end{equation}
which follows from

\begin{equation}
\mathrm{{Var}}\left(\tilde{\mu}^{\textrm{{MIE1}}}\right)=\sum_{j=1}^{M}\frac{N_{j}^{2}}{N^{2}}\mathrm{{Var}}\left(\tilde{\mu}_{j}\right)
\end{equation}
and the approximate variance for self-normalised importance sampling
(SNIS), $\mathrm{{Var}}\left(\tilde{\mu}_{j}\right)$, derived using
the delta method for a ratio of means (see e.g. \citet{owen2013monte}).

\subsubsection{Multiple importance estimator 2\label{subsec:mie2}}

Importance sampling fails when the proposal p.d.f. and the target
p.d.f. have little overlap; in particular, when regions of parameter
space with high posterior density have low proposal density. In this
case most sample weights will be close to zero and some rare samples
will have very large weight, and consequently the estimator $\tilde{\mu}_{j}$
will have high variance. The risk of this increases as the dimension
of the parameter space increases (\citet{mackay2003information}).

Our second estimator combines the local posteriors in a mixture distribution
and uses this for the denominator of the importance weights. This
reflects the intuition that the mixture of local posterior \foreignlanguage{british}{p.d.f.s}
is likely to provide better coverage of the posterior p.d.f., resulting
in more stable sample weights. We define the mixture distribution
with p.d.f.

\begin{equation}
\phi\left(\theta\right):=\sum_{j=1}^{M}q_{j}\pi_{j}\left(\theta\mid\mathbf{x}_{j}\right)\label{eq:mixture-of-local-posteriors}
\end{equation}
and component weights $q_{j}=\frac{N_{j}}{N}$ chosen so that the
pooled samples from the local posteriors can be considered samples
from $\phi$.

The use of a mixture distribution in importance sampling was investigated
by \citet{veach1995optimally} and named the ``balance heuristic''.
However, those authors assumed that all of the component distributions
in the mixture had computable p.d.f.s. In other words, the normalising
constants must be known to use the balance heuristic. The weights
in the balance heuristic can be defined as

\begin{equation}
w_{\phi}\left(\theta\right):=\frac{\pi\left(\theta\mid x_{1:n}\right)}{\phi\left(\theta\right)}.\label{eq:mopp2-normalised-weights}
\end{equation}
Then, as in Equation \ref{eq:normalised-importance-sampling},

\begin{equation}
\mathbb{{E}}\left[w_{\phi}\left(\theta\right)f\left(\theta\right)\right]=\mathbb{{E}}_{\pi}\left[f\left(\theta\right)\right].
\end{equation}
However, if the densities in Equation \ref{eq:mopp2-normalised-weights}
were unnormalised, e.g.

\begin{equation}
\tilde{w}_{\phi}^{\textrm{{BH}}}\left(\theta\right):=\frac{\tilde{\pi}\left(\theta\mid x_{1:n}\right)}{\frac{1}{N}\sum_{j=1}^{M}N_{j}\tilde{\pi}_{j}\left(\theta\mid\mathbf{x}_{j}\right)},
\end{equation}
the average

\begin{equation}
\frac{1}{N}\sum_{j=1}^{M}\sum_{h=1}^{N_{j}}\tilde{w}_{\phi}^{\textrm{{BH}}}\left(\theta_{j,h}\right)
\end{equation}
would not be an unbiased estimator of the normalising constants as
in Equation \ref{eq:estimator-of-normalising-constants}, and therefore
the resulting estimator of $\mathbb{{E}}_{\pi}\left[f\left(\theta\right)\right]$
is biased, even with self-normalised weights and any number of samples.

Our MIE2 estimator fills this gap, i.e. it is an asymptotically unbiased
estimator of $\mathbb{{E}}_{\pi}\left[f\left(\theta\right)\right]$
using a mixture of unnormalised densities. We will use a similar idea
to Equation \ref{eq:estimator-of-normalising-constants} where the
weights provide a Monte Carlo estimate of the ratio of normalising
constants. To this end, define

\begin{equation}
\hat{c}_{j}:=\frac{1}{N_{j}}\sum_{h=1}^{N_{j}}\tilde{w}_{j}\left(\theta_{j,h}\right)\label{eq:estimator-of-normalising-constants-ratio}
\end{equation}
for $j=1,2,\ldots,M$. Let

\begin{equation}
\tilde{\psi}\left(\theta\right):=\sum_{j=1}^{M}q_{j}\hat{c}_{j}\tilde{\pi}_{j}\left(\theta\mid\mathbf{x}_{j}\right)\label{eq:unnormalised-mixture-distribution-1}
\end{equation}
and define the importance weights

\begin{equation}
\tilde{w}_{\phi}\left(\theta\right):=\frac{\tilde{\pi}\left(\theta\mid x_{1:n}\right)}{\tilde{\psi}\left(\theta\right)}\label{eq:unnormalised-mie-2-weights}
\end{equation}
(we deliberately use the notation $\tilde{\psi}$ rather than $\tilde{\phi}$,
despite the resemblance of $\tilde{\psi}$ to the mixture distribution
$\phi$, to avoid misleading the reader into assuming that dividing
$\tilde{\psi}$ by its integral with respect to $\theta$ results
in $\phi$: it does not, as explained above in relation to why the
balance heuristic does not work). Assume that $\tilde{\psi}\left(\theta\right)>0$
for all $\theta$ such that $\pi\left(\theta\mid x_{1:n}\right)>0$
(this is a weaker assumption than in MIE1 because only one of the
local posterior densities needs to be positive for $\tilde{\psi}\left(\theta\right)>0$).
Then we define the estimator

\begin{equation}
\tilde{\mu}^{\textrm{{MIE2}}}:=\frac{1}{N}\sum_{j=1}^{M}\sum_{h=1}^{N_{j}}\tilde{w}_{\phi}\left(\theta_{j,h}\right)f\left(\theta_{j,h}\right).\label{eq:mopp-2-unselfnormalised}
\end{equation}
This is asymptotically unbiased, as shown in Appendix \ref{subsec:Asymptotic-unbiasedness-mie2};
briefly, this is because $\tilde{w}_{\phi}\left(\theta\right)\to w_{\phi}\left(\theta\right)$
as all of the $N_{j}\to\infty$, with the $\hat{c}_{j}$ terms in
Equation \ref{eq:unnormalised-mixture-distribution-1} in the limit
supplying the normalising constants%
.

In \foreignlanguage{british}{practise} we find that self-normalising
the weights is beneficial to this estimator. Define the self-normalised
importance weights

\begin{equation}
\tilde{\mathbf{w}}_{\phi}\left(\theta\right):=\frac{\tilde{w}_{\phi}\left(\theta\right)}{\sum_{j=1}^{M}\sum_{h=1}^{N_{j}}\tilde{w}_{\phi}\left(\theta_{j,h}\right)},\label{eq:selfnormalised-mie-2-weights}
\end{equation}
and the self-normalised estimator

\begin{equation}
\bar{\tilde{\mu}}^{\textrm{{MIE2}}}:=\sum_{j=1}^{M}\sum_{h=1}^{N_{j}}\tilde{\mathbf{w}}_{\phi}\left(\theta_{j,h}\right)f\left(\theta_{j,h}\right).\label{eq:mopp-2-selfnormalised}
\end{equation}

It can be shown (see Appendix \ref{subsec:Finite-sample-bias-mie2})
that the finite sample bias of $\tilde{\mu}^{\textrm{{MIE2}}}$ is
approximately

\[
\mathbb{{E}}\left[\tilde{\mu}^{\textrm{{MIE2}}}\right]-\mathbb{{E}}_{\pi}\left[f\left(\theta\right)\right]\approx\frac{1}{\mathbb{{E}}\left[\phi\left(\theta\right)\right]^{2}}\textrm{{Var}}\left(\varepsilon\left(\theta\right)\right)\frac{\mathbb{{E}}\left[\pi\left(\theta\mid x_{1:n}\right)f\left(\theta\right)\right]}{\mathbb{{E}}\left[\phi\left(\theta\right)\right]},
\]
where $\textrm{{Var}}\left(\varepsilon\left(\theta\right)\right)$,
given by Equation \ref{eq:variance-of-monte-carlo-estimator-mie2}
in Appendix \ref{subsec:Finite-sample-bias-mie2}, is related to the
Monte Carlo errors of the $\hat{c}_{j}$ estimators, which go to zero
with $\sqrt{{N_{j}}}$. The variance of $\tilde{\mu}^{\textrm{{MIE2}}}$
can be approximated using a similar approach; see Appendix \ref{subsec:Variance-of-MIE2}.
The variance of $\bar{\tilde{\mu}}^{\textrm{{MIE2}}}$ can be approximated
using the delta method for the variance of SNIS (see e.g. \citet{owen2013monte}).

\begin{figure}
\begin{centering}
\subfloat[Contours of the posterior and $M=4$ local posteriors (50 outcomes
each).\label{fig:mie2-demo-normal-a}]{\includegraphics[scale=0.55]{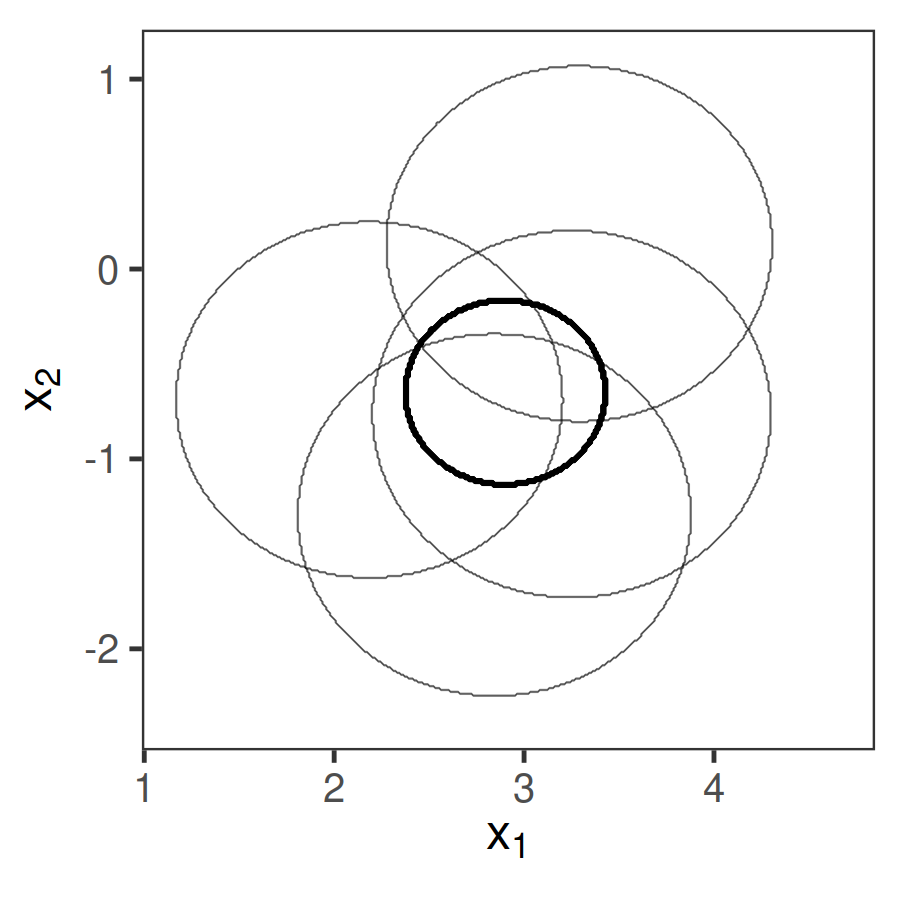}}\quad{}\subfloat[Contours of the same posterior and of KDEs for the naive pooling method
and MIE2.\label{fig:mie2-demo-normal-b}]{\includegraphics[scale=0.55]{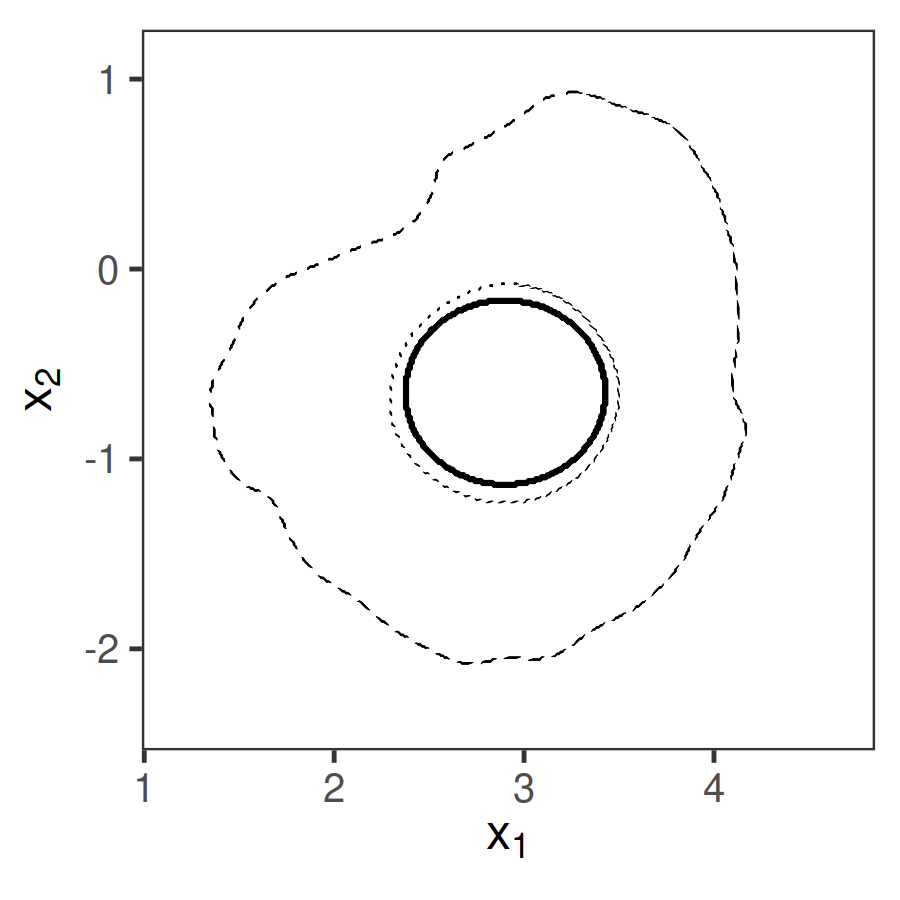}}\quad{}\includegraphics[scale=0.55]{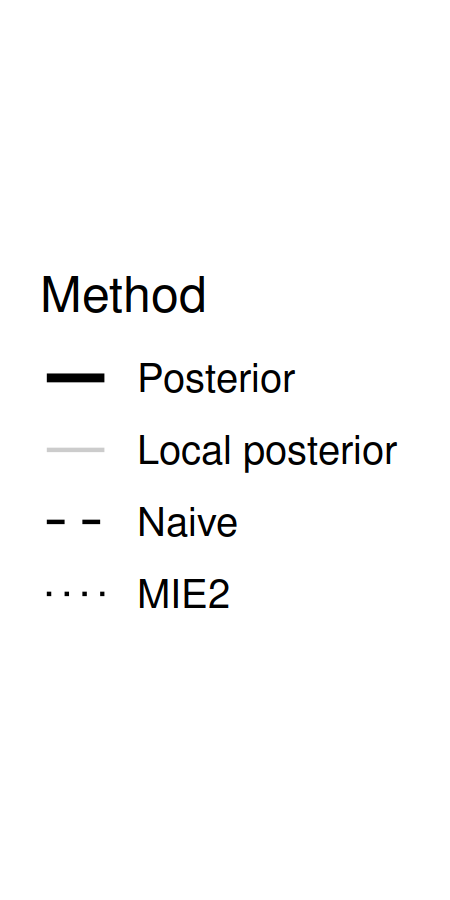}
\par\end{centering}
\caption{Contours of the central 90\% high density region of a 2 dimensional
MVN posterior distribution, from $n=200$ simulated MVN outcomes,
and an approximation using MIE2 from Section \ref{subsec:mie2}.\label{fig:mie2-demo-normal-d2-sigmaknown}}
\end{figure}

Figure \ref{fig:mie2-demo-normal-d2-sigmaknown} presents an example
of how the MIE2 estimator approximates the posterior using samples
from $M=4$ local posteriors. The data for this example are $n=200$
simulated 2 dimensional multivariate normal (MVN) outcomes with mean
vector $\left[3.00,-0.72\right]^{\intercal}$ and covariance matrix
$\left[12.06,10.22\right]I_{2}$ (we use $I_{p}$ for the identity
matrix of dimension $p$). The estimand is the mean vector, with the
covariance matrix known and using an uninformative MVN prior; see
Section \ref{subsec:Multivariate-normal-models} for more details.
Figure \ref{fig:mie2-demo-normal-a} depicts contours of the central
90\% high density region of the posterior and local posteriors. Figure
\ref{fig:mie2-demo-normal-b} depicts the central 90\% for kernel
density estimation (KDE) approximations to the posterior based on
naive pooling of samples (``naive'') and the MIE2 estimator (Section
\ref{subsec:Density-estimation-mie} explains how to perform KDE with
the MIE estimators). The MIE2 KDE approximates the posterior well,
at least the greatest 90\% of the density, and is certainly a much
closer approximation than naive pooling.

\subsubsection{Multiple importance estimator 3\label{subsec:mie3}}

The mixture distribution $\phi$ has component weights $\nicefrac{N_{j}}{N}$
because the importance estimator with proposal $\phi$ uses all of
the samples available and a fraction $\nicefrac{N_{j}}{N}$ of them
are drawn from the $j^{\textrm{{th}}}$ component. This is regardless
of the utility of each component for constructing an approximation
to the posterior: if the $j^{\textrm{{th}}}$ component is a good
approximation to the posterior and the $k^{\textrm{{th}}}$ component
is a poor one, MIE2 still uses all $N_{j}$ samples from the former
and all $N_{k}$ samples from the latter.

One idea for a more efficient estimator is to relax the requirement
that all components have the same weight and use the component weights
to prioritise samples from some local posteriors over others. %
The drawback is we must compromise on using all samples available,
discarding samples from those local posteriors with lower component
weight.

We use the Kullback-Leibler (KL) divergence from the posterior to
a local posterior to measure the ability of the latter to approximate
the former. The KL divergence is defined as

\begin{equation}
D_{\mathrm{{KL}}}\left(q\parallel\pi\right):=H\left(q,\pi\right)-H\left(q\right)\label{eq:kl-divergence-1}
\end{equation}
with 

\begin{equation}
H\left(q,\pi\right):=-\mathbb{{E}}_{q}\left[\log\left(\pi\left(\theta\right)\right)\right],\label{eq:cross-entropy}
\end{equation}
the cross entropy of target $\pi$ relative to an approximating distribution
$q$, and

\begin{equation}
H\left(q\right):=-\mathbb{{E}}_{q}\left[\log\left(q\left(\theta\right)\right)\right],
\end{equation}
the (differential) entropy of $q$. We use $D_{\mathrm{{KL}}}\left(\pi_{j}\parallel\pi\right)$
for the $j^{\textrm{{th}}}$ local posterior. $D_{\mathrm{{KL}}}\left(\pi_{j}\parallel\pi\right)$
is non-negative for all $\pi_{j}$ and $\pi$; when it is zero, the
$j^{\textrm{{th}}}$ local posterior is equal to the posterior, which
means if this component of the mixture is given component weight 1
all samples in the importance estimator will have a weight of 1. On
the other hand, when $D_{\mathrm{{KL}}}\left(\pi_{j}\parallel\pi\right)$
is large there will be regions of $\theta$ where $\pi\left(\theta\right)\gg\pi_{j}\left(\theta\right)$,
so if the $j^{\textrm{{th}}}$ local posterior has a relatively large
component weight the resulting importance weights are likely to be
degenerate. These observations suggest setting mixture component weights
inversely proportional to $D_{\mathrm{{KL}}}\left(\pi_{j}\parallel\pi\right)$.

\begin{sloppypar}The KL divergence cannot be computed exactly because
the normalising constants of the p.d.f.s involved are unknown. It
can be estimated  as

\begin{eqnarray}
D_{\mathrm{{KL}}}\left(\pi_{j}\parallel\pi\right) & = & \mathbb{{E}}\left[\log\left(\frac{\tilde{\pi}_{j}\left(\theta\right)}{\tilde{\pi}\left(\theta\right)}\right)\right]+\log\left(\frac{Z_{\pi}}{Z_{j}}\right)\nonumber \\
 & \approx & -\frac{1}{N_{j}}\sum_{h=1}^{N_{j}}\log\tilde{w}_{j}\left(\theta_{j,h}\right)+\log\left(\frac{1}{N_{j}}\sum_{h=1}^{N_{j}}\tilde{w}_{j}\left(\theta_{j,h}\right)\right)\nonumber \\
 &  & =:\widehat{D_{\mathrm{{KL}}}}\left(\pi_{j}\parallel\pi\right),\label{eq:kl-divergence-estimator}
\end{eqnarray}
using Equations \ref{eq:posterior-expectation} and \ref{eq:estimator-of-normalising-constants}
with samples $\theta_{j,h}\sim\pi_{j}$. We then set $\tilde{q}_{j}:=\nicefrac{1}{\widehat{D_{\mathrm{{KL}}}}\left(\pi_{j}\parallel\pi\right)}$
for each $j$ and use mixture component weights

\begin{equation}
q_{j}:=\frac{\tilde{q}_{j}}{\sum_{i=1}^{M}\tilde{q}_{i}}\label{eq:mopp-3-component-weights}
\end{equation}
in Equation \ref{eq:mixture-of-local-posteriors}. Then we sample
from the set $\left\{ \theta_{j,h}:j=1,2,\ldots,M,h=1,2,\ldots,N_{j}\right\} $
$\min_{j}N_{j}$ times where each $\theta_{j,h}$ has probability
of being sampled $\nicefrac{q_{j}}{N_{j}}$. This can also be achieved
by first sampling an index $j\in\left\{ 1,2,\ldots,M\right\} $ $\bar{N}:=\min_{j}N_{j}$
times where the probability of sampling $j$ is $q_{j}$, then for
each sampled $j$ draw a sample from $\left\{ \theta_{j,h};h=1,2,\ldots,N_{j}\right\} $
uniformly at random. Denote the resulting sample $\tilde{\theta}_{h},h=1,2,\ldots,\bar{N}$.
We compute importance weights for this sample using Equation \ref{eq:unnormalised-mie-2-weights},
using the original $\hat{c}_{j}$ estimates in Equation \ref{eq:unnormalised-mixture-distribution-1},
and define the estimator

\begin{equation}
\tilde{\mu}^{\textrm{{MIE3}}}:=\frac{1}{\bar{N}}\sum_{h=1}^{\bar{N}}\tilde{w}_{\phi}\left(\tilde{\theta}_{h}\right)f\left(\tilde{\theta}_{h}\right).\label{eq:mie-3-unselfnormalised}
\end{equation}
As with $\bar{\tilde{\mu}}^{\textrm{{MIE2}}}$, this benefits from
the weights being normalised, so in practise we prefer to use

\begin{equation}
\bar{\tilde{\mu}}^{\textrm{{MIE3}}}:=\sum_{h=1}^{\bar{N}}\tilde{\mathbf{w}}_{\phi}\left(\tilde{\theta}_{h}\right)f\left(\tilde{\theta}_{h}\right)\label{eq:mie-3-selfnormalised}
\end{equation}
with $\tilde{\mathbf{w}}_{\phi}$ from Equation \ref{eq:selfnormalised-mie-2-weights}.
The finite sample bias and variance of $\tilde{\mu}^{\textrm{{MIE3}}}$
are the same as for $\tilde{\mu}^{\textrm{{MIE2}}}$ but with denominator
$\bar{N}$ instead of $N$.\end{sloppypar}

\subsubsection{Density estimation with MIE\label{subsec:Density-estimation-mie}}

The posterior density $\pi\left(\theta^{\prime}\mid x_{1:n}\right)$
can be estimated using $\tilde{\mu}^{\textrm{{MIE1}}},\bar{\tilde{\mu}}^{\textrm{{MIE2}}}$
or $\bar{\tilde{\mu}}^{\textrm{{MIE3}}}$. For example, using a rectangular
window kernel, this would be, for MIE2:

\begin{equation}
\bar{\tilde{\mu}}^{\textrm{{MIE2}}}\left(K_{\xi}\left(\theta,\theta^{\prime}\right)\right)=\frac{1}{\xi}\sum_{j=1}^{M}\sum_{h=1}^{N}\tilde{\mathbf{w}}_{\phi}\left(\theta_{j,h}\right)\mathbf{{1}}\left(\left|\theta_{j,h}-\theta^{\prime}\right|<\frac{\xi}{2}\right),
\end{equation}
where $\mathbf{{1}}\left(P\right)=1$ if $P$ is true and is zero
otherwise.

\subsection{Additional samples from Laplace approximations\label{subsec:laplace-enrichment}}

We propose an extension to the estimators described in Section \ref{subsec:Multiple-importance-estimation}
which may improve their utility in applications where they may struggle,
such as with large dimension $p$. Our idea is to supplement the samples
from the local posteriors with additional samples drawn from one of
three Laplace approximations to the posterior constructed from the
local posteriors. The intention is that the importance estimators
can be improved by providing better representation of regions of $\theta$
under the posterior where little is provided by the local posteriors.

A Laplace approximation is an MVN with mean and covariance matrix
chosen to approximate a posterior distribution. With such an approximation
we can generate any number of samples much faster, in general, than
generating additional samples from the local posteriors (which may
require many likelihood evaluations).

Samples from a Laplace approximation can be included in the MIE 1,
2 or 3 estimators as an additional importance proposal or mixture
component. We will sample $N_{M+j}$ samples from Laplace approximation
$j=1,2,3$. To enrich MIE1, we define importance weights

\begin{equation}
\tilde{w}_{M+j}\left(\theta\right):=\frac{\tilde{\pi}\left(\theta\mid x_{1:n}\right)}{\varphi\left(\theta;\mu_{j}^{\textrm{{La}}},\Sigma_{j}^{\textrm{{La}}}\right)}
\end{equation}
for $j=1,2,3$, where $\varphi\left(\theta;\mu_{j}^{\textrm{{La}}},\Sigma_{j}^{\textrm{{La}}}\right)$
is the p.d.f. of the Laplace approximation MVN with mean $\mu_{j}^{\textrm{{La}}}$
and covariance matrix $\Sigma_{j}^{\textrm{{La}}}$ (which will be
defined later in this section). Then, with Laplace samples

\[
\theta_{M+j,h}\sim\mathrm{{N}}_{p}\left(\mu_{j}^{\textrm{{La}}},\Sigma_{j}^{\textrm{{La}}}\right),j=1,2,3,h=1,2,\ldots,N_{M+j},
\]
the LEMIE1 estimator is

\begin{equation}
\tilde{\mu}^{\textrm{{LEMIE1}}}:=\sum_{j=1}^{M+3}\frac{N_{j}}{N^{\textrm{{La}}}}\sum_{h=1}^{N_{j}}\tilde{\mathbf{w}}_{j}\left(\theta_{j,h}\right)f\left(\theta_{j,h}\right),\label{eq:lemie-1}
\end{equation}
where $N^{\textrm{{La}}}:=\sum_{j=1}^{M+3}N_{j}$. For LEMIE2, we
can include Laplace samples in the mixture distribution approximation
Equation \ref{eq:unnormalised-mixture-distribution-1} thus:

\begin{equation}
\tilde{\psi}^{\textrm{{La}}}\left(\theta\right):=\sum_{j=1}^{M}q_{j}\hat{c}_{j}\tilde{\pi}_{j}\left(\theta\mid\mathbf{x}_{j}\right)+\sum_{j=1}^{3}q_{M+j}\hat{c}_{M+j}\varphi\left(\theta;\mu_{j}^{\textrm{{La}}},\Sigma_{j}^{\textrm{{La}}}\right),
\end{equation}
where the component weights are now $q_{j}=\frac{N_{j}}{N^{\textrm{{La}}}}$
for $j=1,2,\ldots,M+3$. Then we use

\begin{equation}
\tilde{w}_{\phi}^{\textrm{{La}}}\left(\theta\right):=\frac{\tilde{\pi}\left(\theta\mid x_{1:n}\right)}{\tilde{\psi}^{\textrm{{La}}}\left(\theta\right)},\label{eq:unnormalised-mie-2-weights-1}
\end{equation}

\begin{equation}
\tilde{\mathbf{w}}_{\phi}^{\textrm{{La}}}\left(\theta\right):=\frac{\tilde{w}_{\phi}^{\textrm{{La}}}\left(\theta\right)}{\sum_{j=1}^{M+3}\sum_{h=1}^{N_{j}}\tilde{w}_{\phi}^{\textrm{{La}}}\left(\theta_{j,h}\right)},\label{eq:selfnormalised-mie-2-weights-1}
\end{equation}
and the resulting self-normalised estimator is

\begin{equation}
\bar{\tilde{\mu}}^{\textrm{{LEMIE2}}}:=\sum_{j=1}^{M+3}\sum_{h=1}^{N_{j}}\tilde{\mathbf{w}}_{\phi}^{\textrm{{La}}}\left(\theta_{j,h}\right)f\left(\theta_{j,h}\right).\label{eq:lemie2-unselfnormalised}
\end{equation}
The LEMIE3 estimator is similar except the component weights are

\begin{equation}
q_{j}:=\frac{\tilde{q}_{j}}{\sum_{i=1}^{M+3}\tilde{q}_{i}},
\end{equation}
for $j=1,2,\ldots,M+3$, in place of Equation \ref{eq:mopp-3-component-weights}
with

\begin{equation}
\tilde{q}_{M+j}:=\frac{1}{\widehat{D_{\mathrm{{KL}}}}\left(\varphi_{j}\parallel\pi\right)}
\end{equation}
for $j=1,2,3$ and in which $\varphi_{j}$ is the density function
of the $j^{\textrm{{th}}}$ Laplace approximation. Then samples are
resampled from the $M+3$ components with probabilities $q_{j}$ as
in Section \ref{subsec:mie3}. The estimator $\widehat{D_{\mathrm{{KL}}}}\left(\varphi_{j}\parallel\pi\right)$
in this case can be made more efficient than Equation \ref{eq:kl-divergence-estimator}
because $H\left(\varphi_{j}\right)=\frac{1}{2}\log\det\left(2\pi e\Sigma_{j}^{\textrm{{La}}}\right)$
(\citet{cover2006elements}), so only the cross entropy in Equation
\ref{eq:kl-divergence-1} needs to be estimated.

The three ways to construct a Laplace approximation from the local
posterior samples are as follows.

\subsubsection{Laplace approximation 1: parametric estimator\label{subsec:Laplace-enrichment-1}}

If we assume the local posteriors are MVN with mean $\mu_{j}$ and
covariance matrix $\Sigma_{j}$ for the $j^{\textrm{{th}}}$ local
posterior then the linear combination

\begin{equation}
\left(\sum_{j=1}^{M}\Sigma_{j}^{-1}\right)^{-1}\left(\sum_{j=1}^{M}\Sigma_{j}^{-1}\theta_{j,h}\right)\label{eq:normal-pooled-samples}
\end{equation}
also follows an MVN. Under an additional assumption, this MVN is also
the posterior: this is the motivation for the consensus Monte Carlo
algorithm of \citet{scott2016bayes} and the parametric density product
estimator (PDPE) of \citet{neiswanger2013asymptotically} (see Section
\ref{subsec:Consensus-Monte-Carlo} for more details). We define

\begin{equation}
\Sigma_{1}^{\textrm{{La}}}:=\left(\sum_{j=1}^{M}\hat{\Sigma}_{j}^{-1}\right)^{-1}\label{eq:laplace-1-covariance}
\end{equation}
and

\begin{equation}
\mu_{1}^{\textrm{{La}}}:=\Sigma_{1}^{\textrm{{La}}}\left(\sum_{j=1}^{M}\hat{\Sigma}_{j}^{-1}\hat{\mu}_{j}\right),
\end{equation}
where $\hat{\mu}_{j}$ and $\hat{\Sigma}_{j}$ are respectively the
sample mean and sample covariance matrix for the $j^{\textrm{{th}}}$
local posterior, and

\begin{equation}
\bar{\theta}_{h}:=\Sigma_{1}^{\textrm{{La}}}\left(\sum_{j=1}^{M}\hat{\Sigma}_{j}^{-1}\theta_{j,h}\right),\label{eq:laplace-pooled-samples}
\end{equation}
which is distributed as $\mathrm{{N}}_{p}\left(\mu_{1}^{\textrm{{La}}},\Sigma_{1}^{\textrm{{La}}}\right)$
if the local posteriors are MVN. This is our first Laplace approximation.
We can pool up to $\bar{N}$ samples using Equation \ref{eq:laplace-pooled-samples}
for use in the LEMIE estimators. Additional samples can be generated
from $\mathrm{{N}}_{p}\left(\mu_{1}^{\textrm{{La}}},\Sigma_{1}^{\textrm{{La}}}\right)$
if required.

Numerical problems can arise in the calculation of the $\hat{\Sigma}_{j}^{-1}$,
particularly when $p$ is large. When $\hat{\Sigma}_{j}$ cannot be
inverted we fall back to using the diagonal matrix with the variances
in $\hat{\Sigma}_{j}$ on the diagonal, following \citet{scott2016bayes}.

\subsubsection{Laplace approximation 2: pooled estimated Laplace\label{subsec:laplace-enrichment-2}}

Our second idea is to use the maximum likelihood estimates of the
mean and covariance matrix of an MVN for all the pooled samples from
the local posteriors. These are the sample mean and sample covariance
matrix. I.e.

\begin{equation}
\mu_{2}^{\textrm{{La}}}:=\frac{1}{N}\sum_{j=1}^{M}\sum_{h=1}^{N_{j}}\theta_{j,h},\label{eq:laplace2-pooled-sample-mean}
\end{equation}

\begin{equation}
\Sigma_{2}^{\textrm{{La}}}:=\frac{1}{N-1}\sum_{j=1}^{M}\sum_{h=1}^{N_{j}}\left(\theta_{j,h}-\mu_{2}^{\textrm{{La}}}\right)\left(\theta_{j,h}-\mu_{2}^{\textrm{{La}}}\right)^{\intercal},
\end{equation}
and any number of samples can be drawn from $\mathrm{{N}}_{p}\left(\mu_{2}^{\textrm{{La}}},\Sigma_{2}^{\textrm{{La}}}\right)$.
This approximation is likely to be relatively diffuse, which may help
when local posterior coverage of the posterior p.d.f. is very poor.

\subsubsection{Laplace approximation 3: Bayesian estimated Laplace\label{subsec:laplace-enrichment-3}}

Our third idea is to pool all the samples as above but then to place
an inverse Wishart prior on their covariance matrix,

\begin{equation}
\Sigma\sim\mathrm{{IW}}\left(\Psi^{\textrm{{La3}}},\nu^{\textrm{{La3}}}\right),
\end{equation}
with scale matrix $\Psi^{\textrm{{La3}}}$ and degrees of freedom
$\nu^{\textrm{{La3}}}$, and compute the posterior mean of $\Sigma$
using the samples as data after shifting them to have mean zero. That
is,

\[
\Sigma_{3}^{\textrm{{La}}}:=\frac{1}{N+\nu^{\textrm{{La3}}}-p-1}\left(\sum_{j=1}^{M}\sum_{h=1}^{N_{j}}\left(\theta_{j,h}-\hat{\theta}_{j}\right)\left(\theta_{j,h}-\hat{\theta}_{j}\right)^{\intercal}+\Psi^{\textrm{{La3}}}\right),
\]
where $\hat{\theta}_{j}:=\frac{1}{N_{j}}\sum_{h=1}^{N_{j}}\theta_{j,h}$
for each $j$. Then set the mean vector $\mu_{3}^{\textrm{{La}}}$
to be the pooled sample mean Equation \ref{eq:laplace2-pooled-sample-mean},
and any number of samples can be drawn from $\mathrm{{N}}_{p}\left(\mu_{3}^{\textrm{{La}}},\Sigma_{3}^{\textrm{{La}}}\right)$.
This approximation is similar to that in Section \ref{subsec:laplace-enrichment-2},
but we have some influence over its shape, in particular how diffuse
it is, via the prior parameters $\Psi^{\textrm{{La3}}}$ and $\nu^{\textrm{{La3}}}$.

\subsection{Estimator diagnostics\label{subsec:Performance-indicators}}

\subsubsection{Effective sample size\label{subsec:Effective-sample-size}}

The effective sample size (ESS) of a Monte Carlo estimator, in the
definition of \citet{kong1992note}, is the ratio of the posterior
variance of the estimand to the variance of the estimator, which measures
the efficiency lost due to sampling from the approximation rather
than the posterior itself.

We can derive an estimate of the ESS for each of the estimators in
Section \ref{subsec:Multiple-importance-estimation} with an additional
application of the delta method to the variance estimates. These are,
for MIE 1, 2 and 3 respectively:

\begin{equation}
\textrm{{ESS}}_{1}\approx\frac{1}{\sum_{j=1}^{M}\frac{N_{j}^{2}}{N^{2}}\sum_{h=1}^{N_{j}}\tilde{\mathbf{w}}_{j}\left(\theta_{j,h}\right)^{2}},\label{eq:ess-mie1}
\end{equation}

\begin{equation}
\textrm{{ESS}}_{2}\approx\frac{1}{\sum_{j=1}^{M}\sum_{h=1}^{N_{j}}\tilde{\mathbf{w}}_{\phi}\left(\theta_{j,h}\right)^{2}},\label{eq:ess-mie2}
\end{equation}
and

\begin{equation}
\textrm{{ESS}}_{3}\approx\frac{1}{\sum_{h=1}^{\bar{N}}\tilde{\mathbf{w}}_{\phi}\left(\tilde{\theta}_{h}\right)^{2}}.\label{eq:ess-mie3}
\end{equation}
When Laplace samples are included these become

\begin{equation}
\textrm{{ESS}}_{1}^{\textrm{{La}}}\approx\frac{1}{\sum_{j=1}^{M+3}\frac{N_{j}^{2}}{N^{2}}\sum_{h=1}^{N_{j}}\tilde{\mathbf{w}}_{j}\left(\theta_{j,h}\right)^{2}},\label{eq:ess-mie1-1}
\end{equation}

\begin{equation}
\textrm{{ESS}}_{2}^{\textrm{{La}}}\approx\frac{1}{\sum_{j=1}^{M+3}\sum_{h=1}^{N_{j}}\tilde{\mathbf{w}}_{\phi}^{\textrm{{La}}}\left(\theta_{j,h}\right)^{2}},\label{eq:ess-mie2-1}
\end{equation}
and

\begin{equation}
\textrm{{ESS}}_{3}^{\textrm{{La}}}\approx\frac{1}{\sum_{h=1}^{\bar{N}}\tilde{\mathbf{w}}_{\phi}^{\textrm{{La}}}\left(\tilde{\theta}_{h}\right)^{2}}.\label{eq:ess-mie3-1}
\end{equation}
A derivation of Equation \ref{eq:ess-mie1} can be found in Appendix
\ref{subsec:ESS-derivations}. Equations \ref{eq:ess-mie2} and \ref{eq:ess-mie3}
for MIE2 and MIE3 respectively are equivalent to the approximate ESS
for SNIS; see e.g. \citet{owen2013monte} and the derivation of \citet{kong1992note}.

\subsubsection{Tail distribution shape estimate ($\hat{{k}}$)\label{subsec:Tail-distribution-shape}}

\citet{vehtari2015pareto} introduce a useful diagnostic for importance
sampling, introduced as part of an algorithm to smooth importance
weights. This is used by \citet{vehtari2017practical} for leave-one-out
cross validation (LOO). Their LOO estimator is computed using importance
sampling with the weights smoothed to make them more stable and improve
the estimator's accuracy and reliability. Importance weights have
a tail distribution that is, in the limit, well-approximated by a
generalised Pareto distribution (GPD) under weak conditions (\citet{pickands1975statistical}).
The idea of \citet{vehtari2015pareto} is to replace the largest weights
above a threshold with quantiles from the fitted GPD.

The GPD has shape parameter $k\in\mathbb{{R}}$, which is estimated
as $\hat{{k}}$ using the efficient estimator of \citet{zhang2009new}
(along with the other parameters). \citet{vehtari2015pareto} find
empirically that $\hat{{k}}$ is a useful diagnostic, indicating when
an importance estimator may be unreliable. They find that $\hat{{k}}<0.5$
is an indicator of good performance, but that Pareto smoothed importance
weights will still provide reliable results for $\hat{{k}}<0.7$ and
that importance sampling is unreliable beyond this. The R package
\emph{loo} (\citet{loo}) includes an implementation of the $\hat{{k}}$
estimator.

\subsection{Other approaches\label{subsec:Other-approaches}}

In this section we briefly describe the estimators of \citet{scott2016bayes}
and \citet{neiswanger2013asymptotically}, which are used for comparisons
with our methods in Section \ref{sec:experiments}. Each of these
approaches starts from the observation that

\begin{equation}
\pi\left(\theta\mid x_{1:n}\right)\propto\prod_{j=1}^{M}p\left(\mathbf{x}_{j}\mid\theta\right)\pi\left(\theta\right)^{\nicefrac{1}{M}},\label{eq:fractionated-posterior}
\end{equation}
assuming $\mathbf{x}_{j}$ is conditionally independent of $\mathbf{x}_{k}$,
for all $j\ne k$, given $\theta$. I.e. the posterior p.d.f. is a
product of the local posterior p.d.f.s using the prior with density
$\pi\left(\theta\right)^{\nicefrac{1}{M}}$, which is referred to
as the \emph{fractionated prior}.

\subsubsection{Consensus Monte Carlo\label{subsec:Consensus-Monte-Carlo}}

As explained in Section \ref{subsec:Laplace-enrichment-1}, if $\mu_{j}$
and $\Sigma_{j}$ are the mean and covariance matrix of the $j^{\textrm{{th}}}$
local posterior, and if the posterior and local posteriors are MVN,
then linear combinations of samples Equation \ref{eq:normal-pooled-samples}
also follow an MVN. If the prior used in the local posteriors has
density $\pi\left(\theta\right)^{\nicefrac{1}{M}}$ then that MVN
is in fact the posterior. This can be seen from Equation \ref{eq:fractionated-posterior}
by inductively applying Bayes' theorem with a normal likelihood function
and a normal prior.

In the consensus Monte Carlo algorithm (CMC) of \citet{scott2016bayes}
we define

\begin{equation}
\Sigma^{*}:=\left(\sum_{j=1}^{M}\tilde{\Sigma}_{j}^{-1}\right)^{-1}\label{eq:consensus-covariance-1}
\end{equation}
and

\begin{equation}
\mu^{*}:=\Sigma^{*}\left(\sum_{j=1}^{M}\tilde{\Sigma}_{j}^{-1}\tilde{\mu}_{j}\right),\label{eq:consensus-mean-1}
\end{equation}
where $\tilde{\mu}_{j}$ and $\tilde{\Sigma}_{j}$ are respectively
the sample mean and sample covariance matrix for the $j^{\textrm{{th}}}$
local posterior using the \emph{fractionated} prior, and

\begin{equation}
\theta_{h}^{*}:=\Sigma^{*}\left(\sum_{j=1}^{M}\tilde{\Sigma}_{j}^{-1}\theta_{j,h}\right)\label{eq:consensus-pooled-samples}
\end{equation}
 for $h=1,2,\ldots,\bar{N}$. When the posterior is not MVN, Monte
Carlo estimators based on these samples should still be useful, especially
in big data situations, because of the Bernstein-von Mises theorem
(posterior distributions tend towards a normal distribution as $n\to\infty$,
\citet{van1998asymptotic}). Equation \ref{eq:consensus-pooled-samples}
requires the number of samples from each local posterior to be $\bar{N}$;
samples from a local posterior in excess of this must be discarded.

There are two estimators defined by the choice of weight matrix in
Equation \ref{eq:consensus-pooled-samples}. The first is

\begin{equation}
\tilde{\mu}^{\textrm{{CMC1}}}:=\frac{1}{\bar{N}}\sum_{h=1}^{\bar{N}}f\left(\frac{1}{M}\sum_{j=1}^{M}\theta_{j,h}\right),\label{eq:consensus-estimator-1}
\end{equation}
i.e. the identity matrix $I_{p}$ is used for each of the weight matrices.
The second estimator is

\begin{equation}
\tilde{\mu}^{\textrm{{CMC2}}}:=\frac{1}{\bar{N}}\sum_{h=1}^{\bar{N}}f\left(\Sigma^{*}\left(\sum_{j=1}^{M}\tilde{\Sigma}_{j}^{-1}\theta_{j,h}\right)\right).\label{eq:consensus-estimator-2}
\end{equation}

When the posterior distribution is normal, $\tilde{\mu}^{\textrm{{CMC2}}}$
is an unbiased estimator of the posterior expectation of $f$. One
drawback is that the method is unlikely to perform well when the posterior
deviates greatly from normality. Another is that we must use the fractionated
prior for the local posterior sampling. This can be a problem if the
distribution with density $\pi\left(\theta\right)^{\nicefrac{1}{M}}$
is improper, in which case we may need to compromise on the form or
parameterisation of the prior distribution. As in Section \ref{subsec:Laplace-enrichment-1},
numerical problems can arise in the calculation of the $\tilde{\Sigma}_{j}^{-1}$;
here as well we fall back to using the diagonal matrix with the variances
in $\hat{\Sigma}_{j}$ on the diagonal when $\hat{\Sigma}_{j}$ cannot
be inverted.

There is a small sample bias correction proposed by \citet{scott2016bayes}.
We do not include this in our comparisons in Section \ref{sec:experiments}
because we find the performance to be similar to that of the CMC2
algorithm without it.

\subsubsection{Density product estimator\label{subsec:Density-product-estimator}}

The density product

\begin{equation}
\pi_{1}\cdots\pi_{M}\left(\theta\right):=\prod_{j=1}^{M}\pi_{j}\left(\theta\mid\mathbf{x}_{j}\right),\label{eq:density-product}
\end{equation}
with $\pi_{j}\left(\theta\mid\mathbf{x}_{j}\right)$ using the fractionated
prior, is not necessarily equal to the posterior density but is proportional
to it. \citet{neiswanger2013asymptotically} use kernel density estimation
to approximate Equation \ref{eq:density-product} using the pooled
samples from the local posteriors.The nonparametric density product
estimator (NDPE) is designed to be more robust than CMC to deviations
in the posterior from normality (they also propose a ``parametric
subposterior density product estimator'', PDPE, which is very similar
to CMC so we have not included it). It uses a MVN KDE $\widehat{\pi_{1}\cdots\pi_{M}}$
that is asymptotically unbiased and consistent. The KDE for the $j^{\textrm{{th}}}$
local posterior, assuming all $N_{j}=\bar{N}$, as a function of $\theta$
is

\begin{equation}
\frac{1}{\bar{N}}\sum_{h=1}^{\bar{N}}\mathrm{{N}}_{p}\left(\theta\mid\theta_{j,h},b^{2}I_{p}\right),\label{eq:kde-for-partial-posterior}
\end{equation}
where $b$ is a tuning parameter (bandwidth). The product of KDEs
can be rewritten as a normal mixture density thus:

\begin{eqnarray}
\hat{\pi}^{\mathrm{{NDPE}}}\left(\theta\right) & := & \frac{1}{\bar{N}^{M}}\prod_{j=1}^{M}\sum_{h=1}^{\bar{N}}\mathrm{{N}}_{p}\left(\theta\mid\theta_{j,h},b^{2}I_{p}\right)\nonumber \\
 & \propto & \sum_{h_{1}=1}^{\bar{N}}\cdots\sum_{h_{M}=1}^{\bar{N}}w\left(h_{1},\ldots,h_{M}\right)\mathrm{{N}}_{p}\left(\theta\mid\bar{\theta}\left(h_{1},\ldots,h_{M}\right),b^{2}I_{p}\right),\nonumber \\
\label{eq:ndpe}
\end{eqnarray}
where

\begin{equation}
\bar{\theta}\left(h_{1},\ldots,h_{M}\right):=\frac{1}{M}\sum_{j=1}^{M}\theta_{j,h_{j}}
\end{equation}
and with unnormalised component weights

\begin{equation}
w\left(h_{1},\ldots,h_{M}\right):=\prod_{j=1}^{M}\mathrm{{N}}_{p}\left(\theta_{j,h_{j}}\mid\bar{\theta}\left(h_{1},\ldots,h_{M}\right),b^{2}I_{p}\right).
\end{equation}
There are $\bar{N}^{M}$ terms in the mixture density Equation \ref{eq:ndpe},
so it is not feasible to compute it exactly. Therefore, \citet{neiswanger2013asymptotically}
propose to sample from it using an independent Metropolis-within-Gibbs
algorithm which alternates between sampling $\theta$ from the mixture
distribution given indices $h_{1},\ldots,h_{M}$ and sampling the
indices independently of $\theta$ over $\bar{N}$ iterations. The
resulting samples of $\theta$ can be used in Monte Carlo estimators
of posterior expectations. The bias and variance of the method shrink
as bandwidth $b\to0$, so we set $b$ to $i^{\nicefrac{-1}{p+4}}$
in iteration $i$. This introduces an element of tempering.

The semiparametric density product estimator (SDPE) aims to combine
the PDPE's fast convergence with NDPE's asymptotic properties. SDPE
involves a KDE of $\frac{\pi_{j}\left(\theta\mid\mathbf{x}_{j}\right)}{\tilde{\varphi}_{j}\left(\theta\right)}$,
the ``correction function'' of the normal approximation to the $j^{\textrm{{th}}}$
local posterior, where $\tilde{\varphi}_{j}$ is the density function
of an MVN with parameters $\tilde{\mu}_{j}$ and $\tilde{\Sigma}_{j}$
from Section \ref{subsec:Consensus-Monte-Carlo}. The remaining details
are similar to the NDPE algorithm; see Appendix \ref{appendix:sdpe}
for details.

The MCMC samplers in NDPE and SDPE can suffer from slow mixing due
to a low acceptance rate. \citet{neiswanger2013asymptotically} propose
the sampler be applied separately to subsets of the local posteriors
(e.g., pairwise), generating a new set of samples, then applied recursively
to the output.

NDPE and SDPE can result in samples that are not possible values of
the random variable $\theta$. For example, if $\theta$ is \foreignlanguage{british}{modelled}
to take only positive values, the DPE algorithms can generate pooled
samples of $\theta$ that are negative. This means the algorithm will
be biased near such boundaries, and in practise, it may be necessary
to constrain the result of the algorithm to obtain valid results.
In SDPE, as in CMC and Section \ref{subsec:Laplace-enrichment-1},
we must calculate $\tilde{\Sigma}_{j}^{-1}$ and so problems can also
arise here when $\tilde{\Sigma}_{j}$ cannot be inverted, in which
case we use the diagonal matrix with the variances in $\tilde{\Sigma}_{j}$
on the diagonal.

\subsubsection{Naive pooling}

In our experiments in Section \ref{sec:experiments} we will also
compare the performance of the LEMIE estimators against the naive
pooling estimator introduced in Section \ref{subsec:Sample-based-Bayesian-computatio}.
This is the pooled sample average

\begin{equation}
\tilde{\mu}^{\textrm{{Naive}}}:=\frac{1}{N}\sum_{j=1}^{M}\sum_{h=1}^{N_{j}}f\left(\theta_{j,h}\right).\label{eq:naive-pooling}
\end{equation}

\section{Simulation studies\label{sec:experiments}}

This section presents results from experiments using synthetic data
generated from three models: a beta-Bernoulli model for binary data,
MVN data with normal-inverse Wishart priors for the mean and covariance
matrix, and logistic regression with an MVN prior. The purpose of
these studies is to explore differences in performance between the
methods described in Section \ref{sec:methods}. Where possible, we
also compare them against the $M=1$ case (i.e. using all data together
on one node).

Each of our analyses was performed in R (\citet{R}), version 3.6.3.
Most computation was carried out on a personal computer running a
Linux operating system with an 8 core CPU and 16 GB RAM. Some were
carried out on Amazon Web Services (AWS) Elastic Compute 2 (EC2) instances
(\citet{AWS}) with specifications (CPU cores / RAM): 8 / 16GB, 16
/ 32GB, 32 / 64GB, 64 / 128GB.

In comparisons of point-wise estimation error, we use error function 

\begin{equation}
E\left(f,\hat{f}\right):=\left\Vert \hat{f}-f\right\Vert _{2},\label{eq:2norm-error}
\end{equation}
where $\left\Vert \cdot\right\Vert _{2}$ is the Euclidean 2-norm
and $f,\hat{f}\in\mathbb{{R}}^{p}$. This will be used, for example,
in comparing estimators of a posterior mean. Another measure of performance
we use is the KL divergence between the posterior and the approximating
distribution implied by each estimator.

\paragraph{KL divergence estimation\label{par:Cross-entropy-estimation}}

We define the cross entropy of an approximating distribution $q$
relative to target $\pi$

\begin{equation}
H\left(\pi,q\right):=-\mathbb{{E}}_{\pi}\left[\log\left(q\left(\theta\right)\right)\right].\label{eq:cross-entropy-2}
\end{equation}
In our simulations we are readily able to generate samples from the
posterior, so this form is preferred over Equation \ref{eq:cross-entropy}
since using Equation \ref{eq:posterior-expectation} we can estimate
Equation \ref{eq:cross-entropy-2} as

\begin{equation}
H\left(\pi,q\right)\approx-\frac{1}{N^{*}}\sum_{h=1}^{N^{*}}\log\left(q\left(\theta_{h}\right)\right)\label{eq:cross-entropy-1}
\end{equation}
using samples $\theta_{h}\sim\pi,h=1,2,\ldots,N^{*}$. The standard
error of this estimator can be estimated using the sample standard
deviation of $\log\left(q\left(\theta_{h}\right)\right)$ divided
by $\sqrt{{N^{*}}}$.%

This definition of cross entropy features in the KL divergence from
$q$ to $\pi$,

\begin{equation}
D_{\mathrm{{KL}}}\left(\pi\parallel q\right)=H\left(\pi,q\right)-H\left(\pi\right),\label{eq:kl-divergence}
\end{equation}
in which $H\left(\pi\right):=-\mathbb{{E}}_{\pi}\left[\log\left(\pi\left(\theta\right)\right)\right]$
is the entropy of $\pi$. Equation \ref{eq:cross-entropy-1} is useful
on its own for comparing the methods, but the fact that $D_{\mathrm{{KL}}}\left(\pi\parallel q\right)\ge0$
allows us to assess performance on an absolute scale using Equation
\ref{eq:kl-divergence}. In some simple models, an analytic form is
known for $H\left(\pi\right)$, and in others samples from $\pi$
can be used to estimate it, again using Equation \ref{eq:posterior-expectation}.

Density function $q$ can be estimated using kernel density estimation,
either using the method described in Section \ref{subsec:Density-estimation-mie}
for LEMIE or using standard KDE with the pooled samples from the CMC,
DPE or Naive methods. We use a normal kernel function; for LEMIE this
means an MVN p.d.f. is used for $K_{\xi}\left(\theta,\theta^{\prime}\right)$
in Section \ref{subsec:Density-estimation-mie} with $\xi$ being
the smoothing covariance matrix.

\subsection{Beta-Bernoulli model\label{subsec:Beta-Bernoulli-model}}

We reproduce the example of \citet{scott2016bayes} in which $M=100$,
data $x_{1:n}$ are binary with $n=1,000$ and exactly one of the
outcomes are positive, say $x_{1}=1$, with the remainder being zero.
The model is

\begin{equation}
x_{i}\sim\textrm{{Bernoulli}}\left(\lambda\right),i=1,2,\ldots,n,
\end{equation}
and the estimand is the parameter $\lambda\in\left(0,1\right)$. We
use a prior distribution of the form $\mathrm{{Beta}}\left(\beta_{1},\beta_{2}\right)$.
This is a conjugate prior, and the posterior distribution is $\mathrm{{Beta}}\left(\beta_{1}+\sum_{i=1}^{n}x_{i},\beta_{2}+n-\sum_{i=1}^{n}x_{i}\right)$.
Following \citet{scott2016bayes}, we use $\beta_{1}=\beta_{2}=1$,
which is equivalent to a uniform prior on $\left(0,1\right)$.

For CMC and DPE, following the principle of Equation \ref{eq:fractionated-posterior}
requires $\mathrm{{Beta}}\left(1,1\right)$ also be used as the fractionated
prior. CMC does not perform well with that prior in this example,
as demonstrated by \citet{scott2016bayes}. The fractionated prior
recommended for this example by \citet{scott2016bayes} is $\mathrm{{Beta}}\left(0.01,0.01\right)$
with the justification that using $\mathrm{{Beta}}\left(1,1\right)$
implies an additional ``prior success'', which is too informative
given the data only contain a single success. In fact it is the $\mathrm{{Beta}}\left(0.01,0.01\right)$
that is more informative; $\mathrm{{Beta}}\left(1,1\right)$ is the
beta distribution of maximum differential entropy and in $\mathrm{{Beta}}\left(0.01,0.01\right)$
the density is concentrated close to 0 and 1.

We drew $\bar{N}=10,000$ samples from each local posterior, using
each of the two priors: $\mathrm{{Beta}}\left(1,1\right)$, which
is used by the MIE algorithms, and $\mathrm{{Beta}}\left(0.01,0.01\right)$,
the fractionated prior proposed by \citet{scott2016bayes} and used
by the CMC and DPE algorithms.

\begin{figure}
\subfloat[KDE approximations for Naive, CMC1, NDPE and SDPE from Section \ref{subsec:Other-approaches}.]{\includegraphics[scale=0.38]{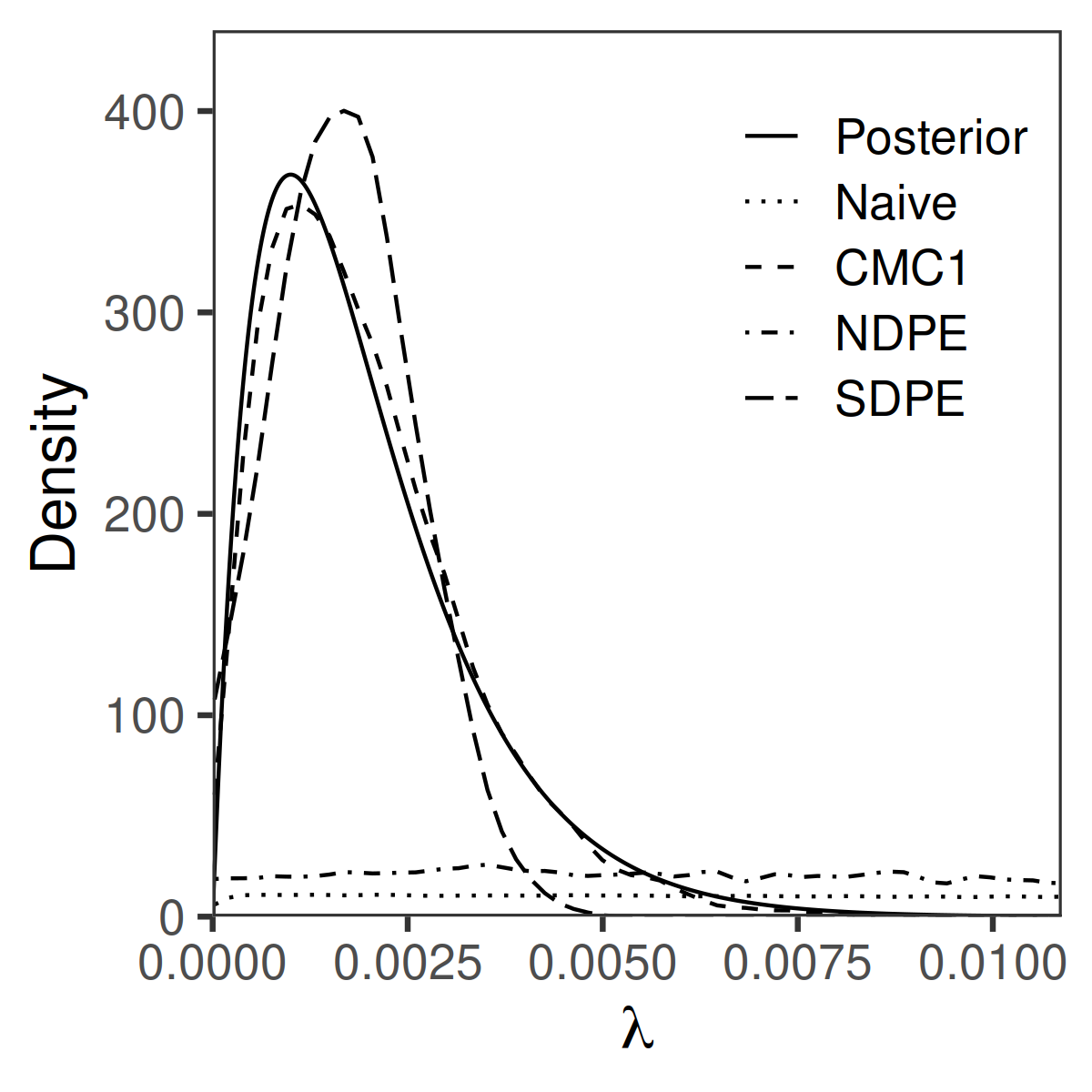}

}\hfill{}\subfloat[KDE approximations for the MIE algorithms from Section \ref{subsec:Multiple-importance-estimation}.]{\includegraphics[scale=0.38]{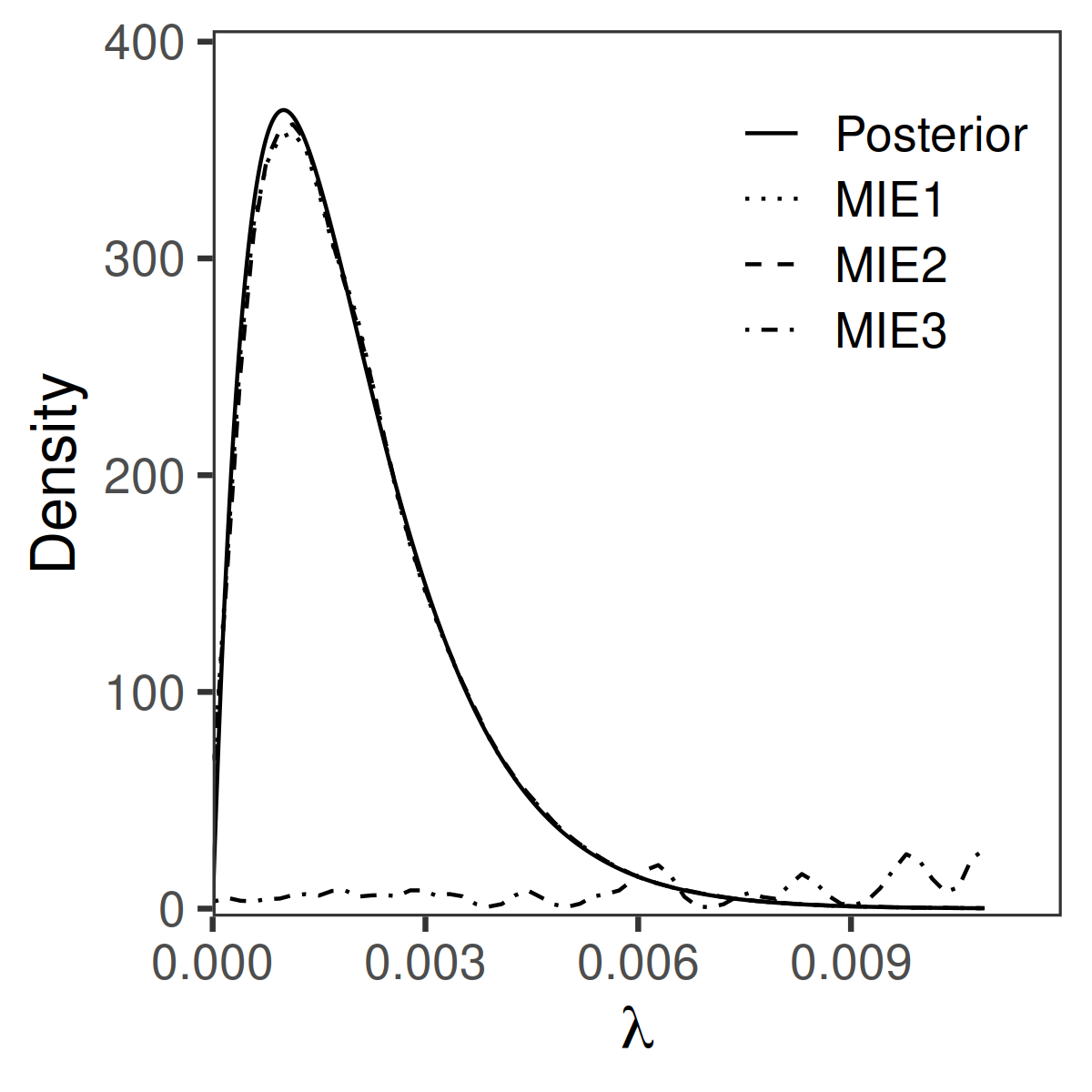}}\hfill{}\subfloat[Posterior quantile-quantile plot for the best of the estimators in
(a) and (b).\label{fig:beta-bernoulli-1-c}]{\includegraphics[scale=0.38]{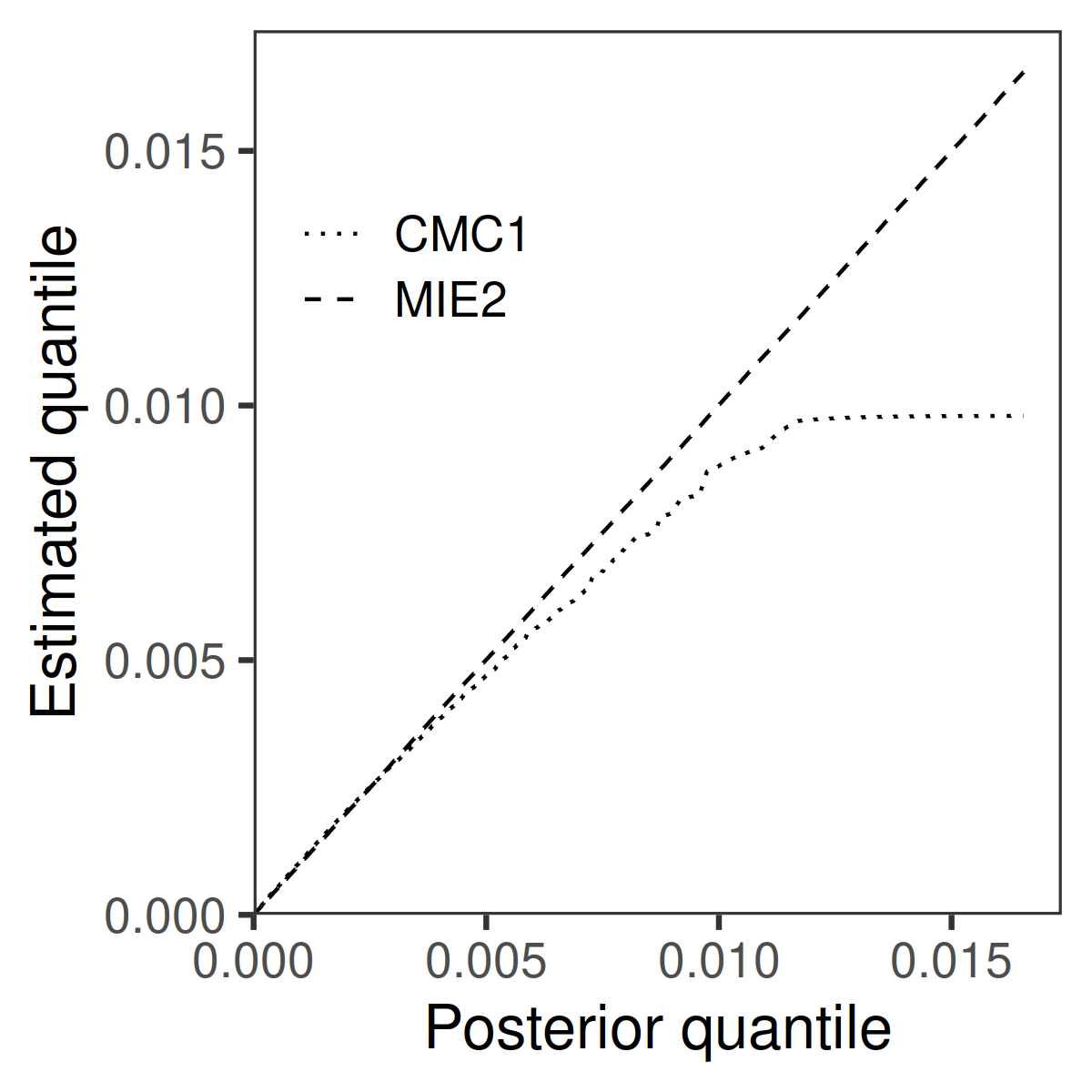}}
\centering{}\caption{KDE (bin width 0.000184) approximations of the beta posterior distribution
in the example of Section \ref{subsec:Beta-Bernoulli-model} with
one positive outcome ($n=1,000$).\label{fig:beta-bernoulli-1}}
\end{figure}

Figure \ref{fig:beta-bernoulli-1} presents the results of posterior
density estimation in this example. As observed by \citet{scott2016bayes},
the CMC1 algorithm using fractionated prior $\mathrm{{Beta}}\left(0.01,0.01\right)$
approximates the posterior p.d.f. quite well. We found that CMC2 is
poor in this example (results not shown), even with the bias correction.
The DPE algorithms perform less well, although better than naive pooling.
MIE1 and MIE2 perform the best, as can be seen for MIE2 in the quantile-quantile
plot of Figure \ref{fig:beta-bernoulli-1-c}, with excellent approximation
of the tail.

{} 

To investigate the role of the prior we looked at another example
in which $\frac{n}{2}$ of the observations are positive and the remaining
$\frac{n}{2}$ are zero, the likelihood of which would be maximised
by $\lambda=\frac{1}{2}$, and we partition the data such that 50
of the $M=100$ parts contain positive outcomes only and the remaining
50 contain negative outcomes only. We used the same $\mathrm{{Beta}}\left(1,1\right)$
and $\mathrm{{Beta}}\left(0.01,0.01\right)$ priors.

\begin{figure}
\subfloat[KDE approximations for CMC1, NDPE and SDPE.]{\includegraphics[scale=0.38]{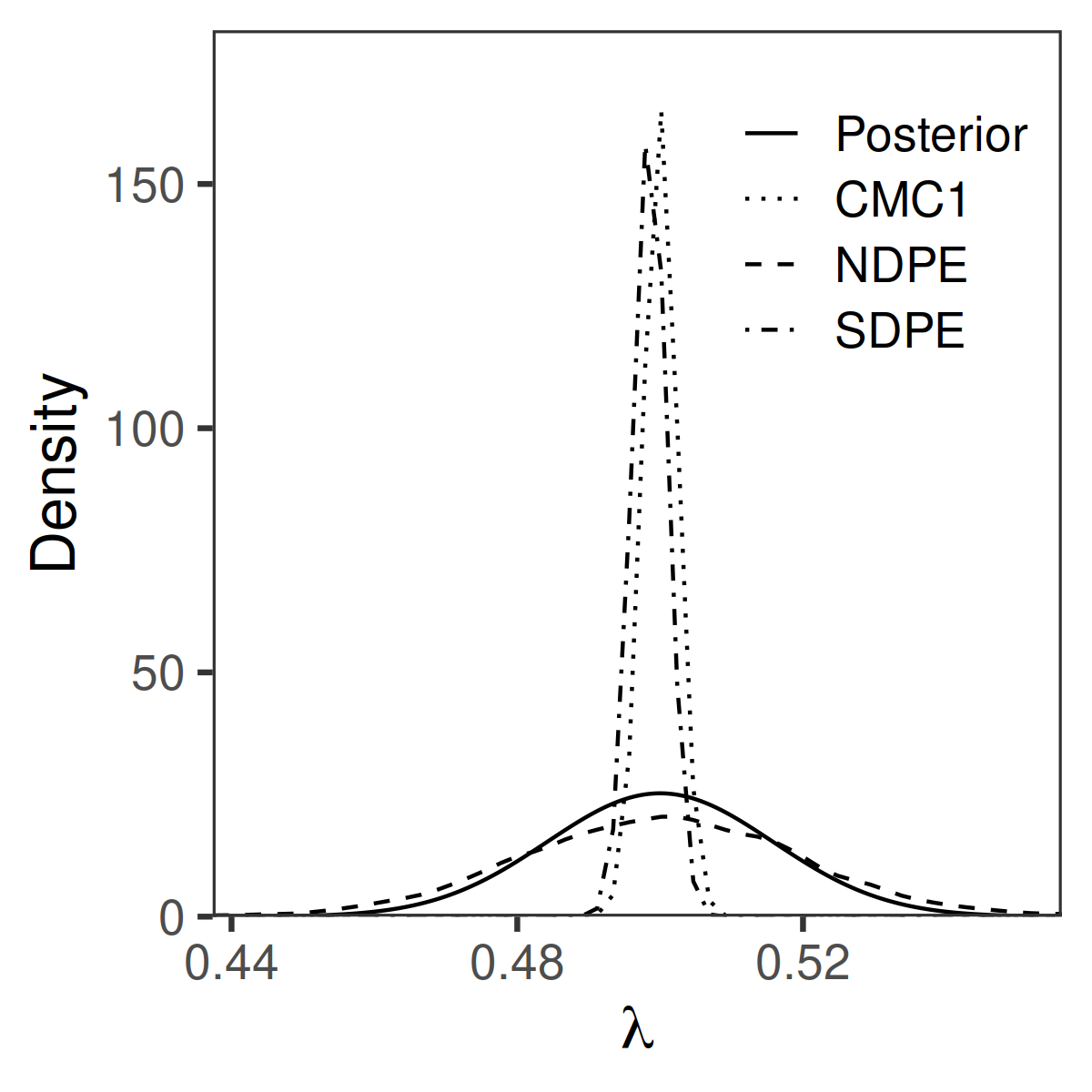}

}\hfill{}\subfloat[KDE approximations for the MIE algorithms.]{\includegraphics[scale=0.38]{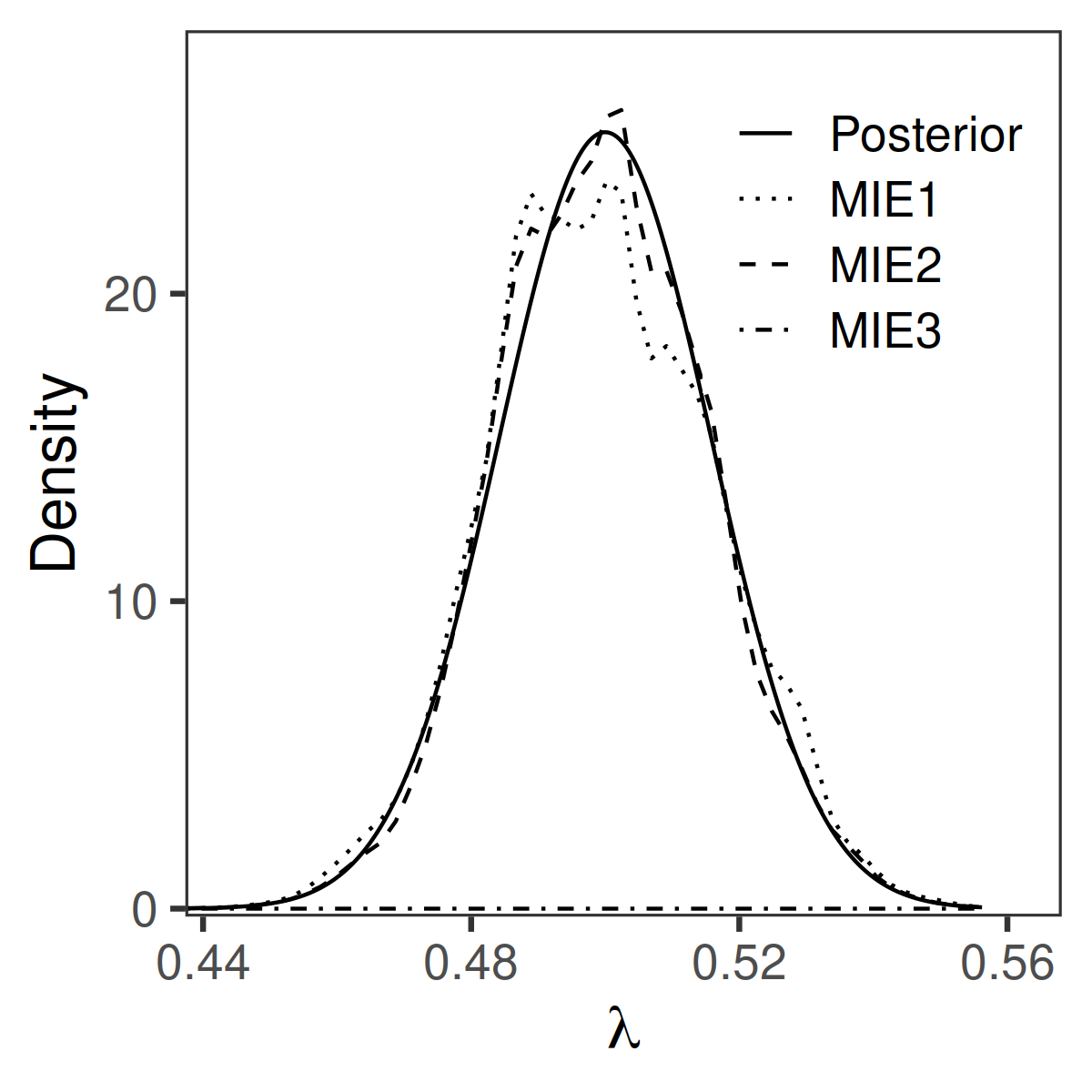}}\hfill{}\subfloat[Posterior quantile-quantile plot for the best of the estimators in
(a) and (b).\label{fig:beta-bernoulli-2-c}]{\includegraphics[scale=0.38]{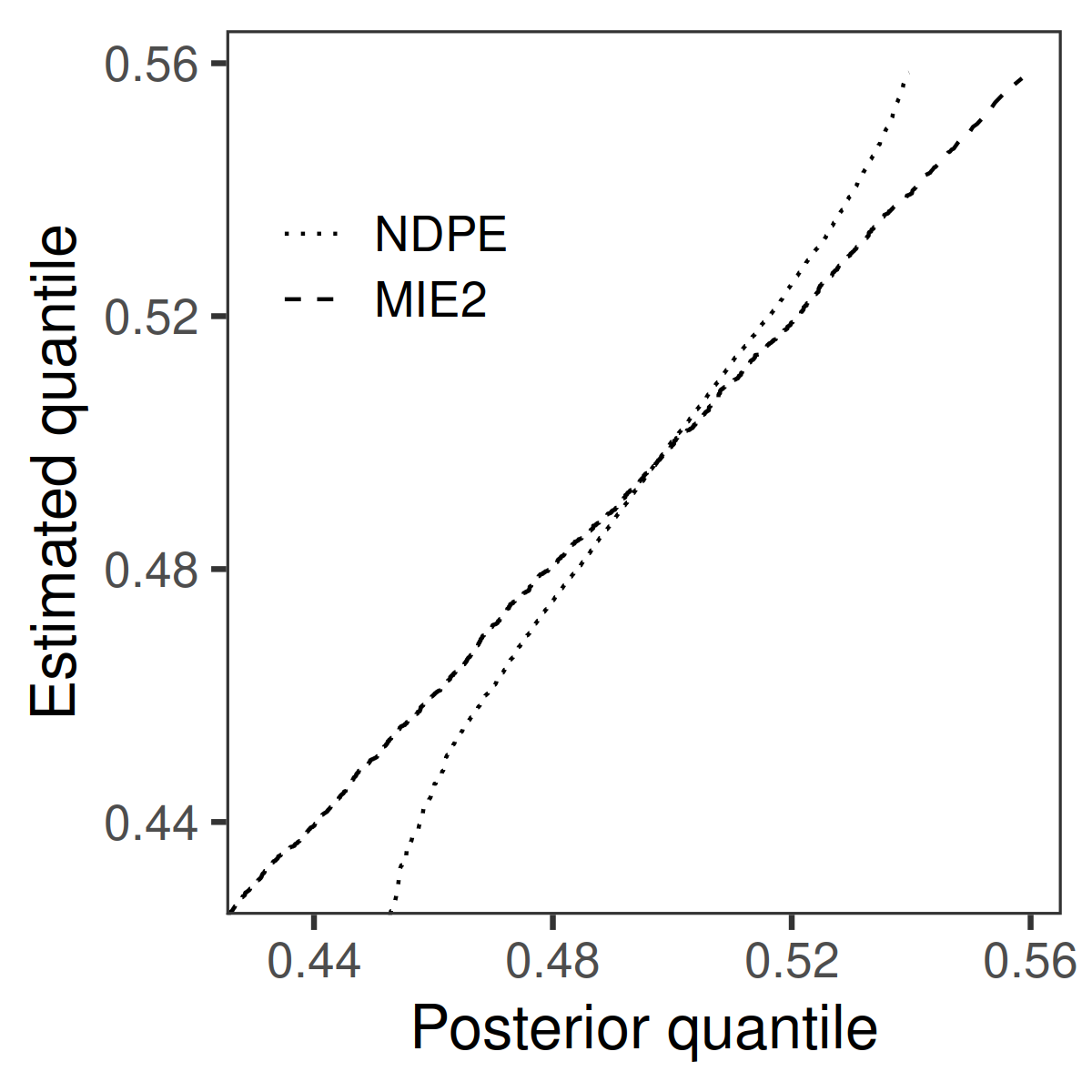}}
\centering{}\caption{Similar to figure \ref{fig:beta-bernoulli-1} but for the example
with 50\% of the observations being positive. KDE bin widths are 0.002243.
Results for Naive are not shown as its estimated density function
appears flat on this scale.\label{fig:beta-bernoulli-2}}
\end{figure}

The results of density estimation in this example are presented in
Figure \ref{fig:beta-bernoulli-2}. Both CMC and DPE struggle in this
example, with CMC1 in particular underestimating the tails, whilst
MIE1 and MIE2 do well. NDPE is the best of the non-MIE algorithms
but is poor at representing the tails of the posterior, as can be
seen in Figure \ref{fig:beta-bernoulli-2-c}. The results of CMC1
and SDPE are symptomatic of using an incorrect prior that is too informative.

MIE3 is poor in both of these examples, doing little better than naive
pooling. This may be because the large $M$ means there are many KL
divergences to estimate and almost all of them are identical (99 of
them in the first example and two sets of 50 in the second). It is
notable that the MIE estimators perform well even when the data are
not partitioned randomly, i.e. the data parts are heterogeneous.

\subsection{Multivariate normal models\label{subsec:Multivariate-normal-models}}

In this section, $x_{i}$ for $i=1,2,\ldots,n$ follows an MVN of
dimension $d$ with parameters $\mu$, the mean vector, and $\Sigma$,
the covariance matrix. We study the performance of the estimators
of Section \ref{sec:methods} in posterior inference for $\mu$ with
$\Sigma$ known, and for both $\mu$ and $\Sigma$, over a range of
$d,n$ and $M$. In each example we simulated each $x_{i}$ from $\textrm{\ensuremath{\mathrm{{N}}}}_{d}\left(\mu,\Sigma\right)$,
then randomly partitioned $x_{1:n}$ into $M$ parts of equal size
(or as close as possible if $n$ is not divisible by $M$).

We restricted these studies to uncorrelated data, i.e. $\Sigma$ is
a diagonal matrix, in order to focus on the effects of $d,n$ and
$M$. The values on the diagonal, $\sigma_{1:d}^{2}$, and $\mu$
were simulated, for each condition, from

\begin{eqnarray}
\sigma_{k}^{2} & \sim & \mathrm{{Gamma}}\left(10,1\right),k=1,2,\ldots,d,\nonumber \\
\mu\mid\sigma_{1:d}^{2} & \sim & \textrm{\ensuremath{\mathrm{{N}}}}_{d}\left(0_{d},\frac{1}{2}\left(\sigma_{1:d}^{2}\right)^{\intercal}I_{d}\right).
\end{eqnarray}

We estimated the posterior mean of $\mu$ and $\Sigma$ and the 2.5\%
and 97.5\% quantiles of the marginals of the posterior for each element
of $\mu$ and $\Sigma$ and calculated the error for each method using
Equation \ref{eq:2norm-error}. We also estimated the KL divergence
between the posterior distribution and the KDE approximation for each
method using the approach explained at the start of Section \ref{sec:experiments}.

In posterior inference for $\mu$ when $\Sigma$ is known, a prior
is needed for $\mu$, for which we used the MVN

\begin{equation}
\mu\sim\textrm{\ensuremath{\mathrm{{N}}}}_{d}\left(\mu_{0},\Sigma_{0}\right),\label{eq:multivariate-normal-prior}
\end{equation}
where $\mu_{0}\in\mathbb{{R}}^{d}$ and $\Sigma_{0}$ is a $d$ by
$d$ positive definite matrix. We used the uninformative prior with
$\mu_{0}=0_{d}$ and $\Sigma_{0}=\mathbf{0}_{d\times d}$, so the
posterior distribution is $\textrm{\ensuremath{\mathrm{{N}}}}_{d}\left(\bar{x},\frac{1}{n}\Sigma\right)$
where $\bar{x}:=\frac{1}{n}\sum_{i=1}^{n}x_{i}$. For CMC and DPE,
the approach to fractionation implied by Equation \ref{eq:fractionated-posterior}
demands an MVN in the density function of which the argument of the
exponential function is

\begin{equation}
-\frac{1}{2M}\left(\mu-\mu_{0}\right)^{\intercal}\Sigma_{0}^{-1}\left(\mu-\mu_{0}\right),\label{eq:fractionated-mvn}
\end{equation}
so the fractionated prior can be parameterised as an MVN by replacing
$\Sigma_{0}$ with $M\Sigma_{0}$. When $\Sigma_{0}=\mathbf{0}_{d\times d}$
this fractionated prior is the same as the prior.

In posterior inference for both $\mu$ and $\Sigma$ we used the normal-inverse
Wishart prior for $\mu$ and $\Sigma$, i.e $\left(\mu,\Sigma\right)\sim\mathrm{{NIW}}\left(\mu_{0},\kappa,\Psi,\nu\right)$
or

\begin{eqnarray}
\Sigma & \sim & \mathrm{{IW}}\left(\Psi,\nu\right),\nonumber \\
\mu\mid\Sigma & \sim & \textrm{\ensuremath{\mathrm{{N}}}}_{d}\left(\mu_{0},\frac{1}{\kappa}\Sigma\right),
\end{eqnarray}
where $\Psi$ is a $d$ by $d$ positive definite matrix, $\nu>0,\kappa>0$
and $\mu_{0}\in\mathbb{{R}}^{d}$. The fractionated prior using the
same approach as above has a density function proportional to

\begin{equation}
\left|\Sigma\right|^{\nicefrac{-\left(\nu+d+1\right)}{2M}}e^{-\frac{1}{2M}\mathrm{{tr}}\left(\Psi\Sigma^{-1}\right)}.\label{eq:normal-inverse-wishart-prior-fractionated}
\end{equation}
We used an uninformative prior, achieved by setting $\kappa=0,\nu=0$
and $\Psi=\mathbf{0}_{d\times d}$ (which results in an improper prior).
Then we can replace $\nu$ with $\nu^{*}=\frac{\nu}{M}-\frac{M-1}{M}d-\frac{M-1}{M}$
to parameterise the fractionated prior as a normal-inverse Wishart
distribution (the exponential function in Equation \ref{eq:normal-inverse-wishart-prior-fractionated}
is the same as in the prior). The posterior distribution obtained
using this fractionated prior is proper so long as $n,M$ and $d$
satisfy $\left\lfloor \frac{n}{M}\right\rfloor >2d$ (see Appendix
\ref{subsec:Fractionated-prior-mvn}), which is a constraint we respect
in our simulations.

For estimation, we drew $\bar{N}=1,000$ samples from each local posterior.
For the LEMIE estimators, we drew an additional 1,000 samples from
each of the Laplace approximations, as explained in Section \ref{subsec:laplace-enrichment}.
In these examples, the naive estimator of the posterior mean of $\mu$
and the CMC1 estimator are optimal because they are equivalent to
the maximum likelihood estimator of $\mu$, which in this example
with an uninformative prior is equal to the posterior mean.

\begin{figure}
\begin{centering}
\subfloat[KL divergence from the approximations to the posterior of $\mu$.]{\includegraphics[scale=0.55]{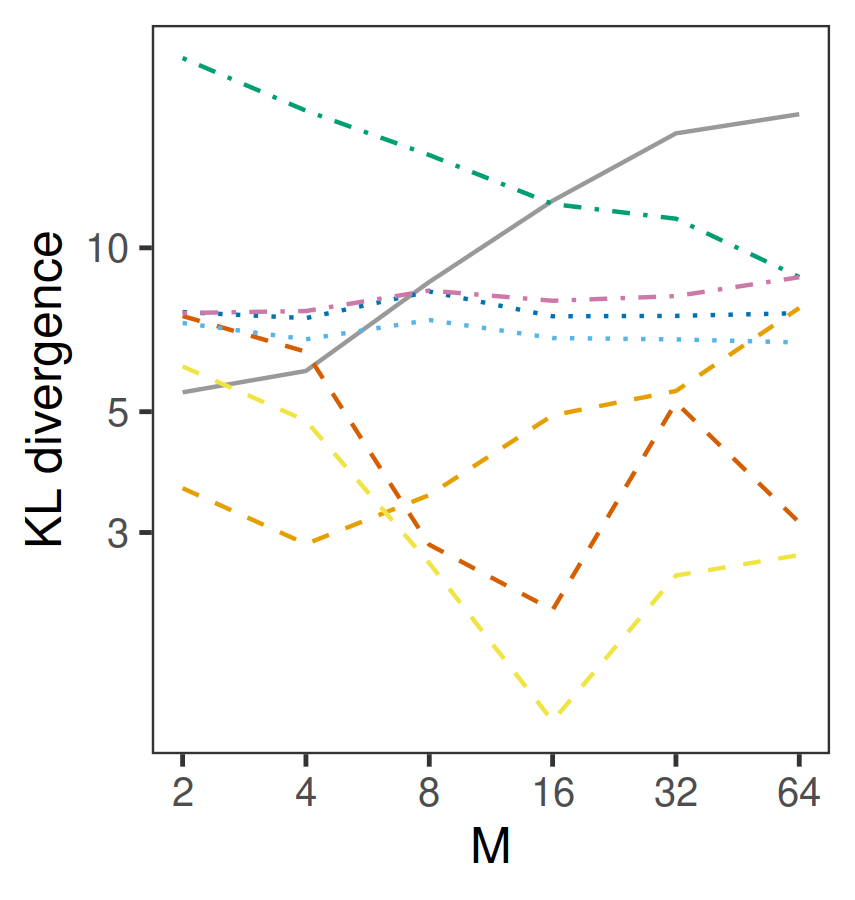}}\quad{}\subfloat[KL divergence from the approximations to the posterior of $\Sigma$.]{\includegraphics[scale=0.55]{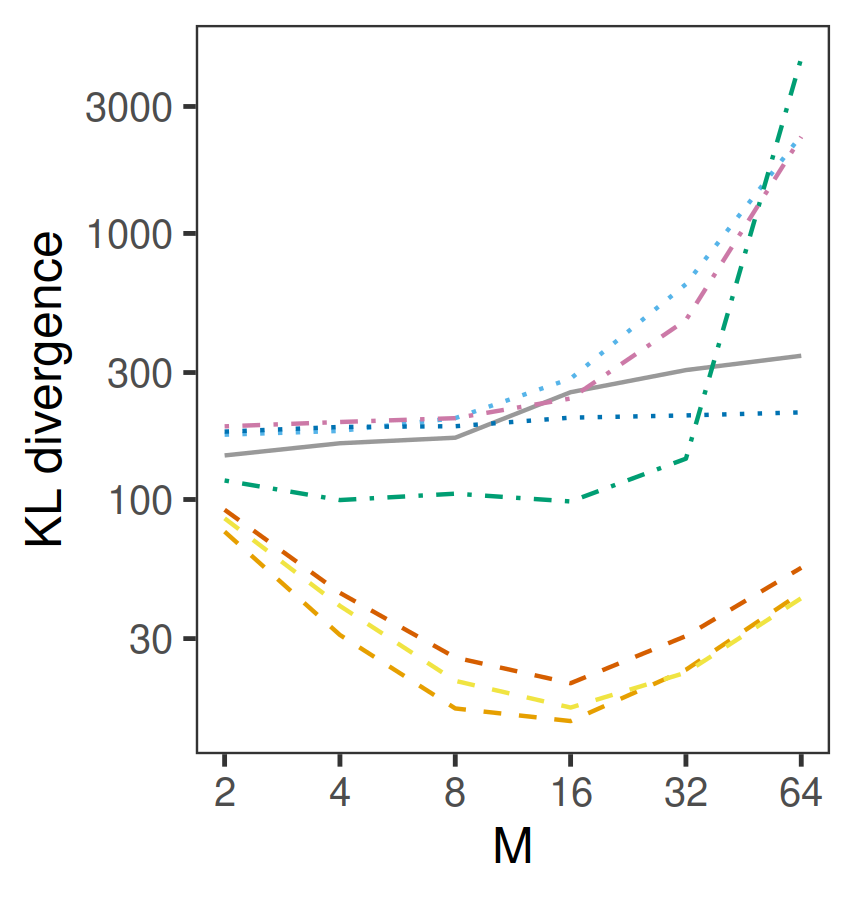}

}\quad{}\includegraphics[scale=0.55]{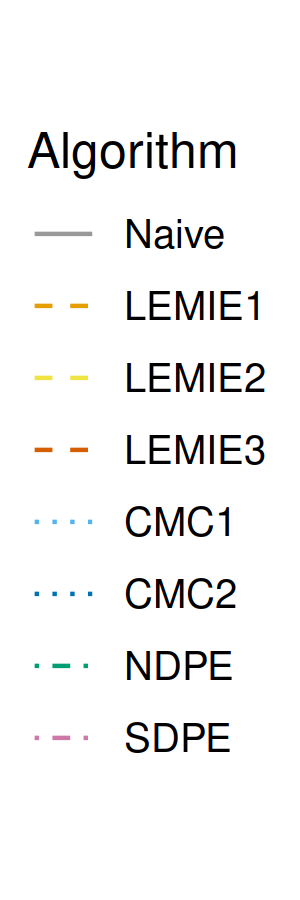}
\par\end{centering}
\begin{centering}
\subfloat[Error in estimating the posterior mean of $\mu$.]{\includegraphics[scale=0.55]{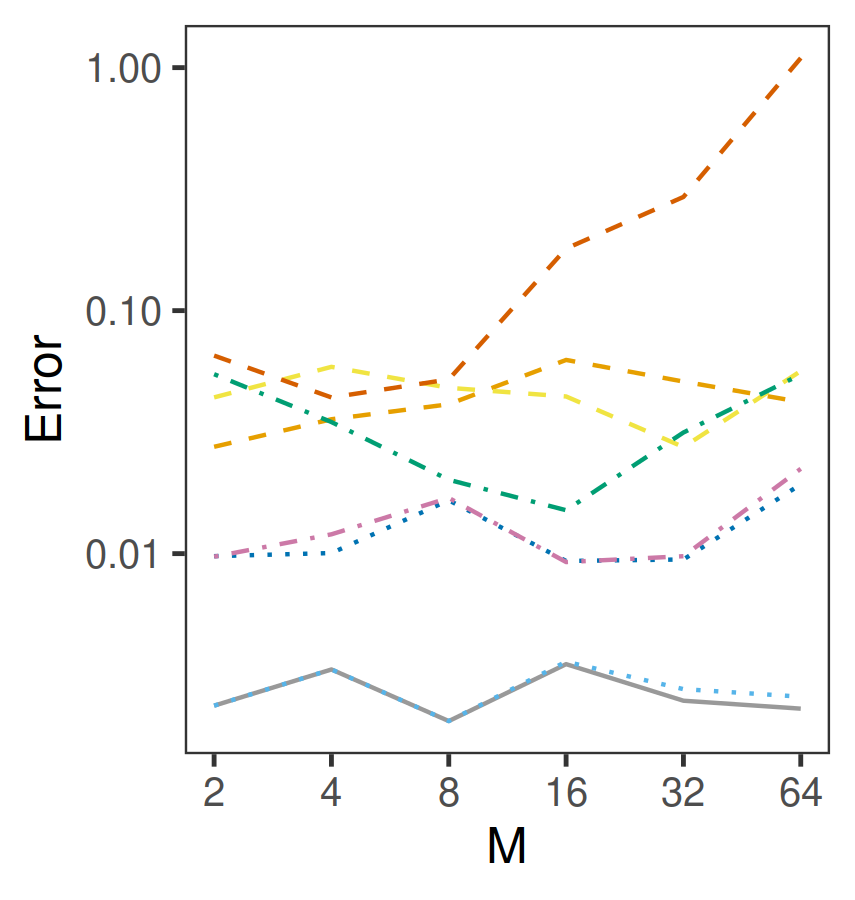}}\quad{}\subfloat[Error in estimating the posterior mean of $\Sigma$.]{\includegraphics[scale=0.55]{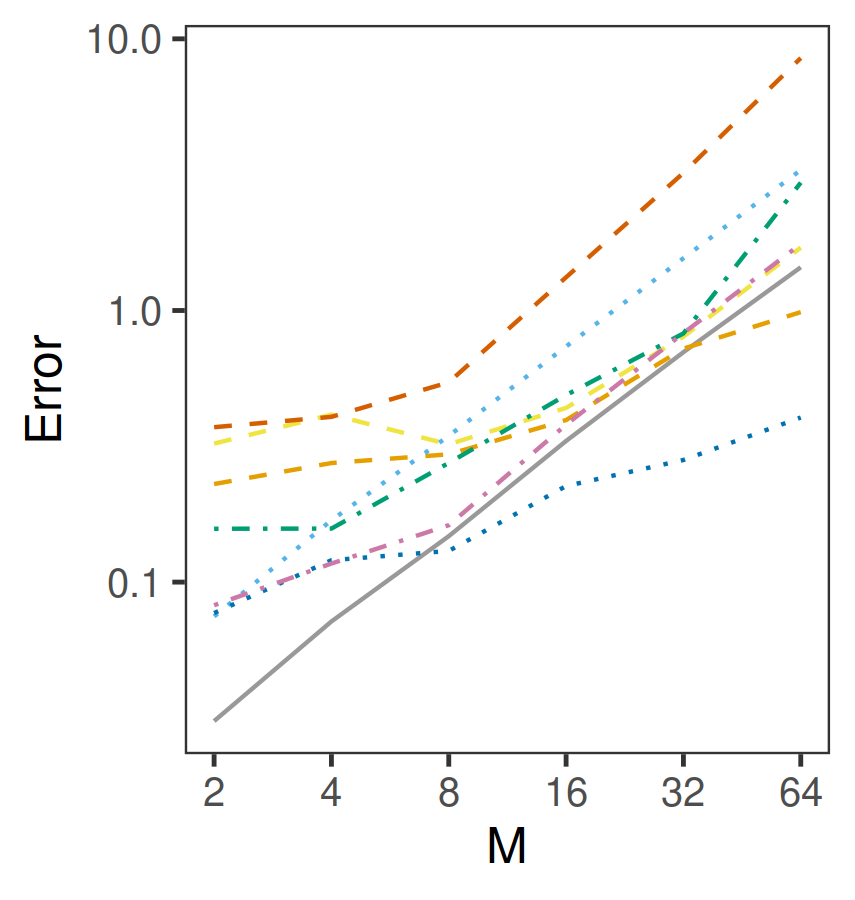}}\qquad{}\qquad{}\enskip{}
\par\end{centering}
\caption{Posterior estimation performance in the MVN example of Section \ref{subsec:Multivariate-normal-models}
with $\mu$ and $\Sigma$ unknown, $d=8$ and $n=10,000$. The LEMIE
methods are enhanced using Laplace samples of all 3 types.\label{fig:normal-sigma-unknown-comparisons-1}}
\end{figure}

In Figure \ref{fig:normal-sigma-unknown-comparisons-1} are plotted
performance metrics for the simulations with $\mu$ and $\Sigma$
unknown, $d=8$ and $n=10,000$, which is one of the more challenging
examples. Of the methods from Section \ref{subsec:Multiple-importance-estimation},
LEMIE using Laplace samples of all 3 types performed best; results
for these are plotted against the naive, CMC and DPE methods. The
approximate KL divergences for the LEMIE algorithms are generally
lower than the other methods, indicating a closer approximation to
the posterior. LEMIE type 1 and 2 approximate the posterior mean of
$\mu$ and $\Sigma$ about as well as the other suboptimal methods.
For $\Sigma$, LEMIE1 outperforms naive at $M=64$ and all other methods
except CMC2. Similar results for the simulations with $d=8$ with
$n=1,088$ and $n=100,000$ can be found in Appendix \ref{appendix:additional-Multivariate-normal-studies},
and the results are similar.

\begin{figure}
\begin{raggedright}
\subfloat[]{\includegraphics[scale=0.45]{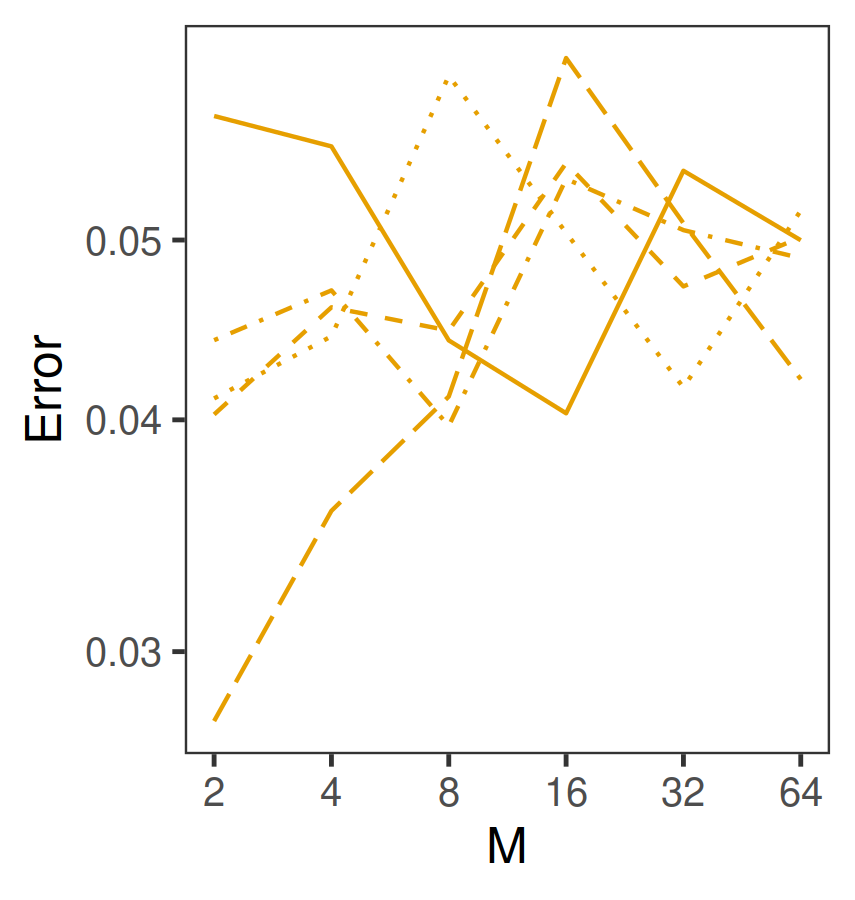}}\enskip{}\subfloat[]{\includegraphics[scale=0.45]{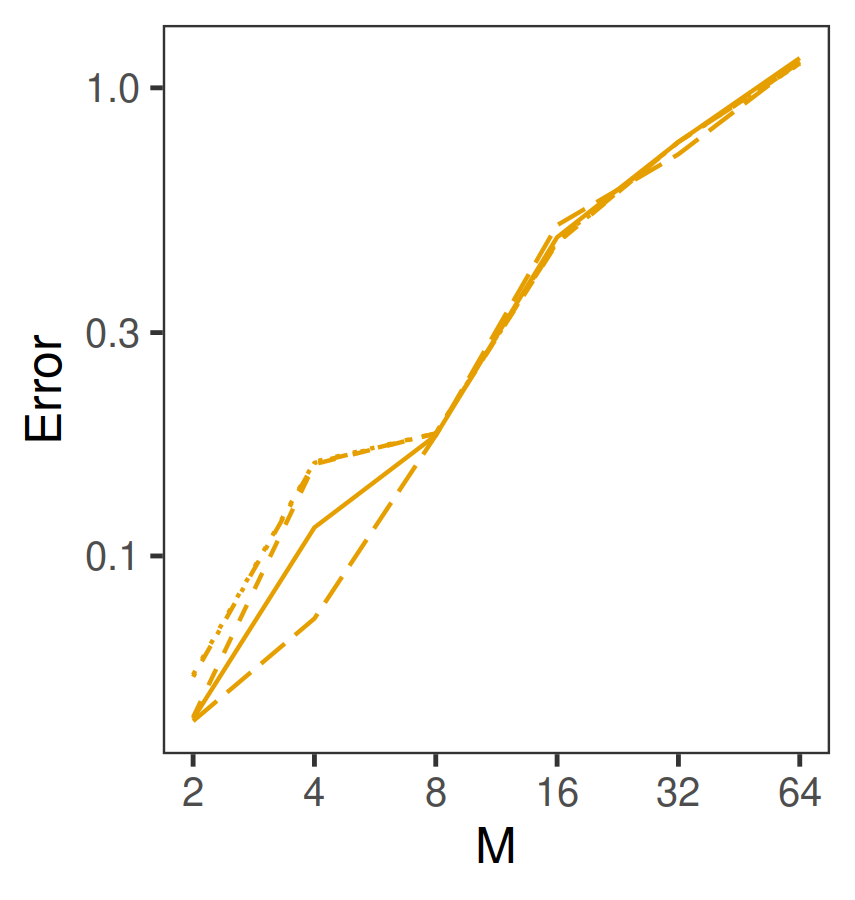}}\enskip{}\subfloat[]{\includegraphics[scale=0.45]{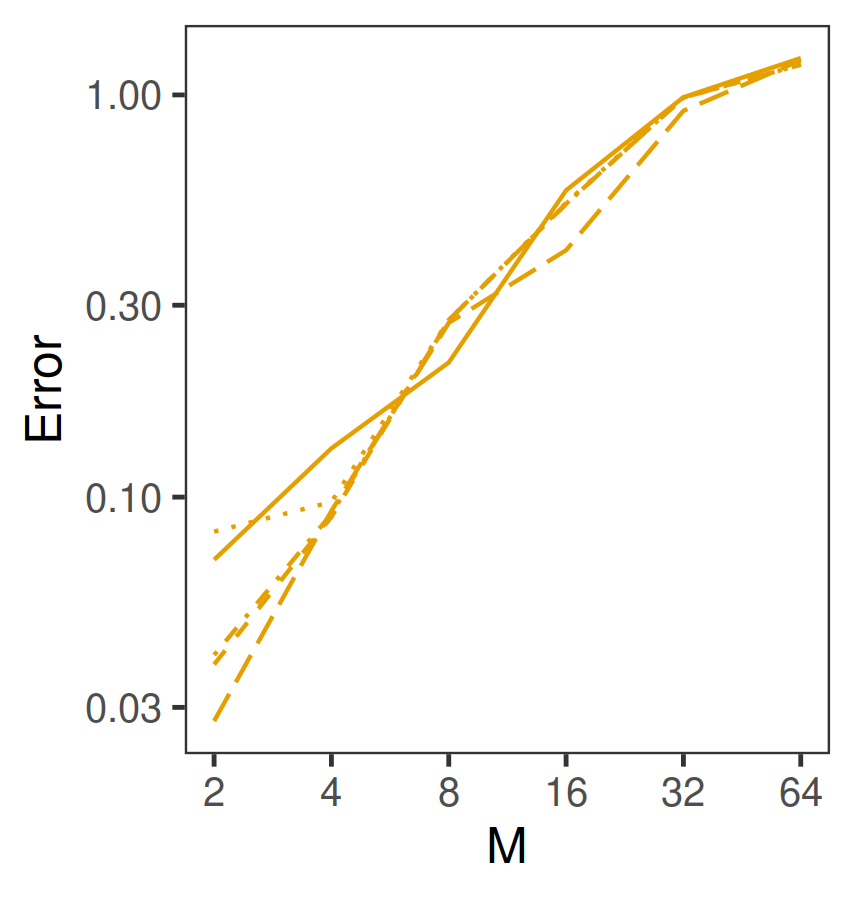}}\enskip{}\includegraphics[scale=0.45]{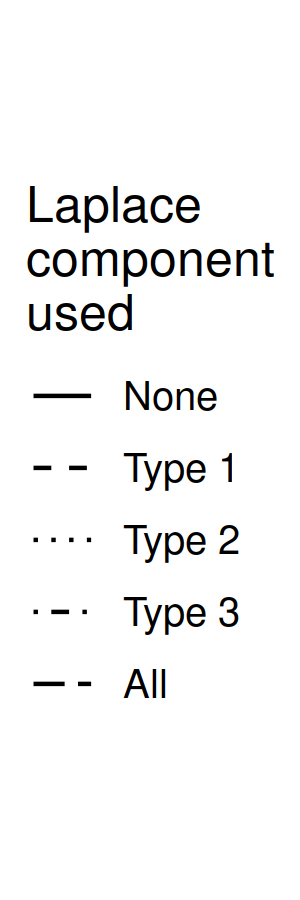}
\par\end{raggedright}
\begin{raggedright}
\subfloat[]{\includegraphics[scale=0.45]{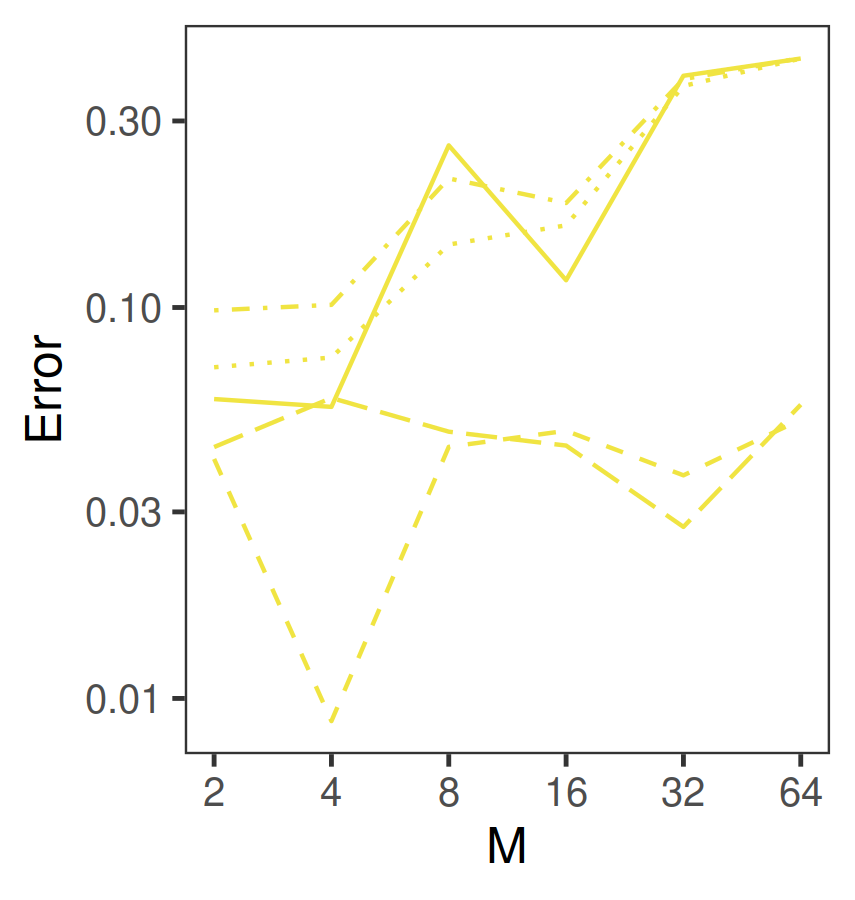}}\enskip{}\subfloat[]{\includegraphics[scale=0.45]{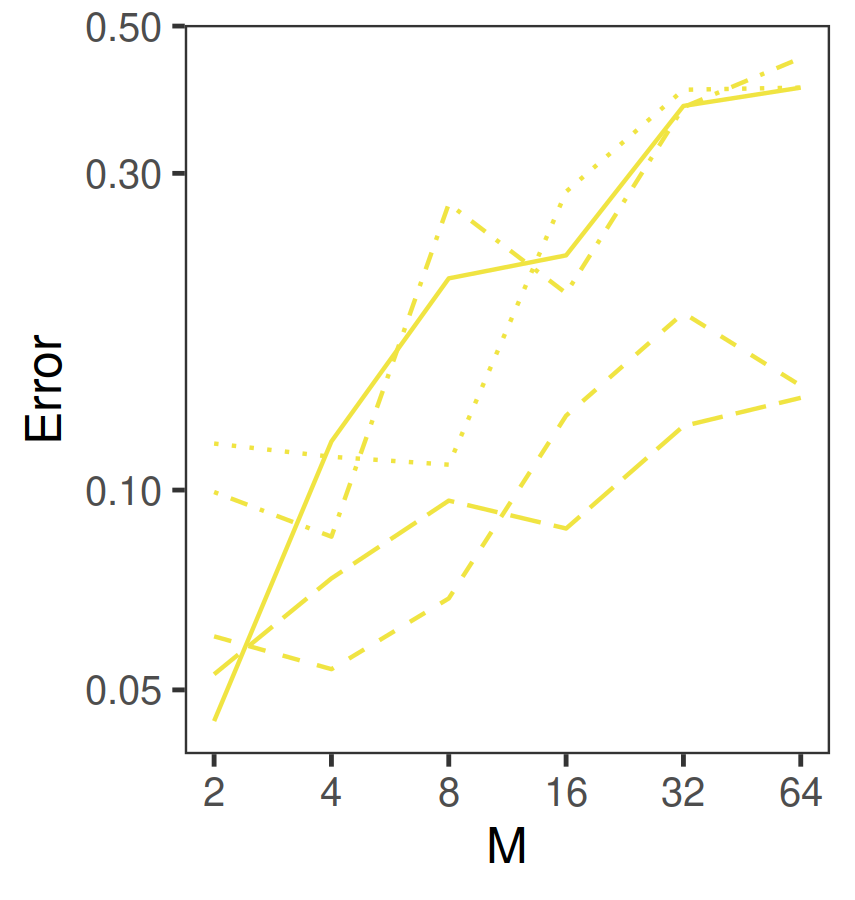}}\enskip{}\subfloat[]{\includegraphics[scale=0.45]{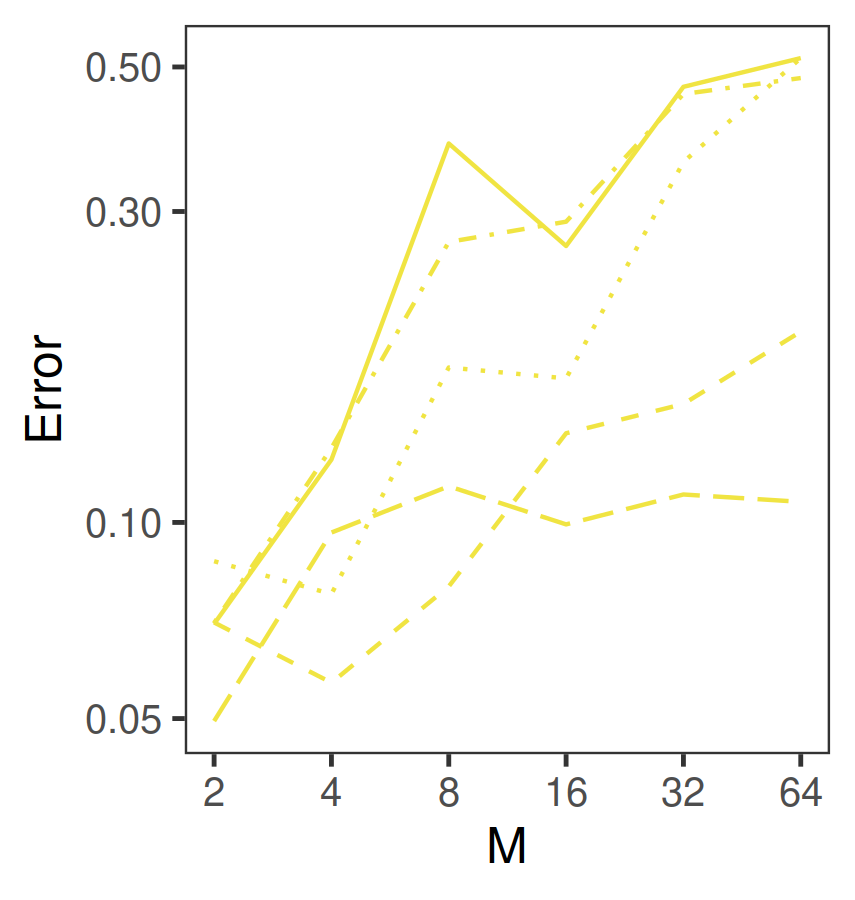}}\enskip{}\includegraphics[scale=0.45]{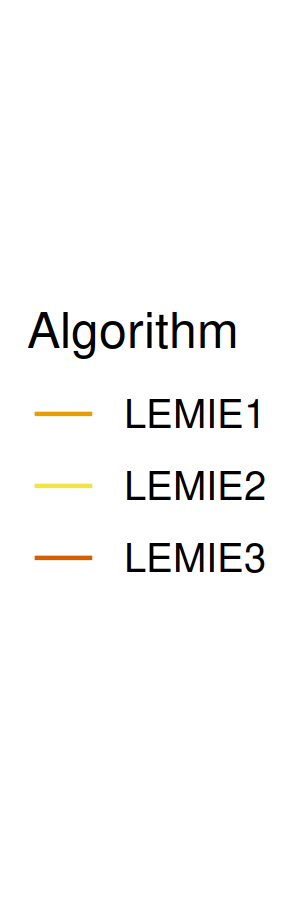}
\par\end{raggedright}
\begin{raggedright}
\subfloat[]{\includegraphics[scale=0.45]{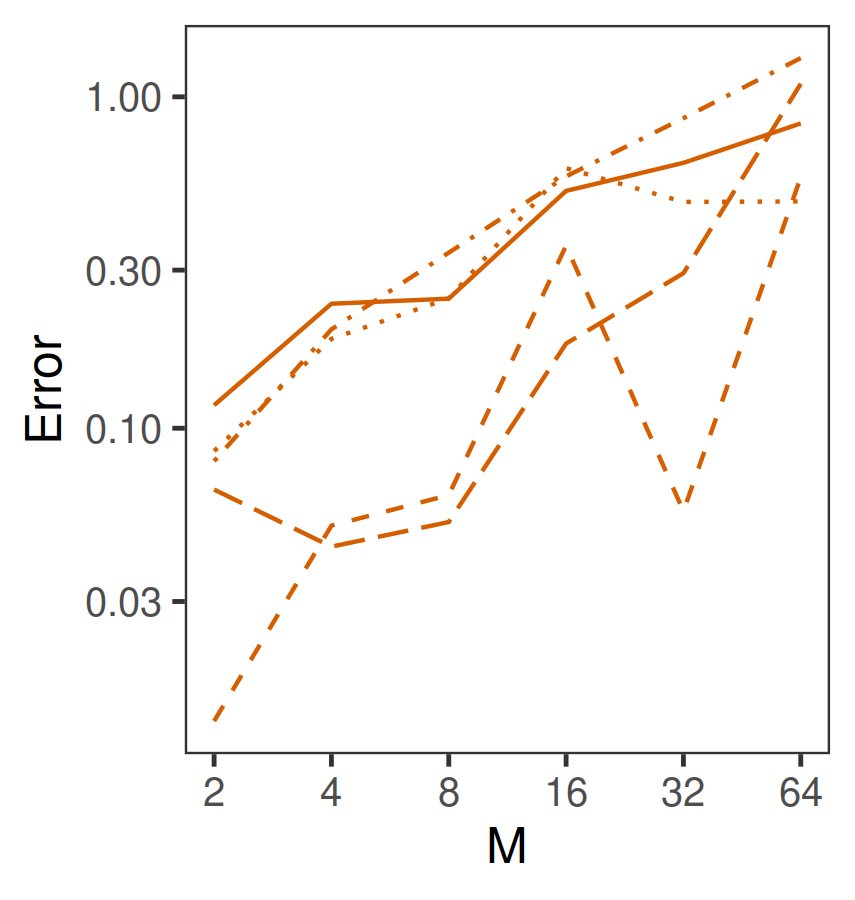}}\enskip{}\subfloat[]{\includegraphics[scale=0.45]{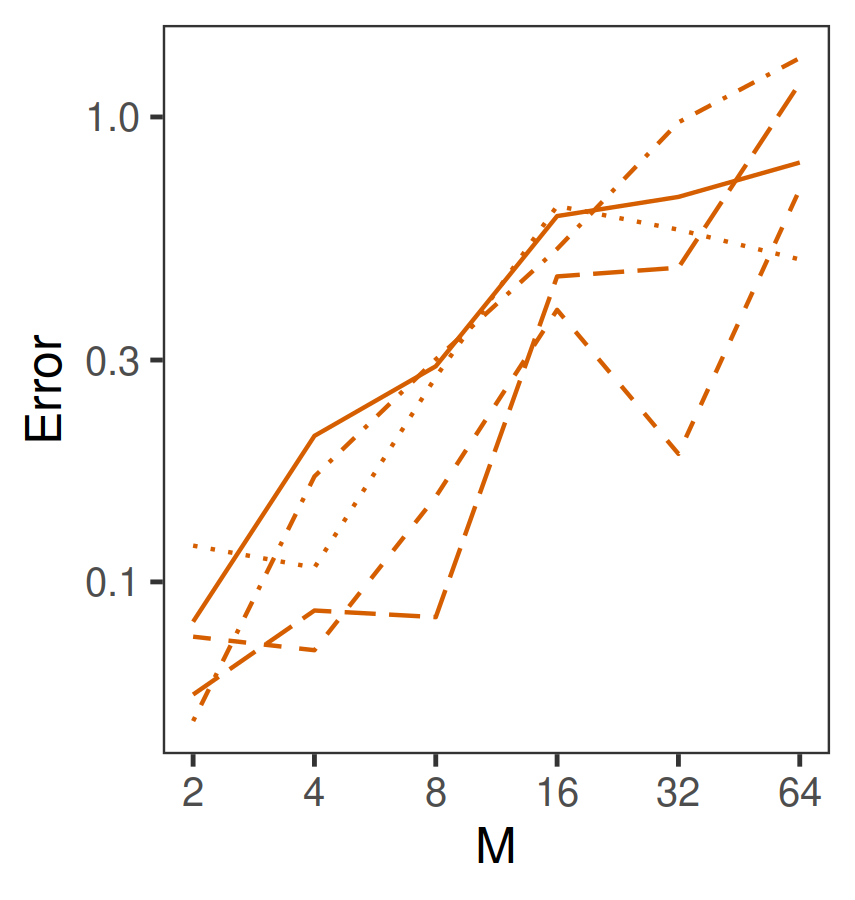}}\enskip{}\subfloat[]{\includegraphics[scale=0.45]{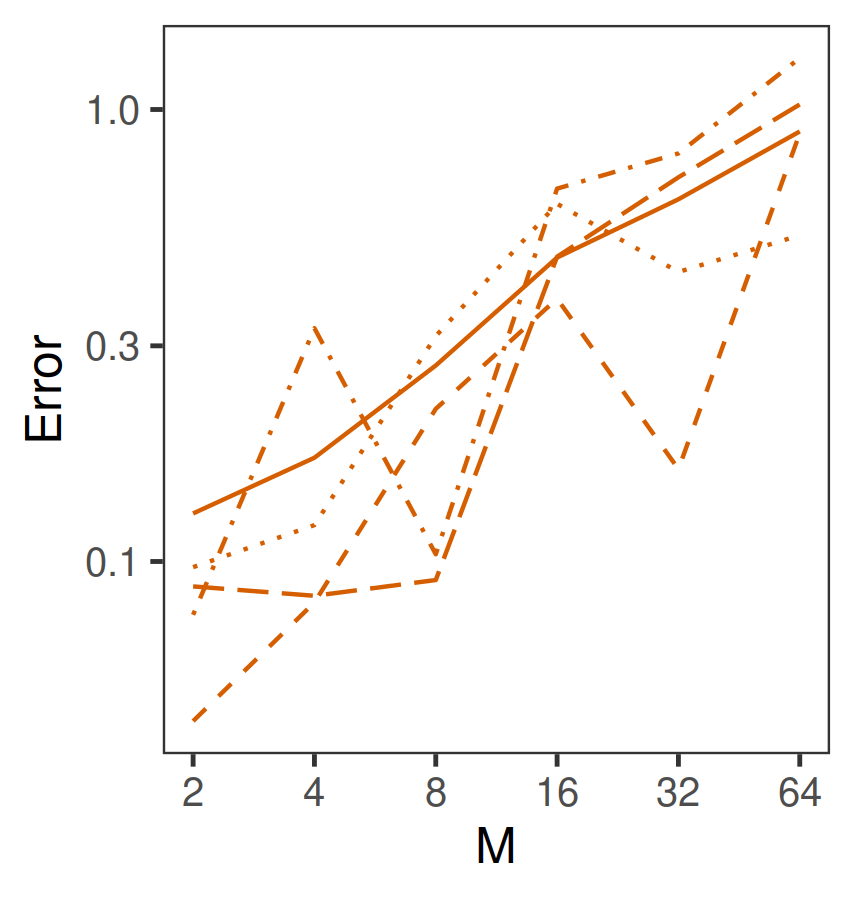}}
\par\end{raggedright}
\begin{raggedright}
\subfloat[]{\includegraphics[scale=0.45]{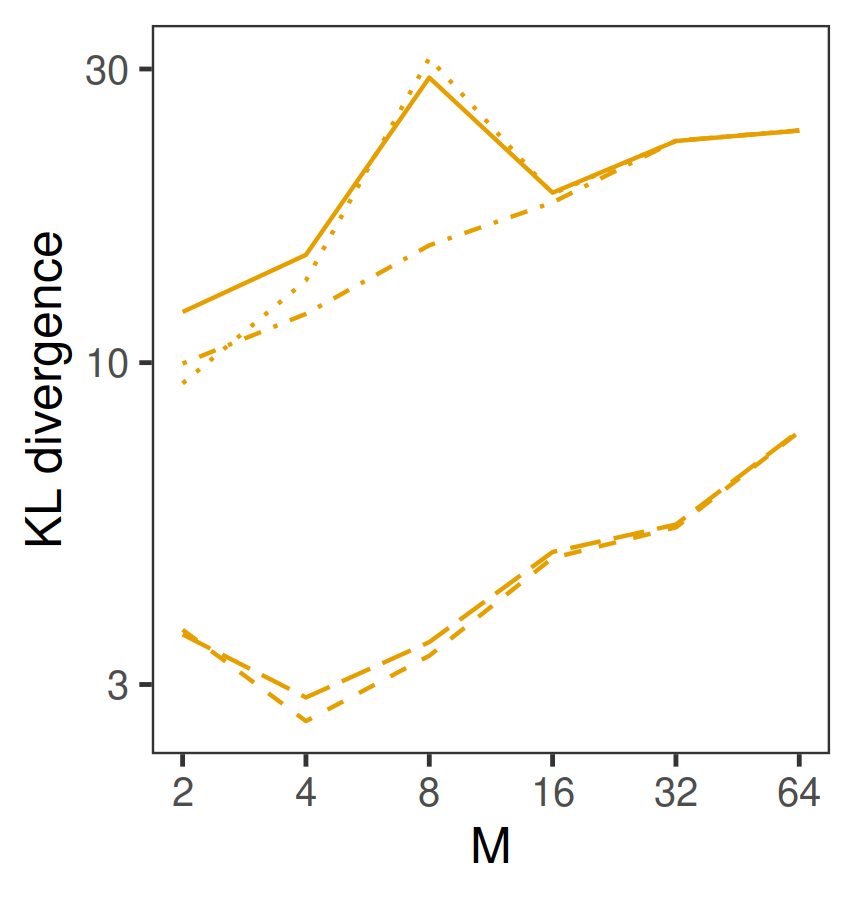}}\enskip{}\subfloat[]{\includegraphics[scale=0.45]{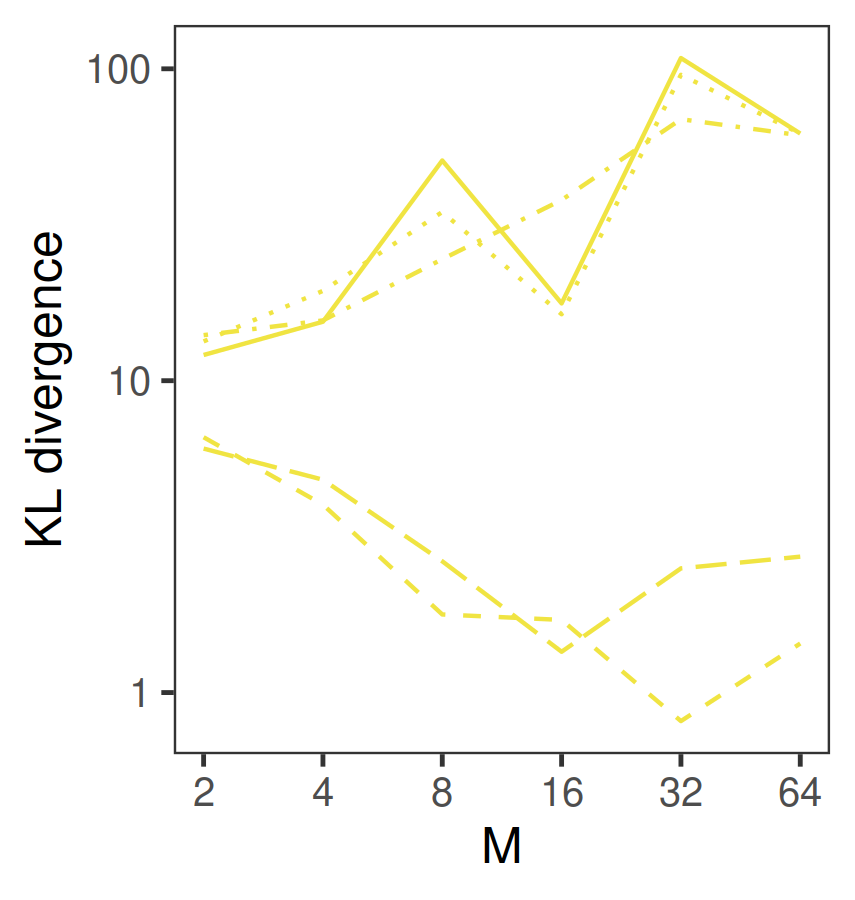}}\enskip{}\subfloat[]{\includegraphics[scale=0.45]{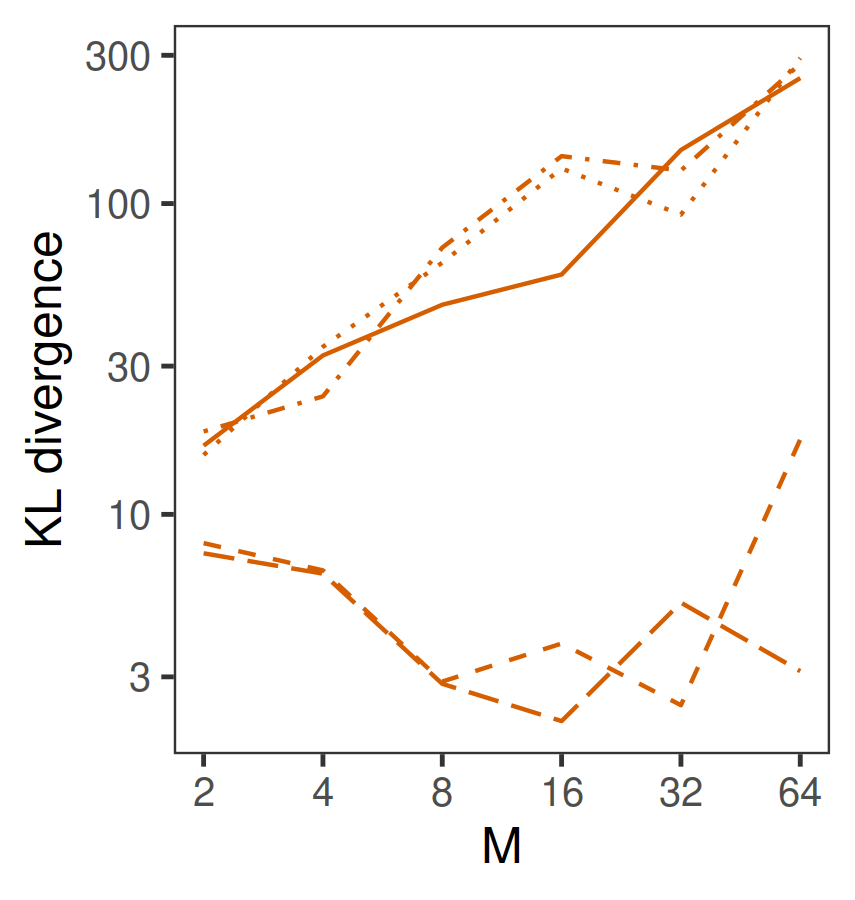}}
\par\end{raggedright}
\caption{Posterior approximation comparisons for the algorithms from Section
\ref{subsec:Multiple-importance-estimation} approximating the $\mu$
marginal of the posterior in the MVN example with $\Sigma$ unknown
and $d=8$ and $n=10,000$ of Section \ref{subsec:Multivariate-normal-models}.
The error in estimating (a)(d)(g) the posterior mean, (b)(e)(h) 2.5\%
quantiles of the marginals of the posterior, (c)(f)(i) 97.5\% quantiles
of the marginals of the posterior. (j)(k)(l) The KL divergence from
the posterior to each approximation.\label{fig:normal-sigma-unknown-comparisons-lemie}}
\end{figure}

Figure \ref{fig:normal-sigma-unknown-comparisons-lemie} compares
across the LEMIE algorithm variants in the same example (results shown
for the $\mu$ marginal of the posterior; see Appendix \ref{appendix:additional-Multivariate-normal-studies}
for a similar plot showing results for the $\Sigma$ marginal). For
LEMIE1, using Laplace samples improves posterior mean estimation only
for low $M$ and not clearly for estimating tail quantiles. For LEMIE2
and LEMIE3, using Laplace samples of type 1 improves estimation across
all $M$; using the other Laplace types may have further benefit but
this is not clear. The KL divergence is improved greatly by using
Laplace samples of type 1; using samples of types 2 and 3 does not
provide any further benefit. We also find that only a small number
of Laplace samples are required to provide benefit and adding additional
samples over a range from 5,000 to 25,000 samples does not seem to
help further (Figure \ref{fig:normal-sigma-known-laplace-extensions}
in Appendix \ref{appendix:additional-Multivariate-normal-studies}).

\begingroup\tabcolsep=0pt
\begin{table}
\begin{centering}
\begin{tabular}{|cc|c|c|c|c|c|c|c|}
\hline 
 &  &  & \multicolumn{3}{c|}{{\scriptsize{}$n=1,088$}} & \multicolumn{3}{c|}{{\scriptsize{}$n=10,000$}}\tabularnewline
 & {\scriptsize{}Laplace} &  & {\scriptsize{}$d=2$} & {\scriptsize{}$d=4$} & {\scriptsize{}$d=8$} & {\scriptsize{}$d=2$} & {\scriptsize{}$d=4$} & {\scriptsize{}$d=8$}\tabularnewline
\hline 
\hline 
\multirow{8}{*}{{\scriptsize{}LEMIE1}} & \multirow{2}{*}{{\scriptsize{}None}} & {\scriptsize{}$\mu$} & {\tiny{}1.221 (0.002) } & {\tiny{}5.892 (0.009) } & {\tiny{}33.43 (0.04) } & {\tiny{}1.422 (0.002) } & {\tiny{}6.95 (0.01) } & {\tiny{}23.83 (0.05) }\tabularnewline
 &  & {\scriptsize{}$\Sigma$} & {\tiny{}1.713 (0.003) } & {\tiny{}29.3 (0.03) } & {\tiny{}175.99 (0.07) } & {\tiny{}1.959 (0.002) } & {\tiny{}27.45 (0.05) } & {\tiny{}341.6 (0.2) }\tabularnewline
\cline{2-9} \cline{3-9} \cline{4-9} \cline{5-9} \cline{6-9} \cline{7-9} \cline{8-9} \cline{9-9} 
 & \multirow{2}{*}{{\scriptsize{}Type 1}} & {\scriptsize{}$\mu$} & {\tiny{}1.318 (0.002) } & {\tiny{}4.25 (0.01) } & {\tiny{}7.28 (0.02) } & {\tiny{}1.214 (0.002) } & {\tiny{}4.671 (0.006) } & {\tiny{}7.72 (0.02) }\tabularnewline
 &  & {\scriptsize{}$\Sigma$} & {\tiny{}1.748 (0.003) } & {\tiny{}29.28 (0.03) } & {\tiny{}175.8 (0.07) } & {\tiny{}2.05 (0.002) } & {\tiny{}10.39 (0.02) } & {\tiny{}44.6 (0.07) }\tabularnewline
\cline{2-9} \cline{3-9} \cline{4-9} \cline{5-9} \cline{6-9} \cline{7-9} \cline{8-9} \cline{9-9} 
 & \multirow{2}{*}{{\scriptsize{}Type 2}} & {\scriptsize{}$\mu$} & {\tiny{}1.352 (0.002) } & {\tiny{}4.604 (0.009) } & {\tiny{}33.45 (0.04) } & {\tiny{}1.256 (0.002) } & {\tiny{}6.96 (0.01) } & {\tiny{}23.84 (0.05) }\tabularnewline
 &  & {\scriptsize{}$\Sigma$} & {\tiny{}1.752 (0.003) } & {\tiny{}29.28 (0.03) } & {\tiny{}176.01 (0.07) } & {\tiny{}2.115 (0.002) } & {\tiny{}27.47 (0.05) } & {\tiny{}341.7 (0.2) }\tabularnewline
\cline{2-9} \cline{3-9} \cline{4-9} \cline{5-9} \cline{6-9} \cline{7-9} \cline{8-9} \cline{9-9} 
 & \multirow{2}{*}{{\scriptsize{}Type 3}} & {\scriptsize{}$\mu$} & {\tiny{}1.352 (0.002) } & {\tiny{}4.604 (0.009) } & {\tiny{}33.45 (0.04) } & {\tiny{}1.229 (0.002) } & {\tiny{}6.93 (0.01) } & {\tiny{}23.84 (0.05) }\tabularnewline
 &  & {\scriptsize{}$\Sigma$} & {\tiny{}1.753 (0.003) } & {\tiny{}29.28 (0.03) } & {\tiny{}176.01 (0.07) } & {\tiny{}2.053 (0.002) } & {\tiny{}27.44 (0.05) } & {\tiny{}265.7 (0.2) }\tabularnewline
\hline 
\multirow{8}{*}{{\scriptsize{}LEMIE2}} & \multirow{2}{*}{{\scriptsize{}None}} & {\scriptsize{}$\mu$} & {\tiny{}0.496 (0.005) } & {\tiny{}6.46 (0.02) } & {\tiny{}38.06 (0.07) } & {\tiny{}0.477 (0.008) } & {\tiny{}16.53 (0.03) } & {\tiny{}62 (0.09) }\tabularnewline
 &  & {\scriptsize{}$\Sigma$} & {\tiny{}0.622 (0.006) } & {\tiny{}28.25 (0.05) } & {\tiny{}537.89 (0.07) } & {\tiny{}0.986 (0.009) } & {\tiny{}23.58 (0.05) } & {\tiny{}438.1 (0.2) }\tabularnewline
\cline{2-9} \cline{3-9} \cline{4-9} \cline{5-9} \cline{6-9} \cline{7-9} \cline{8-9} \cline{9-9} 
 & \multirow{2}{*}{{\scriptsize{}Type 1}} & {\scriptsize{}$\mu$} & {\tiny{}0.602 (0.004) } & {\tiny{}7.39 (0.02) } & {\tiny{}190.8 (0.2) } & {\tiny{}0.275 (0.005) } & {\tiny{}1.76 (0.02) } & {\tiny{}1.44 (0.02) }\tabularnewline
 &  & {\scriptsize{}$\Sigma$} & {\tiny{}0.672 (0.006) } & {\tiny{}30.24 (0.04) } & {\tiny{}1045.1 (0.2) } & {\tiny{}0.469 (0.008) } & {\tiny{}6.93 (0.03) } & {\tiny{}44.69 (0.07) }\tabularnewline
\cline{2-9} \cline{3-9} \cline{4-9} \cline{5-9} \cline{6-9} \cline{7-9} \cline{8-9} \cline{9-9} 
 & \multirow{2}{*}{{\scriptsize{}Type 2}} & {\scriptsize{}$\mu$} & {\tiny{}55.29 (0.08) } & {\tiny{}299.4 (0.2) } & {\tiny{}538 (0.3) } & {\tiny{}1.49 (0.01) } & {\tiny{}17.44 (0.04) } & {\tiny{}62 (0.09) }\tabularnewline
 &  & {\scriptsize{}$\Sigma$} & {\tiny{}266.4 (0.2) } & {\tiny{}230.2 (0.1) } & {\tiny{}3478.7 (0.3) } & {\tiny{}5.36 (0.02) } & {\tiny{}36.59 (0.04) } & {\tiny{}438.5 (0.2) }\tabularnewline
\cline{2-9} \cline{3-9} \cline{4-9} \cline{5-9} \cline{6-9} \cline{7-9} \cline{8-9} \cline{9-9} 
 & \multirow{2}{*}{{\scriptsize{}Type 3}} & {\scriptsize{}$\mu$} & {\tiny{}3.17 (0.02) } & {\tiny{}171.4 (0.1) } & {\tiny{}155.6 (0.2) } & {\tiny{}1.41 (0.01) } & {\tiny{}9.28 (0.03) } & {\tiny{}61.2 (0.09) }\tabularnewline
 &  & {\scriptsize{}$\Sigma$} & {\tiny{}2.56 (0.01) } & {\tiny{}495.9 (0.2) } & {\tiny{}1566.2 (0.2) } & {\tiny{}2.85 (0.01) } & {\tiny{}48.8 (0.05) } & {\tiny{}261.5 (0.2) }\tabularnewline
\hline 
\multirow{8}{*}{{\scriptsize{}LEMIE3}} & \multirow{2}{*}{{\scriptsize{}None}} & {\scriptsize{}$\mu$} & {\tiny{}18.6 (0.04) } & {\tiny{}83.45 (0.05) } & {\tiny{}329.8 (0.2) } & {\tiny{}3 (0.01) } & {\tiny{}92.5 (0.1) } & {\tiny{}253.4 (0.2) }\tabularnewline
 &  & {\scriptsize{}$\Sigma$} & {\tiny{}24.33 (0.03) } & {\tiny{}149.37 (0.08) } & {\tiny{}928.18 (0.1) } & {\tiny{}41.89 (0.06) } & {\tiny{}213.2 (0.2) } & {\tiny{}1132.9 (0.3) }\tabularnewline
\cline{2-9} \cline{3-9} \cline{4-9} \cline{5-9} \cline{6-9} \cline{7-9} \cline{8-9} \cline{9-9} 
 & \multirow{2}{*}{{\scriptsize{}Type 1}} & {\scriptsize{}$\mu$} & {\tiny{}3.64 (0.02) } & {\tiny{}111.92 (0.1) } & {\tiny{}263.1 (0.2) } & {\tiny{}1.523 (0.005) } & {\tiny{}12.17 (0.01) } & {\tiny{}17.43 (0.01) }\tabularnewline
 &  & {\scriptsize{}$\Sigma$} & {\tiny{}22.43 (0.03) } & {\tiny{}206.99 (0.08) } & {\tiny{}537.6 (0.1) } & {\tiny{}1.81 (0.01) } & {\tiny{}24.11 (0.03) } & {\tiny{}64.78 (0.08) }\tabularnewline
\cline{2-9} \cline{3-9} \cline{4-9} \cline{5-9} \cline{6-9} \cline{7-9} \cline{8-9} \cline{9-9} 
 & \multirow{2}{*}{{\scriptsize{}Type 2}} & {\scriptsize{}$\mu$} & {\tiny{}80.58 (0.06) } & {\tiny{}48.86 (0.08) } & {\tiny{}125.9 (0.1) } & {\tiny{}14.3 (0.03) } & {\tiny{}29.9 (0.06) } & {\tiny{}294 (0.2) }\tabularnewline
 &  & {\scriptsize{}$\Sigma$} & {\tiny{}43.45 (0.07) } & {\tiny{}388.3 (0.1) } & {\tiny{}2104.9 (0.1) } & {\tiny{}14.23 (0.03) } & {\tiny{}212.1 (0.2) } & {\tiny{}1020.6 (0.2) }\tabularnewline
\cline{2-9} \cline{3-9} \cline{4-9} \cline{5-9} \cline{6-9} \cline{7-9} \cline{8-9} \cline{9-9} 
 & \multirow{2}{*}{{\scriptsize{}Type 3}} & {\scriptsize{}$\mu$} & {\tiny{}89.12 (0.01) } & {\tiny{}193.6 (0.2) } & {\tiny{}476.7 (0.3) } & {\tiny{}15.35 (0.01) } & {\tiny{}73.96 (0.09) } & {\tiny{}281.8 (0.2) }\tabularnewline
 &  & {\scriptsize{}$\Sigma$} & {\tiny{}97.76 (0.1) } & {\tiny{}329.5 (0.1) } & {\tiny{}1323.4 (0.2) } & {\tiny{}20.31 (0.03) } & {\tiny{}196.69 (0.1) } & {\tiny{}685.8 (0.3) }\tabularnewline
\hline 
\multirow{2}{*}{{\scriptsize{}Vanilla}} & \multirow{2}{*}{} & {\scriptsize{}$\mu$} & {\tiny{}0.03 (0.01) } & {\tiny{}0.52 (0.03) } & {\tiny{}6.71 (0.08) } & {\tiny{}0.02 (0.01) } & {\tiny{}0.49 (0.02) } & {\tiny{}7.3 (0.09) }\tabularnewline
 &  & {\scriptsize{}$\Sigma$} & {\tiny{}0.16 (0.02) } & {\tiny{}13.6 (0.1) } & {\tiny{}177.9 (0.4) } & {\tiny{}0.13 (0.02) } & {\tiny{}14 (0.1) } & {\tiny{}179.9 (0.4) }\tabularnewline
\hline 
\multirow{2}{*}{{\scriptsize{}Naive}} & \multirow{2}{*}{} & {\scriptsize{}$\mu$} & {\tiny{}3.9794 (2e-04) } & {\tiny{}7.7391 (9e-04) } & {\tiny{}21.35 (0.02) } & {\tiny{}3.93 (1e-04) } & {\tiny{}8.066 (0.001) } & {\tiny{}17.6 (0.02) }\tabularnewline
 &  & {\scriptsize{}$\Sigma$} & {\tiny{}5.8749 (8e-04) } & {\tiny{}25.69 (0.02) } & {\tiny{}140.15 (0.05) } & {\tiny{}5.8092 (7e-04) } & {\tiny{}31.59 (0.03) } & {\tiny{}346.9 (0.2) }\tabularnewline
\hline 
\multirow{2}{*}{{\scriptsize{}CMC2}} & \multirow{2}{*}{} & {\scriptsize{}$\mu$} & {\tiny{}0.04 (0.01) } & {\tiny{}0.53 (0.03) } & {\tiny{}7.57 (0.09) } & {\tiny{}0.01 (0.01) } & {\tiny{}0.53 (0.03) } & {\tiny{}7.58 (0.09) }\tabularnewline
 &  & {\scriptsize{}$\Sigma$} & {\tiny{}1.01 (0.01) } & {\tiny{}55.7 (0.1) } & {\tiny{}255.90125 (3e-05) } & {\tiny{}0.35 (0.02) } & {\tiny{}20.7 (0.2) } & {\tiny{}212.5 (0.5) }\tabularnewline
\hline 
\multirow{2}{*}{{\scriptsize{}SDPE}} & \multirow{2}{*}{} & {\scriptsize{}$\mu$} & {\tiny{}3 (0.1) } & {\tiny{}3.96 (0.1) } & {\tiny{}19 (0.2) } & {\tiny{}0.07 (0.01) } & {\tiny{}0.73 (0.03) } & {\tiny{}8.83 (0.1) }\tabularnewline
 &  & {\scriptsize{}$\Sigma$} & {\tiny{}67709 (129) } & {\tiny{}222621 (132) } & {\tiny{}3839.1 (0.8) } & {\tiny{}256 (2) } & {\tiny{}249 (1) } & {\tiny{}2306 (5) }\tabularnewline
\hline 
\end{tabular}
\par\end{centering}
\caption{The KL divergence from the $\mu$ and $\Sigma$ marginals of the posterior
to each approximation in the example of Section \ref{subsec:Multivariate-normal-models}
with $\Sigma$ unknown and $M=64$. Standard errors for each estimate
are shown in parentheses.\label{tab:normal-kl-divergence-m64}}

\end{table}
\endgroup

Table \ref{tab:normal-kl-divergence-m64} presents approximate KL
divergences from the $\mu$ and $\Sigma$ marginals of the posterior
to the approximations, with $M=64$. This includes the Monte Carlo
estimator using $N^{*}=1,000$ samples drawn directly from the posterior,
which we label ``Vanilla'', and the best performing of each of the
CMC and DPE estimators. In approximating the $\mu$ marginal of the
posterior, the normality assumptions of CMC are met so it is unsurprising
that CMC2 does almost as well as Vanilla. We found that it is actually
possible to do better than Vanilla with LEMIE in the most difficult
examples (larger dimension $d$). This is likely because LEMIE uses
all $\bar{N}M$ samples, which in this case is 64,000, in estimators.
We also find that LEMIE can do better than CMC or DPE in approximating
the $\Sigma$ marginal of the posterior than the other methods, which
we suggest is because the CMC and DPE methods struggle with deviations
from normality such as in the inverse-Wishart distribution.

\begin{figure}
\begin{centering}
\includegraphics[scale=0.55]{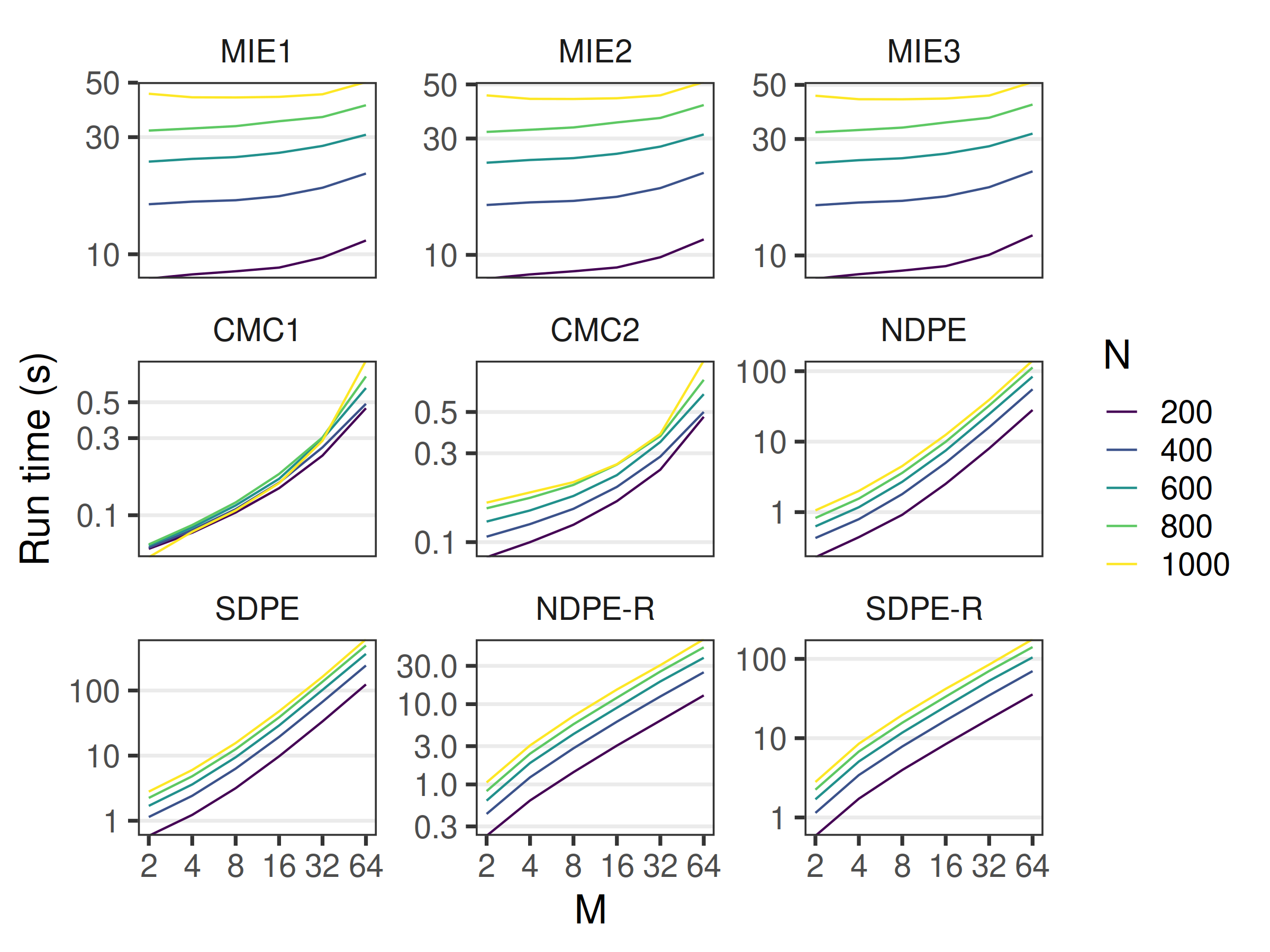}
\par\end{centering}
\caption{Run times for each algorithm from Section \ref{sec:methods} (excluding
Laplace samples) in the MVN example with $\mu$ and $\Sigma$ unknown,
$d=8$ and $n=10,000$. $N$ is the number samples drawn from each
local posterior ($\bar{N}$ in the text). NDPE-R and SPDE-R are the
recursive versions of the algorithms explained in Section \ref{subsec:Density-product-estimator}
using pairwise pooling of the local posteriors. \label{fig:normal-sigma-unknown-runtimes}}
\end{figure}

Figure \ref{fig:normal-sigma-unknown-runtimes} presents the time
taken, in seconds, to run each algorithm across simulations with $d=8$
and $n=10,000$ for a range of $M$ and for different sample sizes
$\bar{N}$. These run times are exclusive of the time taken to draw
samples from the local posteriors, and multiple CPU core parallelisation
was used as much as possible (using the \emph{parallel} R package
from \citet{R} and running on a 64 core AWS EC2 instance). The MIE
algorithms' timings are for the methods of Section \ref{subsec:Multiple-importance-estimation}
without any Laplace samples. The run times of all methods appear to
increase exponentially with $M$. The MIE algorithms have a relatively
gentle exponential increase because the majority of the computation
is in likelihood evaluations which can be performed in parallel when
there are $M$ cores available. The MIE algorithms seem to have more
overhead, unrelated to $M$, but for large $M$ they are faster than
the DPE algorithms except for the recursive version of NDPE. CMC is
very fast over this range of $M$; its hardest computations are calculating
the sample covariance matrices, which can be done in parallel.

\begin{figure}
\begin{centering}
\subfloat[KL divergence vs ESS.]{\includegraphics[scale=0.45]{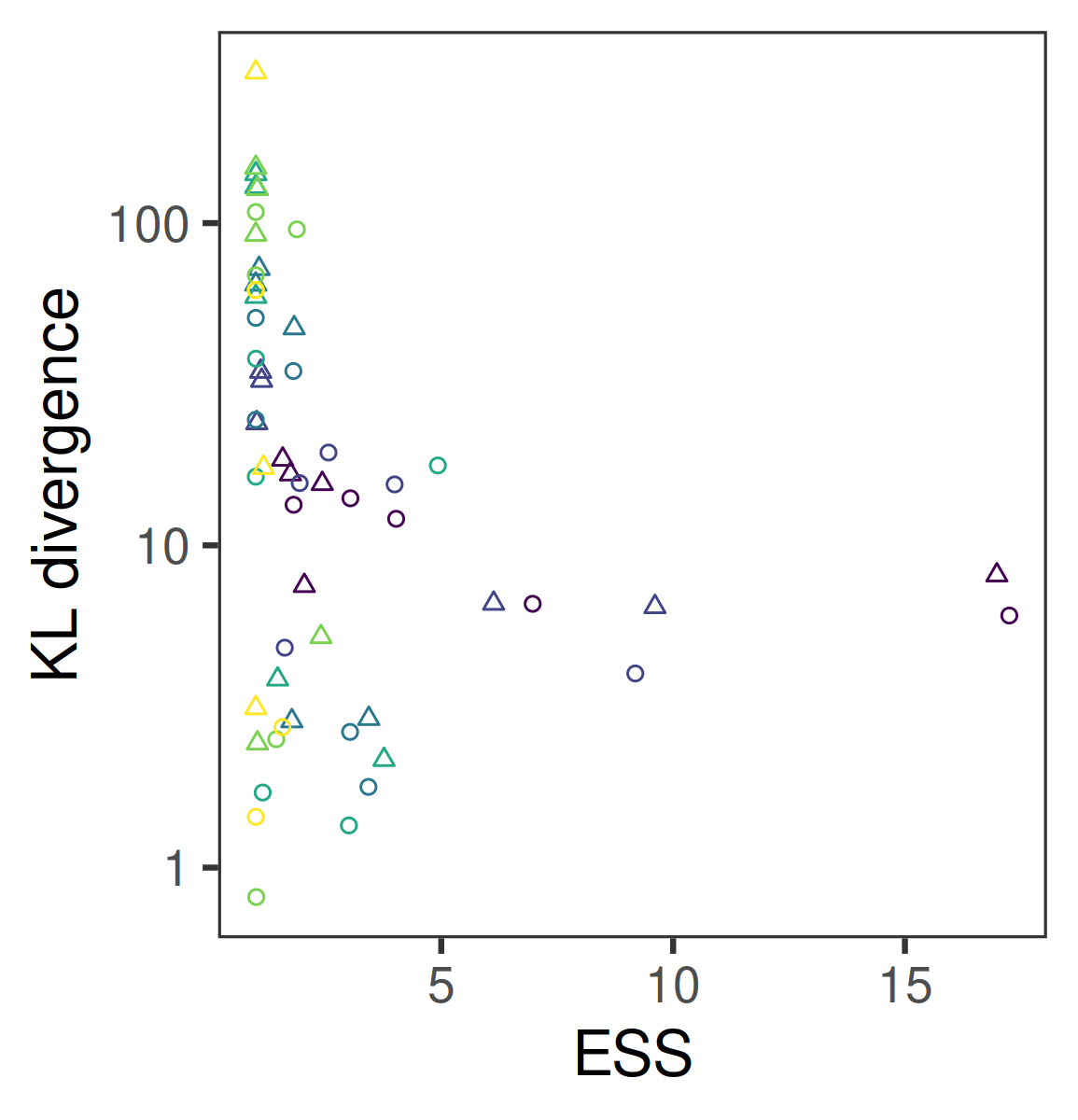}

}\quad{}\subfloat[KL divergence vs $\hat{{k}}$.]{\includegraphics[scale=0.45]{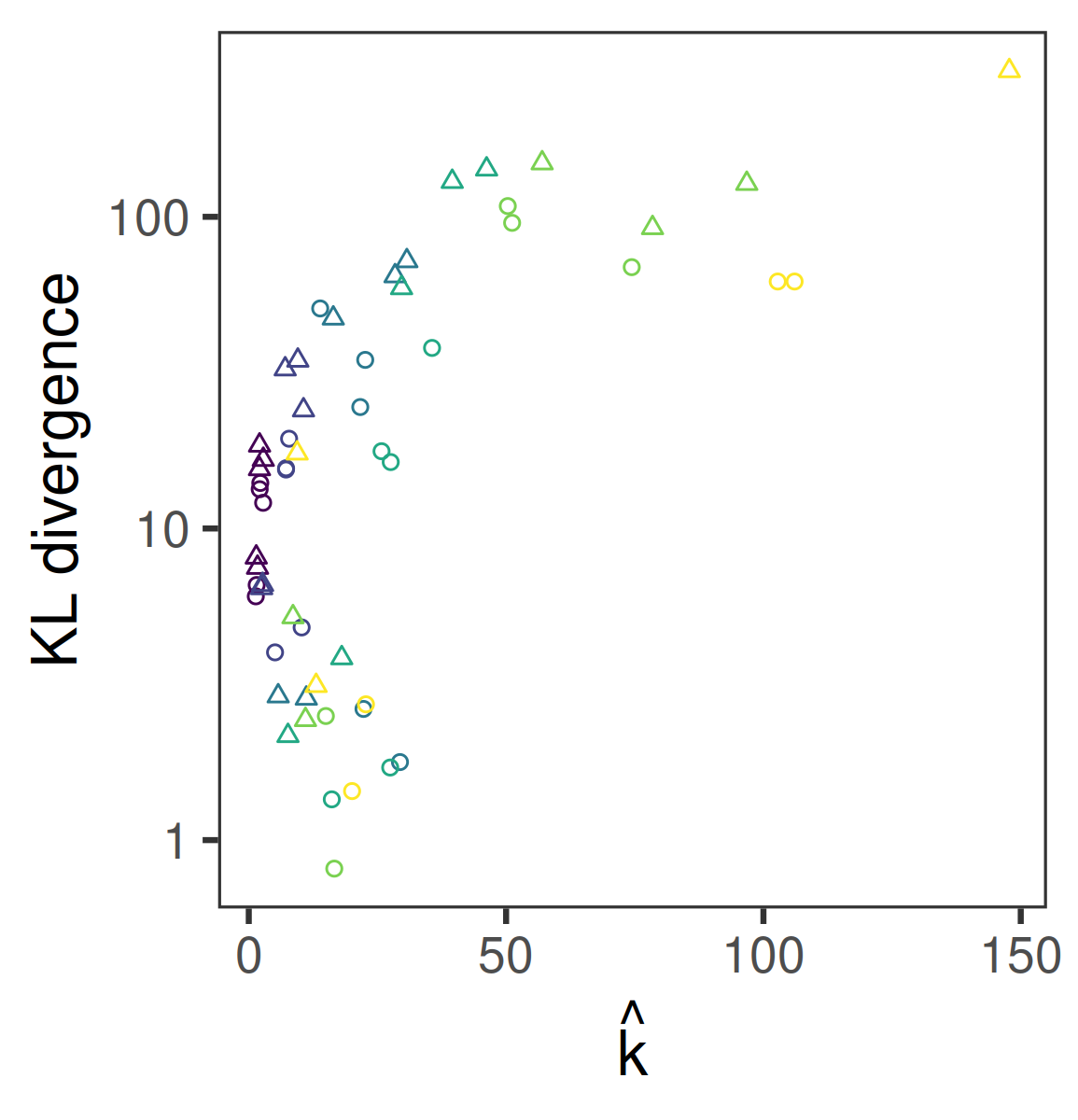}}\quad{}\includegraphics[scale=0.45]{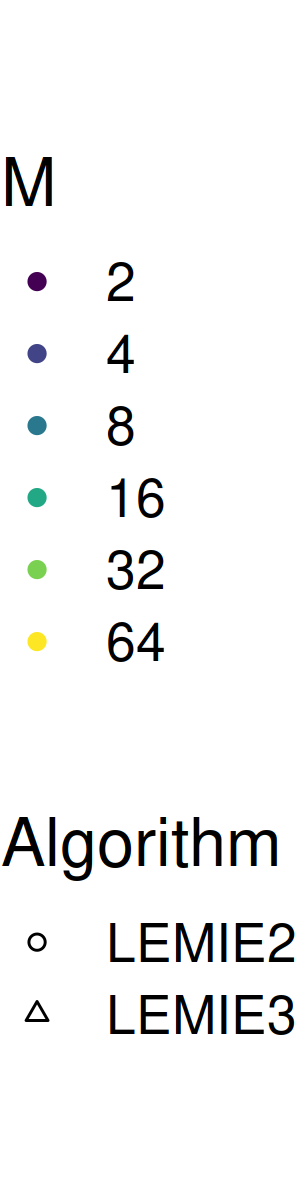}
\par\end{centering}
\caption{For the simulated examples of Section \ref{subsec:Multivariate-normal-models}
with $d=8$ and $n=10,000$, KL divergences Equation \ref{eq:kl-divergence}
from the LEMIE approximations (of all types defined in Section \ref{subsec:Laplace-enrichment-1})
to the posterior of $\mu$ using the approach explained in Section
\ref{par:Cross-entropy-estimation} against the performance metrics
of Section \ref{subsec:Performance-indicators}.\label{fig:normal-sigma-unknown-mu-kl-div-all-no-mie1}}
\end{figure}

\begin{figure}
\begin{centering}
\subfloat[KL divergence vs ESS.]{\includegraphics[scale=0.45]{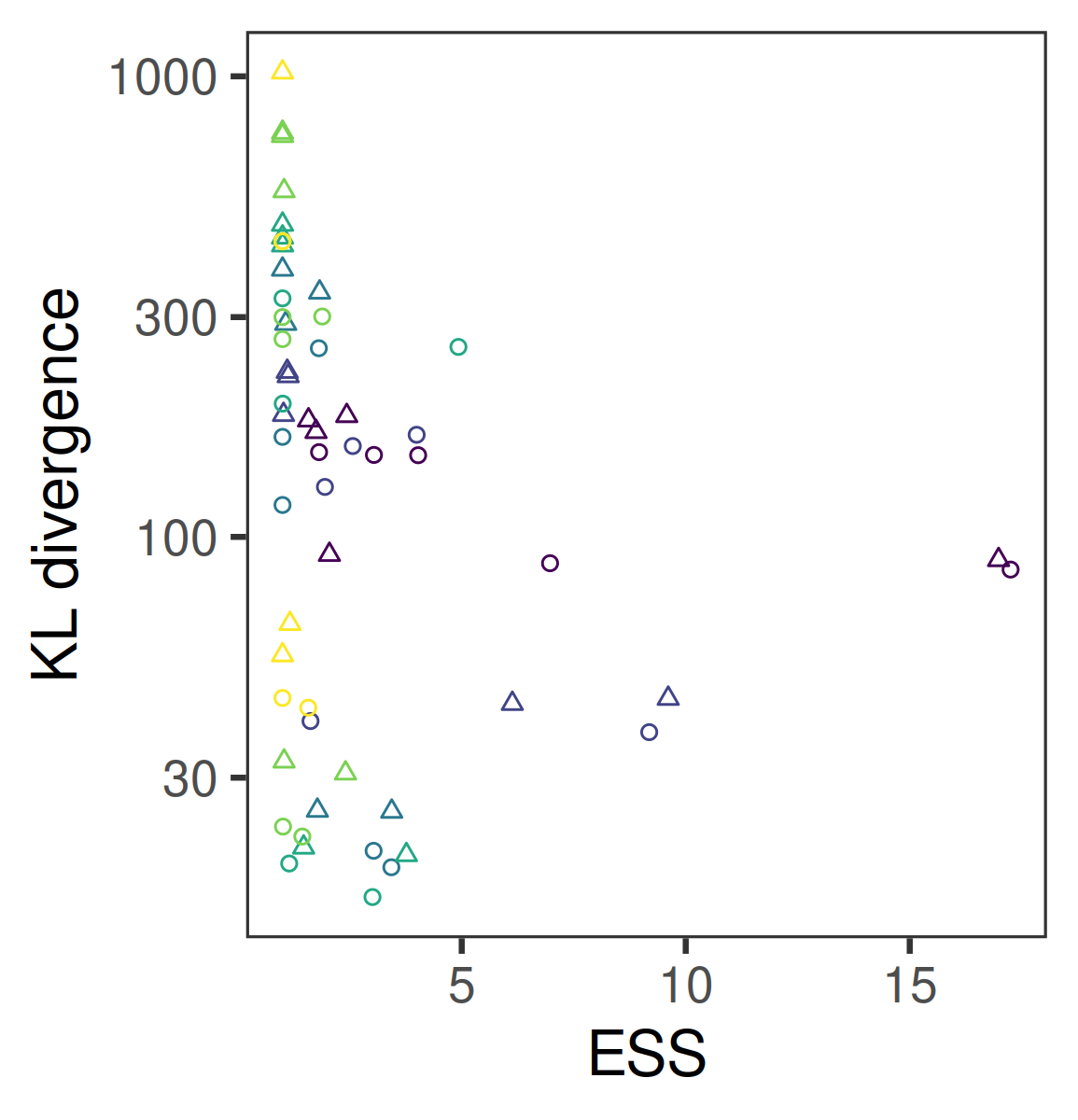}

}\quad{}\subfloat[KL divergence vs $\hat{{k}}$.]{\includegraphics[scale=0.45]{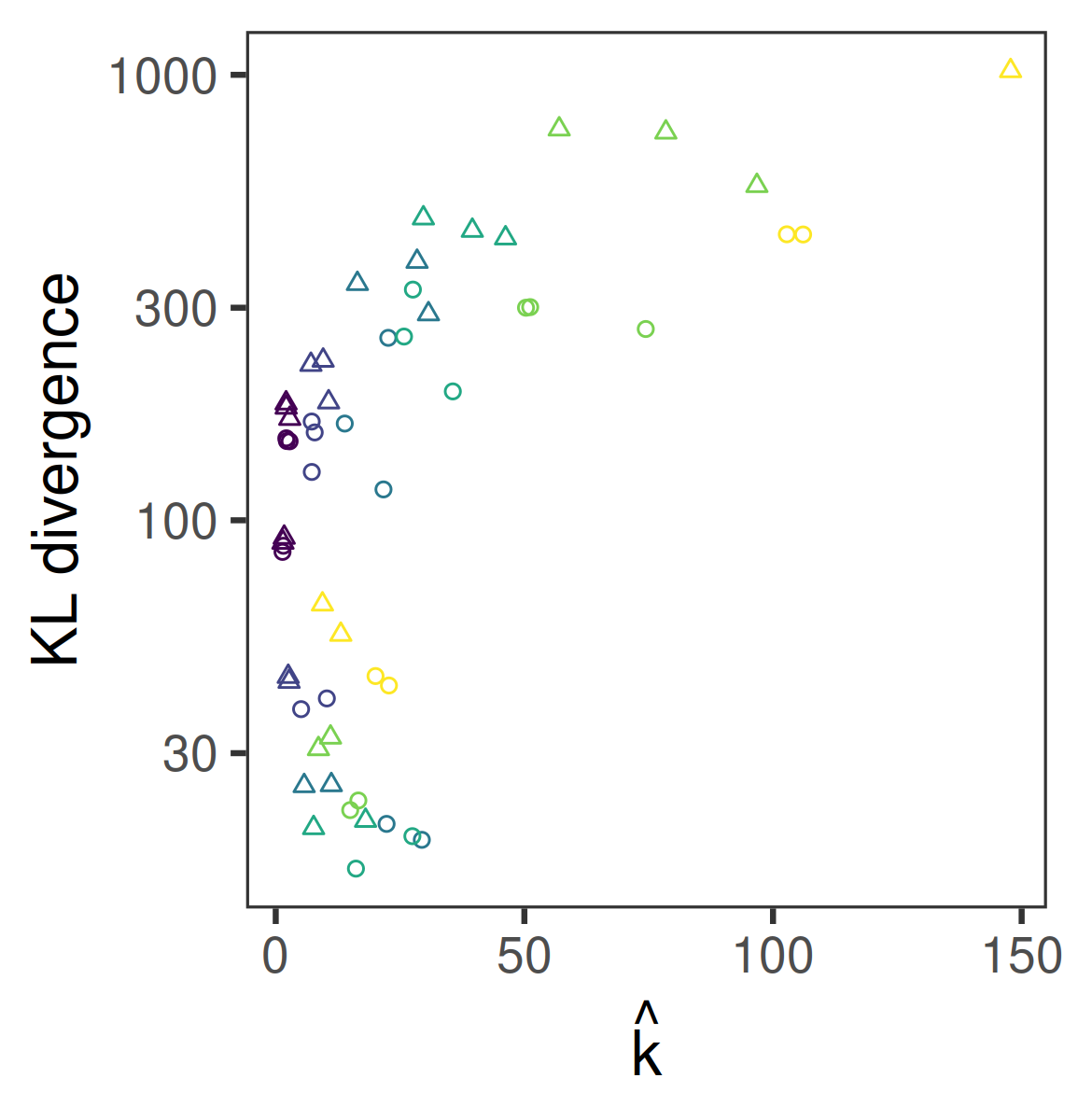}}\quad{}\includegraphics[scale=0.45]{images/experiments-normal-uncorrelated-legend-d8-n10000-log-error-mean-vs-log-khat-no-mie1}
\par\end{centering}
\caption{Similar to Figure \ref{fig:normal-sigma-unknown-mu-kl-div-all-no-mie1}
but of the KL divergences from the LEMIE approximations to the $\Sigma$
marginal of the posterior.\label{fig:normal-sigma-unknown-sigma-kl-div-all-no-mie1}}
\end{figure}

Figures \ref{fig:normal-sigma-unknown-mu-kl-div-all-no-mie1} and
\ref{fig:normal-sigma-unknown-sigma-kl-div-all-no-mie1} plot approximate
KL divergences on a log scale against the estimator diagnostics ESS
and $\hat{{k}}$ for LEMIE 1 and 2 using Laplace types 1, 2, 3, all
3 and none in the $d=8$ and $n=10,000$ examples. By eye, it looks
like these diagnostics may be predictive of performance, at least
relatively, comparing one estimator against another. We did not include
LEMIE1 in Figures \ref{fig:normal-sigma-unknown-mu-kl-div-all-no-mie1}
and \ref{fig:normal-sigma-unknown-sigma-kl-div-all-no-mie1} for clarity,
because ESS does not appear to be a useful predictor of performance
for LEMIE1 (although $\hat{{k}}$ does). Figures including LEMIE1
can be found in Appendix \ref{appendix:additional-Multivariate-normal-studies}.

\begin{figure}
\centering{}\subfloat[Approximating the $\mu$ marginal of the posterior.]{\includegraphics[scale=0.55]{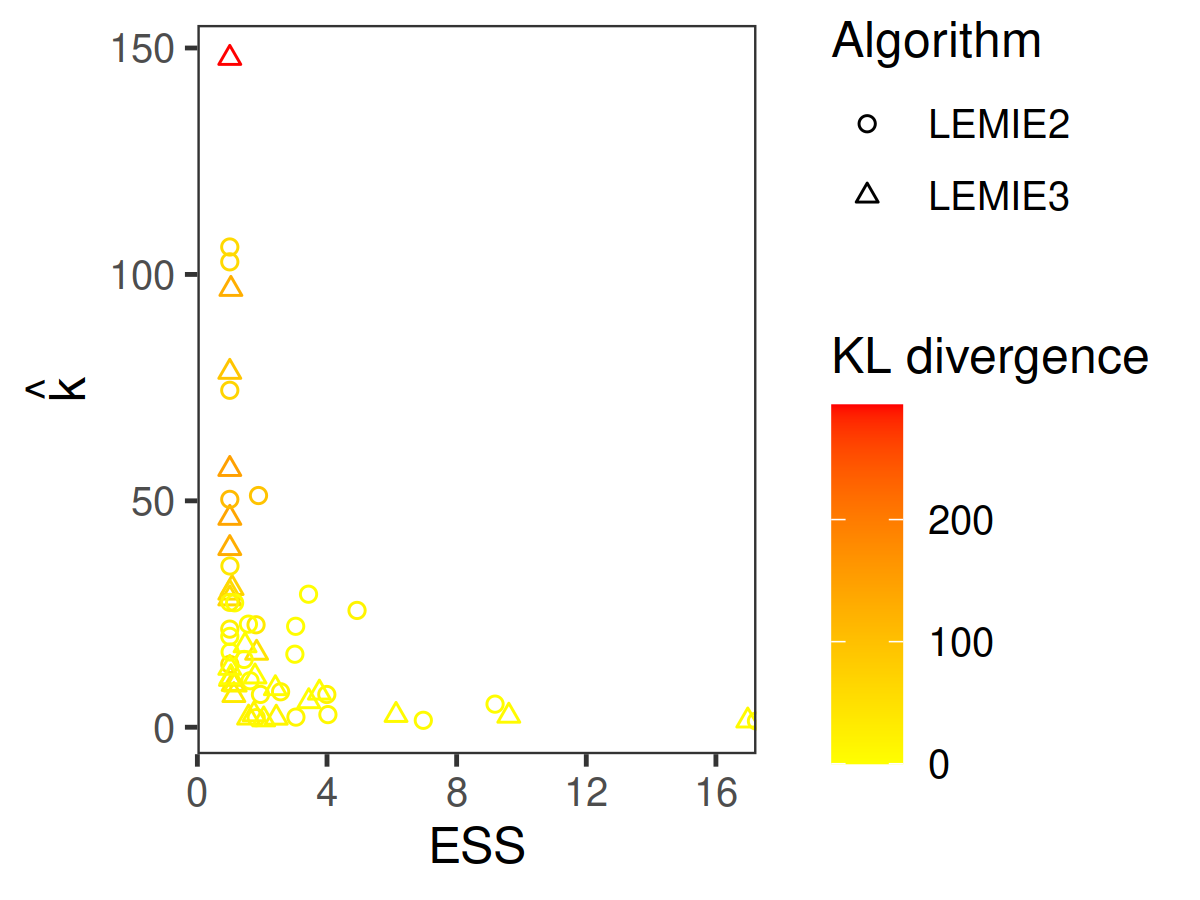}

}\quad{}\subfloat[Approximating the $\Sigma$ marginal of the posterior.]{\includegraphics[scale=0.55]{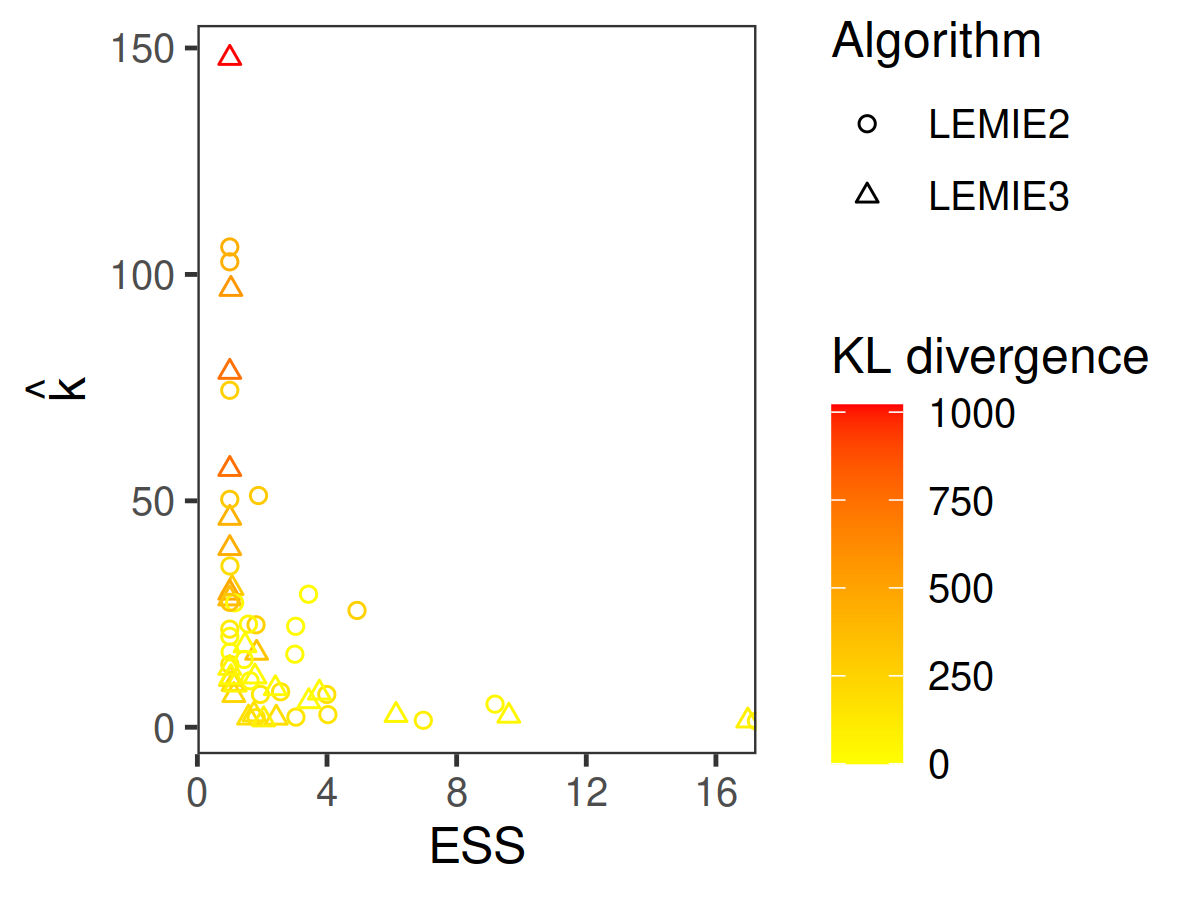}}\caption{For the simulated examples of Section \ref{subsec:Multivariate-normal-models}
with $d=8$ and $n=10,000$, KL divergences from the LEMIE approximations
to the $\mu$ and $\Sigma$ marginals of the posterior using the approach
explained in Section \ref{par:Cross-entropy-estimation} against the
performance metrics of Section \ref{subsec:Performance-indicators}.\label{fig:normal-sigma-unknown-kl-div-vs-ess-khat}}
\end{figure}

As an attempt to quantify the value of ESS and $\hat{{k}}$ as predictors
of performance, we looked to fit a gamma GLM to these  data. Fitted
GLM results using all LEMIE 1, 2 and 3 results, and for LEMIE 2 and
3 only, can be found in Appendix \ref{appendix:normal-studies-Gamma-GLMs}.
The residual deviances suggest there is predictive value in the performance
metrics. For the $\mu$ marginal of the posterior, the log mean of
KL divergence increases 0.032 per unit increase in $\hat{{k}}$ (standard
error 0.003) and decreases -0.057 per unit increase in ESS (s.e. 0.038,
model excludes LEMIE1). For the $\Sigma$ marginal of the posterior,
the log mean of KL divergence increases 0.021 with a unit increase
in $\hat{{k}}$ (s.e. 0.003) and decreases -0.030 with a unit increase
in ESS (s.e. 0.032, model excludes LEMIE1). The existence of a positive
relationship for $\hat{{k}}$ is clear whilst the negative relationship
of ESS, which is clearer without LEMIE1 than with it, is not statistically
significant under this model. The interaction of ESS and $\hat{{k}}$
does not appear to be usefully related to KL divergence (see estimates
in Appendix \ref{appendix:normal-studies-Gamma-GLMs}). These results
provide some validation for the judgement by eye that ESS and $\hat{{k}}$
are useful, but the gamma GLM model may not be the best way to do
this. In particular there is heteroscedasticity visible in Figures
\ref{fig:normal-sigma-unknown-mu-kl-div-all-no-mie1} and \ref{fig:normal-sigma-unknown-sigma-kl-div-all-no-mie1},
although this might be explained between ESS and $\hat{{k}}$ when
both are used as predictors of KL divergence. This joint relationship
is depicted in Figure \ref{fig:normal-sigma-unknown-kl-div-vs-ess-khat}. 

\subsection{Logistic regression\label{subsec:Logistic-regression}}

We look at two logistic regression simulations, replicating those
from \citet{scott2016bayes} and \citet{neiswanger2013asymptotically}.
The model is

\begin{equation}
y_{i}\sim\mathrm{{Binom}}\left(c_{i},\mathrm{{logit}}^{-1}\left(x_{i}^{\intercal}\theta\right)\right),i=1,2,\ldots,n,\label{eq:binomial-logit}
\end{equation}
where $\mathrm{{logit}}^{-1}$ is the inverse logit function, $x_{i}\in\mathbb{{R}}^{p},c_{i}\in\mathbb{{N}},y_{i}\in\mathbb{{N}}$
with $y_{i}\le c_{i}$ and $\theta\in\mathbb{{R}}^{p}$, which is
the posterior estimand of interest given data $\left\{ x_{i},c_{i},y_{i};i=1,2,\ldots,n\right\} $.

For sampling from the local posteriors in these examples, we employ
the Gibbs sampler of \citet{polson2013bayesian}. A brief explanation
of this is provided in Appendix \ref{sec:appendix-Gibbs-sampler-Polson}.
We will compare the methods' estimates of the posterior mean, 2.5\%
and 97.5\% quantiles of the marginals of the posterior  using error
function Equation \ref{eq:2norm-error}. Since the posterior is not
available in analytic form, we must estimate the true values. We do
this using the same MCMC algorithm with the unpartitioned data, running
the sampler for longer than the local posterior samplers for greater
precision in the ground truth: 2,000,000 samples with the first 50\%
discarded as burn-in.

\subsubsection{Simulation of \citet{scott2016bayes}\label{subsec:scott-logistic}}

We use the same data as \citet{scott2016bayes}, which is reproduced
in their Table 1 (a) and consists of 10,000 binary outcomes with $p=5$
binary predictor variables. The $c_{i}$ outcomes with the same combination
of predictor variables $x_{i}$ can be grouped as integer $y_{i}$
to use the model form of Equation \ref{eq:binomial-logit}.

It is not clear what prior distribution or MCMC algorithm is used
by \citet{scott2016bayes}, so we took our own initiative and used
a $\mathrm{{N}}_{p}\left(0_{p},2.5^{2}I_{p}\right)$ prior and sample
$\bar{N}=400,000$ times from each local posterior, discarding the
first 50\% as burn-in. The fractionated prior used for CMC and DPE
is $\mathrm{{N}}_{p}\left(0_{p},2.5^{2}MI_{p}\right)$ (see Equation
\ref{eq:fractionated-mvn}). Whilst \citet{scott2016bayes} use $M=100$,
we looked at performance over a range of $M$ to study the effect
of this on performance, as in the simulations in Section \ref{subsec:Multivariate-normal-models}.
For each $M$ we partitioned the data uniformly at random into $M$
parts and ran the Gibbs sampler using each data part independently.

\begin{figure}
\begin{centering}
\subfloat[Error in estimating the posterior mean of $\theta$.]{\includegraphics[scale=0.48]{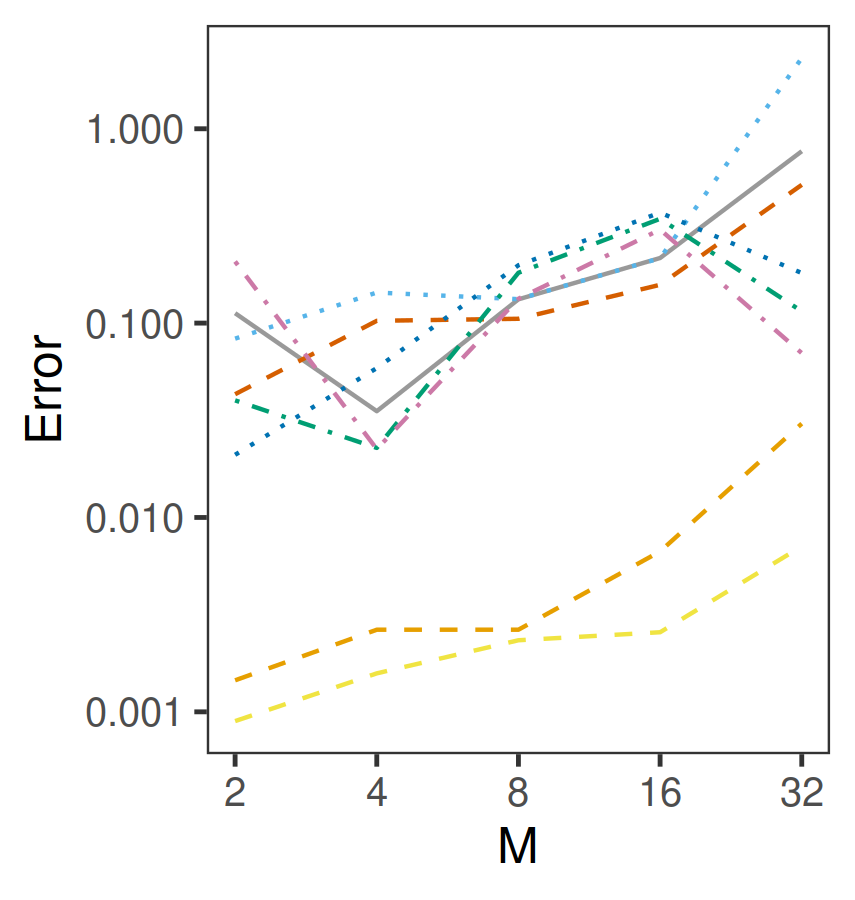}}\hfill{}\subfloat[Error in estimating the 2.5\% quantiles of the marginals of the posterior.]{\includegraphics[scale=0.48]{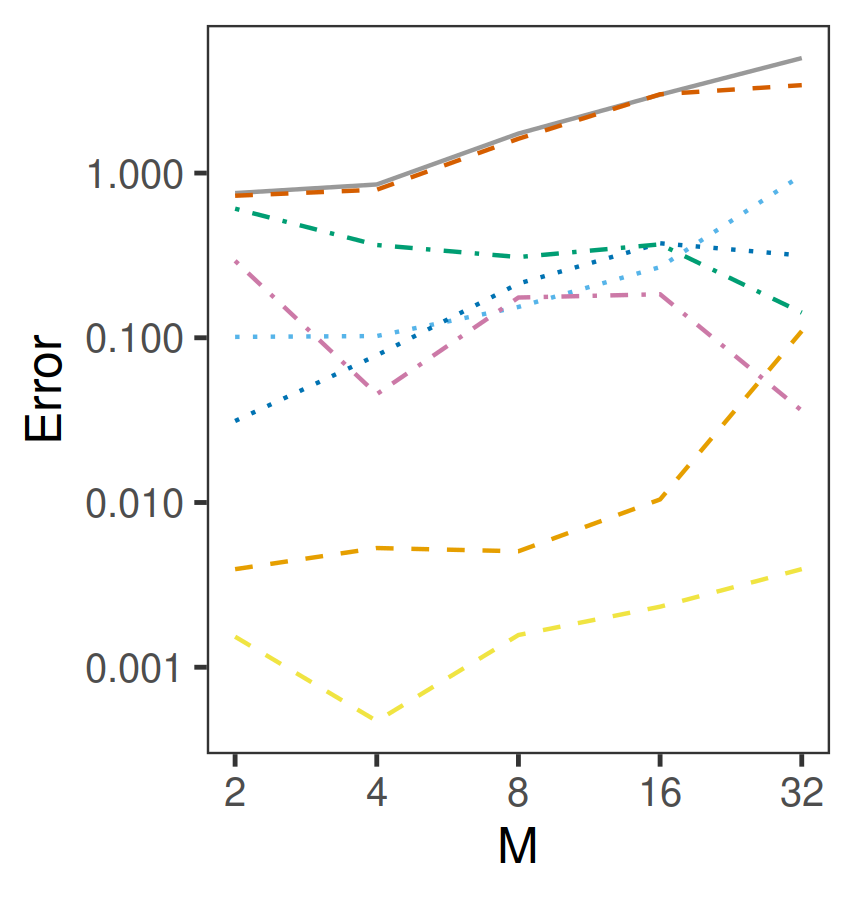}}\hfill{}\subfloat[Error in estimating the 97.5\% quantiles of the marginals of the posterior.]{\includegraphics[scale=0.48]{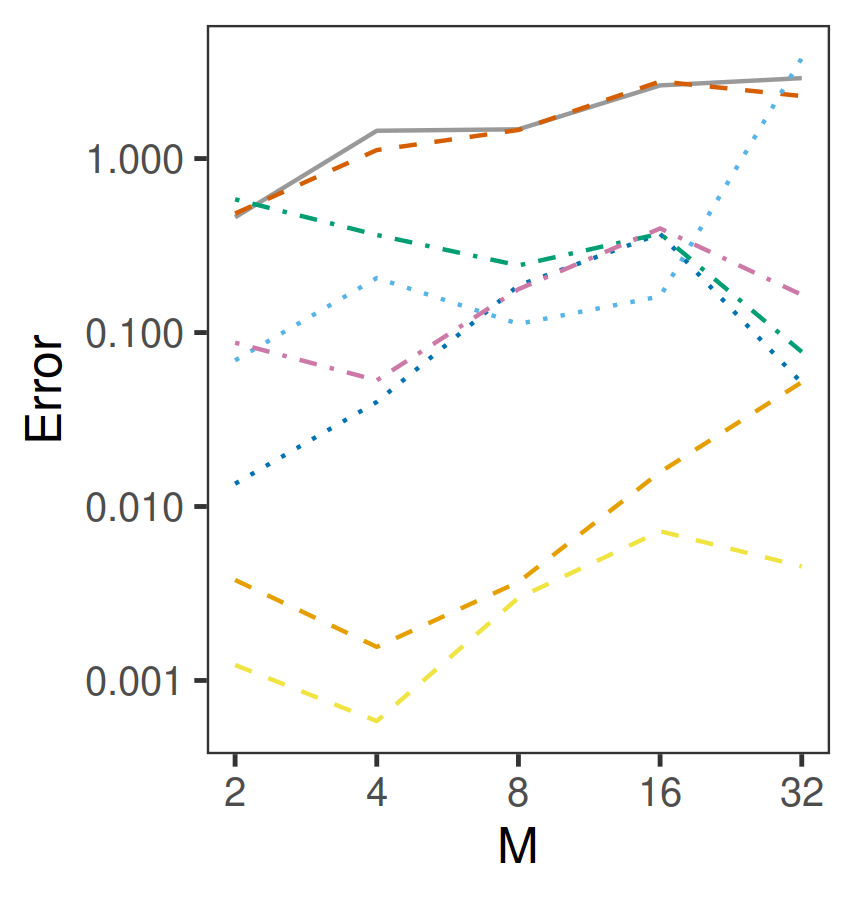}}\hfill{}\includegraphics[scale=0.45]{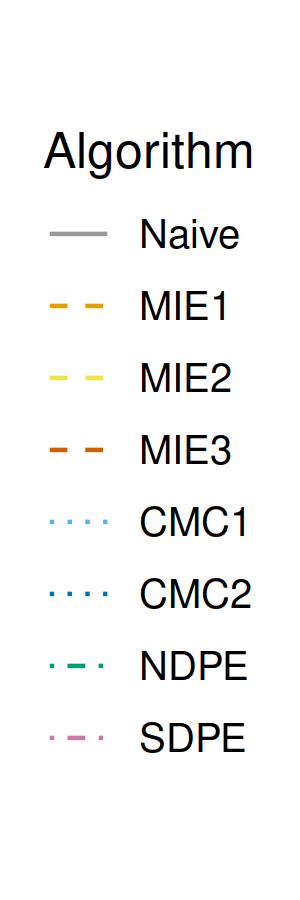}
\par\end{centering}
\caption{Posterior approximation comparisons for $\theta$ in the logistic
regression example of Section \ref{subsec:scott-logistic} due to
\citet{scott2016bayes}.\label{fig:scott-logistic-results}}
\end{figure}

In Figure \ref{fig:scott-logistic-results} are plotted the errors
in estimating the posterior mean of $\theta$ and in estimating the
2.5\% and 97.5\% quantiles of the marginals of the posterior. Using
Laplace samples of any type was not found to provide any benefit to
the estimators of Section \ref{subsec:Multiple-importance-estimation},
so results are plotted for methods without Laplace enrichment. MIE
1 and 2 were found to be more accurate at estimating the posterior
mean and quantiles than any other method across all $M$.

\begin{figure}
\begin{raggedright}
\subfloat[]{\includegraphics[scale=0.45]{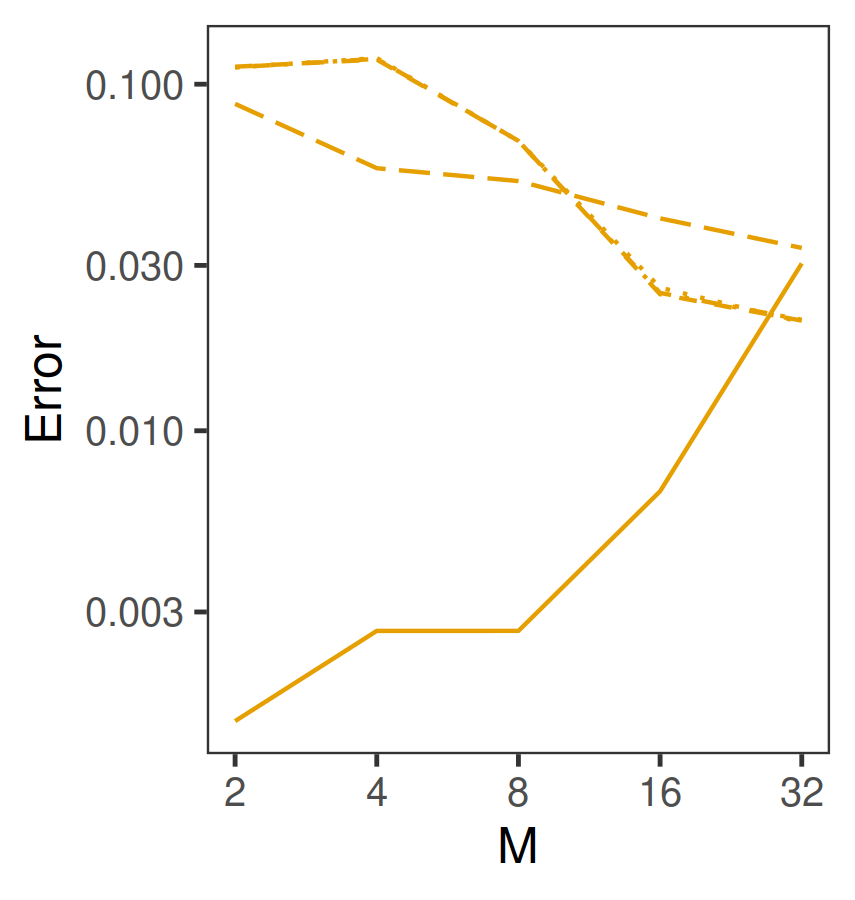}}\enskip{}\subfloat[]{\includegraphics[scale=0.45]{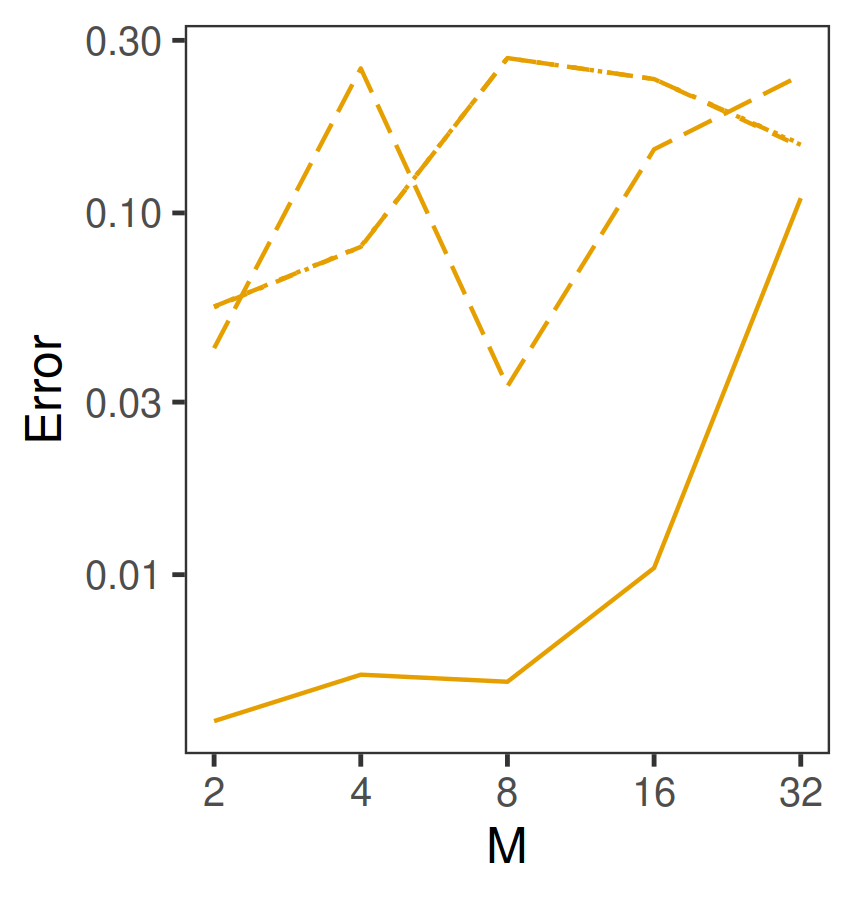}}\enskip{}\subfloat[]{\includegraphics[scale=0.45]{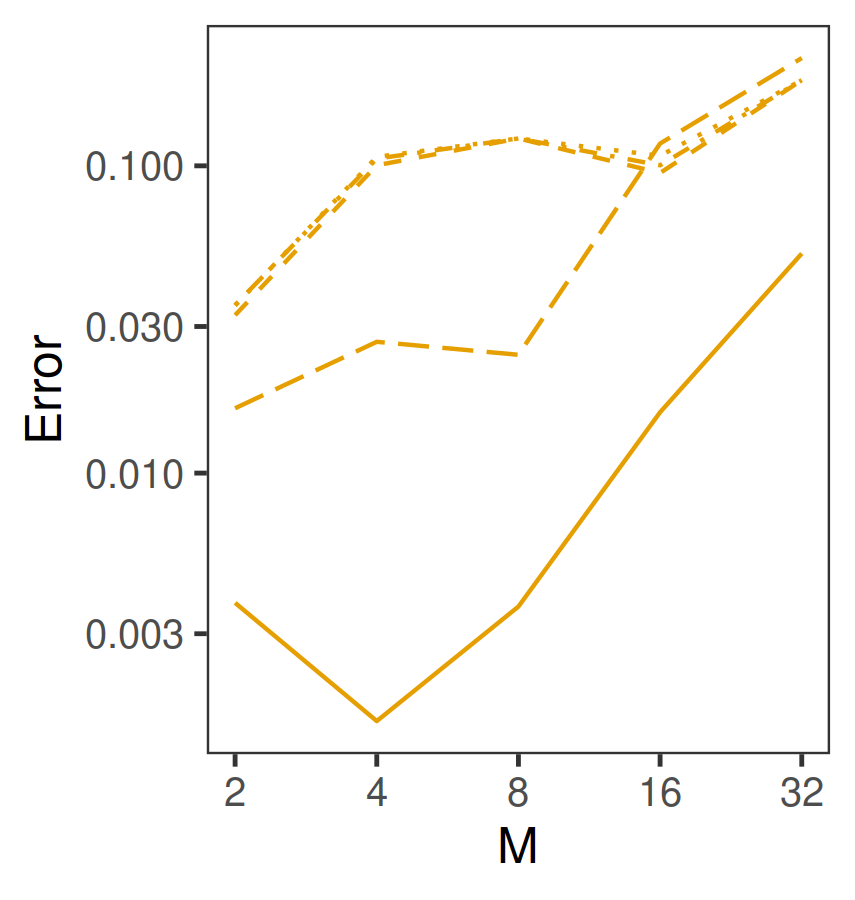}}\enskip{}\includegraphics[scale=0.45]{images/experiments-logistic-scott-legend-m-alt-lemie-presentation}
\par\end{raggedright}
\begin{raggedright}
\subfloat[]{\includegraphics[scale=0.45]{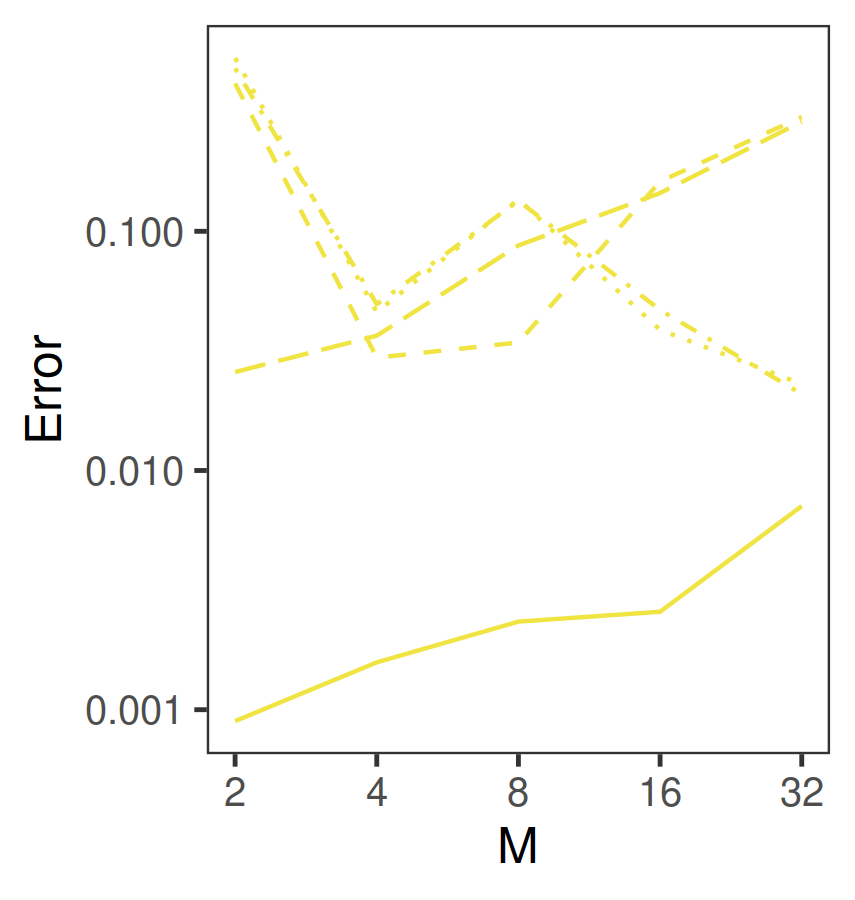}}\enskip{}\subfloat[]{\includegraphics[scale=0.45]{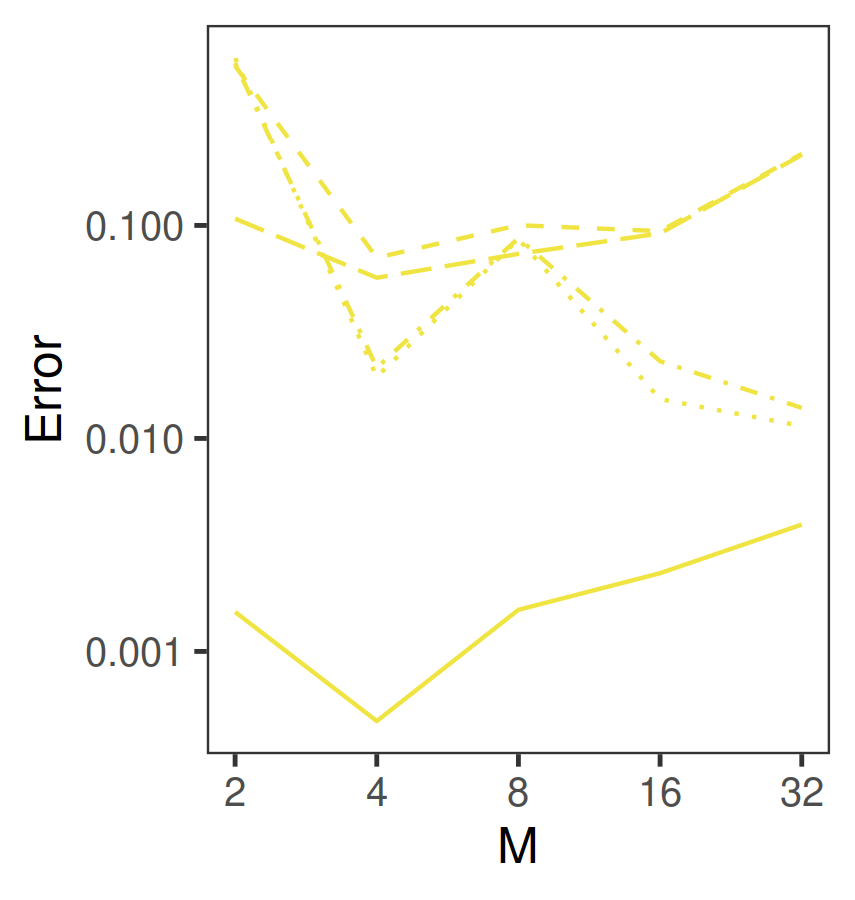}}\enskip{}\subfloat[]{\includegraphics[scale=0.45]{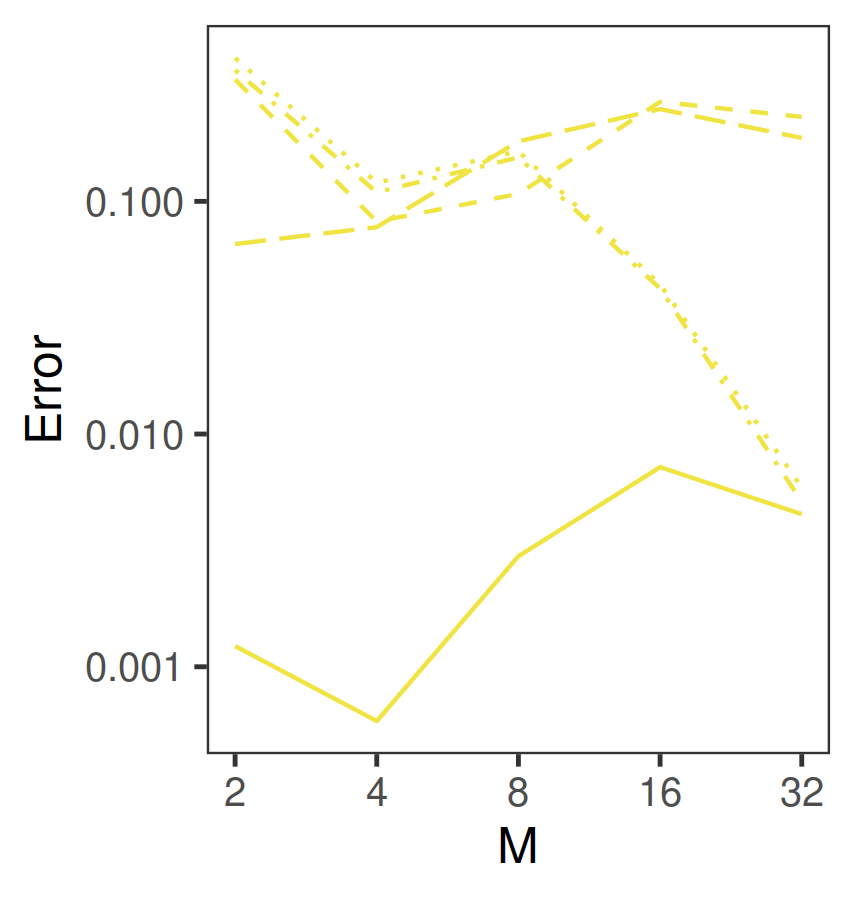}}\enskip{}\includegraphics[scale=0.45]{images/experiments-logistic-scott-class-legend-m-alt-presentation}
\par\end{raggedright}
\begin{raggedright}
\subfloat[]{\includegraphics[scale=0.45]{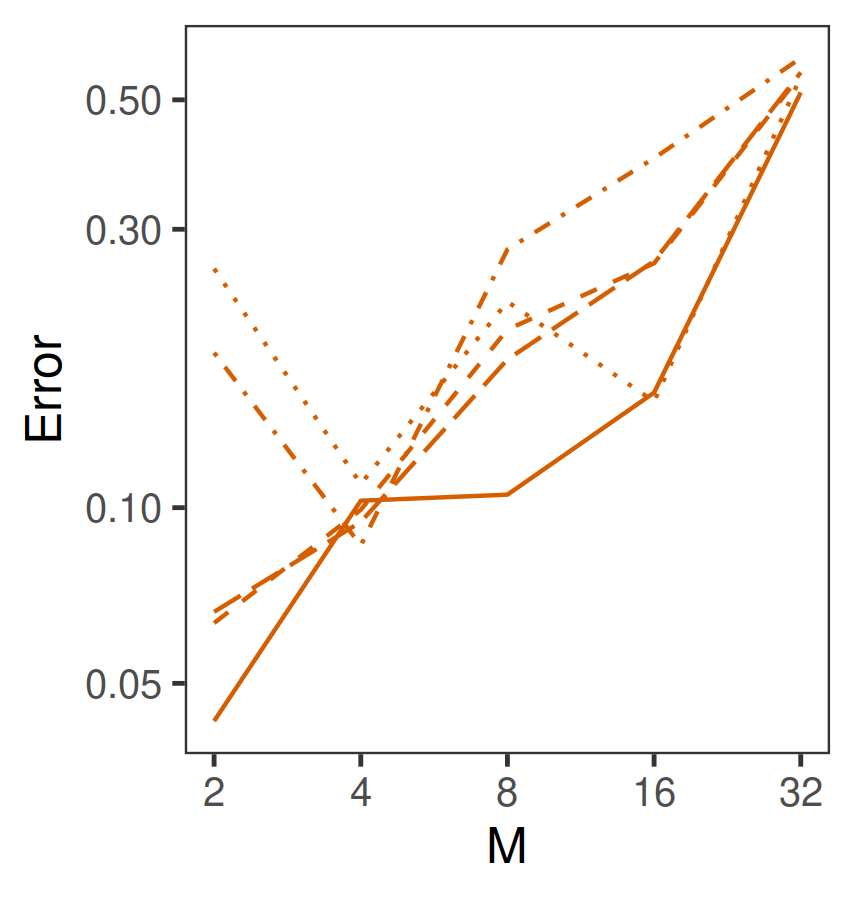}}\enskip{}\subfloat[]{\includegraphics[scale=0.45]{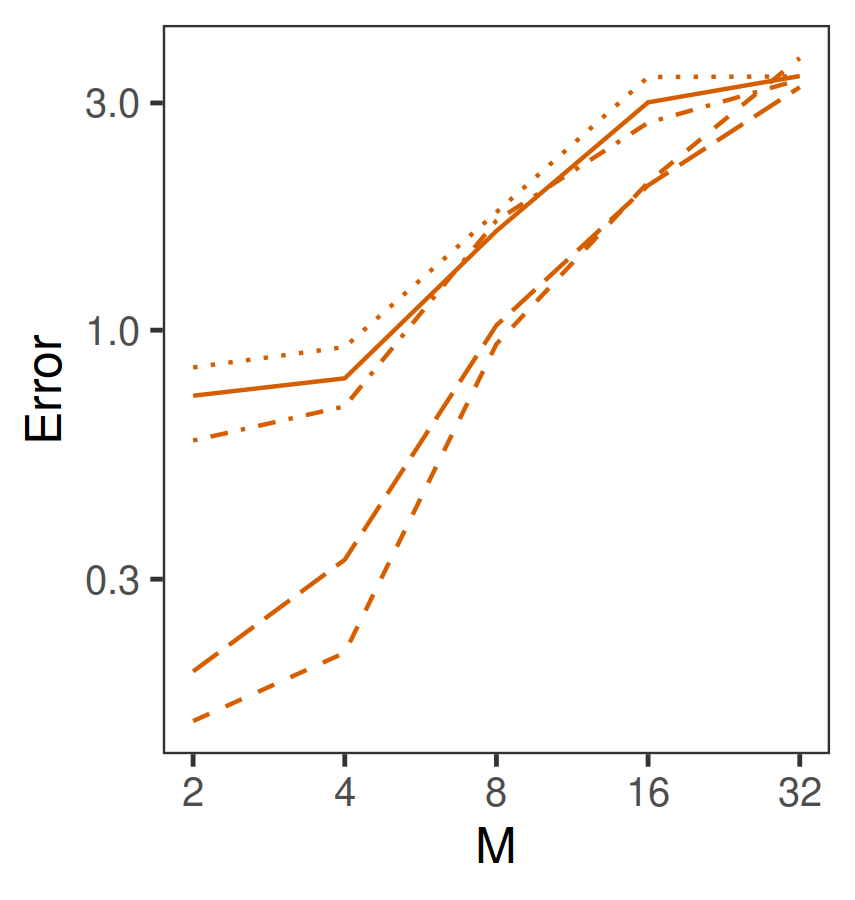}}\enskip{}\subfloat[]{\includegraphics[scale=0.45]{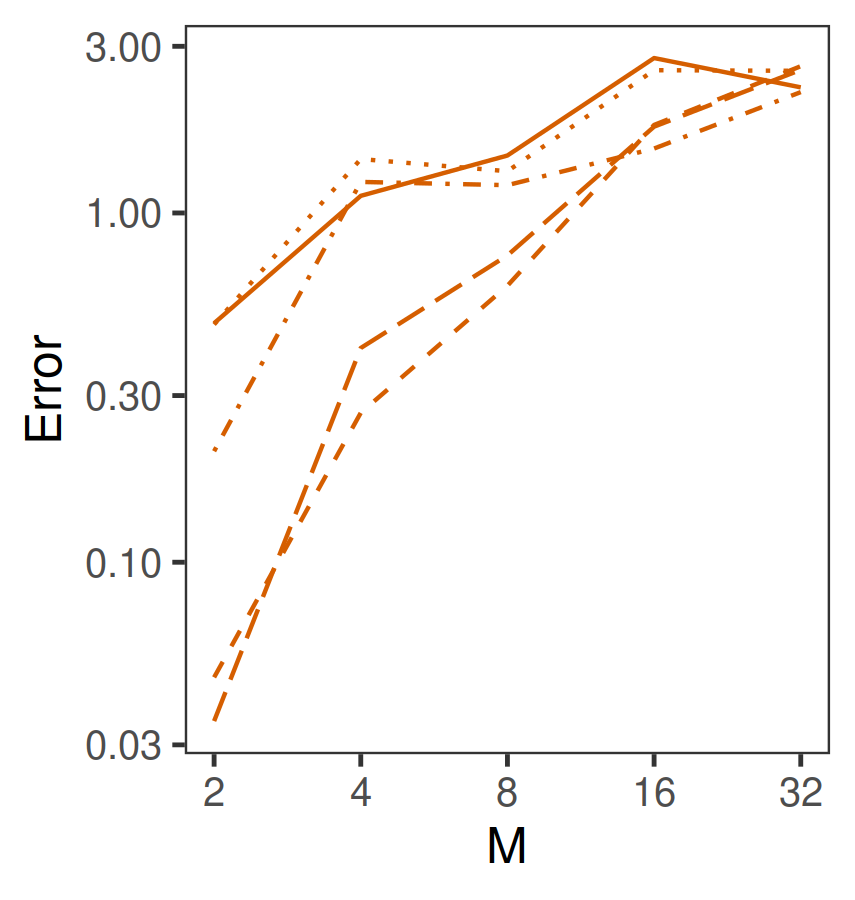}}
\par\end{raggedright}
\caption{Posterior approximation comparisons for the algorithms from Section
\ref{subsec:Multiple-importance-estimation} in the logistic regression
example of Section \ref{subsec:scott-logistic} due to \citet{scott2016bayes}.
The error in estimating (a)(d)(g) the posterior mean, (b)(e)(h) the
2.5\% quantiles of the marginals of the posterior, (c)(f)(i) the 97.5\%
quantiles of the marginals of the posterior.\label{fig:scott-logistic-results-mie-lemie}}
\end{figure}

Figure \ref{fig:scott-logistic-results-mie-lemie} compares the methods
of Section \ref{subsec:Multiple-importance-estimation} in estimating
the posterior mean and tail quantiles with the addition of Laplace
samples. For LEMIE 1 and 2, the trend suggests using Laplace samples
may become more beneficial at large $M$, but up to $M=32$ the estimator
with no Laplace samples is better. LEMIE3 does not perform any better
than the other methods in this example, but does appear to benefit
from using Laplace samples in estimating the tail quantiles, particularly
Laplace samples of type 1. We found that adding Laplace samples numbering
from $1\times10^{6}$ to $2.5\times10^{6}$ samples does not seem
to improve performance (Figure \ref{fig:scott-logistic-laplace-extensions}
in Appendix \ref{subsec:additional-results-scott-logistic}).

\begin{figure}
\begin{centering}
\subfloat[Example of Section \ref{subsec:scott-logistic}.\label{fig:scott-logistic-lemie-diagnostics}]{\includegraphics[scale=0.55]{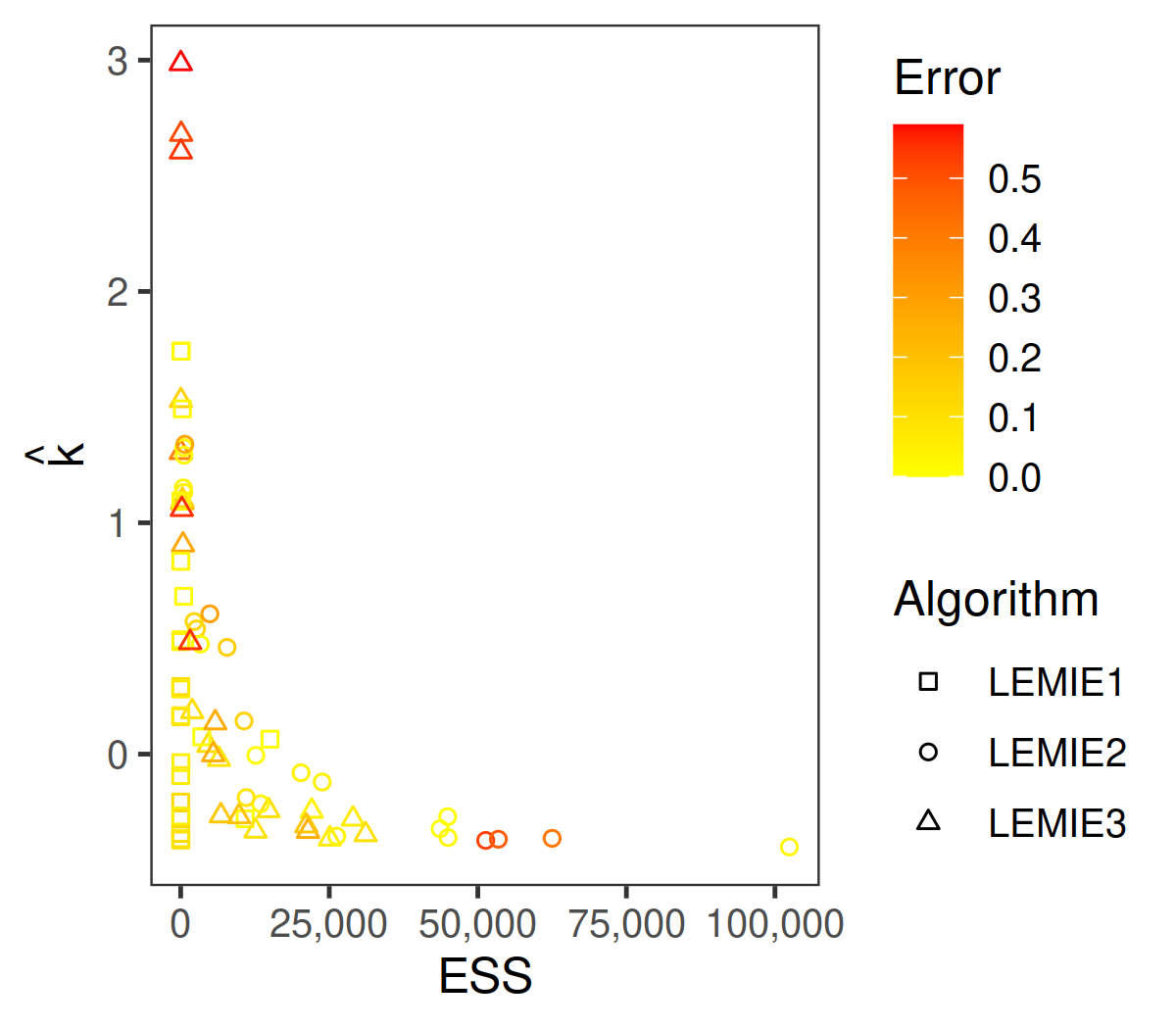}

}\quad{}\subfloat[Example of Section \ref{subsec:logistic-simulations}.\label{fig:simulated-logistic-lemie-diagnostics}]{\includegraphics[scale=0.55]{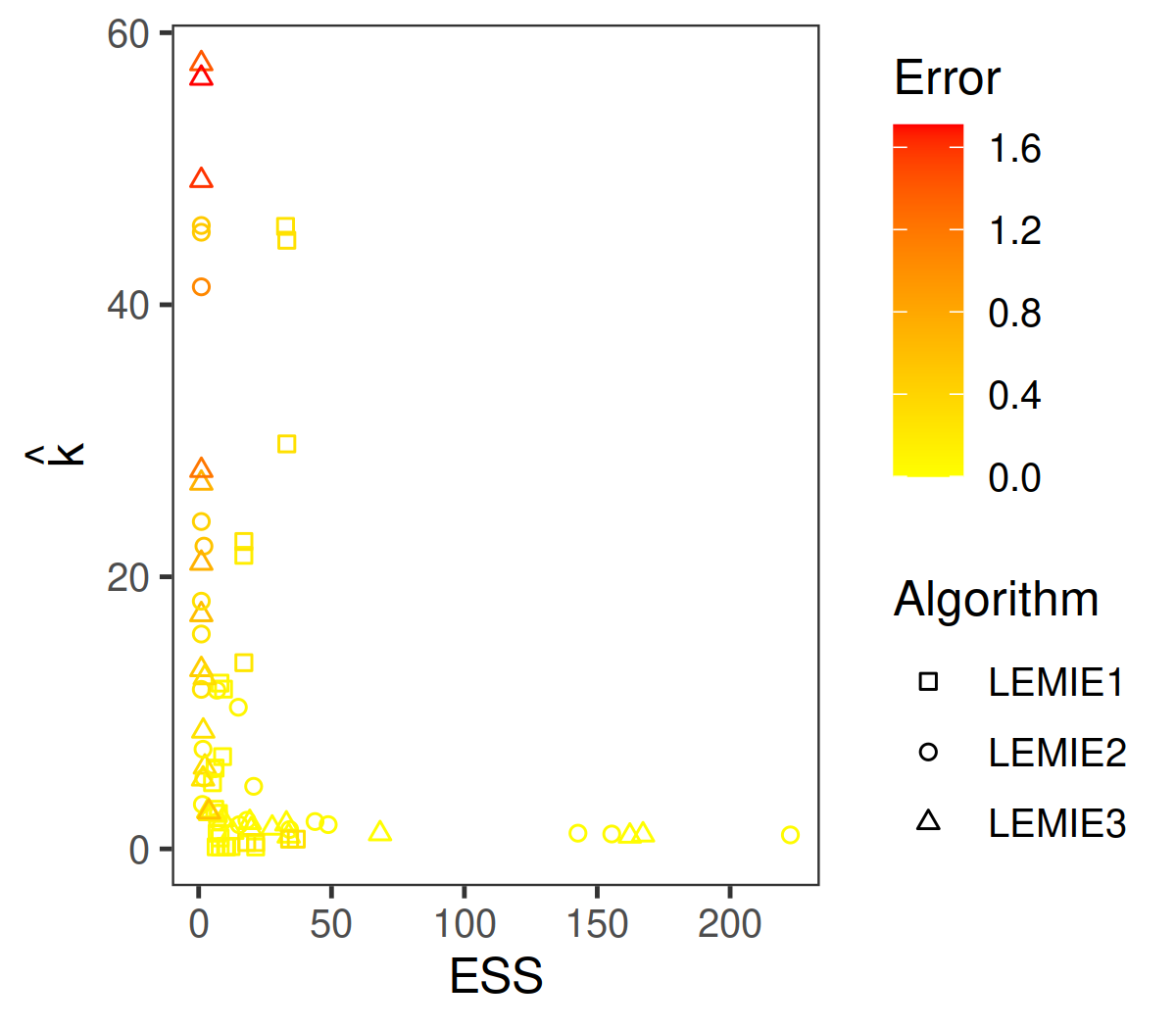}

}
\par\end{centering}
\caption{For the logistic regression examples in Section \ref{subsec:Logistic-regression},
error in estimating the mean of $\theta$ for the LEMIE approximations
(of all types defined in Section \ref{subsec:Laplace-enrichment-1})
and for all $M$ considered against the performance metrics of Section
\ref{subsec:Performance-indicators}. \label{fig:logistic-lemie-diagnostics}}

\end{figure}

Figure \ref{fig:scott-logistic-lemie-diagnostics} plots the error
in estimating $\theta$ for all LEMIE results (all types of Laplace
samples and none) and in all $M$ simulations against the ESS and
$\hat{{k}}$ diagnostics. As in the examples of Section \ref{subsec:Multivariate-normal-models},
ESS and $\hat{{k}}$ are broadly related to performance. However,
there are some results with low ESS and high $\hat{{k}}$ which perform
relatively well, and some with high ESS and low $\hat{{k}}$ which
perform relatively poorly. Those latter results were for the LEMIE2
estimator and $M=2$. The result plotted in the bottom right corner,
having the greatest ESS, low $\hat{{k}}$ and low error, was also
for LEMIE2 and $M=2$ but using no Laplace samples.

With this many samples per local posterior, we run into memory issues
using the LEMIE algorithm when $M$ is large. This is why we only
looked up to $M=32$. This is not an insurmountable barrier to using
LEMIE with large $M$: if the likelihood calculations are implemented
with careful memory management the space requirements can be converted
to additional time requirements (although the time requirement can
be substantial when $\bar{N}M$ is large). If the purpose of parallelising
Bayesian computation is to speed it up, this may present a limit to
the usefulness of the algorithm. We did not look at cross entropy
in this example because the KDE calculations with $\bar{N}M$ samples
is prohibitively slow for large $M$.

\subsubsection{Simulation following \citet{neiswanger2013asymptotically}\label{subsec:logistic-simulations}}

The simulated logistic regression of \citet{neiswanger2013asymptotically}
uses predictors $x_{i}\in\mathbb{{R}}^{p}$ with $p=50$ and $n=50,000$
data realisations ($c_{i}=1$ for all $i$ in the model framework
used at the start of Section \ref{subsec:Logistic-regression}). The
parameters $\theta$ and each realisation $x_{i}$ were simulated
from $\mathrm{{N}}_{p}\left(0_{p},I_{p}\right)$. As in the example
in Section \ref{subsec:scott-logistic} we partitioned the data uniformly
at random into $M$ parts. \citet{neiswanger2013asymptotically} use
the No U-turn Hamiltonian Monte Carlo sampler implemented in Stan
(\citet{stan}). It is not clear what prior distribution they use;
we used a $\mathrm{{N}}_{p}\left(0_{p},2.5^{2}I_{p}\right)$ prior
and sampled $\bar{N}=400,000$ times from each local posterior, discarding
the first 50\% as burn-in, using the Gibbs sampler of \citet{polson2013bayesian}.
As in Section \ref{subsec:scott-logistic}, the fractionated prior
used for CMC and DPE is $\mathrm{{N}}_{p}\left(0_{p},2.5^{2}MI_{p}\right)$.

\begin{figure}
\begin{centering}
\subfloat[Error in estimating the posterior mean of $\theta$.]{\includegraphics[scale=0.48]{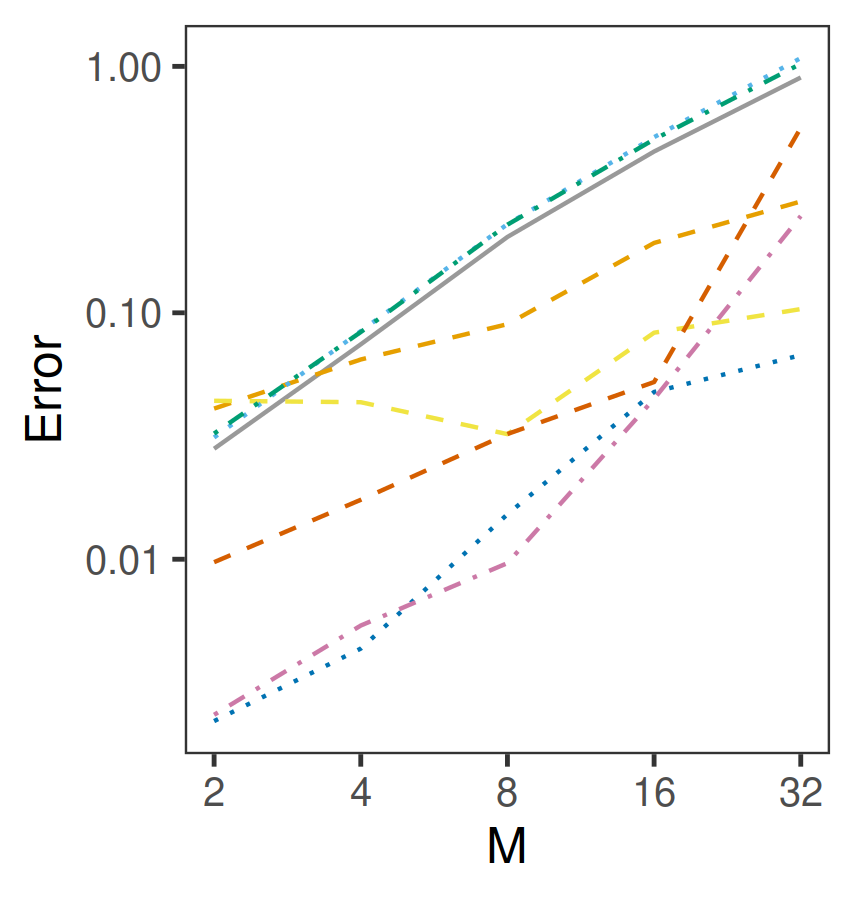}}\hfill{}\subfloat[Error in estimating the 2.5\% quantiles of the marginals of the posterior.]{\includegraphics[scale=0.48]{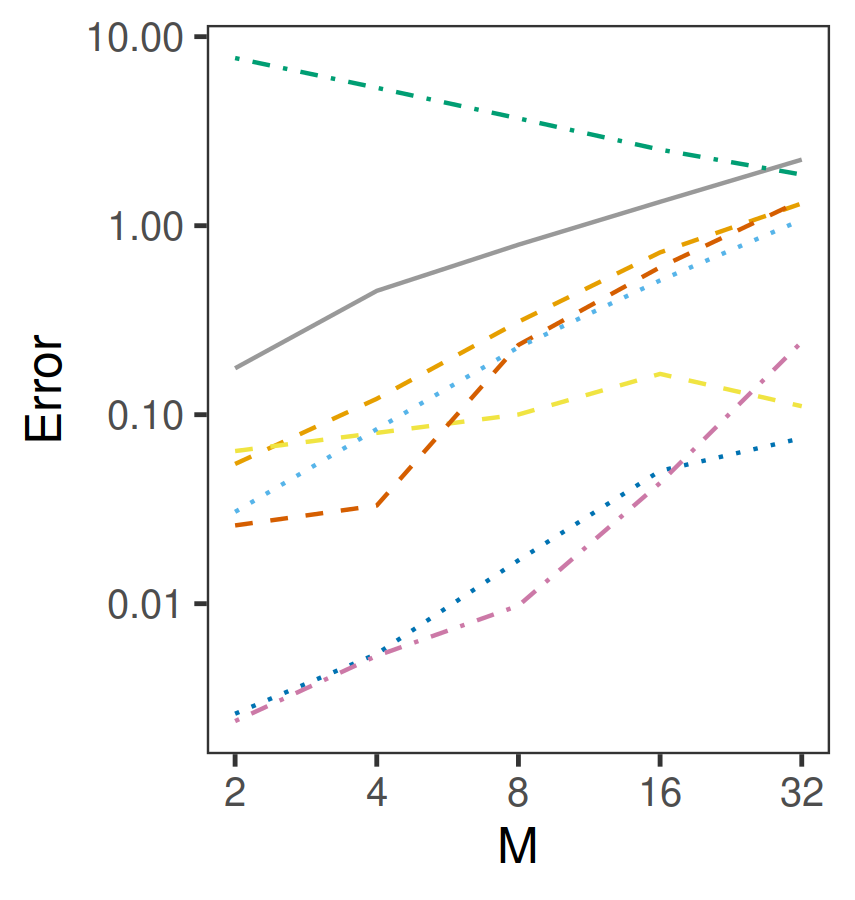}}\hfill{}\subfloat[Error in estimating the 97.5\% quantiles of the marginals of the posterior.]{\includegraphics[scale=0.48]{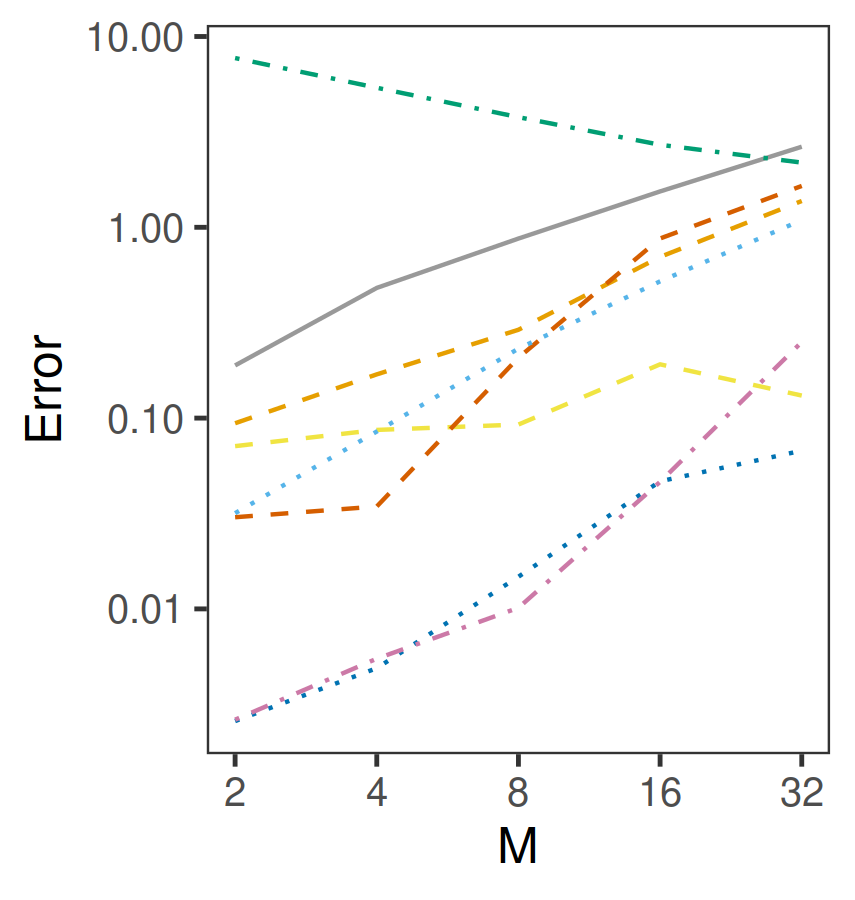}}\hfill{}\includegraphics[scale=0.45]{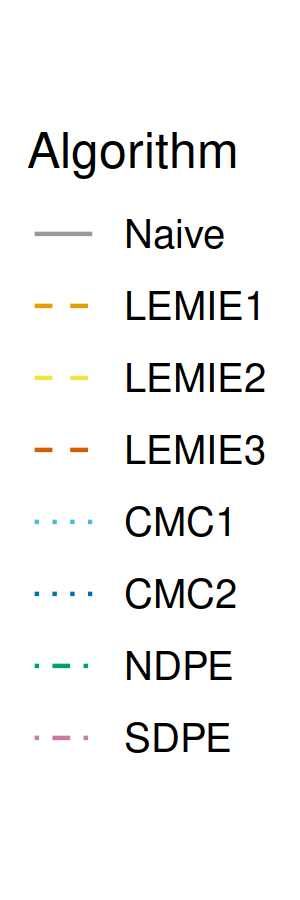}
\par\end{centering}
\caption{Posterior approximation comparisons for $\theta$ in the logistic
regression example of Section \ref{subsec:logistic-simulations}.\label{fig:simulated-logistic-results}}
\end{figure}

Performance results are presented in Figure \ref{fig:simulated-logistic-results}.
CMC2 and SDPE perform consistently well across the range of $M$
considered. The LEMIE algorithms are also fairly reliable across $M$,
and for $M>8$ the best performing LEMIE algorithm does almost as
well as the best of CMC or DPE at estimating the posterior mean and
tail quantiles. CMC and DPE are better suited to this example than
that in Section \ref{subsec:scott-logistic} because of the large
$n$, meaning the posterior is better approximated by an MVN. It is
notable that LEMIE does almost as well as any other method when $M$
is large given the relatively large dimension $p$.

\begin{figure}
\begin{raggedright}
\subfloat[]{\includegraphics[scale=0.45]{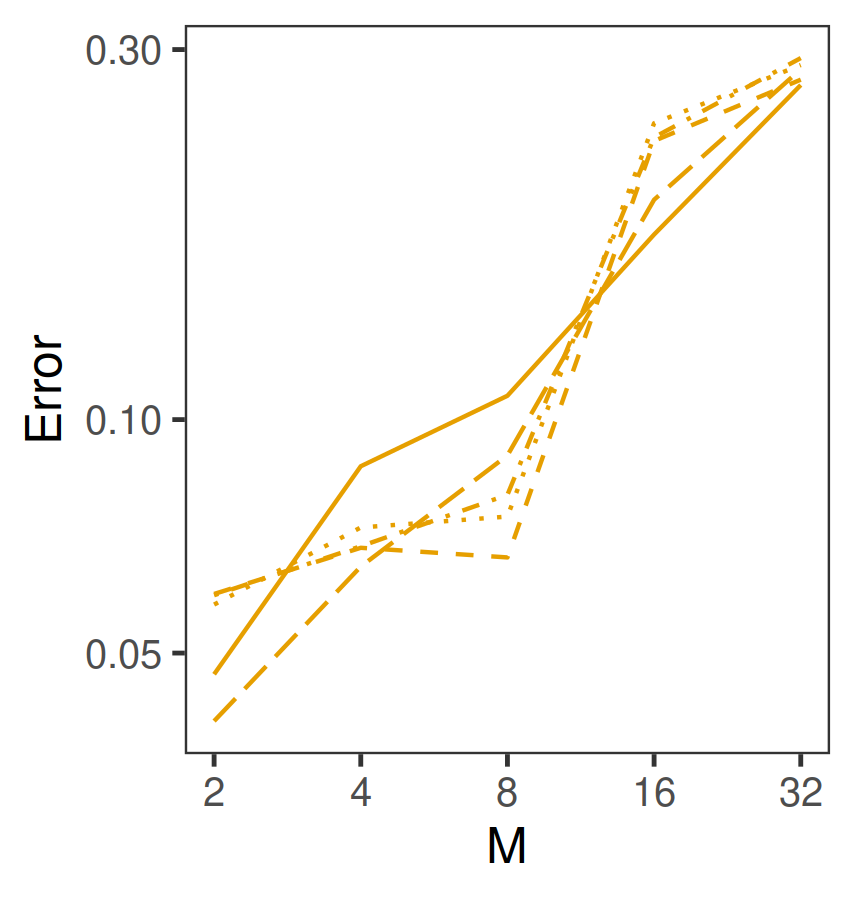}}\enskip{}\subfloat[]{\includegraphics[scale=0.45]{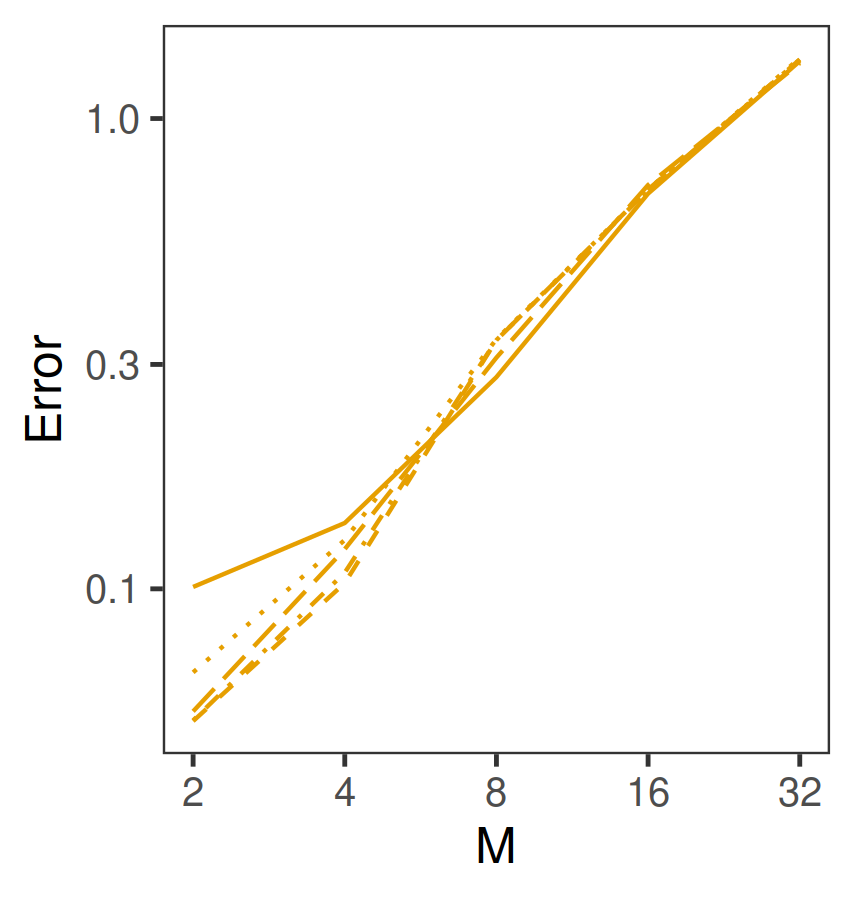}}\enskip{}\subfloat[]{\includegraphics[scale=0.45]{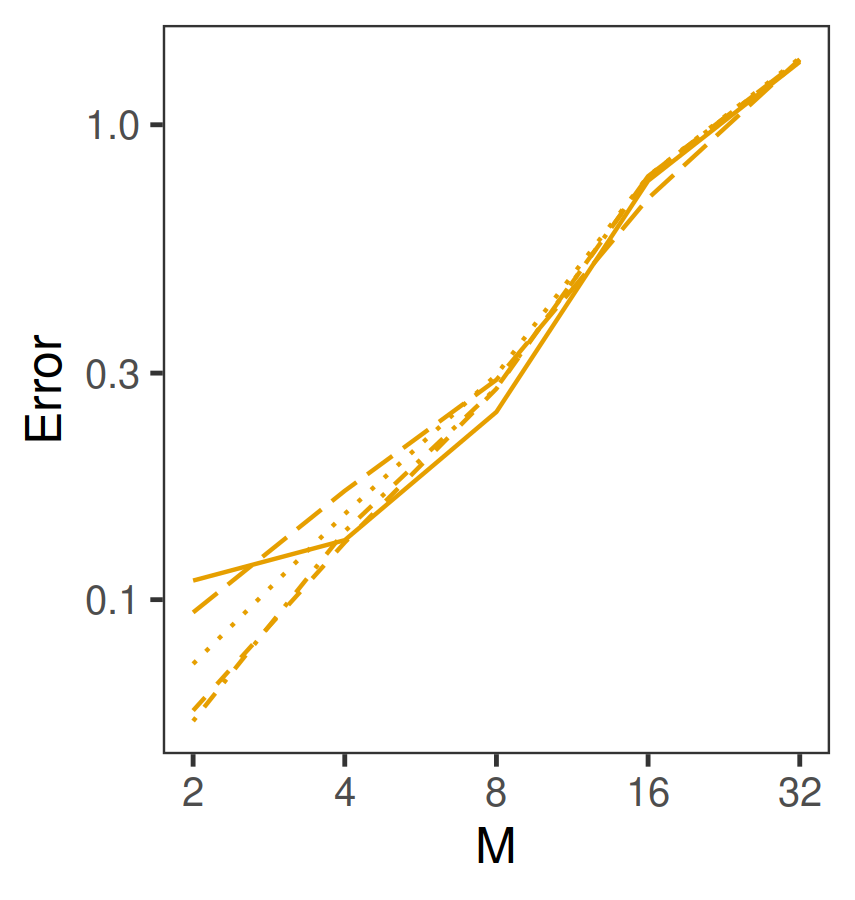}}\enskip{}\includegraphics[scale=0.45]{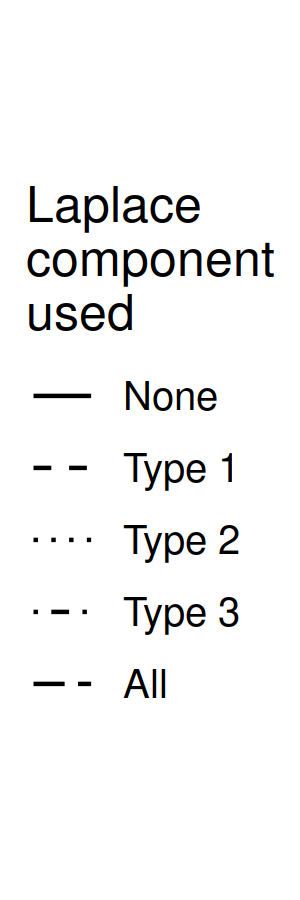}
\par\end{raggedright}
\begin{raggedright}
\subfloat[]{\includegraphics[scale=0.45]{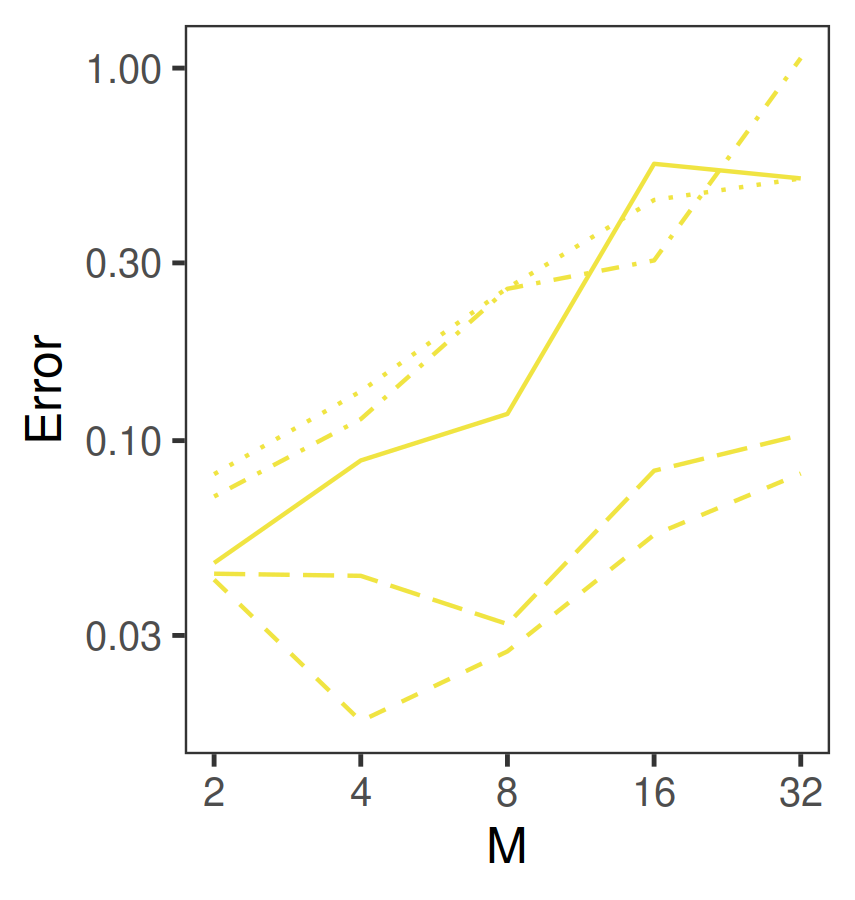}}\enskip{}\subfloat[]{\includegraphics[scale=0.45]{images/experiments-logistic-scott-error-025pc-m-alt-lemie2-presentation}}\enskip{}\subfloat[]{\includegraphics[scale=0.45]{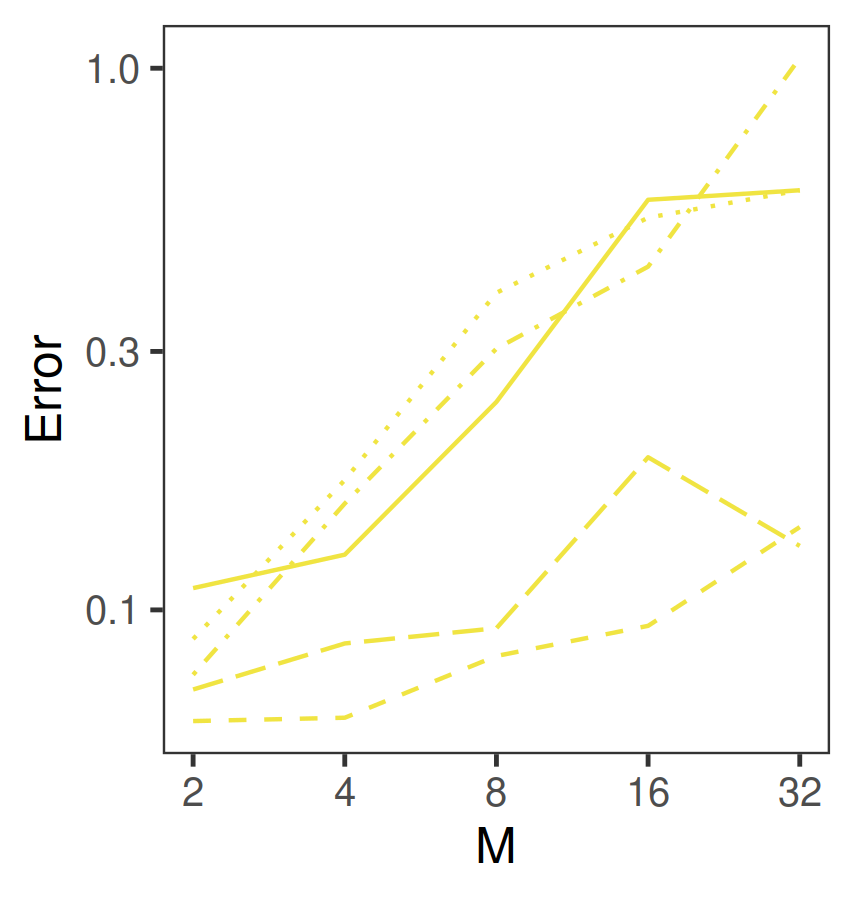}}\enskip{}\includegraphics[scale=0.45]{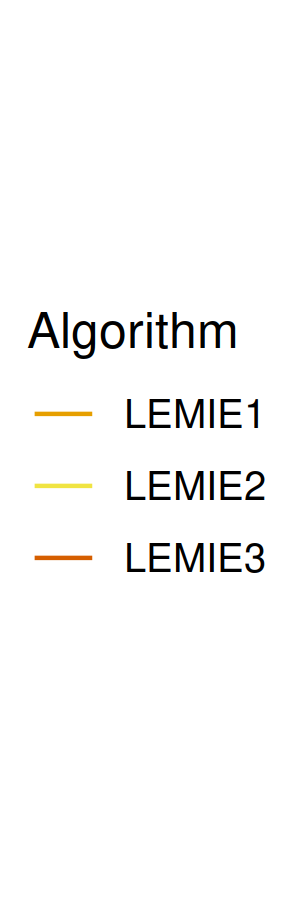}
\par\end{raggedright}
\begin{raggedright}
\subfloat[]{\includegraphics[scale=0.45]{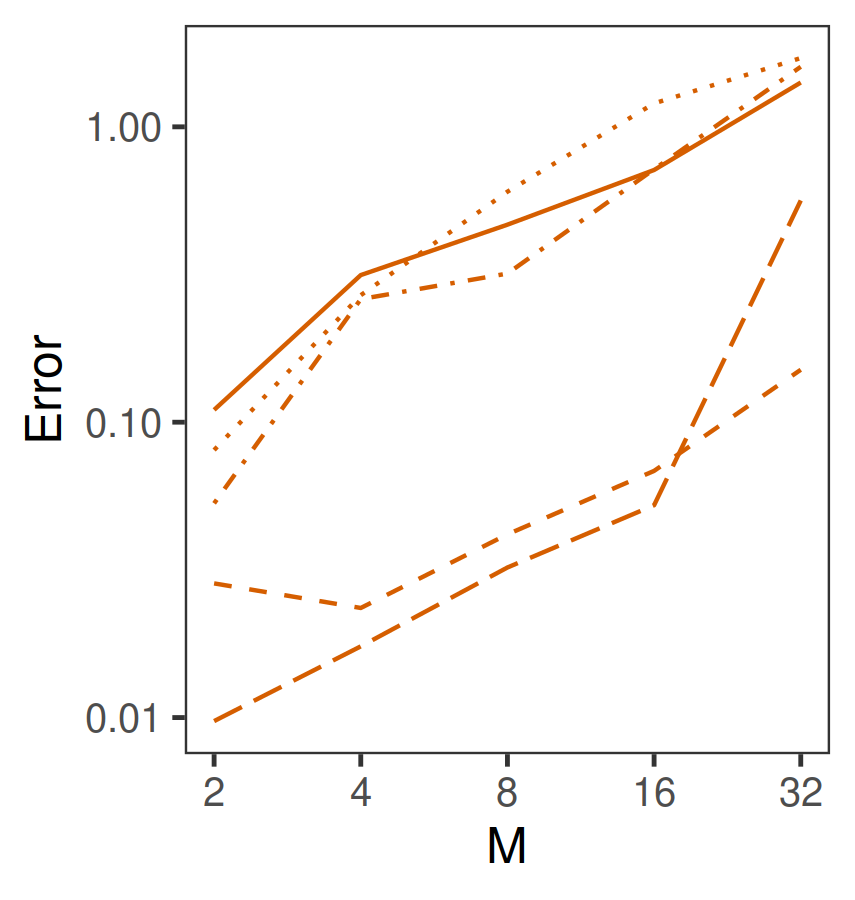}}\enskip{}\subfloat[]{\includegraphics[scale=0.45]{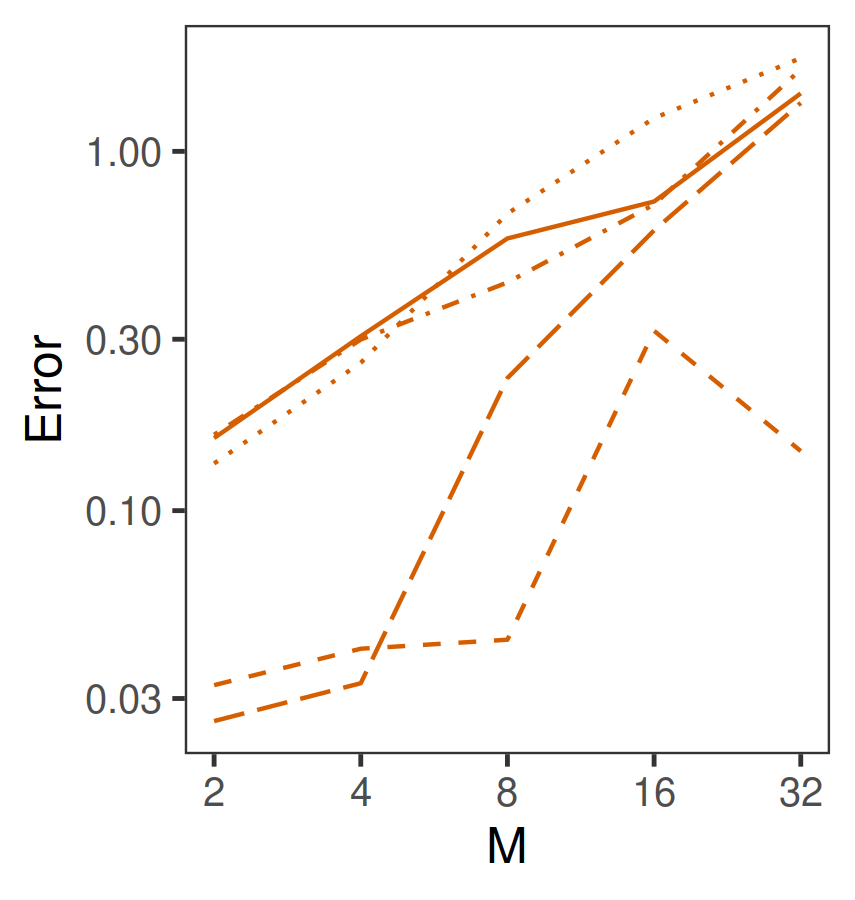}}\enskip{}\subfloat[]{\includegraphics[scale=0.45]{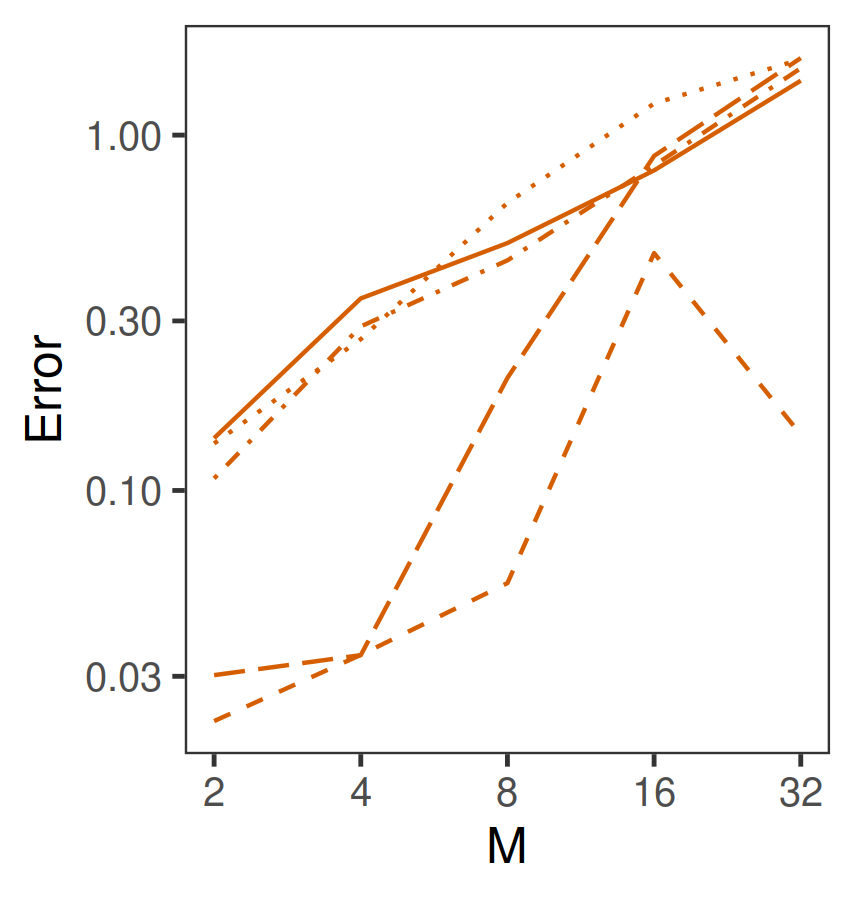}}
\par\end{raggedright}
\caption{Posterior approximation comparisons in the logistic regression example
of Section \ref{subsec:logistic-simulations}. The error in estimating
(a)(d)(g) the posterior mean, (b)(e)(h) the 2.5\% quantiles of the
marginals of the posterior, (c)(f)(i) the 97.5\% quantiles of the
marginals of the posterior.\label{fig:logistic-simulations-results-mie-lemie}}
\end{figure}

Figure \ref{fig:logistic-simulations-results-mie-lemie} shows results
for the LEMIE estimators only using all types of Laplace samples and
none. We find that Laplace samples are beneficial to performance,
but only those of type 1; including type 2 and type 3 Laplace samples
does not seem to help and may make performance worse. LEMIE1 does
not benefit much from adding Laplace samples of any type.

Figure \ref{fig:simulated-logistic-lemie-diagnostics} plots the error
in estimating $\theta$ for all LEMIE results and in all $M$ simulations
against the ESS and $\hat{{k}}$ diagnostics. In these examples, ESS
and $\hat{{k}}$ appear to be useful predictors of performance.

\section{Discussion and future work\label{sec:discussion}}

We have introduced new methodology for estimating posterior expectations
when data are partitioned, the Laplace enriched multiple importance
estimator (LEMIE), which has three variants defined by different importance
weighting schemes. Our method works with parallel sampling from local
posteriors using any unbiased sampling algorithm and weights the samples
obtained for use in Monte Carlo estimators. This is accomplished with
new importance weighting schemes that allow for unnormalised proposal
and posterior densities.

The performance of the LEMIE estimators in terms of KL divergence
and error in estimating the posterior mean and tail quantiles appears
to be generally good, almost always better than naive pooling of samples
and sometimes better than CMC (\citet{scott2016bayes}) or DPE (\citet{neiswanger2013asymptotically}).
It seems to be particularly good relative to these methods in small
data, non-normal examples, as seen in the results with the beta-Bernoulli
model in Section \ref{subsec:Beta-Bernoulli-model}, estimation of
the $\Sigma$ marginal of the posterior in the MVN example of Section
\ref{subsec:Multivariate-normal-models} and the logistic regression
example of Section \ref{subsec:scott-logistic}.

The larger logistic example of Section \ref{subsec:logistic-simulations}
poses a tougher challenge for LEMIE, but the methods still do no worse
than some of the other methods, and for large $M$ become competitive.
This is remarkable for importance sampling-based methods used in a
50 dimensional parameter space given the curse of dimensionality.
This seems to be thanks to including samples from the Laplace approximation
of type 1 (Section \ref{subsec:Laplace-enrichment-1}). We have seen
that a small number of Laplace samples of the right type, in particular
type 1, can improve LEMIE a lot in complex examples. This is likely
because the approximation is closer to the posterior than any of the
local posteriors. The type 1 Laplace approximation is very similar
to CMC2 (and the PDPE of \citet{neiswanger2013asymptotically}), so
this is unsurprisingly good in normal examples or examples with large
$n$. However, we have also seen that further improvements cannot
be easily obtained by increasing the number of Laplace samples.

In our methods, any prior distribution can be used, unlike in some
other methods such as CMC or DPE, which need to use a $\nicefrac{1}{M}^{\textrm{{th}}}$
power of the prior p.d.f. for the local sampling. This often entails
a compromise in the form or parameterisation of the prior used, as
we saw in examples in Sections \ref{subsec:Beta-Bernoulli-model}
and \ref{subsec:Multivariate-normal-models}. Incorrectly specified
priors may not be an issue in big data situations (CMC and DPE are
designed with big data applications in mind), since the influence
of the prior on the posterior diminishes with $n$, but it may well
be consequential in those scenarios with restrictions on data sharing
where $n$ is not very large. It is not an issue when the prior is
uninformative, but uninformative priors can be hard to interpret or
lead to unstable estimates (\citet{gelman2008weakly}).

Having useful performance diagnostics - ESS and $\hat{{k}}$ - is
another advantage of our methods. There are several variants of LEMIE
to consider, including the Laplace approximation options, and some
may work better than others in particular situations. These diagnostics
can be useful in identifying which one is likely to perform better
when there is no other means of calibration. Whilst we have some
empirical evidence that the ESS is a useful diagnostic, it does not
carry any guarantees. It is also based on two approximations of the
estimator variance using the delta method, the second in particular
having a non-negligible error term (see Appendix \ref{subsec:ESS-derivations}).
There are other issues in the ESS, discussed in \citet{elvira2018rethinking},
who also suggest some alternatives that we could use.

There are ways to estimate an ESS for the other methods, for instance
the multivariate ESS for MCMC of \citet{vats2019multivariate}. This
would not permit comparisons across methods, including LEMIE, to predict
relative performance because the methods have different biases and
the ESS is related to the variance of the estimators and not the bias.

One limitation of our methods is on how large we can make $M$. We
found in the logistic regression examples of Section \ref{subsec:Logistic-regression}
that for a given number of samples from each local posterior, $\bar{N}$,
there is an $M$ large enough that the necessary pooling and weighting
operations on the master node of the $\bar{N}M$ samples may exceed
the available memory resources. To get around this it may be possible
to cap the number of samples drawn by each worker, but this may not
be possible if the local MCMC samplers have not converged by that
point, or to discard samples, although this will have a detrimental
effect on the estimator variance.

We think it is likely that a random partition of the data is better
for our methods than a non-random partition, or \foreignlanguage{british}{heterogeneous}
data, because that would make the local posteriors less likely to
be good approximations to the posterior, as required for importance
sampling to work well. This is a limitation for applications with
real data if the data partition cannot be controlled. We did see
in Section \ref{subsec:Beta-Bernoulli-model} that the methods perform
well with heterogeneous data in a simple, 1 dimensional example; however,
we are yet to investigate the impact of heterogeneous data in more
complex examples.

In future work, it would be interesting to compare performance against
methods outside of \citet{scott2016bayes} and \citet{neiswanger2015embarrassingly},
particularly methods which approach the problem in a different way
such as \citet{xu2014distributed}, \citet{jordan2018communication},
\citet{nemeth2018merging}, \citet{park2020variational} and \citet{rendell2020global},
and identify situations where one may perform better than another
or where methods may complement each other. Our methods apply at the
sample collection stage. This means they could be used in conjunction
with some other methods, notably those of \citet{xu2014distributed}
or \citet{nemeth2018merging}, in the latter case using our multiple
importance weighting schemes in place of theirs. This may result in
estimators that perform better in higher dimensions than either method
on its own.

Another idea for improving our methods is to employ the importance
weight smoothing used in the Pareto smoothed importance sampling (PSIS)
of \citet{vehtari2015pareto}. This would apply after computing the
weights of type 1, 2 or 3 from Section \ref{subsec:Multiple-importance-estimation}.
We have already investigated the use of $\hat{{k}}$ from the PSIS
algorithm as a diagnostic (Section \ref{subsec:Tail-distribution-shape}).
\citet{vehtari2015pareto} find that importance estimator performance
can be improved with PSIS when $\hat{{k}}<0.7$. This is something
to explore in future work.

The simulation studies used in Section \ref{sec:experiments} to evaluate
the performance of our methods are limited in scope. In order to better
understand the strengths and limitations of our methods, experiments
should be conducted on more complex models such as hierarchical models,
large data sets and especially with real data. It would also be valuable
to investigate performance under different conditions, such as non-random
partitions of heterogeneous data, i.e. different data distributions
in each node, multimodal posterior distributions, and the choice of
MCMC algorithm employed for local posterior sampling.

\appendix

\section{Derivations}

\subsection{Asymptotic results\label{subsec:appendix-Asymptotic-results}}

\subsubsection{Asymptotic unbiasedness of self-normalised importance sampling\label{subsec:Asymptotic-unbiasedness-snis}}

Equation \ref{eq:estimator-of-normalising-constants} is derived as

\begin{eqnarray}
\lim_{N_{j}\to\infty}\frac{1}{N_{j}}\sum_{h=1}^{N_{j}}\tilde{w}_{j}\left(\theta_{j,h}\right) & = & \mathbb{{E}}_{\pi_{j}}\left[\tilde{w}_{j}\left(\theta\right)\right]\nonumber \\
 & = & \frac{Z_{\pi}}{Z_{j}}\int w_{j}\left(\theta\right)\pi_{j}\left(\theta\mid\mathbf{x}_{j}\right)\mathrm{{d}}\theta\nonumber \\
 & = & \frac{Z_{\pi}}{Z_{j}}\int\pi\left(\theta\mid x_{1:n}\right)\mathrm{{d}}\theta\nonumber \\
 & = & \frac{Z_{\pi}}{Z_{j}},
\end{eqnarray}
with the first line due to the strong law of large numbers. The importance
estimator in Equation \ref{eq:single-importance-estimator} of $\mathbb{{E}}_{\pi}\left[f\left(\theta\right)\right]$
has zero bias in the limit of $N_{j}\to\infty$ because

\begin{eqnarray}
\lim_{N\to\infty}\tilde{\mu}_{j} & = & \lim_{N\to\infty}\frac{\frac{1}{N_{j}}\sum_{h=1}^{N_{j}}\tilde{w}_{j}\left(\theta_{h}\right)f\left(\theta_{h}\right)}{\frac{1}{N_{j}}\sum_{h=1}^{N_{j}}\tilde{w}_{j}\left(\theta_{h}\right)}\nonumber \\
 & = & \frac{\mathbb{{E}}_{\pi_{j}}\left[\tilde{w}_{j}\left(\theta\right)f\left(\theta\right)\right]}{\mathbb{{E}}_{\pi_{j}}\left[\tilde{w}_{j}\left(\theta\right)\right]}\label{eq:importance-estimator-proof-2}\\
 & = & \mathbb{{E}}_{\pi_{j}}\left[\frac{Z_{j}}{Z_{\pi}}\tilde{w}_{j}\left(\theta\right)f\left(\theta\right)\right]\label{eq:importance-estimator-proof-3}\\
 & = & \mathbb{{E}}_{\pi_{j}}\left[w_{j}\left(\theta\right)f\left(\theta\right)\right]\label{eq:importance-estimator-proof-4}\\
 & = & \mathbb{{E}}_{\pi}\left[f\left(\theta\right)\right],\label{eq:importance-estimator-proof-5}
\end{eqnarray}
in which line \ref{eq:importance-estimator-proof-2} holds almost
surely by the strong law of large numbers, line \ref{eq:importance-estimator-proof-3}
is by Equation \ref{eq:estimator-of-normalising-constants} and line
\ref{eq:importance-estimator-proof-5} is by Equation \ref{eq:normalised-importance-sampling}.

\subsubsection{Asymptotic unbiasedness of MIE2\label{subsec:Asymptotic-unbiasedness-mie2}}

The MIE2 estimator, Equation \ref{eq:mopp-2-unselfnormalised}, has
limit

\begin{eqnarray}
\lim_{N_{1},\ldots,N_{M}\to\infty}\tilde{\mu}^{\textrm{{MIE2}}} & = & \lim_{N_{1},\ldots,N_{M}\to\infty}\frac{1}{N}\sum_{j=1}^{M}\sum_{h=1}^{N_{j}}\frac{\pi\left(\theta_{j,h}\mid x_{1:n}\right)f\left(\theta_{j,h}\right)}{\frac{1}{N}\sum_{k=1}^{M}N_{k}\frac{Z_{k}}{Z_{\pi}}\hat{c}_{k}\pi_{k}\left(\theta_{j,h}\mid\mathbf{x}_{k}\right)}\nonumber \\
 & = & \lim_{N_{1},\ldots,N_{M}\to\infty}\frac{1}{N}\sum_{j=1}^{M}\sum_{h=1}^{N_{j}}\pi\left(\theta_{j,h}\mid x_{1:n}\right)f\left(\theta_{j,h}\right)\nonumber \\
 &  & \times\frac{1}{\lim_{N_{1},\ldots,N_{M}\to\infty}\frac{1}{N}\sum_{k=1}^{M}N_{k}\frac{Z_{k}}{Z_{\pi}}\hat{c}_{k}\pi_{k}\left(\theta_{j,h}\mid\mathbf{x}_{k}\right)}\label{eq:asymptotic-unbiasedness-mie2-line2}\\
 & = & \lim_{N_{1},\ldots,N_{M}\to\infty}\frac{1}{N}\sum_{j=1}^{M}\sum_{h=1}^{N_{j}}\frac{\pi\left(\theta_{j,h}\mid x_{1:n}\right)f\left(\theta_{j,h}\right)}{\phi\left(\theta_{j,h}\right)}\label{eq:asymptotic-unbiasedness-mie2-line3}\\
 & = & \mathbb{{E}}\left[w_{\phi}\left(\theta\right)f\left(\theta\right)\right]\label{eq:asymptotic-unbiasedness-mie2-line4}\\
 & = & \mathbb{{E}}_{\pi}\left[f\left(\theta\right)\right].\label{eq:asymptotic-unbiasedness-mie2-line5}
\end{eqnarray}
Line \ref{eq:asymptotic-unbiasedness-mie2-line2} is possible because
the limit of the denominator is not zero, line \ref{eq:asymptotic-unbiasedness-mie2-line3}
evaluates the limit for the denominator by replacing $\hat{c}_{k}$
with its estimand $\frac{Z_{\pi}}{Z_{k}}$, line \ref{eq:asymptotic-unbiasedness-mie2-line4}
is by the strong law of large numbers and line \ref{eq:asymptotic-unbiasedness-mie2-line5}
follows from Equation \ref{eq:normalised-importance-sampling}.

\subsection{Bias and variance of estimators}

\subsubsection{Finite-sample bias of MIE2\label{subsec:Finite-sample-bias-mie2}}

The MIE2 estimator without self-normalising can be written as

\begin{equation}
\tilde{\mu}^{\textrm{{MIE2}}}=\sum_{j=1}^{M}\sum_{h=1}^{N_{j}}\frac{\pi\left(\theta_{j,h}\mid x_{1:n}\right)f\left(\theta_{j,h}\right)}{\frac{1}{N}\sum_{k=1}^{M}N_{k}\bar{\hat{c}}_{k}\pi_{k}\left(\theta_{j,h}\mid\mathbf{x}_{k}\right)}.
\end{equation}
where

\begin{equation}
\bar{\hat{c}}_{k}:=\frac{1}{N_{k}}\sum_{i=1}^{N_{k}}\frac{\pi\left(\theta_{k,i}\mid x_{1:n}\right)}{\pi_{k}\left(\theta_{k,i}\mid\mathbf{x}_{k}\right)}.\label{eq:mie2-bias-mc-estimator}
\end{equation}
The bias can be attributed to the Monte Carlo average Equation \ref{eq:mie2-bias-mc-estimator},
as can be seen in Equation \ref{eq:asymptotic-unbiasedness-mie2-line5}.
Equation \ref{eq:mie2-bias-mc-estimator} converges to 1 in the limit
$N_{k}\to\infty$. Write the denominator of the weights as

\begin{eqnarray}
\tilde{\phi}\left(\theta\right) & := & \frac{1}{N}\sum_{k=1}^{M}N_{k}\bar{\hat{c}}_{k}\pi_{k}\left(\theta\mid\mathbf{x}_{k}\right)\nonumber \\
 & = & \frac{1}{N}\sum_{k=1}^{M}N_{k}\left[1+\varepsilon_{k}\right]\pi_{k}\left(\theta\mid\mathbf{x}_{k}\right)\nonumber \\
 & = & \phi\left(\theta\right)+\varepsilon\left(\theta\right),
\end{eqnarray}
where $\varepsilon_{k}\sim\textrm{{N}}\left(0,\frac{\sigma_{k}^{2}}{N_{k}}\right)$
by the central limit theorem with 

\begin{equation}
\sigma_{k}^{2}:=\textrm{{Var}}\left(\frac{\pi\left(\theta\mid x_{1:n}\right)}{\pi_{k}\left(\theta\mid\mathbf{x}_{k}\right)}\right)
\end{equation}
and

\begin{equation}
\varepsilon\left(\theta\right):=\frac{1}{N}\sum_{k=1}^{M}N_{k}\varepsilon_{k}\pi_{k}\left(\theta\mid\mathbf{x}_{k}\right).
\end{equation}
 Then we can write

\begin{eqnarray}
\mathbb{{E}}\left[\tilde{\mu}^{\textrm{{MIE2}}}\right] & = & \mathbb{{E}}\left[\frac{\frac{1}{N}\sum_{j=1}^{M}\sum_{h=1}^{N_{j}}\pi\left(\theta\mid x_{1:n}\right)f\left(\theta\right)}{\phi\left(\theta\right)+\varepsilon\left(\theta\right)}\right],
\end{eqnarray}
which we can estimate using the delta method. Let

\begin{eqnarray}
\bar{V} & := & \frac{1}{N}\sum_{j=1}^{M}\sum_{h=1}^{N_{j}}\pi\left(\theta\mid x_{1:n}\right)f\left(\theta\right),\nonumber \\
U & := & \phi\left(\theta\right)+\varepsilon\left(\theta\right),
\end{eqnarray}
then the expectation of the Taylor series of $g\left(U,\bar{V}\right)=\frac{\bar{V}}{U}$
truncated at the 2nd order term (required to estimate bias, \citet{owen2013monte})
gives us

\begin{eqnarray}
\mathbb{{E}}\left[\tilde{\mu}^{\textrm{{MIE2}}}\right] & \approx & g\left(\mu_{U},\mu_{V}\right)+\frac{1}{2}\left[\sigma_{U}^{2}\frac{\partial^{2}}{\partial U^{2}}g\left(\mu_{U},\mu_{V}\right)\right.\nonumber \\
 &  & +\frac{1}{N}\sigma_{V}^{2}\frac{\partial^{2}}{\partial V^{2}}g\left(\mu_{U},\mu_{V}\right)\nonumber \\
 &  & \left.+2\textrm{{Cov}}\left(U,V\right)\frac{\partial^{2}}{\partial U\partial V}g\left(\mu_{U},\mu_{V}\right)\right],\label{eq:expectation-delta-method-mie2}
\end{eqnarray}
where $\mu_{U},\sigma_{U}^{2}$ and $\mu_{V},\sigma_{V}^{2}$ are
the expectations and variances of $U$ and $V$ respectively and $\textrm{{Cov}}\left(U,V\right)$
is their covariance. Since $\mathbb{{E}}\left[\varepsilon\left(\theta\right)\right]=0$
and $\varepsilon_{k}$ is independent of $\theta$, Equation \ref{eq:expectation-delta-method-mie2}
can be simplified to

\begin{eqnarray}
\mathbb{{E}}\left[\tilde{\mu}^{\textrm{{MIE2}}}\right] & \approx & \frac{\mu_{V}}{\mu_{U}}+\frac{1}{\mu_{U}^{2}}\left(\textrm{{Var}}\left(\phi\left(\theta\right)\right)\frac{\mu_{V}}{\mu_{U}}-\mathrm{{Cov}}\left(U,V\right)\right)\label{eq:expectation-delta-method-mie2-1}\\
 &  & +\frac{1}{\mu_{U}^{2}}\textrm{{Var}}\left(\varepsilon\left(\theta\right)\right)\frac{\mu_{V}}{\mu_{U}}.\label{eq:expectation-delta-method-mie2-2}
\end{eqnarray}
If we had applied the delta method to the estimator using weights
Equation \ref{eq:mopp2-normalised-weights} we would have gotten line
\ref{eq:expectation-delta-method-mie2-1} with the same values of
$\mu_{U},\mu_{V}$ and $\mathrm{{Cov}}\left(U,V\right)$. But we know
that that estimator is unbiased, so

\begin{eqnarray}
\mathbb{{E}}\left[\tilde{\mu}^{\textrm{{MIE2}}}\right] & \approx & \mathbb{{E}}_{\pi}\left[f\left(\theta\right)\right]+\frac{1}{\mu_{U}^{2}}\textrm{{Var}}\left(\varepsilon\left(\theta\right)\right)\frac{\mu_{V}}{\mu_{U}}\nonumber \\
 & = & \mathbb{{E}}_{\pi}\left[f\left(\theta\right)\right]+\frac{1}{\mathbb{{E}}\left[\phi\left(\theta\right)\right]^{2}}\textrm{{Var}}\left(\varepsilon\left(\theta\right)\right)\frac{\mathbb{{E}}\left[\pi\left(\theta\mid x_{1:n}\right)f\left(\theta\right)\right]}{\mathbb{{E}}\left[\phi\left(\theta\right)\right]}
\end{eqnarray}
where

\begin{eqnarray}
\textrm{{Var}}\left(\varepsilon\left(\theta\right)\right) & = & \frac{1}{N^{2}}\sum_{k=1}^{M}N_{k}^{2}\textrm{{Var}}\left(\varepsilon_{k}\pi_{k}\left(\theta_{j,h}\mid\mathbf{x}_{k}\right)\right)\nonumber \\
 & = & \frac{1}{N^{2}}\sum_{k=1}^{M}N_{k}^{2}\textrm{{Var}}\left(\varepsilon_{k}\right)\mathbb{{E}}\left[\pi_{k}\left(\theta_{j,h}\mid\mathbf{x}_{k}\right)^{2}\right].\label{eq:variance-of-monte-carlo-estimator-mie2}
\end{eqnarray}

\subsubsection{Variance of MIE2\label{subsec:Variance-of-MIE2}}

Using the notation of Section \ref{subsec:Finite-sample-bias-mie2}
we have

\begin{eqnarray}
\textrm{{Var}}\left(\tilde{\mu}^{\textrm{{MIE2}}}\right) & \approx & \frac{\mu_{V}^{2}}{\mu_{U}^{4}}\sigma_{U}^{2}+\frac{1}{N}\frac{1}{\mu_{U}^{2}}\sigma_{V}^{2}-\frac{\mu_{V}}{\mu_{U}^{3}}\textrm{{Cov}}\left(U,V\right)\nonumber \\
 & = & \frac{1}{\mu_{U}^{2}}\left(\frac{\mu_{V}^{2}}{\mu_{U}^{2}}\sigma_{U}^{2}+\frac{1}{N}\sigma_{V}^{2}-\frac{\mu_{V}}{\mu_{U}}\textrm{{Cov}}\left(U,V\right)\right)\nonumber \\
 & = & \frac{1}{\mu_{U}^{2}}\left(\frac{\mu_{V}^{2}}{\mu_{U}^{2}}\textrm{{Var}}\left(\phi\left(\theta\right)\right)+\frac{1}{N}\sigma_{V}^{2}-\frac{\mu_{V}}{\mu_{U}}\textrm{{Cov}}\left(U,V\right)\right)\nonumber \\
 &  & +\frac{1}{\mu_{U}^{2}}\frac{\mu_{V}^{2}}{\mu_{U}^{2}}\textrm{{Var}}\left(\varepsilon\left(\theta\right)\right).
\end{eqnarray}
Following the same reasoning as above, the $\frac{1}{\mu_{U}^{2}}\frac{\mu_{V}^{2}}{\mu_{U}^{2}}\textrm{{Var}}\left(\varepsilon\left(\theta\right)\right)$
term is the additional variance due to the Monte Carlo estimate of
the ratio of normalising constants. The rest is the variance of the
normalised estimator.

\subsection{ESS of MIE1\label{subsec:ESS-derivations}}

\begin{sloppypar}The approximate ESS for MIE1, Equation \ref{eq:ess-mie1},
is derived from an additional application of the delta method to Equation
\ref{eq:mie-1-variance}. We follow the reasoning of \citet{kong1992note}.
Equation \ref{eq:mie-1-variance} can be written as

\begin{eqnarray}
\mathrm{{Var}}\left(\tilde{\mu}^{\textrm{{MIE1}}}\right) & \approx & \sum_{j=1}^{M}\frac{N_{j}}{N^{2}}\left(\textrm{{Var}}\left(w_{j}\left(\theta\right)\right)\mathbb{{E}}_{\pi}^{2}\left[f\left(\theta\right)\right]+\textrm{{Var}}\left(w_{j}\left(\theta\right)f\left(\theta\right)\right)\right.\nonumber \\
 &  & \left.-2\textrm{{Cov}}\left(w_{j}\left(\theta\right),w_{j}\left(\theta\right)f\left(\theta\right)\right)\mathbb{{E}}_{\pi}\left[f\left(\theta\right)\right]\right).\label{eq:mie-1-variance-2}
\end{eqnarray}
The delta method is used to approximate the $\mathbb{{E}}\left[w_{j}\left(\theta\right)^{2}f\left(\theta\right)^{2}\right]$
term in $\textrm{{Var}}\left(w_{j}\left(\theta\right)f\left(\theta\right)\right)$.
This term is equal to $\mathbb{{E}}_{\pi}\left[w_{j}\left(\theta\right)f\left(\theta\right)^{2}\right]$.
The Taylor expansion of $g^{\prime}\left(U,V\right):=UV^{2}$ evaluated
at the expectation of $U$ and $V$ and truncated after the second
order term leads to the approximation

\begin{eqnarray}
\mathbb{{E}}_{\pi}\left[w_{j}\left(\theta\right)f\left(\theta\right)^{2}\right] & \approx & \mathbb{{E}}_{\pi}\left[w_{j}\left(\theta\right)\right]\mathbb{{E}}_{\pi}^{2}\left[f\left(\theta\right)\right]+\textrm{{Var}}_{\pi}\left(f\left(\theta\right)\right)\mathbb{{E}}_{\pi}\left[w_{j}\left(\theta\right)\right]\nonumber \\
 &  & +2\textrm{{Cov}}_{\pi}\left(w_{j}\left(\theta\right),f\left(\theta\right)\right)\mathbb{{E}}_{\pi}\left[f\left(\theta\right)\right].
\end{eqnarray}
The error term of this ``is not necessarily small'', according to
\citet[, p36]{liu2001monte}. Substituting into Equation \ref{eq:mie-1-variance-2}
and simplifying leads us to

\begin{equation}
\mathrm{{Var}}\left(\tilde{\mu}^{\textrm{{MIE1}}}\right)\approx\sum_{j=1}^{M}\frac{N_{j}}{N^{2}}\textrm{{Var}}_{\pi}\left(f\left(\theta\right)\right)\left(\textrm{{Var}}\left(w_{j}\left(\theta\right)\right)+1\right).
\end{equation}
This is rearranged to give $\textrm{{ESS}}_{1}$,

\begin{eqnarray}
\frac{\textrm{{Var}}_{\pi}\left(f\left(\theta\right)\right)}{\mathrm{{Var}}\left(\tilde{\mu}^{\textrm{{MIE1}}}\right)} & \approx & \frac{1}{\sum_{j=1}^{M}\frac{N_{j}}{N^{2}}\mathbb{{E}}_{\pi}\left[w_{j}\left(\theta\right)^{2}\right]},
\end{eqnarray}
which is estimated as Equation \ref{eq:ess-mie1}.\end{sloppypar}

\subsection{Fractionated prior in MVN\label{subsec:Fractionated-prior-mvn}}

The marginal prior density for $\Sigma$ is proportional to

\begin{equation}
\left|\Sigma\right|^{\frac{-\left(\nu+d+1\right)}{2}}\exp\left\{ -\frac{1}{2}\mathrm{{tr}}\left(\Psi\Sigma^{-1}\right)\right\} .
\end{equation}
For fractionation this must be raised to the $\nicefrac{1}{M}^{\textrm{{th}}}$
power, so to reparameterise we define $\nu^{*}$ such that

\begin{equation}
\nu^{*}+d+1=\frac{1}{M}\left(\nu+d+1\right)
\end{equation}
and solve for $\nu^{*}$:

\begin{equation}
\nu^{*}=\frac{\nu}{M}-\frac{M-1}{M}d-\frac{M-1}{M}.
\end{equation}
In the uninformative prior we put $\nu=0$. This implies that

\begin{equation}
\lim_{M\to\infty}\nu^{*}=-d-1.
\end{equation}
We could therefore use $\nu^{*}=-d-1$ for a fractionated prior that
is invariant with respect to $M$ and therefore can be used in all
our simulations. This would be an improper prior for $\Sigma$, but
the posterior distribution is proper under the following condition
for positive integer $n$:

\begin{eqnarray}
 & \nu^{*}+n & >d-1\nonumber \\
\implies & -d-1+n & >d-1\nonumber \\
\implies & n & >2d.
\end{eqnarray}
This must also be the case for every local posterior, so if data are
distributed evenly we must have $n$ such that $\left\lfloor \frac{n}{M}\right\rfloor >2d$.

\section{Semiparametric density product estimator (SDPE)\label{appendix:sdpe}}

As explained in Section \ref{subsec:Density-product-estimator}, the
SDPE algorithm of \citet{neiswanger2013asymptotically} is similar
to the NDPE algorithm but involves a KDE of $\frac{\pi_{j}\left(\theta\mid\mathbf{x}_{j}\right)}{\tilde{\varphi}_{j}\left(\theta\right)}$.
The product of these $M$ KDE approximations and the normal approximations
is a density product estimator of the posterior. Similarly to NDPE,
this can be expressed as the mixture distribution

\begin{eqnarray}
\hat{\pi}^{\mathrm{{SDPE}}}\left(\theta\right) & := & \frac{1}{\bar{N}^{M}}\prod_{j=1}^{M}\sum_{h=1}^{\bar{N}}\frac{\mathrm{{N}}_{p}\left(\theta\mid\theta_{j,h},bI_{p}\right)\mathrm{{N}}_{p}\left(\theta\mid\tilde{\mu}_{j},\tilde{\Sigma}_{j}\right)}{b^{p}\mathrm{{N}}_{p}\left(\theta_{j,h}\mid\tilde{\mu}_{j},\tilde{\Sigma}_{j}\right)}\nonumber \\
 & \propto & \sum_{h_{1}=1}^{\bar{N}}\cdots\sum_{h_{M}=1}^{\bar{N}}W\left(h_{1},\ldots,h_{M}\right)\nonumber \\
 &  & \times\mathrm{{N}}_{p}\left(\theta\mid\mu\left(h_{1},\ldots,h_{M}\right),\Sigma\left(h_{1},\ldots,h_{M}\right)\right)\label{eq:sdpe}
\end{eqnarray}
where

\begin{equation}
\Sigma\left(h_{1},\ldots,h_{M}\right):=\left(\frac{M}{b}I_{p}+\left(\Sigma^{*}\right)^{-1}\right)^{-1}
\end{equation}
and

\begin{equation}
\mu\left(h_{1},\ldots,h_{M}\right):=\Sigma\left(h_{1},\ldots,h_{M}\right)\left(\frac{M}{b}I_{p}\bar{\theta}\left(h_{1},\ldots,h_{M}\right)+\mu^{*}\right),
\end{equation}
with unnormalised weights

\begin{equation}
W\left(h_{1},\ldots,h_{M}\right):=\frac{w\left(h_{1},\ldots,h_{M}\right)\mathrm{\mathrm{{N}}_{p}}\left(\bar{\theta}\left(h_{1},\ldots,h_{M}\right)\mid\mu^{*},\Sigma^{*}\right)}{\prod_{j=1}^{M}\mathrm{{N}}_{p}\left(\theta_{j,h_{j}}\mid\tilde{\mu}_{j},\tilde{\Sigma}_{j}\right)}.
\end{equation}
$\mu^{*}$ and $\Sigma^{*}$ are from Equations \ref{eq:consensus-mean-1}
and \ref{eq:consensus-covariance-1} respectively. A similar independent
Metropolis-within-Gibbs algorithm is used to sample from the mixture
distribution Equation \ref{eq:sdpe} as described in Section \ref{subsec:Density-product-estimator}.

\section{Gibbs sampler for logistic regression from \citet{polson2013bayesian}\label{sec:appendix-Gibbs-sampler-Polson}}

The algorithm of \citet{polson2013bayesian} uses an augmented variable
approach to generate samples from the posterior of $\theta$ conditional
on data $x_{i},c_{i},y_{i}$ for $i=1,2,\ldots,n$ using Gibbs sampling
without any approximations or Metropolis steps, making it computationally
efficient. The likelihood is characterised as a scale mixture of normal
distributions. The marginal distribution of the scale parameter is
the Pólya-gamma distribution, which is constructed so that the full
conditional distribution for $\theta$ with prior $\mathrm{{N}}_{p}\left(b,B\right)$
is also MVN.

\begin{sloppypar}The augmentation variables are $\omega_{i}>0$ for
$i=1,2,\ldots,n$, and the Gibbs sampler consists of sampling iteratively
from the full conditionals
\begin{eqnarray*}
\omega_{i}\mid x_{i},c_{i},y_{i},\theta & \sim & \mathrm{{PG}}\left(c_{i},x_{i}^{\intercal}\theta\right),i=1,2,\ldots,n,\\
\theta\mid x_{1:n},c_{1:n},y_{1:n},\omega_{1:n} & \sim & \mathrm{{N}}_{p}\left(m_{\omega},V_{\omega}\right)
\end{eqnarray*}
where 

\begin{eqnarray*}
V_{\omega} & = & \left(x_{1:n}^{\intercal}\Omega x_{1:n}+B^{-1}\right)^{-1},\\
m_{\omega} & = & V_{\omega}\left(x_{1:n}^{\intercal}\kappa+B^{-1}b\right),
\end{eqnarray*}
in which $\Omega$ is the diagonal matrix with $\omega_{1:n}$ on
the diagonal and $\kappa=\left(y_{1}-\frac{c_{1}}{2},\dots,y_{n}-\frac{c_{n}}{2}\right)$.
$\mathrm{{PG}}\left(b,a\right)$ is the Pólya-gamma distribution,
for details on which see \citet{polson2013bayesian}, who devise an
efficient sampling algorithm implemented in R package \emph{BayesLogit}
(\citet{polson2013bayesian}).\end{sloppypar}

\section{Additional results}

\subsection{Multivariate normal studies\label{appendix:additional-Multivariate-normal-studies}}

\begin{figure}
\begin{centering}
\subfloat[KL divergence vs ESS.]{\includegraphics[scale=0.45]{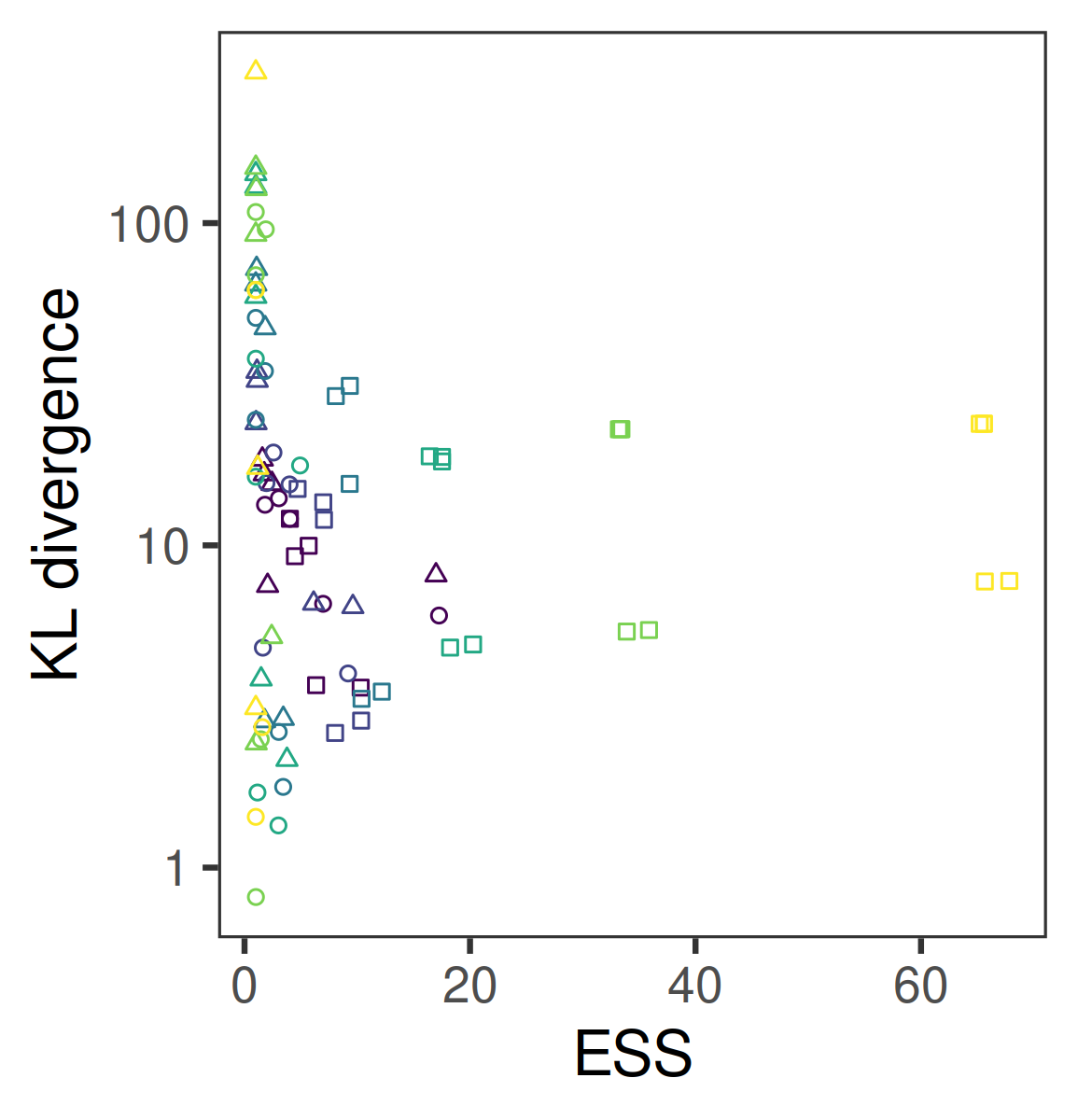}

}\quad{}\subfloat[KL divergence vs $\hat{{k}}$.]{\includegraphics[scale=0.45]{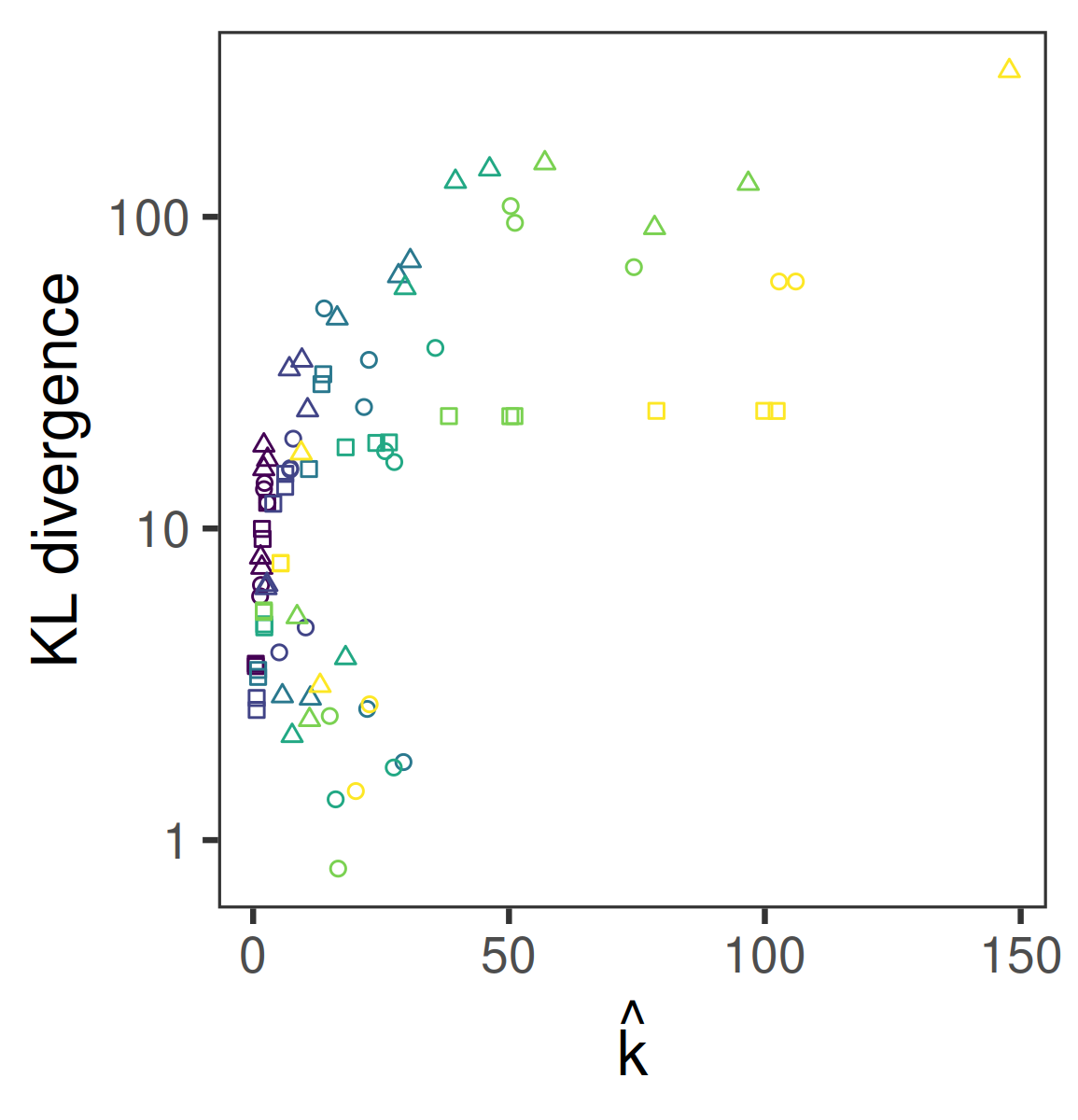}}\quad{}\includegraphics[scale=0.45]{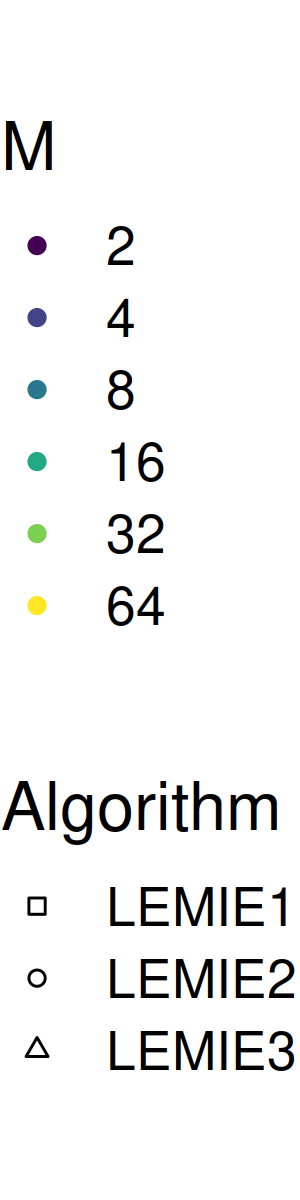}
\par\end{centering}
\caption{For the simulated examples of Section \ref{subsec:Multivariate-normal-models}
with $d=8$ and $n=10,000$, KL divergences from the $\mu$ marginal
of the posterior to the LEMIE approximations using the approach explained
in Section \ref{par:Cross-entropy-estimation} against the performance
metrics of Section \ref{subsec:Performance-indicators}.\label{fig:normal-sigma-known-mu-kl-div-all}}

\end{figure}

\begin{figure}
\begin{centering}
\subfloat[KL divergence vs ESS.]{\includegraphics[scale=0.45]{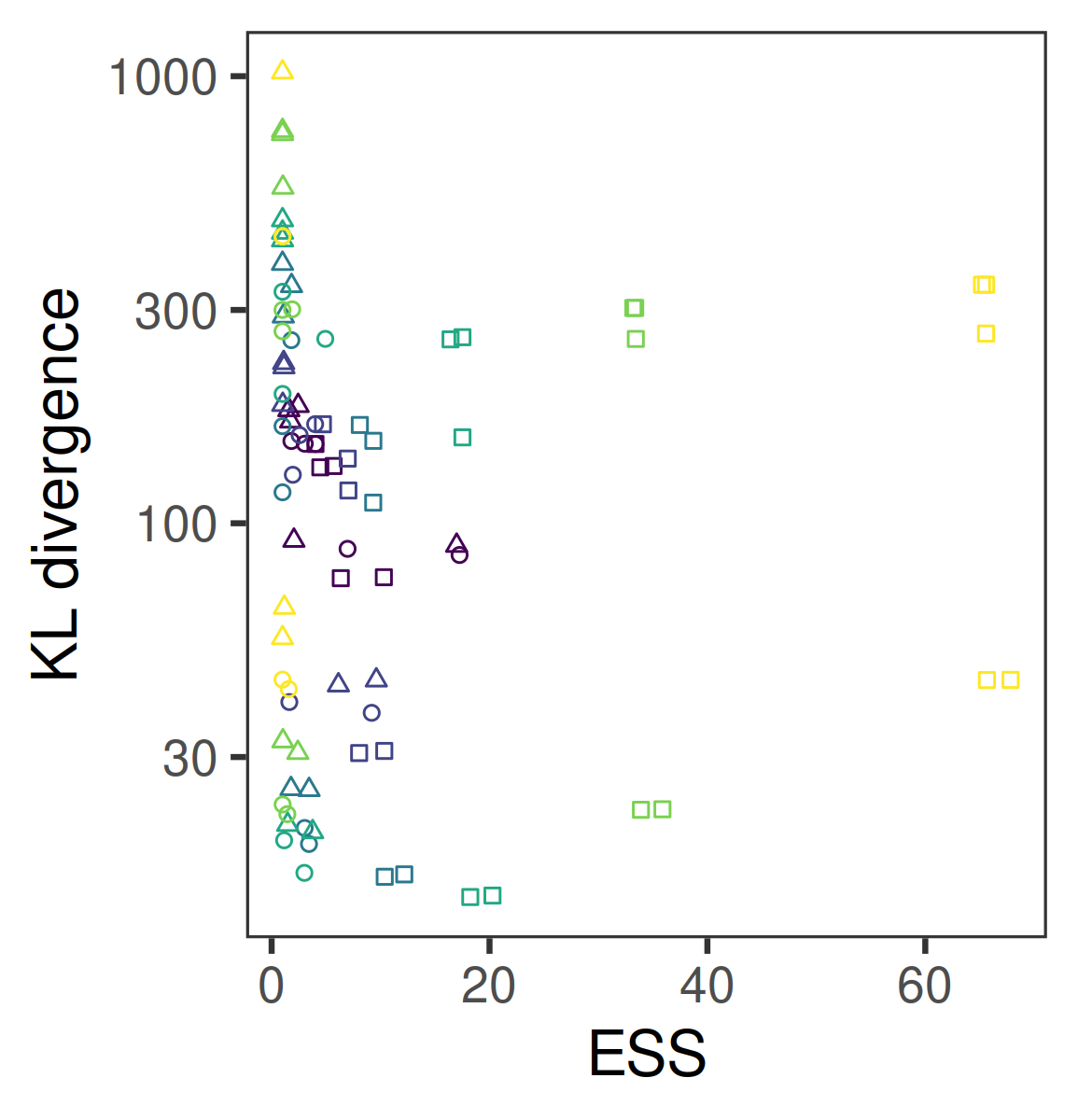}

}\quad{}\subfloat[KL divergence vs $\hat{{k}}$.]{\includegraphics[scale=0.45]{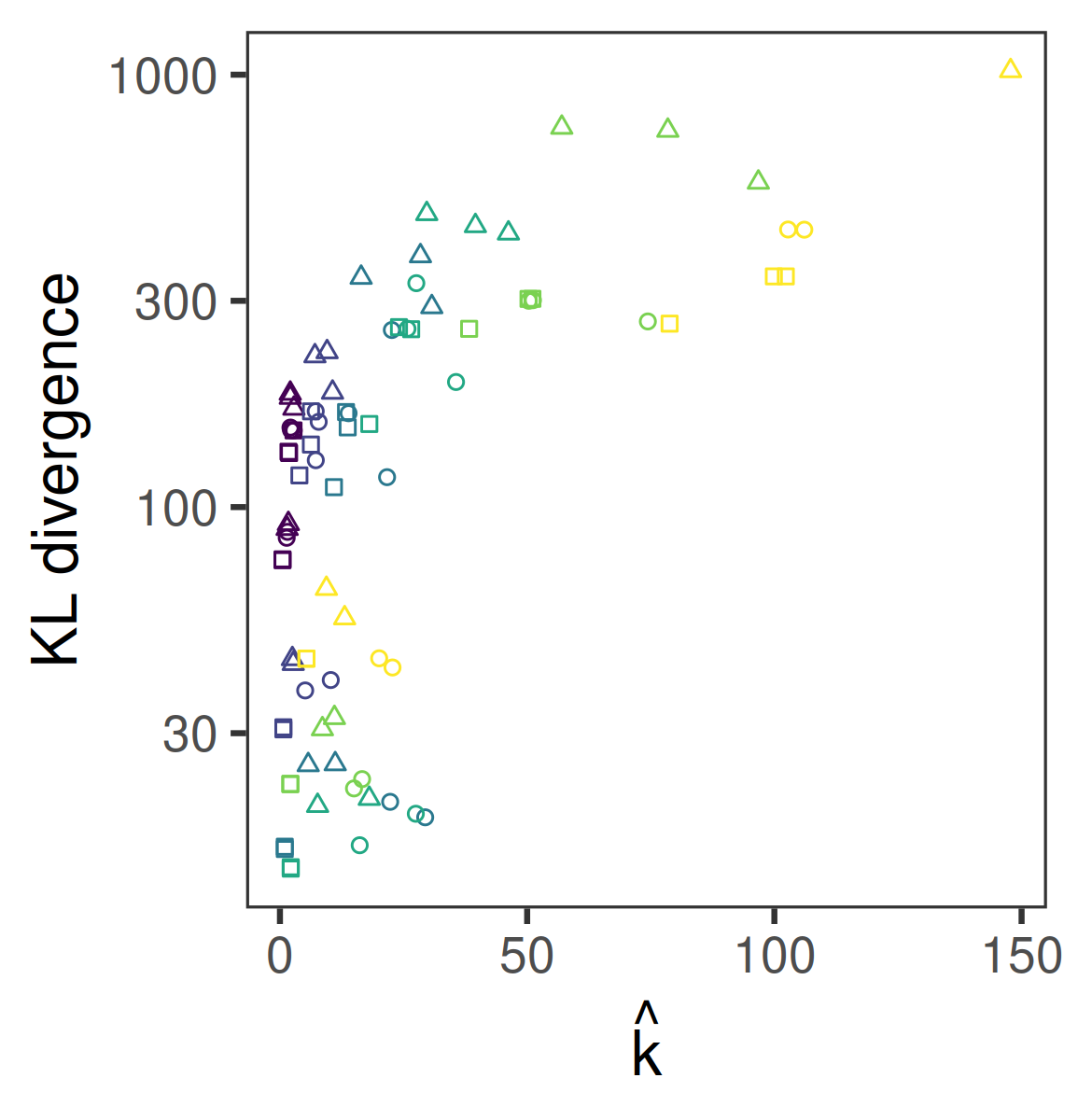}}\quad{}\includegraphics[scale=0.45]{images/experiments-normal-uncorrelated-legend-d8-n10000-log-error-mean-vs-log-khat}
\par\end{centering}
\caption{Similar to Figure \ref{fig:normal-sigma-known-mu-kl-div-all} but
of the KL divergences from the $\Sigma$ marginal of the posterior
to the LEMIE approximations.}
\end{figure}

\begin{figure}
\centering{}\subfloat[Approximating the $\mu$ marginal of the posterior.]{\includegraphics[scale=0.45]{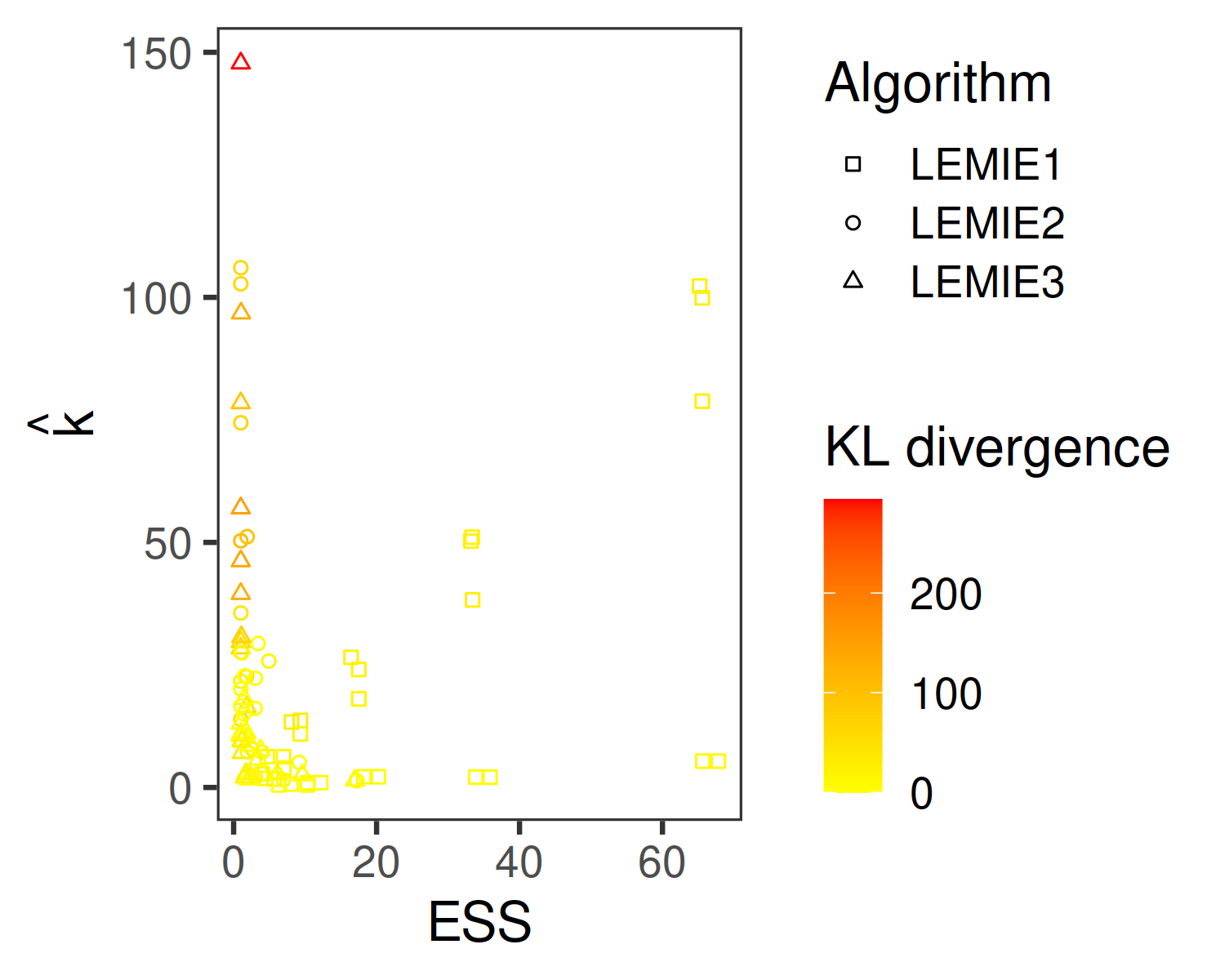}

}\quad{}\subfloat[Approximating the $\Sigma$ marginal of the posterior.]{\includegraphics[scale=0.45]{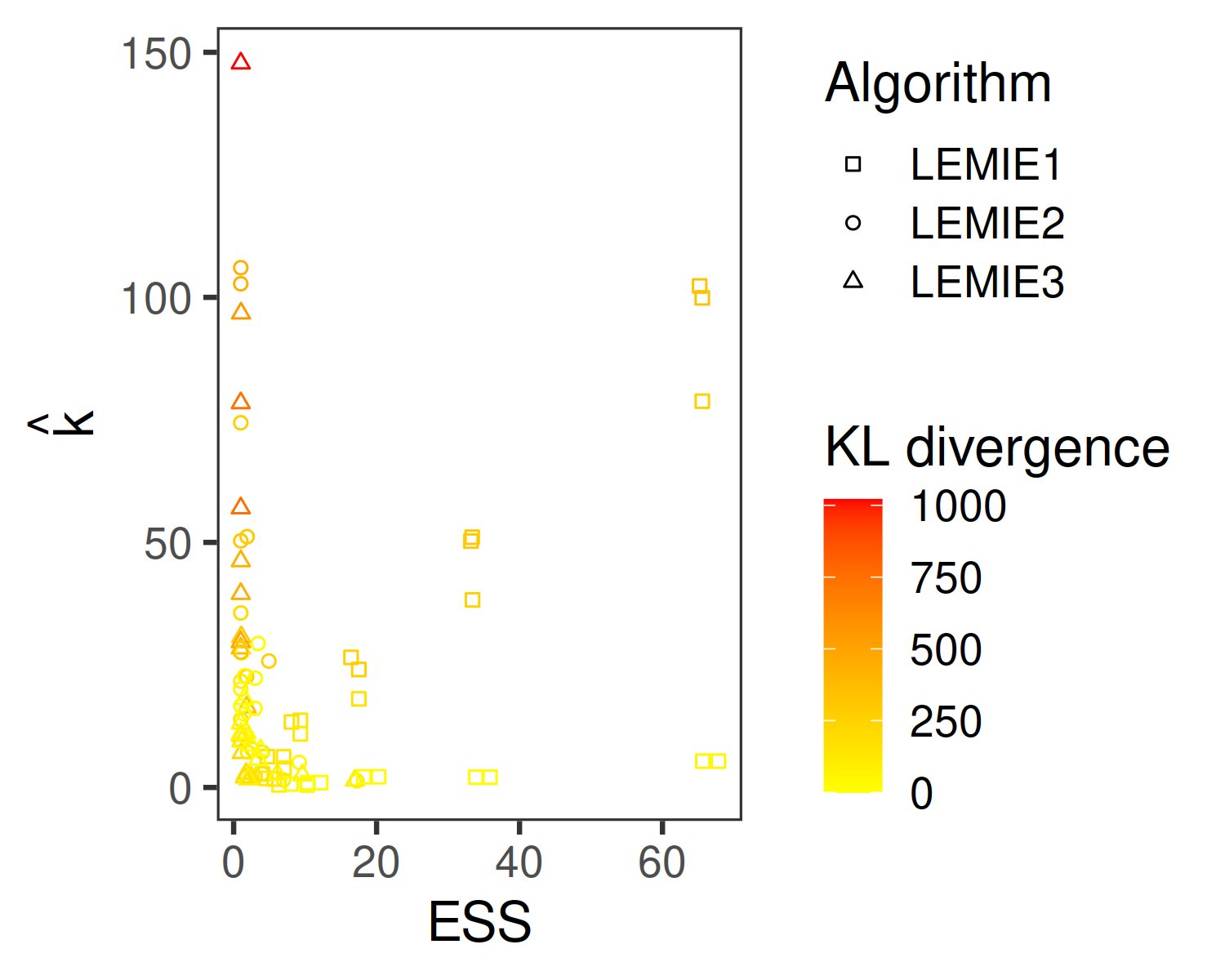}}\caption{For the simulated examples of Section \ref{subsec:Multivariate-normal-models},
KL divergences Equation \ref{eq:kl-divergence} from the posterior
to the LEMIE approximations using the approach explained in Section
\ref{par:Cross-entropy-estimation} against the performance metrics
of Section \ref{subsec:Performance-indicators}.}
\end{figure}

\begin{figure}
\begin{centering}
\subfloat[Error in estimating the 2.5\% quantiles of the $\mu$ marginal of
the posterior.]{\includegraphics[scale=0.55]{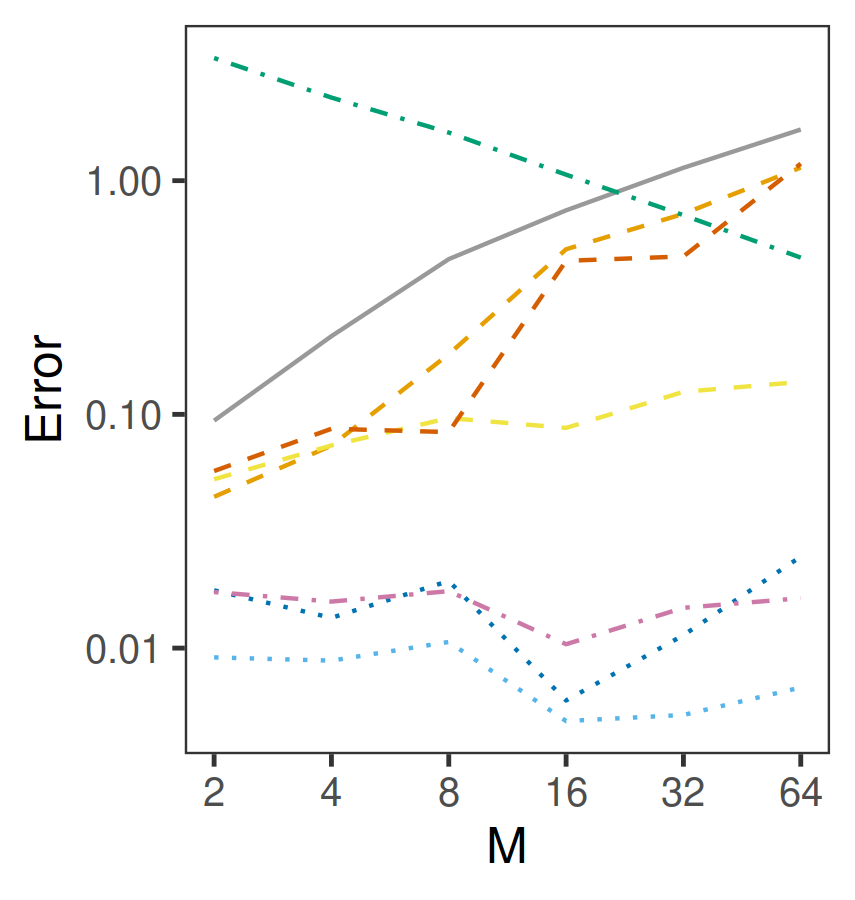}}\quad{}\subfloat[Error in estimating the 97.5\% quantiles of the $\mu$ marginal of
the posterior.]{\includegraphics[scale=0.55]{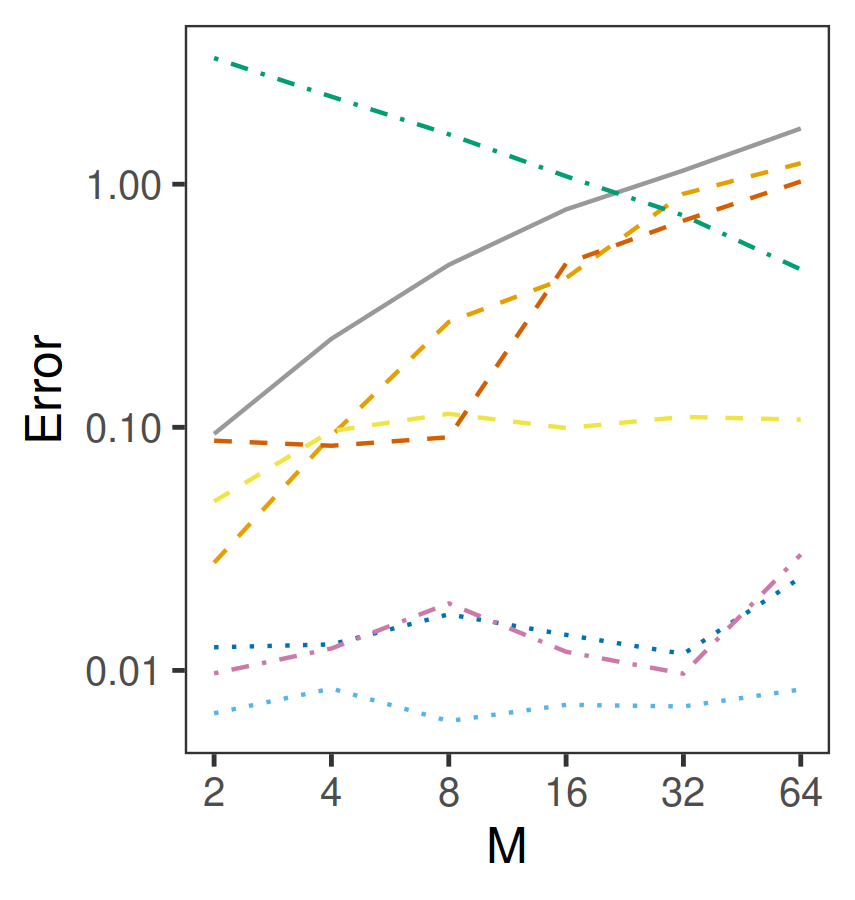}

}\quad{}\includegraphics[scale=0.55]{images/experiments-normal-uncorrelated-legend-d8-n10000-m-vs-n-alt-presentation}
\par\end{centering}
\begin{centering}
\subfloat[Error in estimating the 2.5\% quantiles of the $\Sigma$ marginal
of the posterior.]{\includegraphics[scale=0.55]{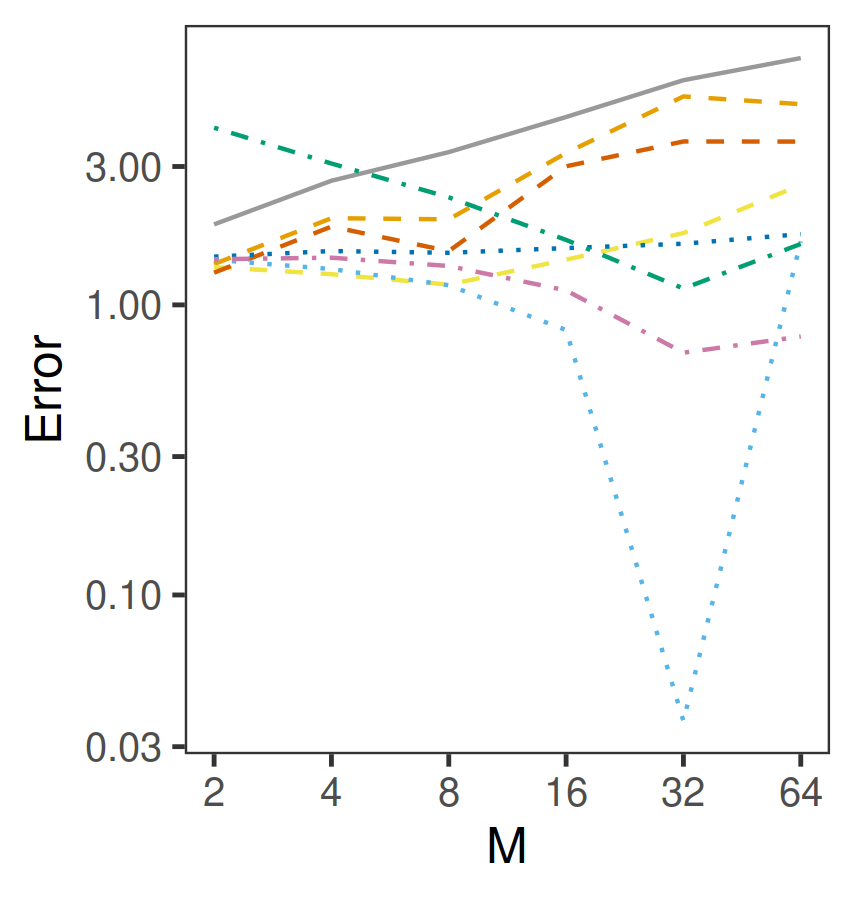}}\quad{}\subfloat[Error in estimating the 97.5\% quantiles of the $\Sigma$ marginal
of the posterior.]{\includegraphics[scale=0.55]{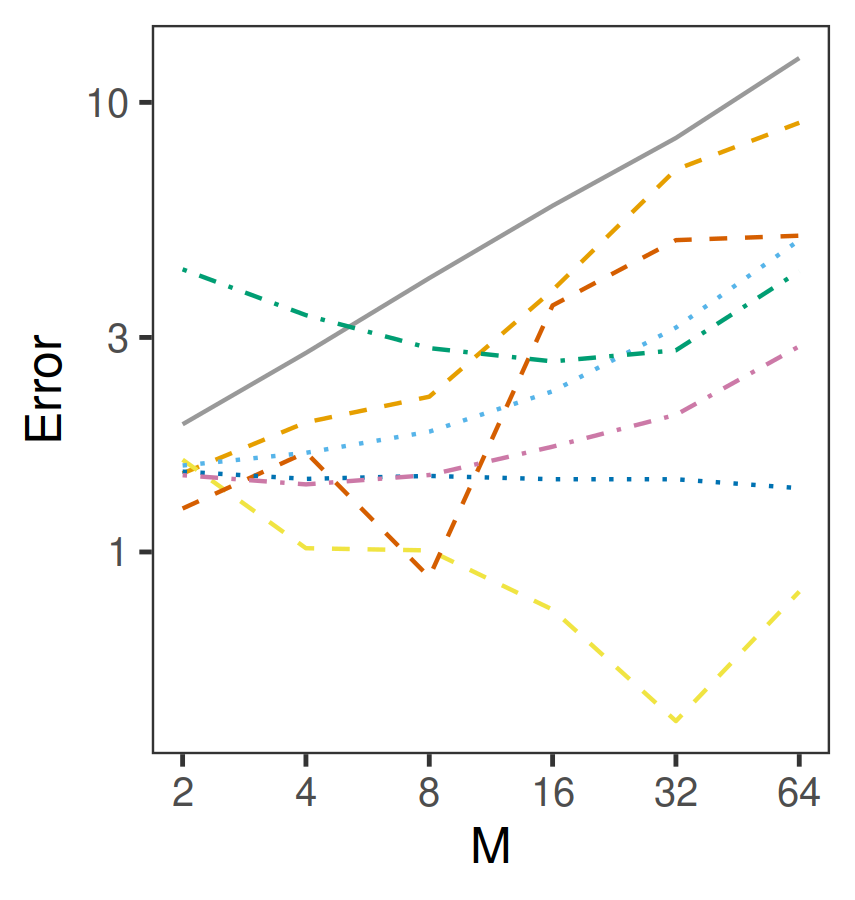}}\qquad{}\qquad{}\enskip{}
\par\end{centering}
\caption{Posterior approximation comparisons in the MVN example of Section
\ref{subsec:Multivariate-normal-models} with $\mu$ and $\Sigma$
unknown, $d=8$ and $n=10,000$. \label{fig:normal-sigma-unknown-comparisons-2}}
\end{figure}

\begin{figure}
\begin{centering}
\subfloat[KL divergence from the approximations to the posterior of $\mu$.]{\includegraphics[scale=0.55]{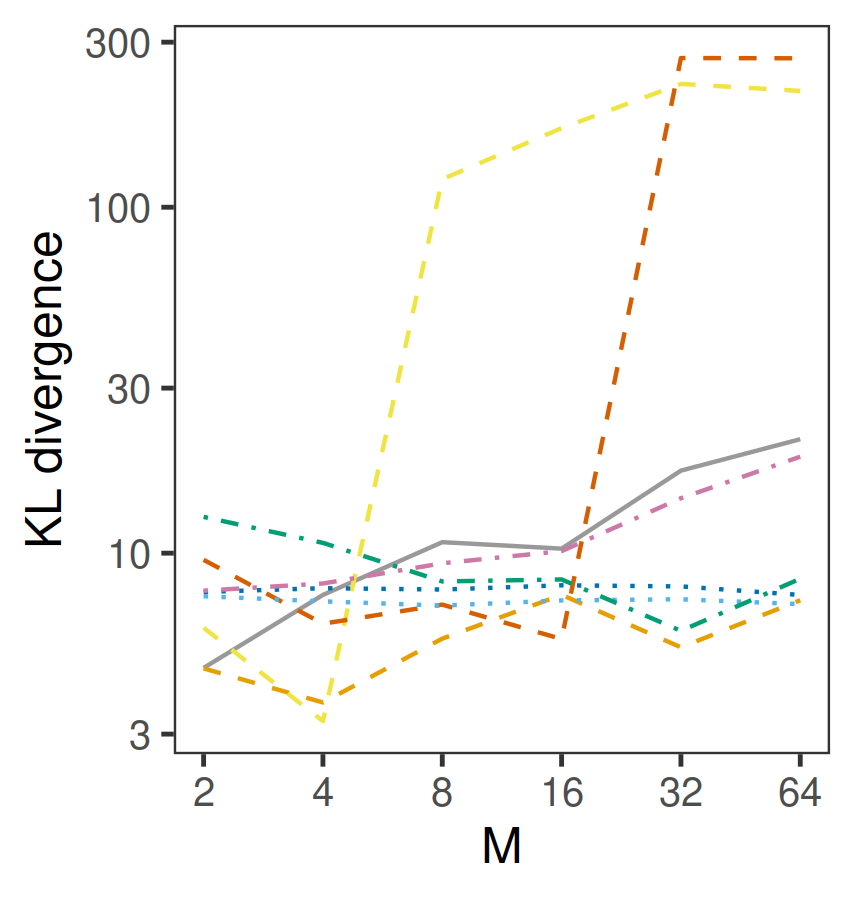}}\quad{}\subfloat[KL divergence from the approximations to the posterior of $\Sigma$.]{\includegraphics[scale=0.55]{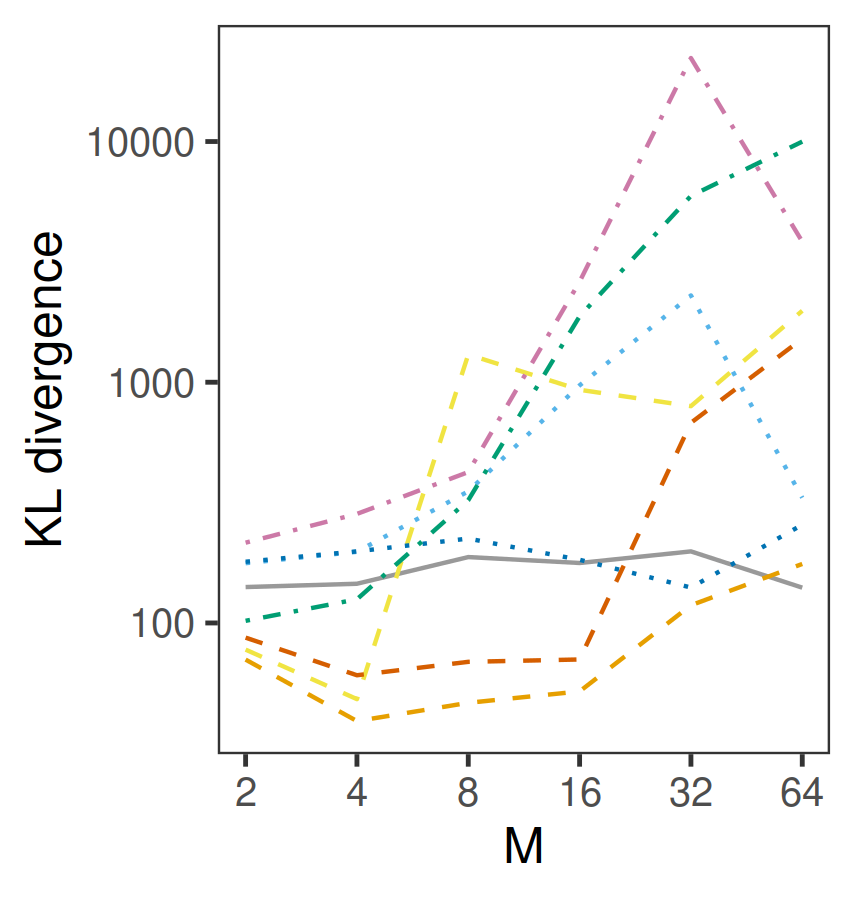}

}\quad{}\includegraphics[scale=0.55]{images/experiments-normal-uncorrelated-legend-d8-n10000-m-vs-n-alt-presentation}
\par\end{centering}
\begin{centering}
\subfloat[Error in estimating the posterior mean of $\mu$.]{\includegraphics[scale=0.55]{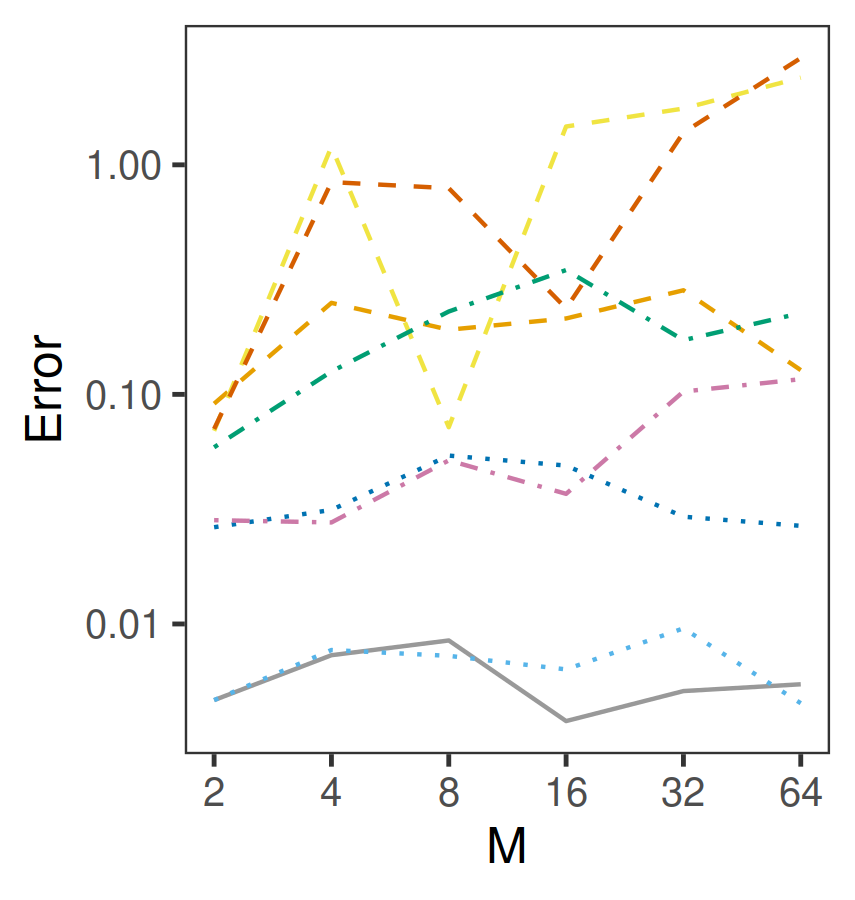}}\quad{}\subfloat[Error in estimating the posterior mean of $\Sigma$.]{\includegraphics[scale=0.55]{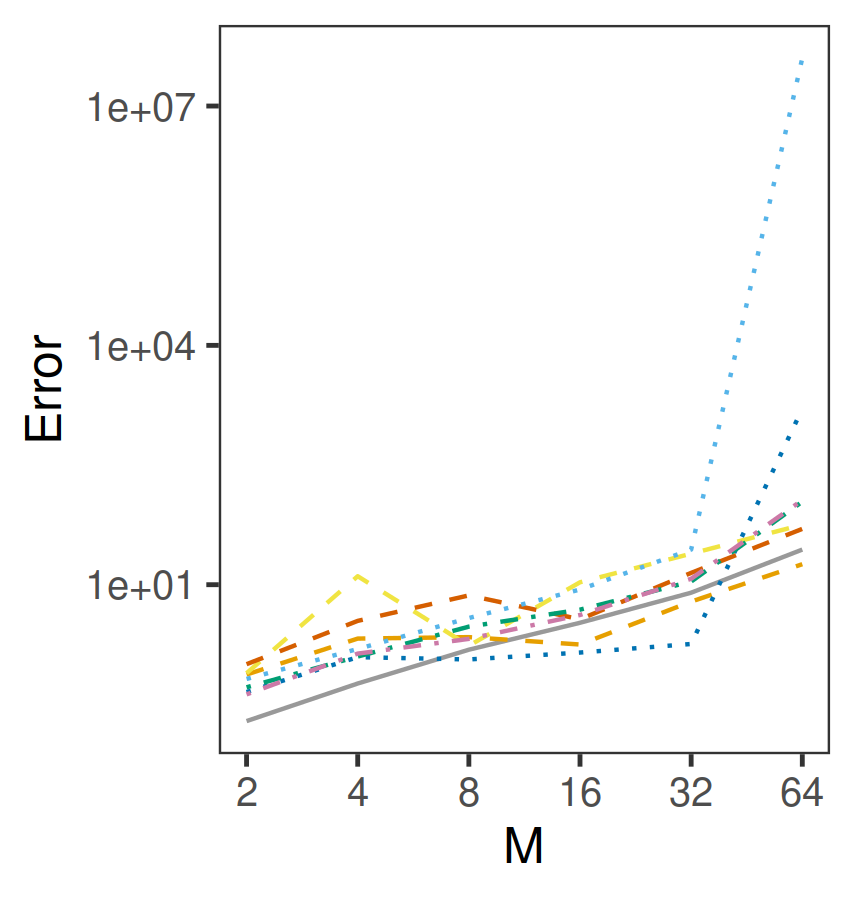}}\qquad{}\qquad{}\enskip{}
\par\end{centering}
\caption{Results similar to Figure \ref{fig:normal-sigma-unknown-comparisons-1}
but with $d=8$ and $n=1,088$.\label{fig:normal-sigma-unknown-comparisons-1-2}}
\end{figure}

\begin{figure}
\begin{centering}
\subfloat[KL divergence from the approximations to the posterior of $\mu$.]{\includegraphics[scale=0.55]{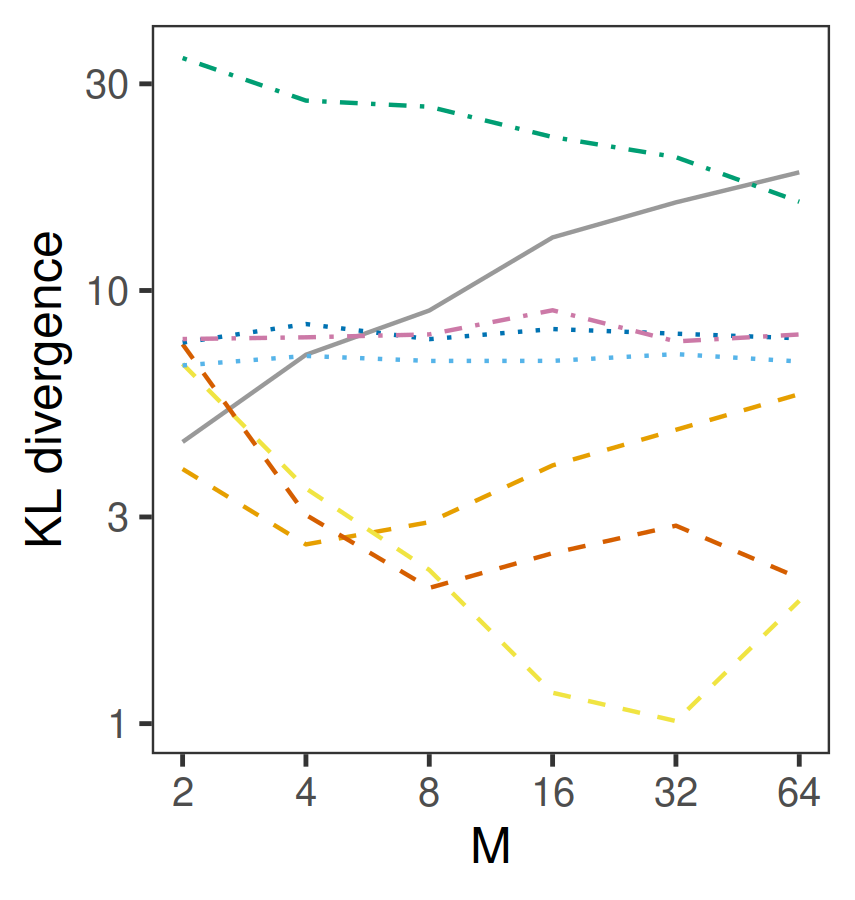}}\quad{}\subfloat[KL divergence from the approximations to the posterior of $\Sigma$.]{\includegraphics[scale=0.55]{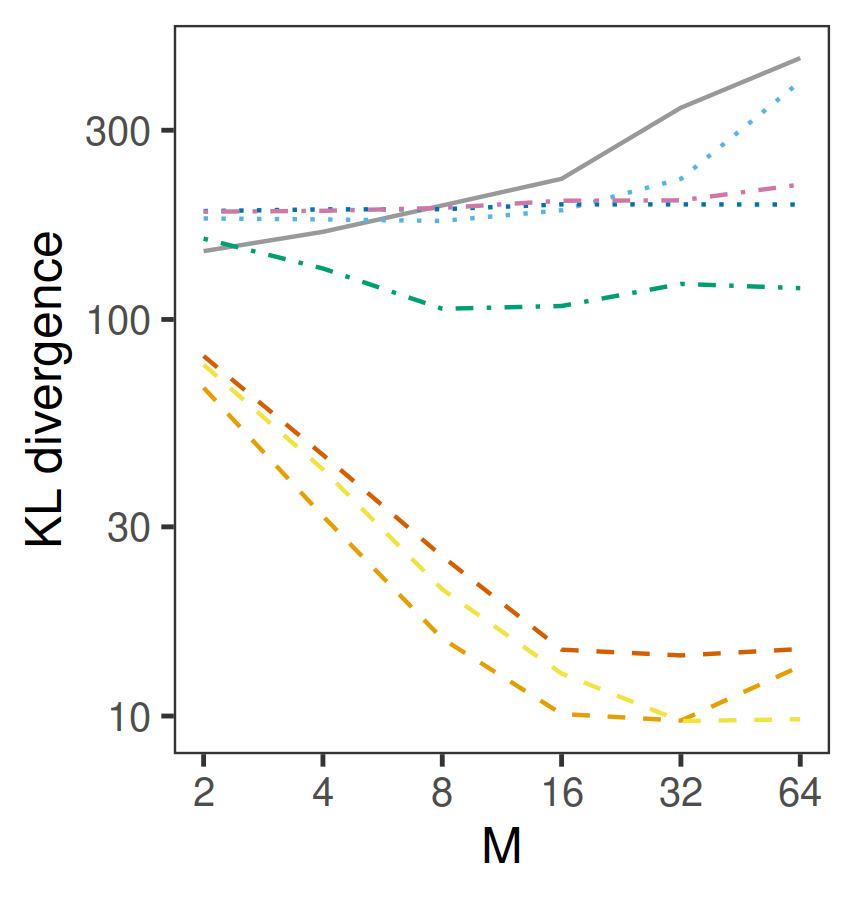}

}\quad{}\includegraphics[scale=0.55]{images/experiments-normal-uncorrelated-legend-d8-n10000-m-vs-n-alt-presentation}
\par\end{centering}
\begin{centering}
\subfloat[Error in estimating the posterior mean of $\mu$.]{\includegraphics[scale=0.55]{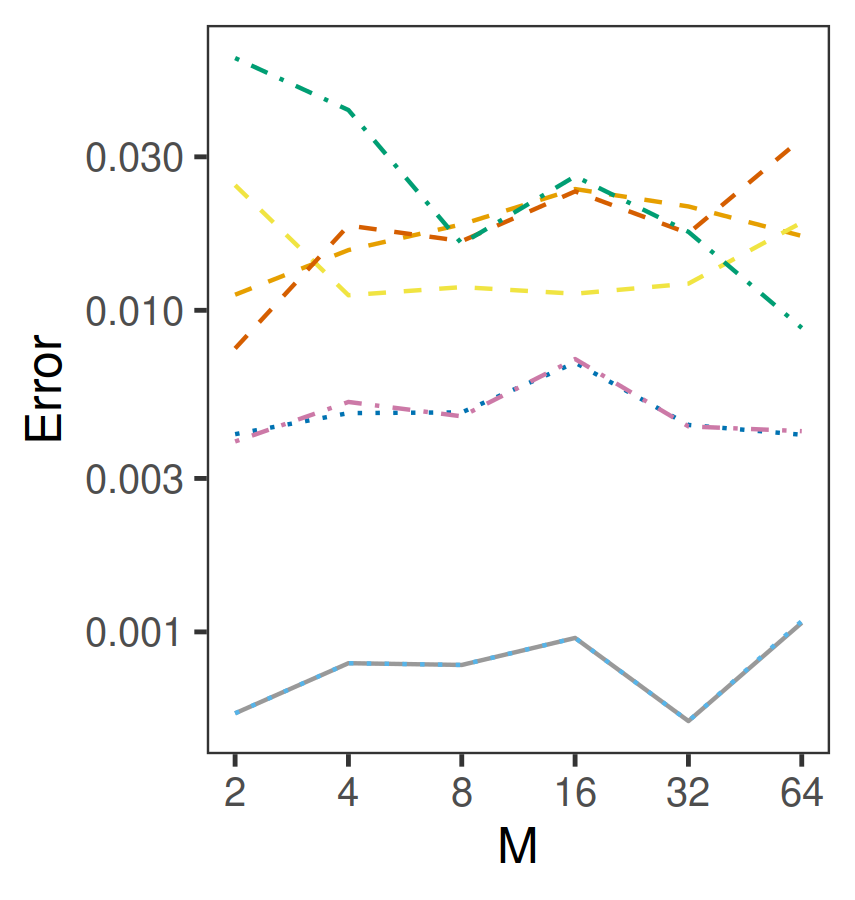}}\quad{}\subfloat[Error in estimating the posterior mean of $\Sigma$.]{\includegraphics[scale=0.55]{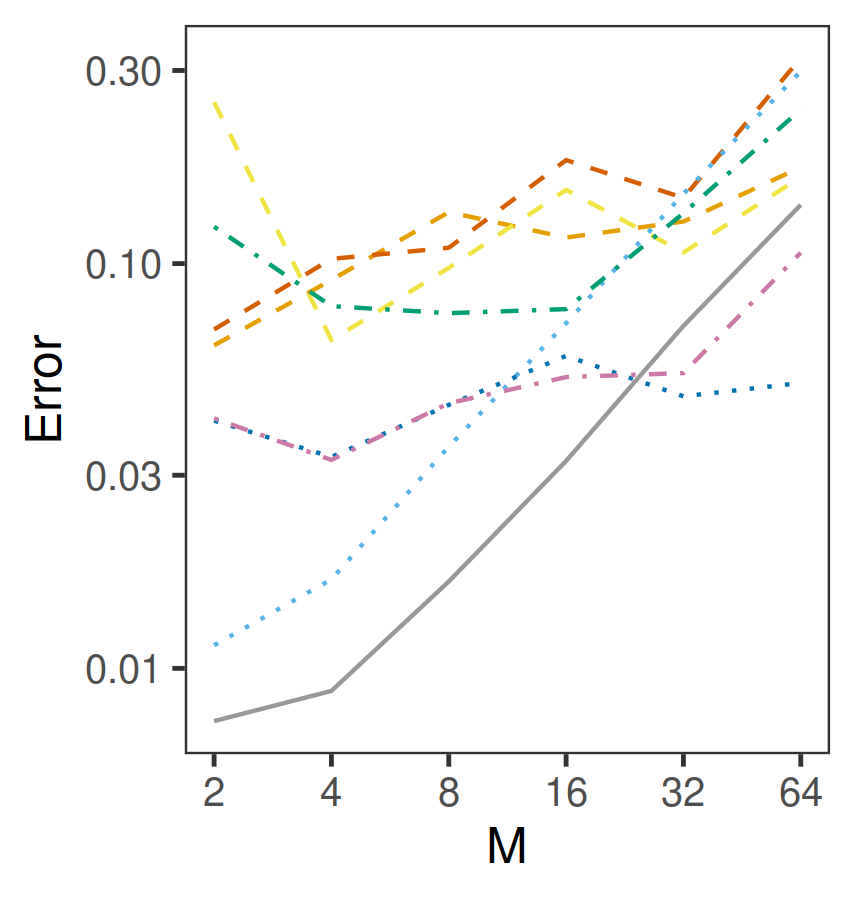}}\qquad{}\qquad{}\enskip{}
\par\end{centering}
\caption{Results similar to Figure \ref{fig:normal-sigma-unknown-comparisons-1}
but with $d=8$ and $n=100,000$.\label{fig:normal-sigma-unknown-comparisons-1-3}}
\end{figure}

\begin{figure}
\begin{raggedright}
\subfloat[]{\includegraphics[scale=0.45]{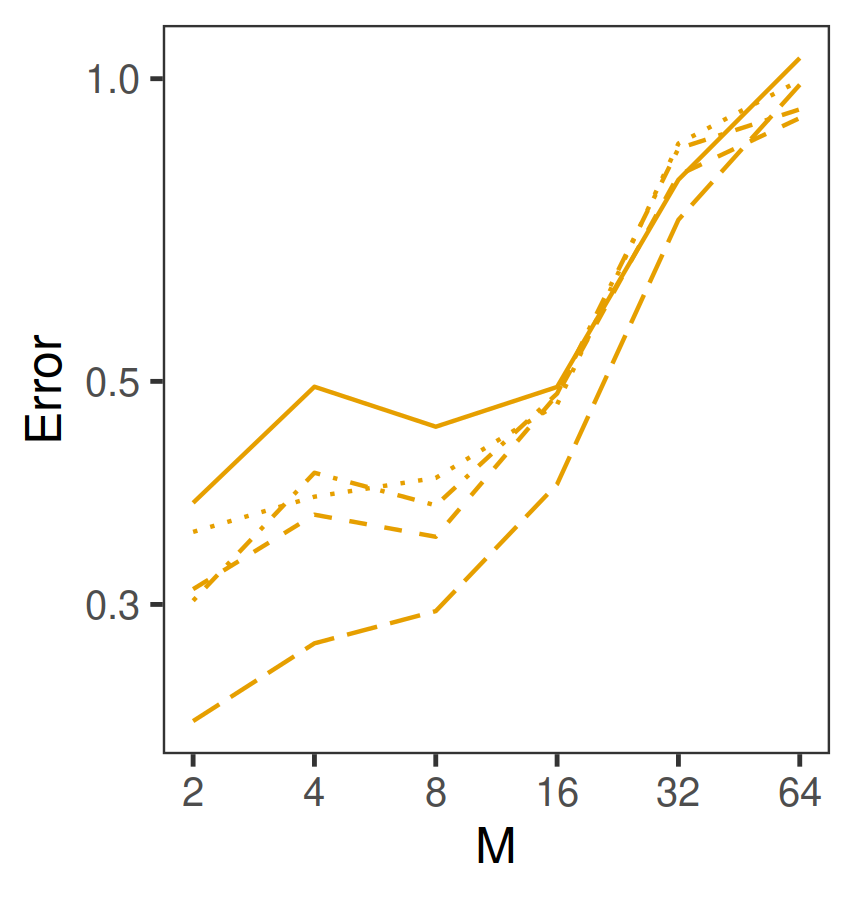}}\enskip{}\subfloat[]{\includegraphics[scale=0.45]{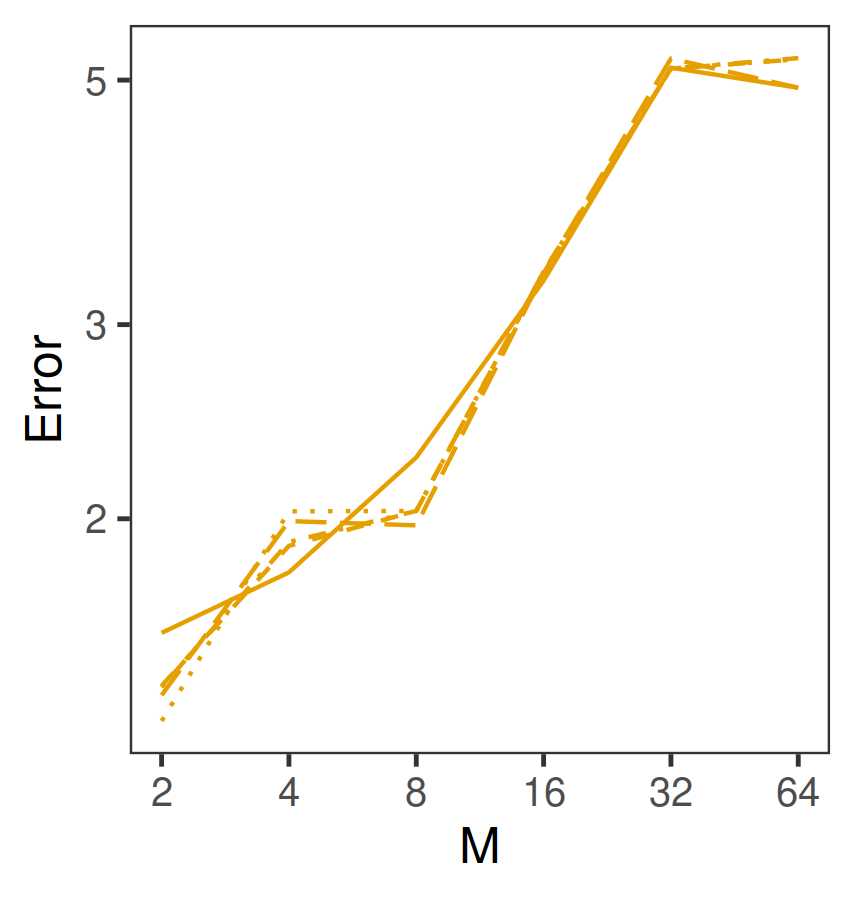}}\enskip{}\subfloat[]{\includegraphics[scale=0.45]{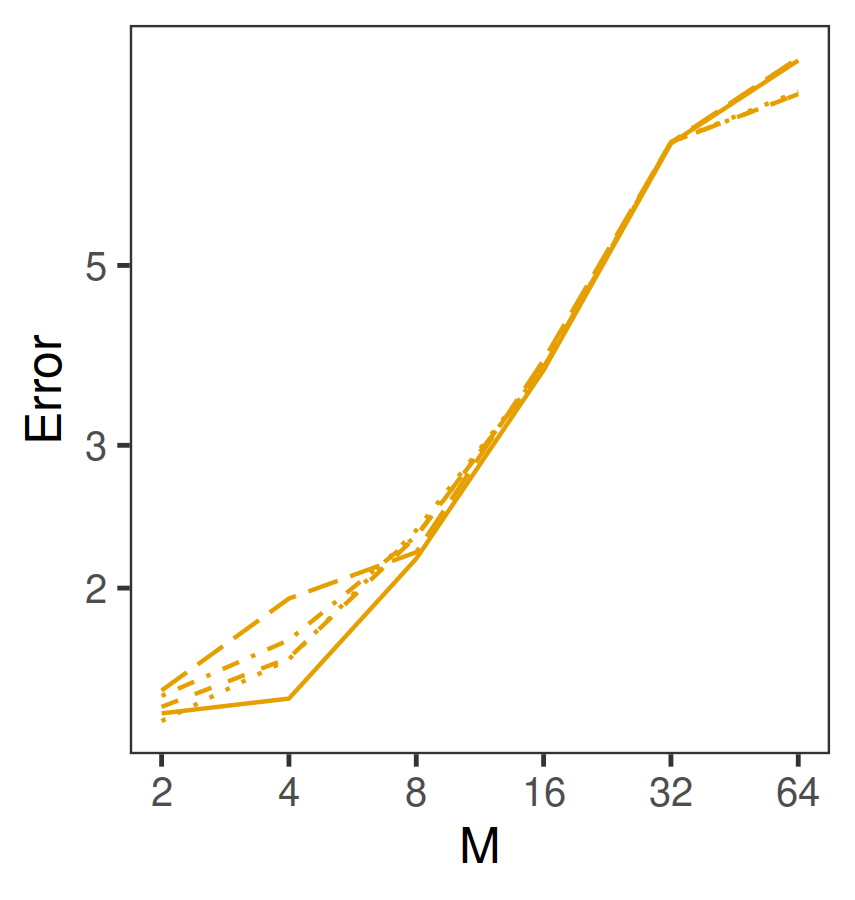}}\enskip{}\includegraphics[scale=0.45]{images/experiments-logistic-scott-legend-m-alt-lemie-presentation}
\par\end{raggedright}
\begin{raggedright}
\subfloat[]{\includegraphics[scale=0.45]{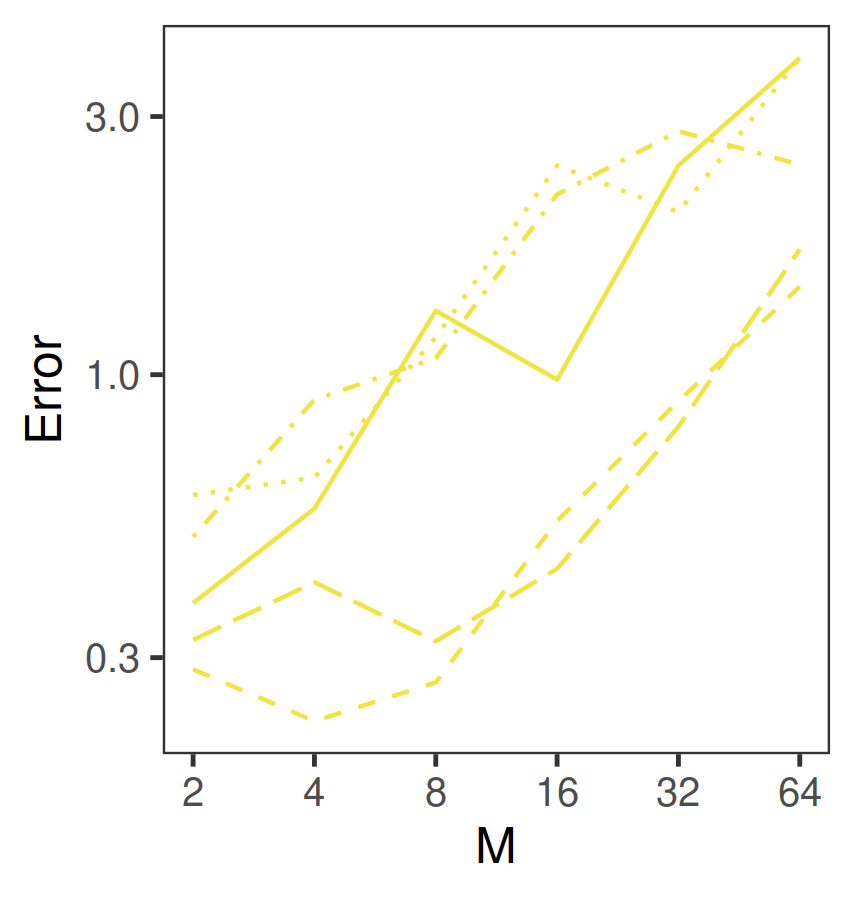}}\enskip{}\subfloat[]{\includegraphics[scale=0.45]{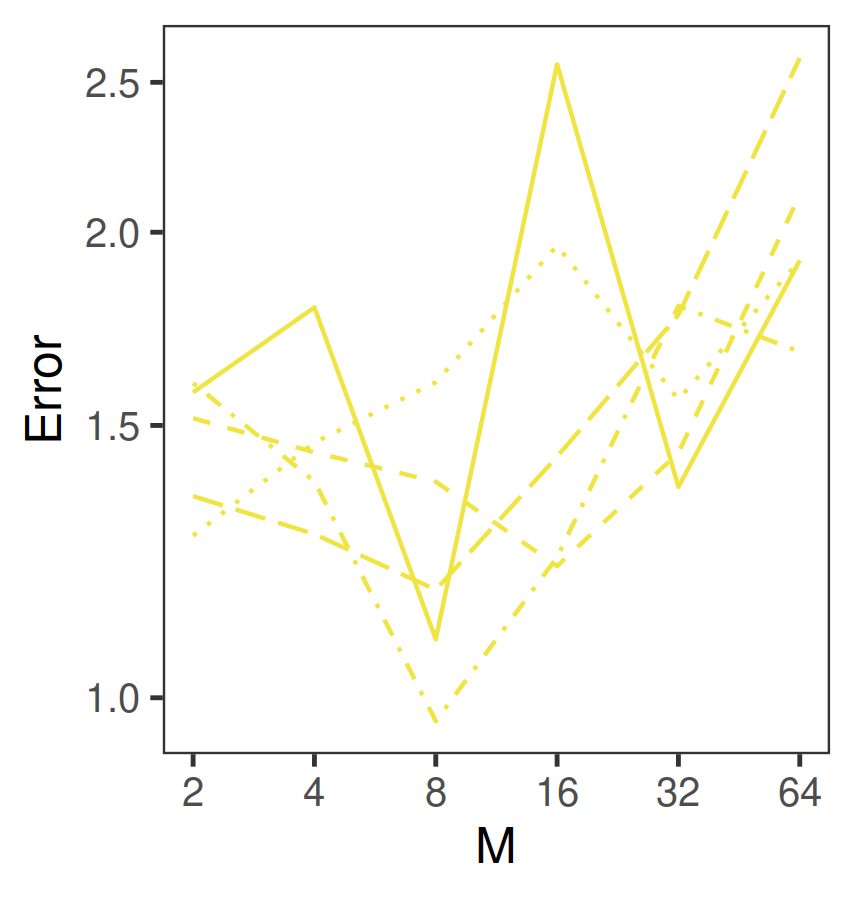}}\enskip{}\subfloat[]{\includegraphics[scale=0.45]{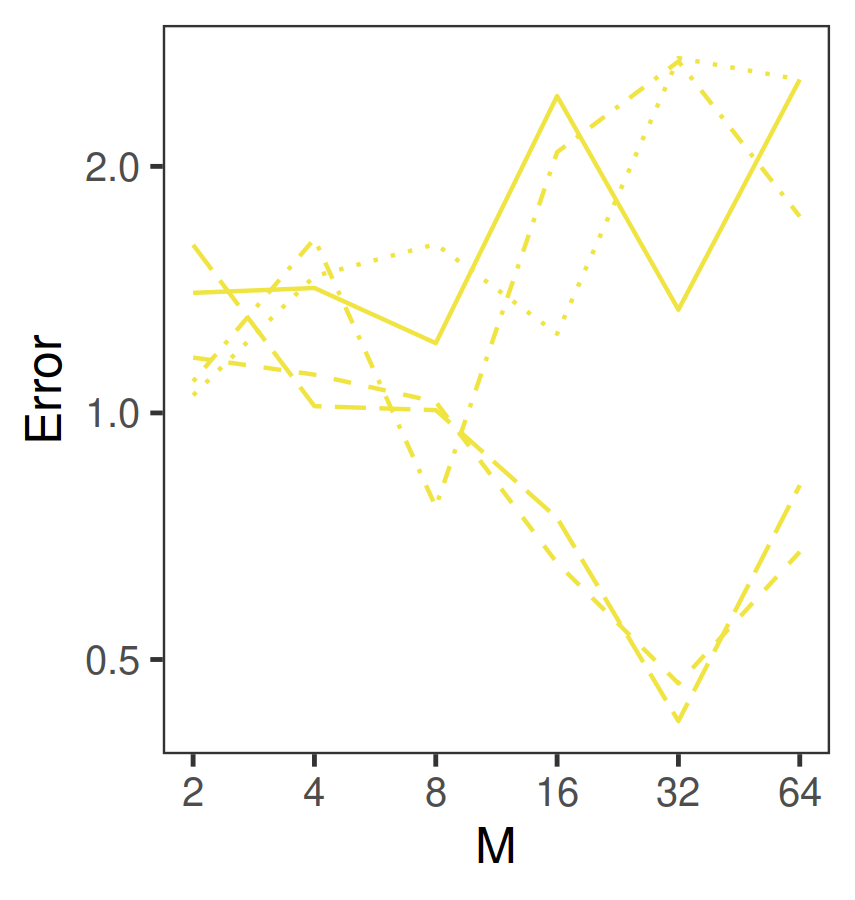}}\enskip{}\includegraphics[scale=0.45]{images/experiments-logistic-scott-class-legend-m-alt-presentation}
\par\end{raggedright}
\begin{raggedright}
\subfloat[]{\includegraphics[scale=0.45]{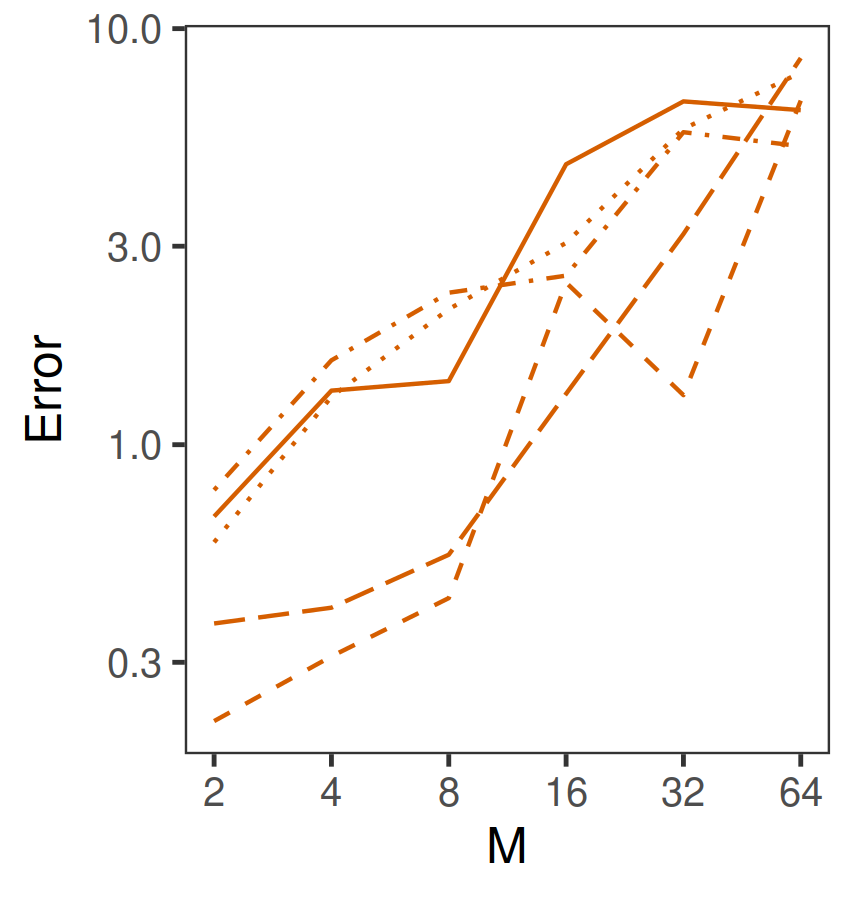}}\enskip{}\subfloat[]{\includegraphics[scale=0.45]{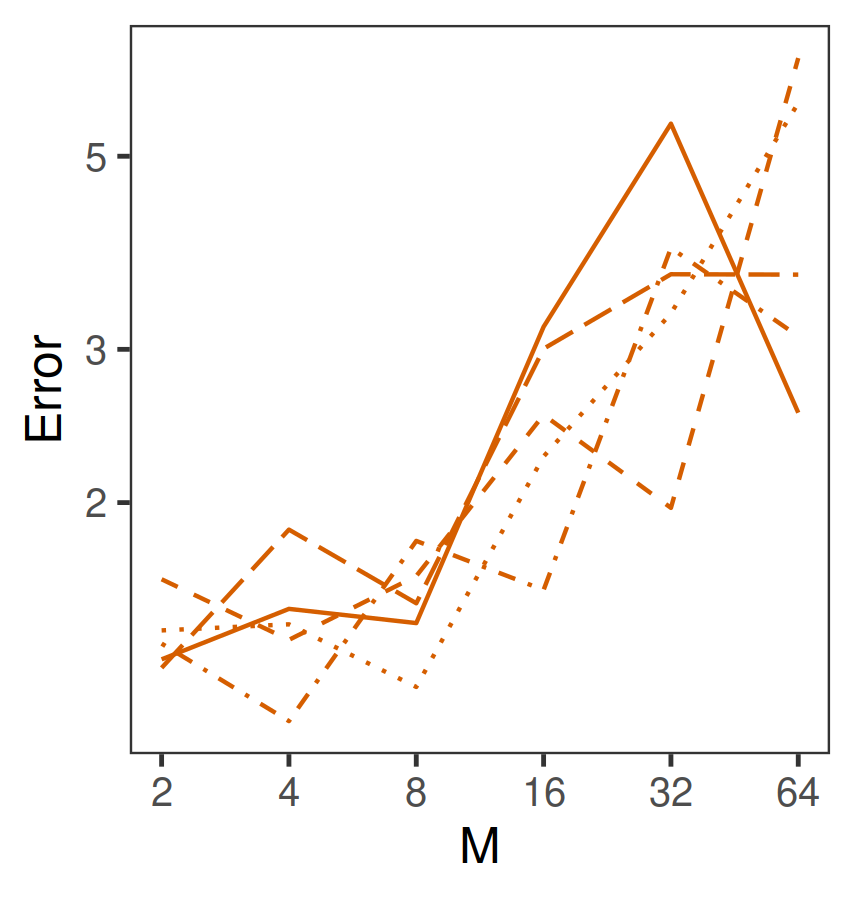}}\enskip{}\subfloat[]{\includegraphics[scale=0.45]{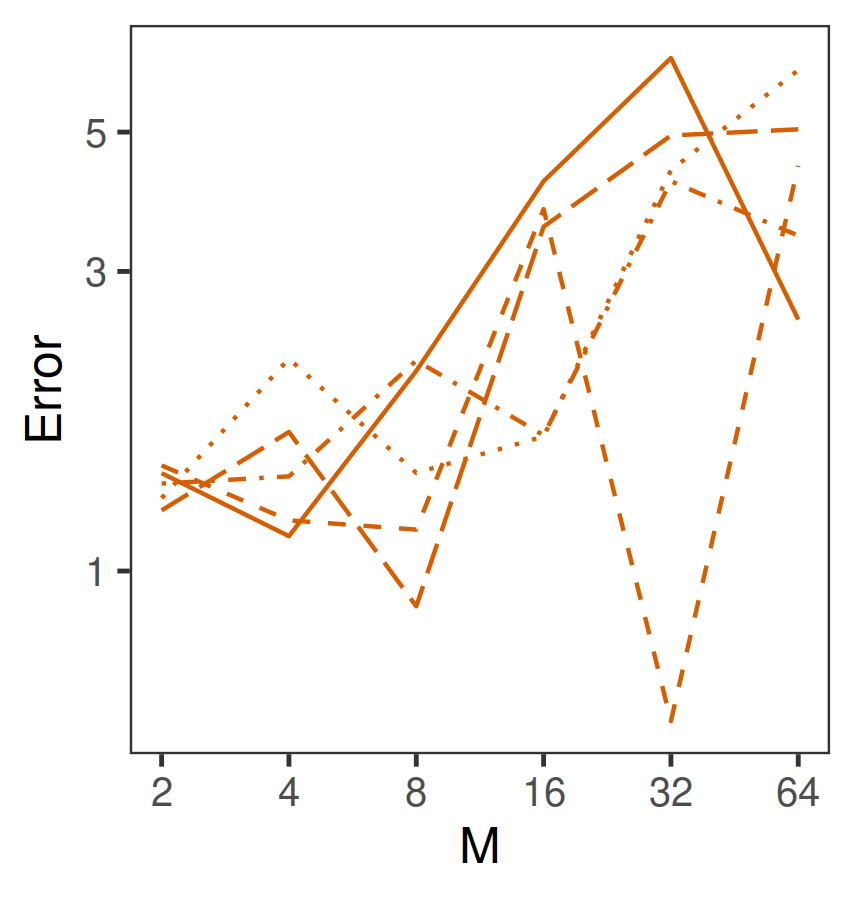}}
\par\end{raggedright}
\begin{raggedright}
\subfloat[]{\includegraphics[scale=0.45]{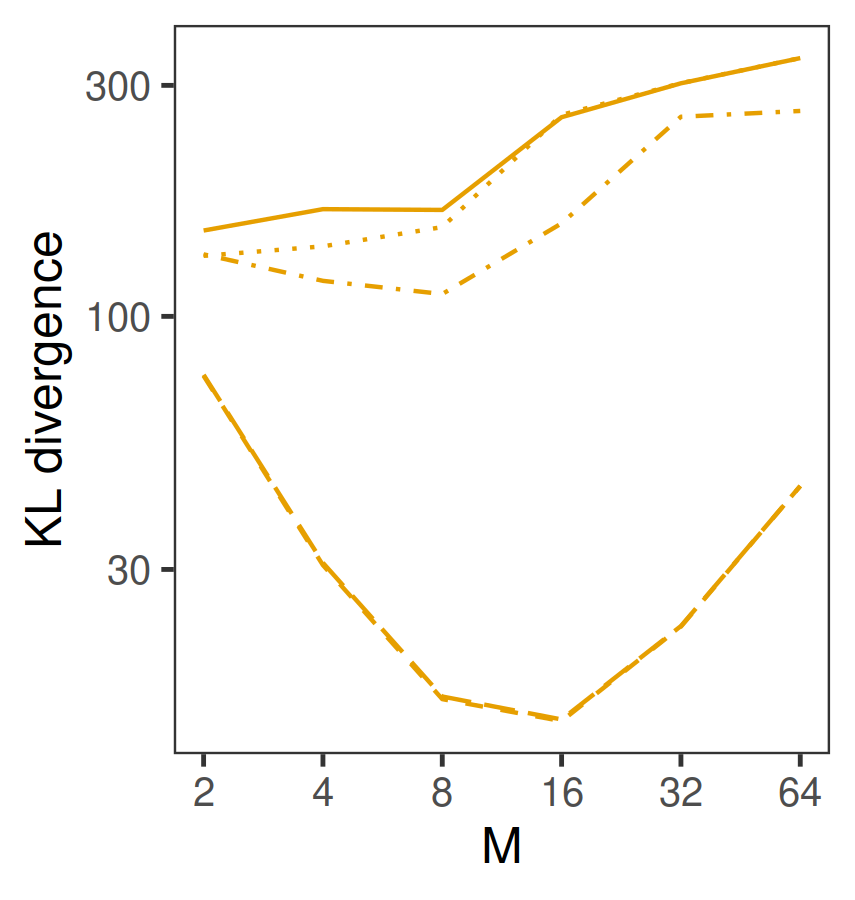}}\enskip{}\subfloat[]{\includegraphics[scale=0.45]{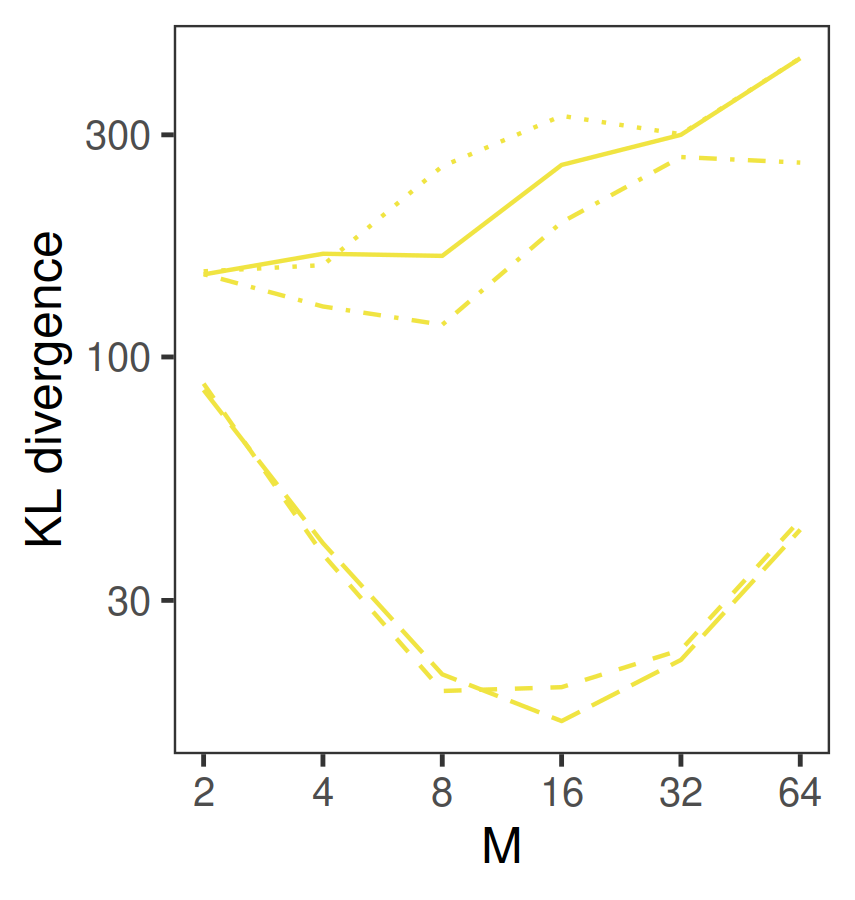}}\enskip{}\subfloat[]{\includegraphics[scale=0.45]{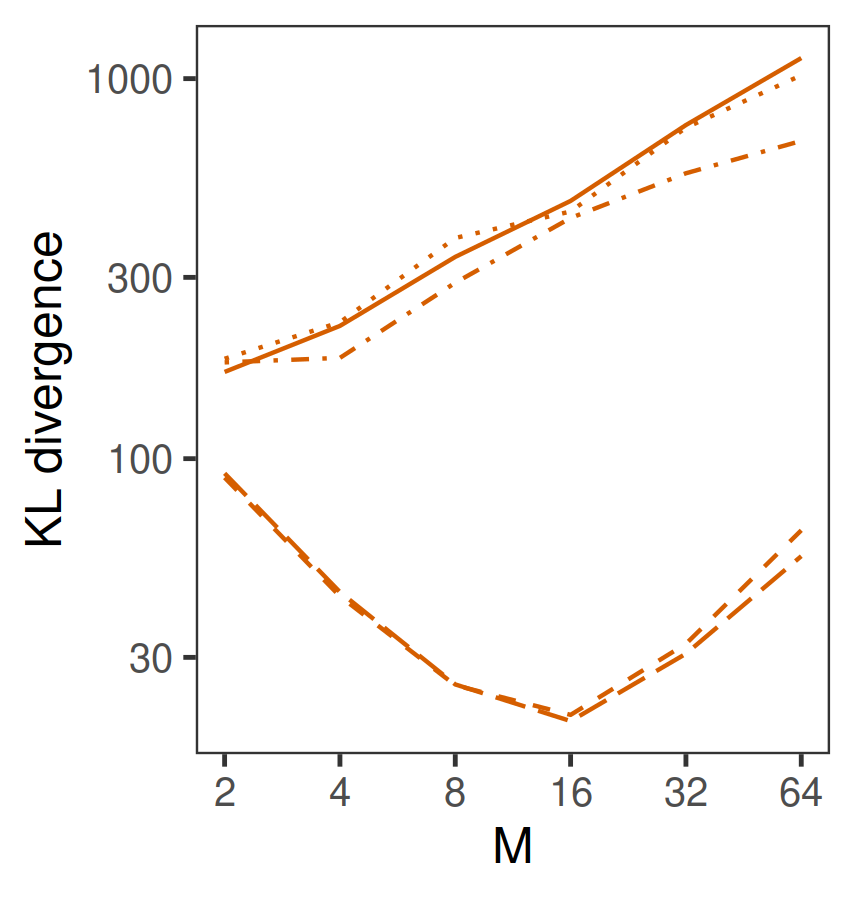}}
\par\end{raggedright}
\caption{Posterior approximation comparisons for the $\Sigma$ marginal of
the posterior in the MVN example of Section \ref{subsec:Multivariate-normal-models}
with $\Sigma$ unknown and $d=8$ and $n=10,000$. The error in estimating
(a)(d)(g) the posterior mean, (b)(e)(h) the 2.5\% quantiles, (c)(f)(i)
the 97.5\% quantiles. (j)(k)(l) The KL divergence from the marginal
of the posterior to each approximation.\label{fig:normal-sigma-unknown-comparisons-lemie-1}}
\end{figure}

\begin{figure}
\begin{centering}
\subfloat[KL divergence from the approximations to the posterior of $\mu$.]{\includegraphics[scale=0.55]{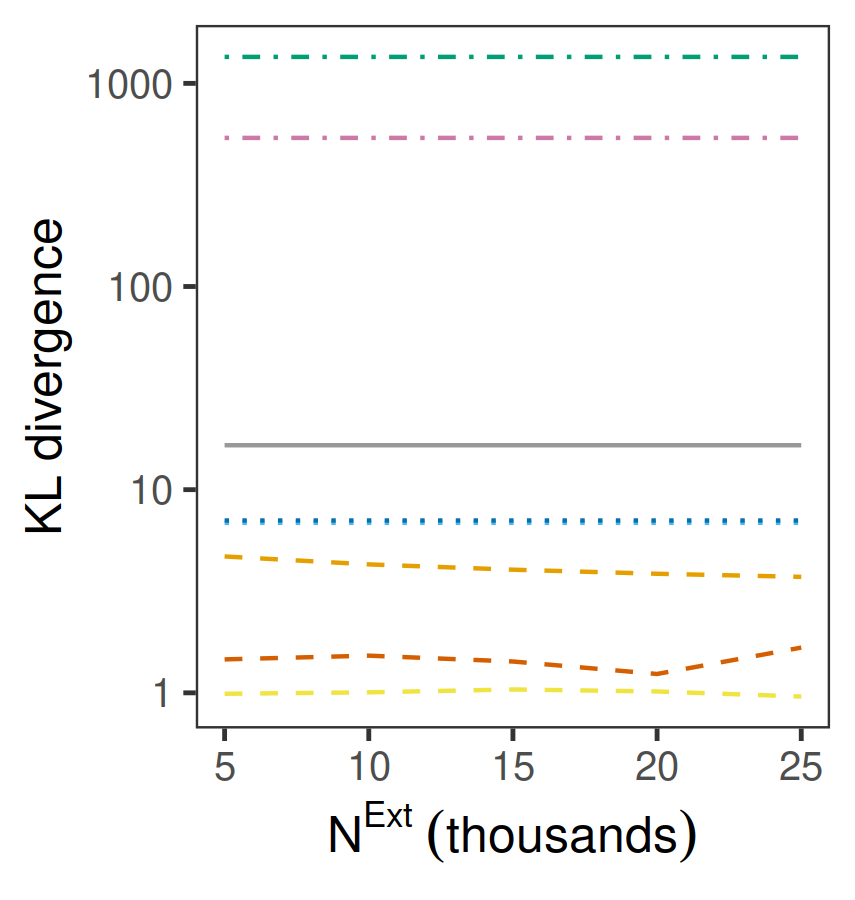}}\quad{}\subfloat[Error in estimating the mean of $\mu$.]{\includegraphics[scale=0.55]{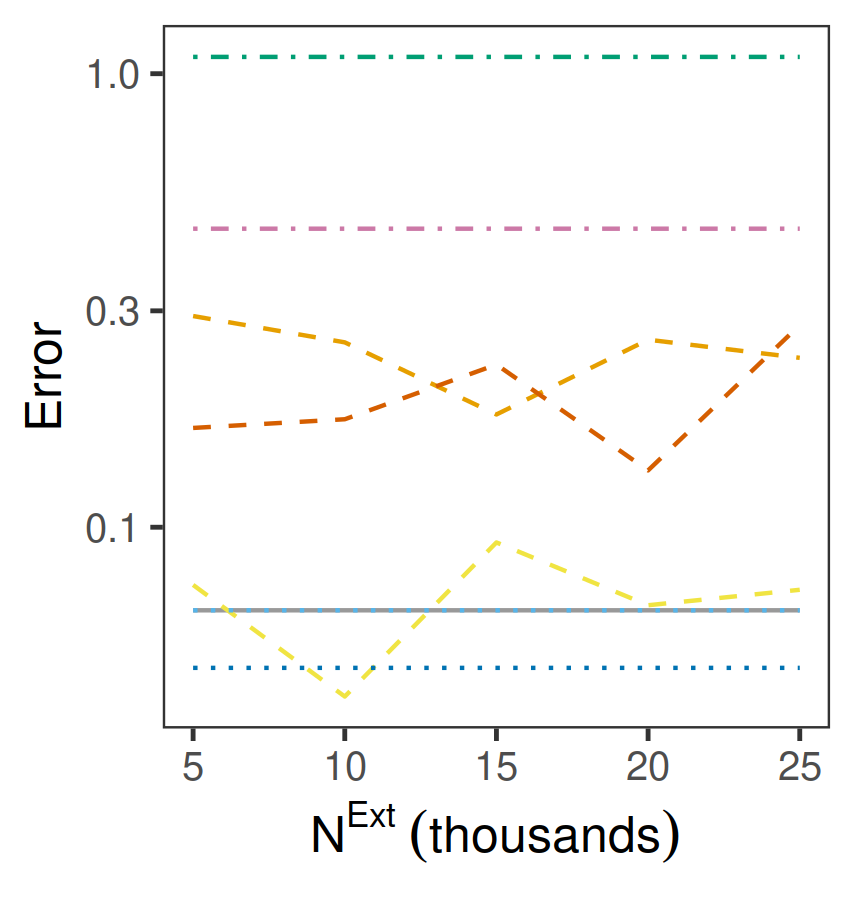}

}\quad{}\includegraphics[scale=0.55]{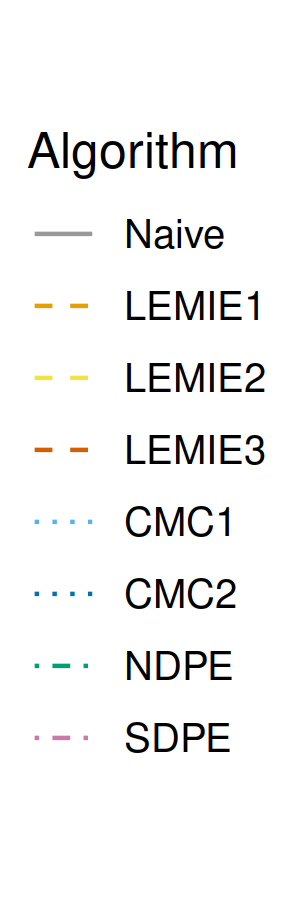}
\par\end{centering}
\begin{centering}
\subfloat[Error in estimating the 2.5\% quantiles of the $\mu$ marginal of
the posterior.]{\includegraphics[scale=0.55]{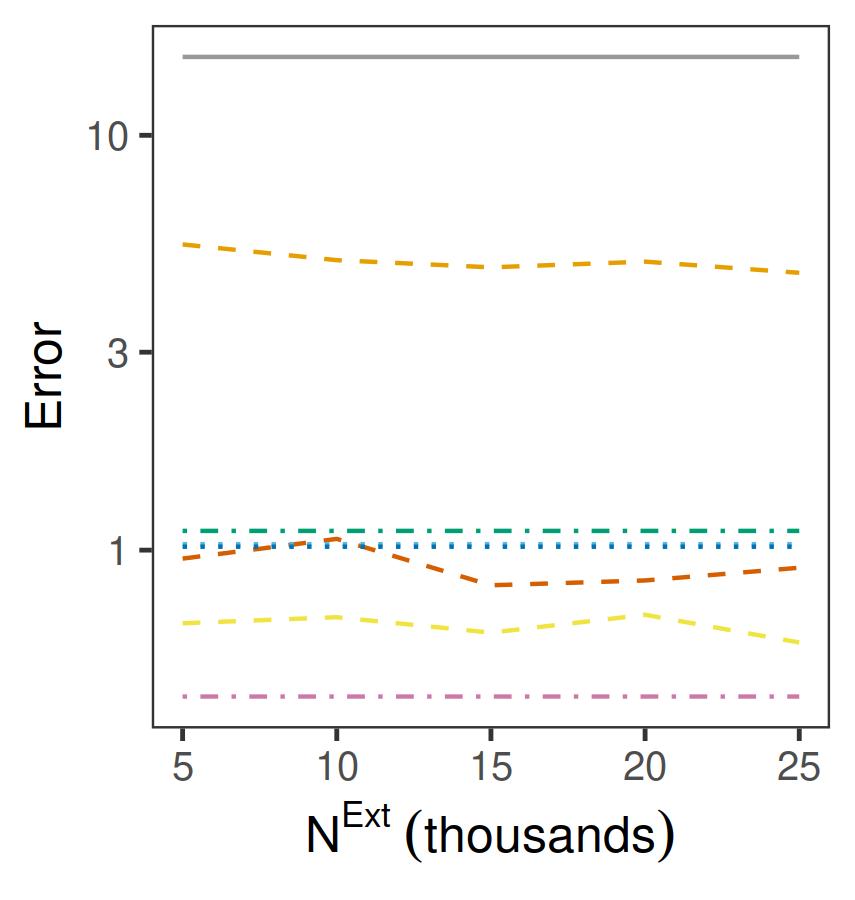}}\quad{}\subfloat[Error in estimating the 97.5\% quantiles of the $\mu$ marginal of
the posterior.]{\includegraphics[scale=0.55]{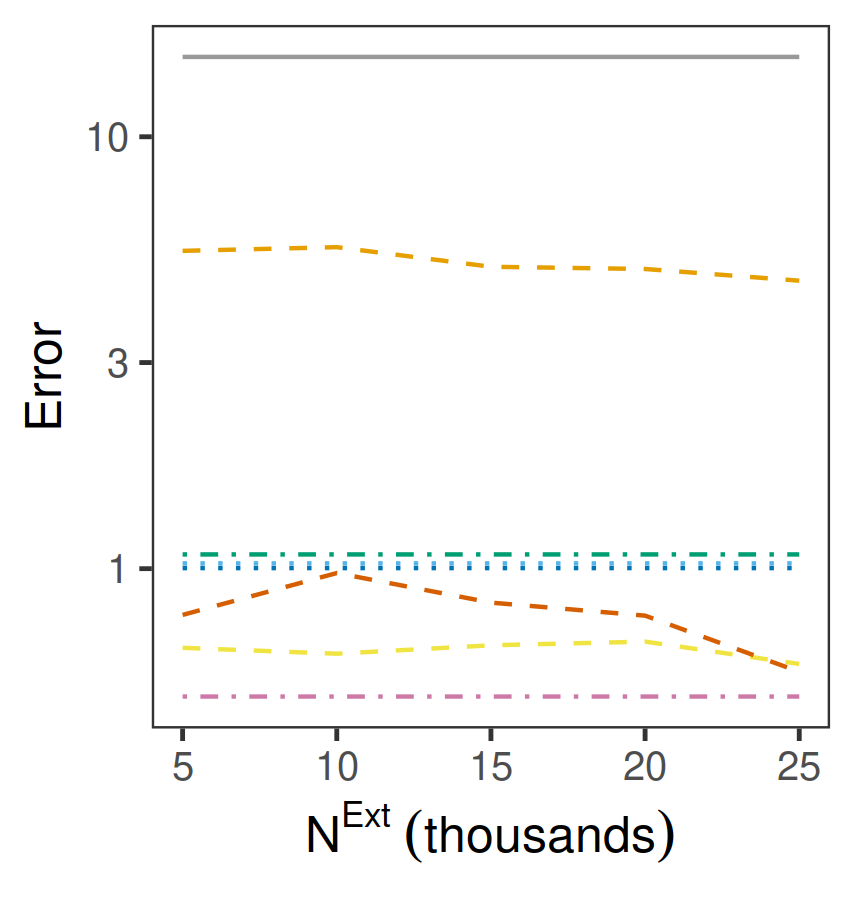}}\qquad{}\qquad{}\enskip{}
\par\end{centering}
\caption{Effect on performance metrics of adding more Laplace samples (of all
types from Section \ref{subsec:laplace-enrichment}) in the MVN example
from Section \ref{subsec:Multivariate-normal-models} with $\Sigma$
known. $N^{\textrm{{Ext}}}$ is the total number of Laplace samples
added.\label{fig:normal-sigma-known-laplace-extensions}}
\end{figure}

\subsubsection{Gamma GLM for diagnostics\label{appendix:normal-studies-Gamma-GLMs}}

All gamma GLMs with the log link, MVN studies with $d=8,n=10,000$.

\paragraph{KL divergence for $\mu$, all LEMIE models\\}

Null deviance: 139.201 (86 dof), residual deviance: 68.056 (83 dof).

\begin{tabular}{|c|c|c|c|c|}
\hline 
 & Estimate & Std. Error & t value & Pr(>|t|)\tabularnewline
\hline 
\hline 
(Intercept) & 2.4118422 & 0.1294829 & 18.627 & < 2e-16\tabularnewline
\hline 
ess & -0.0134534 & 0.0079883 & -1.684 & 0.0959\tabularnewline
\hline 
khat & 0.0316572 & 0.0034003 & 9.31 & 1.57E-14\tabularnewline
\hline 
ess:khat & -0.0002122 & 0.0001274 & -1.666 & 0.0995\tabularnewline
\hline 
\end{tabular}

\paragraph{KL divergence for $\Sigma$, all LEMIE models\\}

Null deviance: 91.982 (86 dof), residual deviance: 53.109 (83 dof).

\begin{tabular}{|c|c|c|c|c|}
\hline 
 & Estimate & Std. Error & t value & Pr(>|t|)\tabularnewline
\hline 
\hline 
(Intercept) & 4.65E+00 & 1.07E-01 & 43.253 & < 2e-16\tabularnewline
\hline 
ess & -1.83E-02 & 6.63E-03 & -2.762 & 0.00707\tabularnewline
\hline 
khat & 2.07E-02 & 2.82E-03 & 7.347 & 1.27E-10\tabularnewline
\hline 
ess:khat & 9.68E-05 & 1.06E-04 & 0.916 & 0.36242\tabularnewline
\hline 
\end{tabular}

\paragraph{KL divergence for $\mu$, LEMIE2 and LEMIE3 only\\}

Null deviance: 105.701 (56 dof), residual deviance: 55.662 (53 dof).

\begin{tabular}{|c|c|c|c|c|}
\hline 
 & Estimate & Std. Error & t value & Pr(>|t|)\tabularnewline
\hline 
\hline 
(Intercept) & 2.745086 & 0.198363 & 13.839 & < 2e-16\tabularnewline
\hline 
ess & -0.057005 & 0.038233 & -1.491 & 0.142\tabularnewline
\hline 
khat & 0.03612 & 0.007583 & 4.763 & 1.52E-05\tabularnewline
\hline 
ess:khat & -0.009408 & 0.006542 & -1.438 & 0.156\tabularnewline
\hline 
\end{tabular}

\paragraph{KL divergence for $\Sigma$, LEMIE2 and LEMIE3 only\\}

Null deviance: 63.574 (56 dof), residual deviance: 38.942 (53 dof).

\begin{tabular}{|c|c|c|c|c|}
\hline 
 & Estimate & Std. Error & t value & Pr(>|t|)\tabularnewline
\hline 
\hline 
(Intercept) & 4.769317 & 0.167882 & 28.409 & <2e-16\tabularnewline
\hline 
ess & -0.029969 & 0.032358 & -0.926 & 0.3586\tabularnewline
\hline 
khat & 0.022828 & 0.006418 & 3.557 & 0.0008\tabularnewline
\hline 
ess:khat & -0.004623 & 0.005537 & -0.835 & 0.4075\tabularnewline
\hline 
\end{tabular}

\subsection{Logistic regression - simulation of \citet{scott2016bayes}\label{subsec:additional-results-scott-logistic}}

\begin{figure}
\begin{centering}
\subfloat[Error in estimating the posterior mean of $\theta$.]{\includegraphics[scale=0.45]{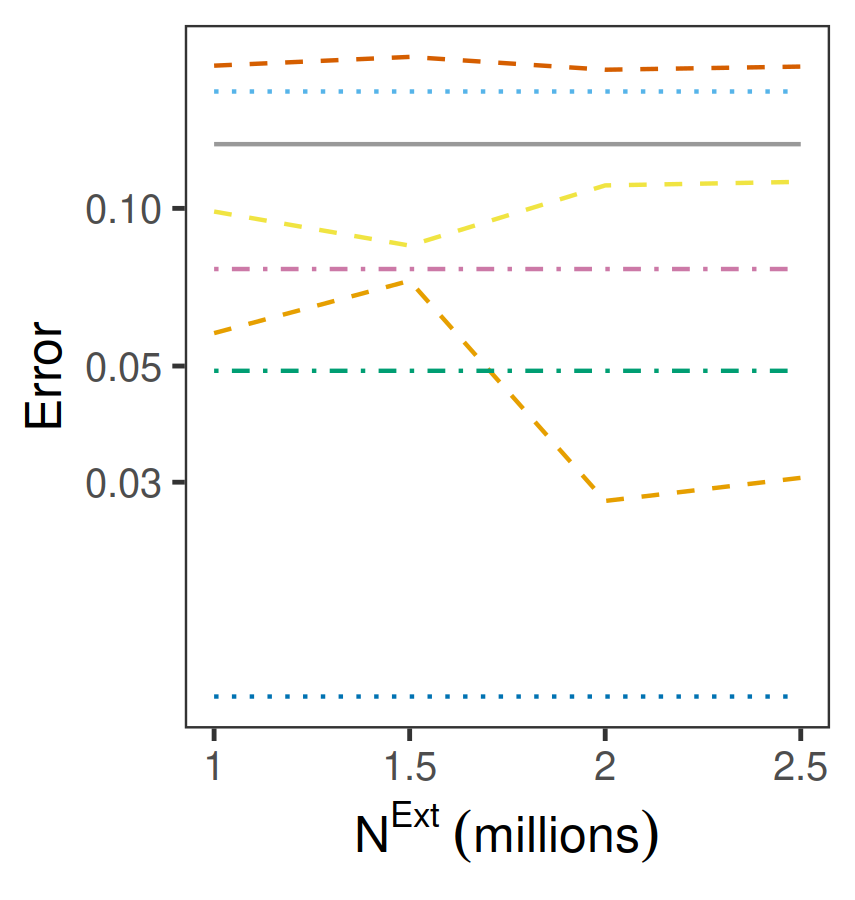}}\hfill{}\subfloat[Error in estimating the 2.5\% quantiles of the marginals of the posterior.]{\includegraphics[scale=0.45]{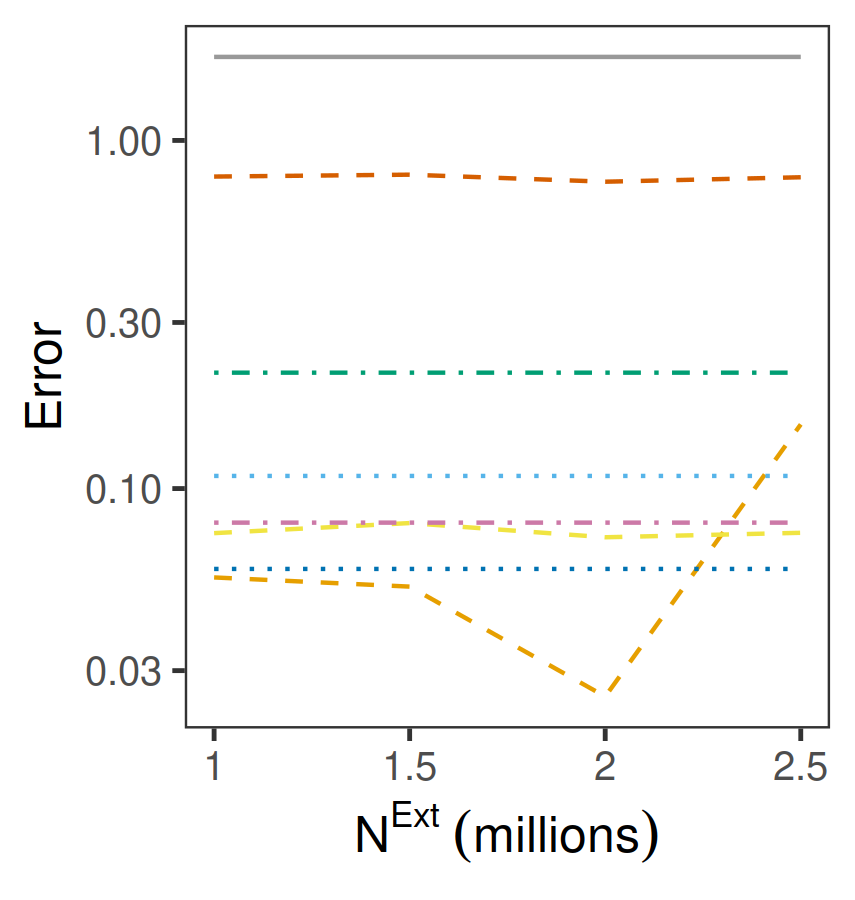}}\hfill{}\subfloat[Error in estimating the 97.5\% quantiles of the marginals of the posterior.]{\includegraphics[scale=0.45]{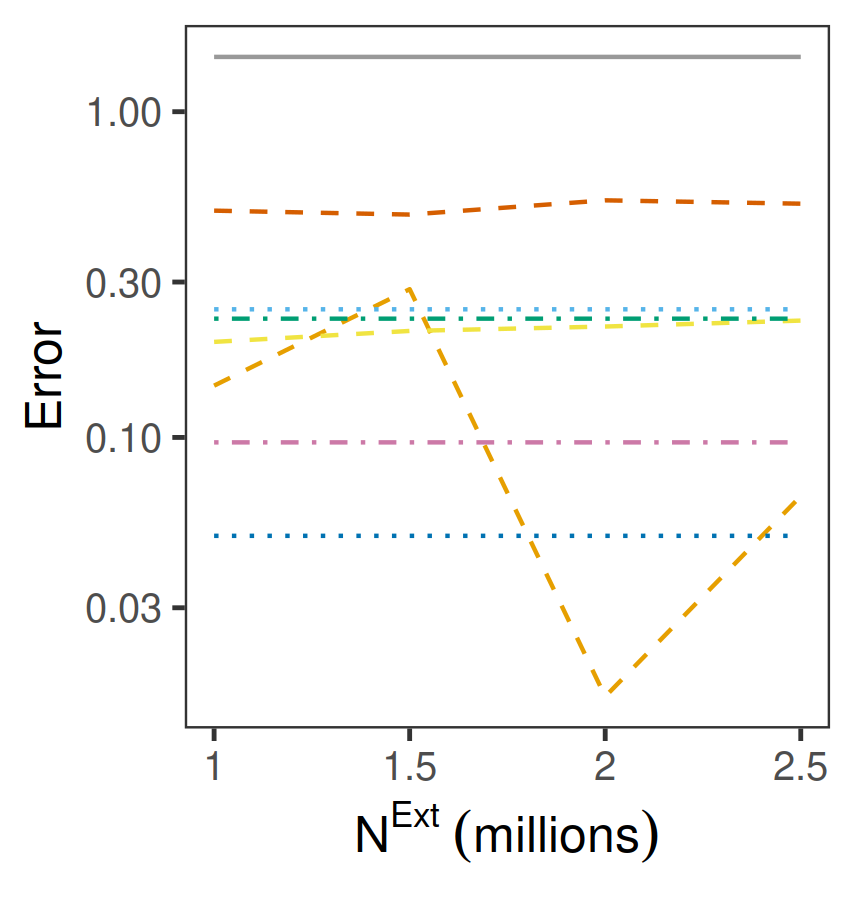}}\hfill{}\includegraphics[scale=0.45]{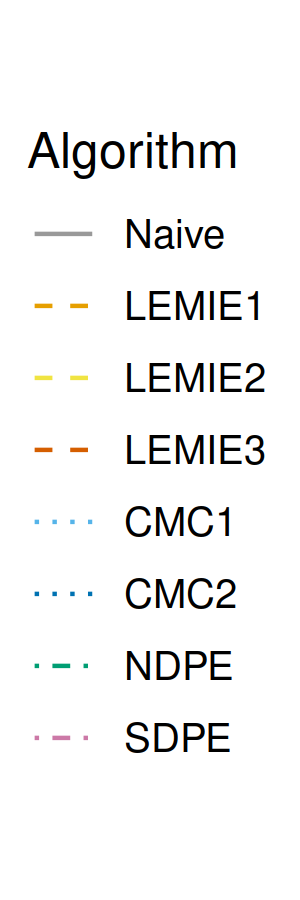}
\par\end{centering}
\caption{Effect on performance metrics of adding more Laplace samples (of all
types from Section \ref{subsec:laplace-enrichment}) in the logistic
example of Section \ref{subsec:scott-logistic}. $N^{\textrm{{Ext}}}$
is the total number of Laplace samples added.\label{fig:scott-logistic-laplace-extensions}}
\end{figure}

See Figure \ref{fig:scott-logistic-laplace-extensions} for the results
from Section \ref{subsec:laplace-enrichment}, showing what happens
to performance when we add more Laplace samples up to $2.5\times10^{6}$.

\bibliographystyle{apalike}
\bibliography{references}

\end{document}